\documentclass[manuscript, nonacm]{acmart}

\usepackage{latexsym}
\usepackage{amsmath}
\usepackage{amsfonts}
\usepackage{amsthm}
\usepackage{graphicx}
\usepackage{subfig}
\usepackage{tikz,tikz-qtree,tikz-qtree-compat}
\usepackage{color}
\usepackage{xspace}
\usepackage{colortbl}
\usepackage{algorithmicx}
\usepackage{algorithm}
\usepackage[noend]{algpseudocode}

\usepackage{todonotes}

\renewcommand{\setminus}{-}

\colorlet{RED}{red!40}
\colorlet{LightViolet}{violet!40}
\colorlet{LightRed}{red!40}
\colorlet{LightOrange}{orange!40}
\colorlet{LightGreen}{green!40}
\colorlet{LightBlue}{blue!40}
\colorlet{DarkGreen}{green!50!black}
\colorlet{DarkRed}{red!70!black}
\colorlet{DarkCyan}{red!70!black}
\colorlet{DarkBlue}{blue!80!black}
{\definecolor{DarkOrange}{rgb}{1.0, 0.49, 0.0}
\definecolor{Airforceblue}{rgb}{0.36, 0.54, 0.66}

\newcommand{\DarkBlue}[1]{{\color{DarkBlue} #1}}

\newcommand{\DarkGreen}[1]{{\color{DarkGreen} #1}}

\renewcommand{\vec}[1]{\ensuremath\boldsymbol{#1}}
\newcommand{\repsize}[1]{\|#1\|}

\newcommand{\Assumption}{Conditional query assumption\xspace}
\newcommand{\ProdMar}{Product-marginalization assumption\xspace}
\newcommand{\conQuery}{conditional query\xspace}
\newcommand{\conQueries}{conditional queries\xspace}

\newcommand{\faqcs}{\text{\sf FAQ-SS}}
\newcommand{\bcp}{\text{\sf BCP}}
\newcommand{\faq}{\text{\sf FAQ}}

\newcommand{\complexityclass}{\mathbf}
\newcommand{\sharpP}{\#\complexityclass{P}}

\newcommand{\np}{\complexityclass{NP}}

\newcommand{\fpt}{\complexityclass{FPT}}
\newcommand{\wone}{\complexityclass{W}[1]}

\newcommand\myortikz{%
  \begin{tikzpicture}[baseline=(X.base), inner sep=0, outer sep=0]
    \node[draw,circle] (X)  {${\scriptstyle\mv{01}}$};
  \end{tikzpicture}%
}

\newcommand\bigmyortikz{%
  \begin{tikzpicture}[baseline=(X.base), inner sep=1.5pt, outer sep=0]
    \node[draw,circle] (X)  {$\displaystyle \mv{01}$};
  \end{tikzpicture}%
}
\newcommand{\myor}{\mathop{\myortikz}}
\newcommand{\bigmyor}{\mathop{\bigmyortikz}}
\newcommand{\functionname}[1]{\text{\sf #1}}

\newcommand{\Idem}{\functionname{Idem}}
\newcommand{\tw}{\functionname{tw}}

\newcommand{\faqw}{\functionname{faqw}}

\newcommand{\Lss}{\functionname{$L$-ss}}
\newcommand{\Fss}{\functionname{$F$-ss}}
\newcommand{\htw}{\functionname{htw}}
\newcommand{\fhtw}{\functionname{fhtw}}

\newcommand{\atoms}{\functionname{atoms}}
\newcommand{\Dom}{\functionname{Dom}}

\newcommand{\vars}{\functionname{vars}}

\newcommand{\weight}{\functionname{weight}}
\newcommand{\clr}{\functionname{color}}

\newcommand{\true}{\functionname{true}}
\newcommand{\false}{\functionname{false}}
\newcommand{\nan}{\functionname{NaN}}

\newcommand{\InsideOut}{\functionname{InsideOut}}

\newcommand{\OI}{\functionname{OutsideIn}}

\newcommand{\avg}{\functionname{avg}}
\newcommand{\unique}{\functionname{unique}}
\newcommand{\xor}{\functionname{XOR}}

\newcommand{\problemname}[1]{\text{\sf #1}}
\newcommand{\mcm}{\problemname{MCM}}

\newcommand{\sumprod}{\problemname{SumProd}}

\newcommand{\csp}{\problemname{CSP}}
\newcommand{\pgm}{\problemname{PGM}}
\newcommand{\cqe}{\problemname{CQE}}
\newcommand{\bcq}{\problemname{BCQ}}
\newcommand{\sat}{\problemname{SAT}}
\newcommand{\ssat}{\problemname{\#SAT}}
\newcommand{\wssat}{\problemname{\#WSAT}}

\newcommand{\scq}{\problemname{\#CQ}}
\newcommand{\sqcq}{\problemname{\#QCQ}}
\newcommand{\mpf}{\problemname{MPF}}
\newcommand{\FFT}{\problemname{FFT}}
\newcommand{\EVO}{\problemname{EVO}}
\newcommand{\LE}{\problemname{LinEx}}
\newcommand{\CW}{\problemname{CW}}
\newcommand{\CWE}{\problemname{CWE}}
\newcommand{\qcq}{\problemname{QCQ}}
\newcommand{\sacq}{\problemname{\#ACQ}}
\newcommand{\cq}{\problemname{CQ}}

\newcommand{\agm}{\text{\sf AGM}\xspace}
\newcommand{\tetris}{\text{\sf Tetris}\xspace}
\newcommand{\ms}{\text{\sf Minesweeper}\xspace}

\newcommand{\calC}{\mathcal C}
\newcommand{\calU}{\mathcal U}

\newcommand{\calE}{\mathcal E}
\newcommand{\calF}{\mathcal F}

\newcommand{\calH}{\mathcal H}
\newcommand{\calV}{\mathcal V}

\newcommand{\F}{\mathbb F} 
\newcommand{\Z}{\mathbb Z} 
\newcommand{\N}{\mathbb N} 
\newcommand{\R}{\mathbb R} 
\newcommand{\D}{\mathbf D} 

\newcommand{\perm}{\text{perm}}

\newcommand{\dft}{\textnormal{DFT}}
\newcommand{\poly}{\textnormal{poly}}

\newcommand{\opt}{\textnormal{\sc opt}}

\DeclareMathOperator*{\argmin}{arg\,min}
\DeclareMathOperator*{\argmax}{arg\,max}

\newcommand{\be}{\begin{enumerate}}
\newcommand{\ee}{\end{enumerate}}
\newcommand{\bi}{\begin{itemize}}
\newcommand{\ei}{\end{itemize}}
\newcommand{\beq}{\begin{equation}}
\newcommand{\eeq}{\end{equation}}

\newcommand{\bp}{\begin{proof}}
\newcommand{\ep}{\end{proof}}
\newcommand{\bcor}{\begin{cor}}
\newcommand{\ecor}{\end{cor}}
\newcommand{\bthm}{\begin{thm}}
\newcommand{\ethm}{\end{thm}}
\newcommand{\blmm}{\begin{lmm}}
\newcommand{\elmm}{\end{lmm}}
\newcommand{\bdefn}{\begin{defn}}
\newcommand{\edefn}{\end{defn}}
\newcommand{\bprop}{\begin{prop}}
\newcommand{\eprop}{\end{prop}}
\newcommand{\bconj}{\begin{conj}}
\newcommand{\econj}{\end{conj}}
\newcommand{\bopm}{\begin{opm}}
\newcommand{\eopm}{\end{opm}}
\newcommand{\brmk}{\begin{rmk}}
\newcommand{\ermk}{\end{rmk}}

\newcommand{\suchthat}{\ | \ }
\newcommand{\inner}[1]{\langle #1 \rangle}

\newcommand{\mv}[1]{\mathbf{#1}}

\theoremstyle{plain}                   
\newtheorem{thm}{Theorem}[section]
\newtheorem{lmm}[thm]{Lemma}
\newtheorem{prop}[thm]{Proposition}
\newtheorem{cor}[thm]{Corollary}

\theoremstyle{definition}              

\newtheorem{assumption}{Assumption}
\newtheorem{opm}[thm]{Open Problem}
\newtheorem{conj}[thm]{Conjecture}
\newtheorem{ex}[thm]{Example}

\newtheorem{defn}[thm]{Definition}

\newtheorem{rmk}[thm]{Remark}
\newtheorem{claim}{Claim}

\definecolor{Red}{RGB}{255,204,204}
\definecolor{Green}{RGB}{204,255,204}
\definecolor{Blue}{RGB}{204,204,255}

\usepackage{etoolbox}
\usepackage{textcase}
\makeatletter
\patchcmd{\@sect}{\uppercase}{\MakeTextUppercase}{}{}
\patchcmd{\@sect}{\uppercase}{\MakeTextUppercase}{}{}
\makeatother

\usepackage{booktabs} 

\usepackage{listings} 

\allowdisplaybreaks

\title[Functional Aggregate Queries]{{\sf FAQ}: Questions Asked Frequently}
\author{Mahmoud Abo Khamis}
\orcid{0000-0003-3894-6494}
\affiliation{%
  \institution{Relational\underline{AI}}
  \city{Berkeley}
  \state{CA}
  \country{USA}
}
\author{Hung Q. Ngo}
\affiliation{%
  \institution{Relational\underline{AI}}
  \city{Berkeley}
  \state{CA}
  \country{USA}
}
\author{Atri Rudra}
\affiliation{%
  \institution{University at Buffalo (SUNY)}
  \city{Buffalo}
  \state{NY}
  \country{USA}
}

\thanks{An extended abstract of this manuscript appeared in the Proceedings of the 35th ACM Symposium on Principles of Database Systems {\em (PODS '16)} \cite{faq-pods16}.}

\begin{document}

\begin{abstract}
We define and study the {\bf F}unctional {\bf A}ggregate {\bf Q}uery ($\faq$)
problem, which encompasses many frequently asked questions in constraint 
satisfaction, databases, matrix operations, probabilistic graphical models 
and logic. This is our main conceptual contribution.

We then present a simple algorithm called $\InsideOut$ 
to solve this general problem. $\InsideOut$ is a variation of the 
traditional dynamic programming approach for constraint programming based on 
variable elimination. Our variation adds a couple of simple twists to 
basic variable elimination in order to deal with the generality of
$\faq$, to take full advantage of Grohe and Marx's
fractional edge cover framework, and of the analysis of
recent worst-case optimal relational join algorithms.

As is the case with constraint programming and graphical model inference, 
to make $\InsideOut$ run efficiently we need to solve an optimization problem 
to compute an appropriate {\em variable ordering}.
The main technical contribution of this work is a precise characterization of
when a variable ordering is `semantically equivalent' to the variable 
ordering given by the input $\faq$ expression.
Then, we design an approximation algorithm to find an equivalent variable 
ordering that has the best `fractional $\faq$-width'.
Our results imply a host of known and a few new results in graphical model 
inference, matrix operations, relational joins, and logic.

We also briefly explain how recent algorithms on beyond 
worst-case analysis for joins and those for solving $\sat$ and $\ssat$ can be viewed
as variable elimination to solve $\faq$ over compactly represented input 
functions.
\end{abstract}

\maketitle

\section{Introduction}
\label{sec:intro}

\subsection{Motivating examples}

The following fundamental problems from three diverse domains share
a common algebraic structure.

\begin{ex}($\problemname{Matrix Chain Multiplication}$ ($\mcm$))
\label{ex:matrix mult}
Given a series of matrices $\mv A_1, \dots, \mv A_n$ over some field $\F$, 
where the dimension of $\mv A_i$ is $p_i \times p_{i+1}$, $i \in [n]$ (where $[n]$ denotes $\{1,\ldots,n\}$ and $[0]$ denotes $\emptyset$),
we wish to compute the product $\mv A = \mv A_1 \cdots \mv A_n$.
The problem can be reformulated as follows.
There are $n+1$ variables $X_1,\dots,X_{n+1}$ with domains 
$\Dom(X_i) = [p_i]$, for $i\in[n+1]$.
For $i \in [n]$, matrix $\mv A_i$ can be viewed as a function of two variables
\[ \psi_{i,i+1}: \Dom(X_i)\times \Dom(X_{i+1}) \to \F, \]
where $\psi_{i,i+1}(x,y) = (\mv A_i)_{xy}$.
The $\mcm$ problem is to compute the output function
\[ \varphi(x_1,x_{n+1}) =
   \sum_{x_2 \in\Dom(X_2)} \cdots \sum_{x_n \in \Dom(X_n)} \prod_{i=1}^n 
   \psi_{i,i+1}(x_i,x_{i+1}).
\]
\end{ex}

\begin{ex}(Maximum A Posteriori ($\problemname{MAP}$) queries in 
      probabilistic graphical models ($\pgm$))
   Consider a discrete graphical model represented by a hypergraph 
   $\calH = (\calV, \calE)$.
   There are $n$ discrete random variables $\calV = \{X_1,\dots,X_n\}$ on finite 
   domains $\Dom(X_i)$, $i\in[n]$, and $m=|\calE|$ {\em factors} 
   \[ \psi_S : \prod_{i \in S} \Dom(X_i) \to \R_+, \ S \in \calE. \]
   A typical inference task is to compute the marginal MAP estimates, 
   written in the form
   \[ \varphi(x_1,\dots,x_f) = 
      \max_{x_{f+1}\in\Dom(X_{f+1})} \cdots
      \max_{x_{n}\in\Dom(X_{n})} \prod_{S\in\calE}\psi_S(\mv x_S).
   \]
\end{ex}

\begin{ex}($\problemname{\# Quantified Conjunctive Query}$ ($\sqcq$))
\label{ex:sqcq}
Let $\Phi$ be a first-order formula of the form
\[ \Phi(X_1,\dots,X_f) = \DarkBlue{Q_{f+1}} X_{f+1} \cdots 
   \DarkGreen{Q_n} X_n 
    \left( \bigwedge_{R\in\atoms(\Phi)} R \right), \]
where $Q_i \in \{\exists, \forall\}$, for $i>f$. 
The $\sqcq$ problem is to {\em count} the number of tuples in relation $\Phi$ 
on the free variables $X_1,\dots,X_f$.
To reformulate $\sqcq$,
construct a hypergraph $\calH = (\calV, \calE)$ as follows: $\calV$ is the set
of all variables $X_1,\dots,X_n$, and for each $R \in \atoms(\Phi)$
there is a hyperedge $S = \vars(R)$ consisting of all variables in $R$.
The atom $R$ can be viewed as a function indicating
whether an assignment $\mv x_S$ to its variables is satisfied by the atom;
namely $\psi_S(\mv x_S) = 1$ if $R(\mv x_S)$ is {\sf true} and $0$ otherwise. 

Now, for each $i \in \{f+1,\dots,n\}$ we define an aggregate operator
\[ \textstyle{\bigoplus^{(i)} = 
    \begin{cases}
    \max & \text{ if } Q_i = \exists, \\
    \times & \text{ if } Q_i = \forall.
    \end{cases}}
\]
Then, the $\sqcq$ problem above is to compute the {\em constant} function
\[ \varphi = \sum_{x_1\in\Dom(X_1)} \cdots \sum_{x_f\in\Dom(X_f)} 
   \mathop{\textstyle{\DarkBlue{\bigoplus^{(f+1)}}}}_{x_{f+1}\in\{0,1\}}
   \cdots
   \mathop{\textstyle{\DarkGreen{\bigoplus^{(n)}}}}_{x_{n}\in\{0,1\}}
   \prod_{S\in\calE}\psi_S(\mv x_S).
\]
\end{ex}

It turns out that these and dozens of other fundamental problems from 
constraint satisfaction ($\csp$), 
databases, 
matrix operations, 
$\pgm$ inference, 
logic,
coding theory,
and complexity theory
can be viewed as special instances
of a generic problem we call the 
{\bf F}unctional {\bf A}ggregate {\bf Q}uery, or
the $\faq$ problem, which we define next.
The first two columns in Table~\ref{tab:results} present eight of these 
problems. 
See~\cite{DBLP:journals/ai/Dechter99, AM00, KW08} and 
Appendix~\ref{app:sec:reductions} for many more examples.

\subsection{The $\faq$ problem}
\label{sec:faq-pbm}

Throughout the paper, we use the following convention.
Uppercase $X_i$ denotes a variable, and lowercase $x_i$ denotes a value in 
the domain $\Dom(X_i)$ of the variable. 
Furthermore, for any subset $S\subseteq [n]$, define
\begin{align*}
    \mv X_S &= (X_i)_{i\in S},
            & \mv x_S = (x_i)_{i\in S} \in \prod_{i\in S}\Dom(X_i).
\end{align*}
In particular, $\mv X_S$ is a tuple of variables and $\mv x_S$ is a
tuple of specific values with support $S$.
The input to $\faq$ is a set of functions and the output is a function
computed using a series of aggregates over the variables and
input functions.
More specifically, for each $i\in [n]$, let $X_i$ be a variable on some discrete 
domain $\Dom(X_i)$, where $|\Dom(X_i)|\ge 2$.
The $\faq$ problem is to compute the following function
\begin{equation}
\label{eqn:gen:faq}
    \varphi(\mv x_{[f]}) =
    \mathop{\DarkBlue{\textstyle{\bigoplus^{(f+1)}}}}_{x_{f+1}\in\Dom(X_{f+1})}
       \cdots
    \mathop{\DarkGreen{\textstyle{\bigoplus^{(n)}}}}_{x_{n}\in\Dom(X_{n})}
    \mathop{\textstyle{\bigotimes}}_{S\in\calE}\psi_S(\mv x_S),
\end{equation}
where 
\bi
 \item $\calH=(\calV,\calE)$ is a multi-hypergraph. $\calV=[n]$ is the index
    set of the variables $X_i$, $i\in [n]$. Overloading notation, $\calV$ is
    also referred to as the set of variables.
 \item The set $F = [f]$ is the set of {\em free variables} for some 
    integer $0\le f\le n$. Variables in $\calV-F$ are called 
    {\em bound variables}.
 \item $\D$ is a fixed domain, such as $\{\textsf{true}, \textsf{false}\}$, 
    $\{0,1\}$ or $\R^+$.
 \item For every hyperedge $S\in\calE$, 
    $\psi_S:\prod_{i\in S} \Dom(X_i) \to \D$ is an 
    {\em input} function (also called a {\em factor}).
    There are $m=|\calE|$ hyperedges. 
 \item For every bound variable $i>f$, $\oplus^{(i)}$ is a binary (aggregate) 
    operator on the domain $\D$. Different bound variables may have different
    aggregate operators.
 \item Finally, for each bound variable $i>f$ either $\oplus^{(i)}=\otimes$ or 
    $(\D,\oplus^{(i)},\otimes)$ forms a commutative semiring \footnote{A triple 
   $(\D,\oplus,\otimes)$ is a {\em commutative semiring} if $\oplus$
and $\otimes$ are commutative binary operators over $\D$ satisfying the
following:
  (1) $(\D, \oplus)$ is a commutative monoid with an additive
                   identity, denoted by $\mv 0$.
  (2) $(\D, \otimes)$ is a commutative
monoid with a multiplicative
                   identity, denoted by $\mv 1$.
 (In the usual
                   semiring definition, we do not need the multiplicative monoid
                   to be commutative.)
  (3) $\otimes$ distributes over $\oplus$.
  (4) For any element $e \in \mathbf D$, we have $e \otimes \mv 0
                       = \mv 0 \otimes e = \mv 0$.
} (with the same additive identity $\mv 0$ and multiplicative identity $\mv 1$).
If $\oplus^{(i)} = \otimes$, then $\oplus^{(i)}$ is called a 
{\em product aggregate};
otherwise, it is a {\em semiring aggregate}. 
\ei
To avoid triviality, we assume that there is at least one semiring aggregate.
(The semiring requirement is not as much of a restriction as one might think at
first glance. In Appendix~\ref{sec:appendix:noCS}, we describe several methods for
`turning' non-semiring aggregates into semiring aggregates.)
Because for $i > f$ every variable $X_i$ has its own aggregate $\oplus^{(i)}$ 
over all values $x_i \in \Dom(X_i)$, in the rest of the paper we will 
write $\bigoplus^{(i)}_{x_i}$ to mean 
$\displaystyle{\mathop{\textstyle{\bigoplus^{(i)}}}_{x_{i}\in\Dom(X_{i})}}$.
Also for brevity, we use ``semiring'' to refer to a ``commutative semiring'' unless otherwise stated. In particular, we don't use non-commutative semirings in this paper.

We will often refer to $\varphi$ as an {\em $\faq$-query}.
We use $\faqcs$\footnote{$\faq$ with a Single Semiring.}
to denote
the special case of $\faq$ when there is only {\em one} variable aggregate,
i.e. $\oplus^{(i)}=\oplus, \forall i>f$, and $(\D, \oplus, \otimes)$ is a 
commutative semiring.
The special case of $\faqcs$ when there is no free variable is called
the $\sumprod$ problem.
As shall be further discussed in Section~\ref{sec:related}, $\sumprod$ and $\faqcs$ are
well-studied problems.

\subsection{Input and output representation}

To make the problem definition complete, we will also have to specify how the
input and output functions of an $\faq$ instance are represented. As we shall 
see in Section~\ref{sec:representation}, this is a subtle issue that vastly 
affects the landscape of tractability of the problem.

To streamline the presentation, in the first part of this paper we will assume 
that both the input and output factors are represented using the {\em listing 
representation}:
each factor is a table of all tuples of
the form $\inner{\mv x_S, \psi_S(\mv x_S)}$,
such that $\psi_S(\mv x_S)\neq \mv 0$.
(In particular, entries not in the table are $\mv 0$-entries.)
This representation is commonly used in the $\csp$, databases, and sparse
matrix computation domains.

Our algorithms are in fact more generic, they work for a general class of
input and output representations, as discussed in 
Section~\ref{sec:representation}.

\subsection{Paper organization}
\label{subsec:ext}

Section~\ref{sec:summary:contributions} summarizes the contributions of the paper and 
sketches the line of attack.
Related works are discussed in Section~\ref{sec:related}.
Section~\ref{sec:prelim} defines notations, terminologies, and establishes a few
facts used throughout the paper.
Section~\ref{sec:insideout} discusses the main ideas behind $\InsideOut$ and
analyzes its runtime {\em given} a variable ordering. 
Section~\ref{sec:evo} explains how to characterize variable orderings that are ``semantically-equivalent" to the original ordering in the given $\faq$-query.
Section~\ref{sec:approx} explains how to \emph{efficiently} search through
all those equivalent variable orderings to find the ``best" one (i.e. the one that allows $\InsideOut$ to run the fastest). 
Finally, Section~\ref{sec:representation} presents the effect of input and
output representations; in particular, it shows how $\InsideOut$ is still
useful for problems such as $\sat$ and $\ssat$.

\section{Summary of contributions}
\label{sec:summary:contributions}

\subsection{Conceptual contribution}

The formulation of $\faq$ has its roots in the $\problemname{SumProd}$ and more generally $\faqcs$ problems, which have been studied by
 by Dechter \cite{DBLP:journals/ai/Dechter99},  Aji and McEliece~\cite{AM00} and Kohlas and Wilson~\cite{KW08}.
The
$\sumprod$ problem is {\em exactly} 
the special case of $\faq$ when all variable aggregates are
semiring aggregates over {\em the same} semiring, and there is no free variable.
We will discuss more of the history of this problem in Section~\ref{sec:related}.

$\faq$ substantially generalizes $\sumprod$, as $\faq$
can now capture problems in logic such as $\qcq$ (quantified conjunctive query) 
or $\sqcq$ (sharp quantified conjunctive query).
We argue that $\faq$ is a very powerful way of thinking about these problems and
related issues. 
$\faq$ can be thought of as a declarative query language over functions.
For example, 
we show in Section~\ref{sec:representation} how different input
representations can vastly affect the landscape of tractability of the problem,
and how the output representation is related to 
the notion of factorized databases~\cite{OZ15}.

\subsection{Algorithmic contribution}

We present a single algorithm, called $\InsideOut$, to solve $\faq$.
$\InsideOut$ is a variation of the variable elimination algorithm 
\cite{DBLP:journals/ai/Dechter99,MR1426261,zhangpoole94}.
In $\pgm$, variable elimination was first proposed by Zhang and 
Poole \cite{zhangpoole94}. Then Dechter \cite{DBLP:journals/ai/Dechter99} 
observed that this strategy can be applied to
problems on other semirings such as constraint satisfaction and $\sat$
solving. In the database literature, Yannakakis'
algorithm~\cite{dblp:conf/vldb/yannakakis81} can also be cast as variable
elimination under the set semiring or Boolean semiring.\footnote{It is 
well-known~\cite{AM00,MR2778120} that variable elimination and 
message passing are equivalent in the special case of $\faqcs$.}

$\InsideOut$ adds three minor twists to the basic variable elimination strategy.
First, we use a backtracking-search strategy 
called $\OI$ to compute the intermediate results. This strategy allows us to 
use recent worst-case optimal join algorithms 
\cite{leapfrog, NPRR12, skew, anrr} to compute intermediate results within
the fractional edge cover bound \cite{AGM08, GM06}. 
Second, we introduce the idea of an {\em indicator projection} of a function
onto a given set of variables to obtain the fractional hypertree width style 
of runtime guarantee \cite{DBLP:journals/talg/GroheM14}.
Third, in addition to making use of the distributive law to `fold'
common factors \cite{AM00} when we face a semiring aggregate, we apply a
swap between an aggregate and the inside product when that aggregate is also a 
product.

We show that $\InsideOut$ runs in time $\tilde O(N^{\faqw(\sigma)}+\repsize{\varphi})$, 
where $\sigma$ is a variable ordering that we choose to run the algorithm on, 
$N$ is the input size, and $\repsize{\varphi}$ is the output size (under the `listing 
representation' of input and output factors), and $\faqw(\sigma)$ is a parameter called the (fractional) 
{\em $\faq$-width} of $\sigma$. 
$\faq$-width is the $\faq$-analog of the induced fractional hypertree
width of a variable ordering. (See Definition~\ref{defn:faqw}.)
In this paper, we use $\tilde O$ to hide a logarithmic factor in data
complexity and a polynomial factor in query complexity.

In fact, Section~\ref{sec:representation} shows
that the variable elimination framework is still powerful
in cases when the fractional hypertree width bounds are no longer applicable.
These are special cases of $\faq$ where the input functions are compactly
represented.
In particular, we explain how -- with a suitable modification -- $\InsideOut$
can be used to recover recently known beyond worst-case results in join algorithms
($\ms$ \cite{nnrr}, and $\tetris$ \cite{anrr}),
and results on the tractability of $\sat$ and $\ssat$ for $\beta$-acyclic formulas
\cite{ordyniak_et_al:LIPIcs:2010:2855, braultbaron_et_al:LIPIcs:2015:4910}.

\subsection{Main technical contributions}

\subsubsection{``Width'' of an $\faq$}

In light of $\InsideOut$ running in time $\tilde
O(N^{\faqw(\sigma)}+\repsize{\varphi})$
for a given variable order $\sigma$,
the key technical problem  is choosing a $\sigma$
that minimizes $\faqw(\sigma)$. This is where the generality of $\faq$ requires
new techniques and results. Traditional variable elimination for $\csp$s or $\pgm$ inference
also requires computing a good variable ordering to minimize the  (induced)
{\em treewidth} \cite{MR2778120} or {\em fractional
hypertree width} \cite{DBLP:journals/talg/GroheM14} of the variable ordering.
However, in those cases all variable orderings are valid; 
hence, all we have to do in this traditional setting is to compute a tree 
decomposition whose maximum bag size (or maximum fractional edge cover number over the bags)
is minimized;
then, the GYO-elimination procedure will produce a good variable ordering
(see Section~\ref{sec:prelim}).
In the general setting of $\faq$, just like in logic where there are 
alternating quantifiers, the set of
semantically equivalent variable orderings depends on both the scoping 
structure specified by the input query expression {\em and} the connectivity
structure of the query's hypergraph. 

To see how the query's hypergraph affects the set of equivalent variable
orderings, consider the following simple example.
A natural class of valid permutations to consider are those that only permute 
aggregates in a maximal block of identical aggregates in the query expression. 
However, taking the query hypergraph into account, one can do much better.
Consider, for example, 
the $\faq$-query 
$$\varphi = \max_{x_1}\sum_{x_2}\max_{x_3} \cdots \sum_{x_{2k}} 
\psi_{\{1,3,\dots,2k-1\}}\psi_{\{2,4,\dots,2k\}}$$ 
where both factors have range $\R_+$. In this case, even though $\max$ and
$\sum$ do not commute with one another, we can rewrite $\varphi$
using any of the $(2k)!$ variable orderings and still obtain the same result.
(The aggregates
have to be permuted along with the variables to which they are attached.)

Even for the special case of $\faqcs$, where there is only one type of semiring
aggregates hence all permutations are valid, computing the optimal variable 
ordering is already $\np$-hard in query complexity, because 
computing the (fractional hyper) treewidth of the query hypergraph is $\np$-hard~\cite{Marx:2010:AFH:1721837.1721845,DBLP:journals/corr/FischlGP16}.
(See also Gottlob et al~\cite{DBLP:conf/pods/GottlobGLS16} 
for a survey.)
Hence, the extra complication of only considering `valid' orderings for 
$\faq$ seems to make our task much harder. 
Somewhat surprisingly, we are able to show that the complexity of computing the 
optimal ordering for general $\faq$ is essentially the same as the complexity 
of computing the optimal ordering for $\faqcs$ instances.
Figure~\ref{fig:contrib:summary} presents a schematic summary of our main
technical contributions, described in more details below.
\begin{figure}   
\centering{
\usetikzlibrary{arrows,shapes,positioning,matrix}
\definecolor{MidnightBlue}{HTML}{006795}
\tikzset{>=stealth}
\begin{tikzpicture}
      [every text node part/.style={align=center},
       et node/.style={rectangle, draw,rounded corners=3pt}]

\begin{scope}[scale=0.8, every node/.style={scale=.8}]
   \coordinate (input) at (0,6);
   \coordinate (et) at (1,1);
   \coordinate (td1) at (3,1);
   \coordinate (td2) at (5.5,1);

{
   \node[] (sigma) at (-1,6) {$\faq$-expr. $\sigma$ \\ for $\varphi$, hypergraph $\calH$};
}

{
   \node[draw,cloud] at (2.2,6) (evo) {$\EVO(\varphi)$};
   \path[->,thick,color=DarkOrange] (input) edge (evo);
}
{
   \node[color=DarkOrange] at (2, 7.5) {$\EVO(\varphi) = $ set of expressions \\ ``semantically equivalent'' to $\varphi$};
}
{
   \node[] at (5.5,6) (sigmastar) {$\faq$-expr. $\sigma^*$ \\ for $\varphi$};
   \path[->,thick,color=DarkOrange] (evo) edge (sigmastar);
}
{
   \node[color=DarkOrange] at (5.5, 5.3) {$\sigma^* = \argmin_{\tau \in \EVO(\varphi)} \faqw(\tau)$};
}

{
   \node[draw,rectangle] at (8.5,6) (algo) {$\InsideOut$};
   \path[->,thick,color=DarkOrange] (sigmastar) edge (algo);
}
{
   \node[color=DarkOrange,anchor=west] at (6.5, 7.5) {Runtime = $\tilde O(N^{\faqw(\sigma^*)} + \repsize{\varphi})$};
   \node[color=DarkOrange,anchor=west] at (7.8, 6.8) {= $\tilde O(N^{\opt} + \repsize{\varphi})$};
}

{
   \node[] at (10.5,6) (phi) {$\varphi$};
   \path[->,thick] (algo) edge (phi);
}

{
   \draw (-1,2) node[et node] (a) {$\sum_{x_1,x_4}$};
   \draw (-2,1) node[et node] (b) {$\max_{x_3}$};
   \draw (0,1) node[et node] (c) {$\prod_{x_2,x_7}$};
   \draw (-1,0) node[et node] (d) {$\sum_{x_5}$};
   \draw (1,0) node[et node] (e) {$\max_{x_6}$};
   \draw (a) -- (b);
   \draw (a) -- (c);
   \draw (c) -- (d);
   \draw (c) -- (e);
   \path[->,thick,color=DarkGreen] (sigma) edge node[left] {{\sf poly}($|\calH|$)} (a);
   \node[color=DarkGreen] at (-1,-1) {Expression Tree \\ Precedence Poset $P$};
}

{
   \draw[implies-implies,color=MidnightBlue,double equal sign distance, line width=1pt] (evo) -- (a);
   \node[color=MidnightBlue] at (2.5,4.2) {$\LE(P) \subseteq \EVO(\varphi)$};
   \node[color=MidnightBlue] at (2.4,3.5) {$\EVO(\varphi) = \CWE(\LE(P))$};
}

{
   \node[color=MidnightBlue] at (3.1,2.8) {$\min_{\tau \in \LE(P)}\faqw(\tau) = \min_{\tau \in \EVO(\varphi)} \faqw(\tau)$};
}

{
   \path[->,thick,color=DarkGreen] (et) edge node[above] {{\sf poly}$(|\calH|)$} (td1);
   \node[color=DarkGreen] (td) at (4.5,-1) {Tree Decomposition of $\calH$};
   \node[cloud,draw] (1) at (4,1) {};
   \node[cloud,draw] (2) at (4.5,2) {};
   \node[cloud,draw] (3) at (3,1.5) {};
   \node[cloud,draw] (4) at (3,0.5) {};
   \node[cloud,draw] (5) at (5,1) {};
   \node[cloud,draw] (6) at (4.5,0) {};
   \node[cloud,draw] (7) at (5.5,0) {};
   \node[cloud,draw] (8) at (6,1.5) {};
   \draw (1) -- (2);
   \draw (1) -- (3);
   \draw (1) -- (4);
   \draw (1) -- (5);
   \draw (5) -- (6);
   \draw (5) -- (7);
   \draw (5) -- (8);
}

{
   \node[] at (8.5,1) (sigmabar) {$\faq$-expr. $\bar \sigma$ \\ for $\varphi$};
   \path[->,thick,color=DarkGreen] (td2) edge node[below] {{\sf poly}$(|\calH|)$} (sigmabar);
   \node[color=DarkGreen] at (9.5,0) {$\faqw(\bar \sigma) \leq \opt+g(\opt)$};
   \node[color=DarkGreen] at (9.5,-1) {$g = $ approx. factor \\ for
   fractional \\ hypertree width of $\calH$};
}

{
   \path[->,thick,color=DarkGreen] (sigmabar) edge (algo);
   \node[color=DarkGreen,anchor=west] at (6.5, 4.5) {Runtime = $\tilde O(N^{\opt+g(\opt)} + \repsize{\varphi})$};
}
\end{scope}

\end{tikzpicture}
}
\caption{Summary of our technical contributions}
\Description{Summary of our technical contributions}
\label{fig:contrib:summary}
\end{figure}

\bi
 \item Given an $\faq$-query $\varphi$, we define the set
$\EVO(\varphi)$ of all variable orderings {\em semantically equivalent} to 
$\varphi$. Roughly, for any $\sigma \in \EVO(\varphi)$,
if we rewrite the expression for $\varphi$ using the ordering $\sigma$
(with all aggregates permuted along), 
then we obtain a function identical to $\varphi$, {\bf no matter} what the input 
factors are.
The $\faq$-width of $\varphi$ is
$\faqw(\varphi) = \min_{\sigma\in\EVO(\varphi)} \faqw(\sigma)$.
(If the $\faq$ instance was a $\sumprod$ instance, then $\faqw(\varphi)$
is exactly the fractional hypertree width of the query.)

In Figure~\ref{fig:contrib:summary}, suppose we are looking at the top/orange 
path: from the input expression $\varphi$ with variable order $\sigma$, if 
we somehow were able to compute the optimal variable ordering
$\sigma^* = \argmin_{\tau \in \EVO(\varphi)} \faqw(\tau)$, then
we would have an algorithm running in time $\tilde
O(N^{\faqw(\varphi)}+\repsize{\varphi})$.
Here, $\repsize{\varphi}$ denotes the output size.
However, even if we were willing to spend an exponential time
in query complexity, in order to take the orange path and find $\sigma^*$ we need to be able to
characterize the set $\EVO(\varphi)$. In particular, given a variable ordering
$\tau$, how do we know whether $\tau \in \EVO(\varphi)$ or not? Our answer
to this question comes next.
 \item We describe how to (in poly-time, query complexity) 
 construct an {\em expression tree} for
 the input $\faq$-query. The expression tree induces a partially ordered set
 on the variables called the {\em precedence poset}. 
 By defining a notion called {\em component-wise equivalence} ($\CWE$),
 we completely characterize $\EVO(\varphi)$:
 $\sigma \in \EVO(\varphi)$ if and only if it is component-wise equivalent to 
 some linear extension of the precedence poset.
 In fact, we also show that checking whether $\sigma \in \EVO(\varphi)$ can be 
 done in polynomial time in query complexity.
 \item While membership in $\EVO(\varphi)$ is verifiable in polynomial time,
    as aforementioned the optimization problem
    $\sigma^* = \argmin_{\tau \in \EVO(\varphi)} \faqw(\tau)$ is $\np$-hard.
    In some applications such as $\pgm$-inference, we cannot sweep
    query complexity under the rug as we do in database applications. It is thus
    natural to design approximation algorithms for $\faqw(\varphi)$. 
    To this end, we prove another technical result, that going through
    all orderings in $\EVO(\varphi)$ is {\em not} necessary. The set of linear 
    extensions of the precedence poset is sufficient:
    every linear extension of the precedence poset is 
    semantically equivalent to $\varphi$,
    and every $\sigma \in \EVO(\varphi)$ has the same
    $\faq$-width as some linear extension of this poset.\footnote{If the 
    instance is an $\faqcs$ instance, then the poset imposes no order
 on the non-free variables, i.e. the set of linear extensions is the set 
 of all possible variable orderings of non-free variables.}
 \item 
 Finally, using an approximation algorithm for the fractional hypertree width 
($\fhtw$) with approximation ratio $g(\fhtw)$ as a blackbox,\footnote{The best known such algorithm 
due to Marx~\cite{Marx:2010:AFH:1721837.1721845} has $g(\fhtw) = O(\fhtw^3)$ and runs in polynomial time, given $\fhtw$ is bounded by a constant.} 
and using the expression tree as a guide,
we give an approximation algorithm computing an 
ordering $\sigma$ such that $\faqw(\sigma)  \leq \opt + g(\opt)$,
where $\opt = \faqw(\varphi)$.
(In the $\faqcs$ case, our approximation guarantee is slightly better:
$\faqw(\sigma) \leq g(\opt)$.)
\ei

\subsubsection{The effect of input and output representation}
\label{subsubsec:ext}

The input to $\faq$ is a collection of functions.
We observe that the representation of the input functions has a huge effect 
on the computational complexity of the problem. 

We begin with the representation of the input factors. One option is the truth
table representation for each factor involved in the $\faq$ instance $\varphi$
(e.g. the {\em conditional probability table} in $\pgm$ 
\cite{MR2778120} or the usual 2D-array representation of matrices). 
This representation is wasteful when the input factors have many $\mv 0$-entries. 
Another option, which is commonly used in the $\csp$, databases, and sparse
matrix computation domains is to list pairs $\inner{\mv x_S, \psi_S(\mv x_S)}$ for 
a given factor $\psi_S$ such that $\psi_S(\mv x_S)\neq \mv 0$.
The results listed in Table~\ref{tab:results} implicitly assumed the listing 
representation. 

However, our results can handle even more succinct representations such as
GDNFs and decision diagrams~\cite{DBLP:journals/jcss/ChenG10a} in $\csp$
literature and algebraic decision diagrams in the 
$\pgm$ literature~\cite{DBLP:journals/fmsd/BaharFGHMPS97}. 
We present a common view of these representations: an input factor
$\psi_S(\mv x_S)$ might itself be the output of some other $\faq$ instance
$\varphi'_S$ on the (free) variables $\mv x_S$.
The succinctness in the 
representation comes from representing factors of $\varphi'_S$ in the 
listing format (instead of listing $\psi_S$). 
Technically, we analyze what happens when one {\em composes} an
$\faq$ instance with another. More interestingly, this framework is 
general enough to present a class of structured matrices (including the 
DFT matrices) for which we can quantify how much better our algorithm 
runs than the na\"ive quadratic time algorithm. 

Furthermore, we observe that when the input factors are too ``compact'', then the 
class of tractable $\faq$-queries is much smaller, and the fractional 
hypertree width framework no longer applies.  Nevertheless, the variable 
elimination strategy is still a powerful approach. We will, however, have to 
change the algorithm used to solve sub-problems: it can no longer be 
backtracking-search-style of algorithm.
We give four examples where we explain how recent beyond worst-case results in 
join algorithms ($\ms$ \cite{nnrr}, and $\tetris$ \cite{anrr}),
and the tractability of $\sat$ and $\ssat$ for $\beta$-acyclic formulas
\cite{ordyniak_et_al:LIPIcs:2010:2855, braultbaron_et_al:LIPIcs:2015:4910},
all of which are special cases of $\faq$, 
can be explained using this idea.

Last but not least, we make several observations regarding the representations
of the {\em output} function $\varphi$.
For the case when the $\faq$ instance $\varphi$ has no free variables, the
algorithm needs to output a single element from $\D$. However, when 
$\varphi$ has free variables, then we have to also represent the output 
somehow. The default option is to also use the listing representation for 
the output. However, $\InsideOut$ is general enough to be able to output 
an $\faq$ instance as the output. This issue is slightly subtle as 
$\varphi$ is {\em already} an $\faq$ instance that represents the output 
but it is not an ``interesting'' representation. However, the generality of 
our algorithm allows the input factors and output both to be 
represented as $\faq$ instances.
Our observations here are very close in spirit to the recent results on 
database factorization of Olteanu and Z\'{a}vodn\'{y}~\cite{OZ15}.

\subsection{Highlights of our corollaries}

\begin{table*}[th!]
{\small
\centering
{\renewcommand{\arraystretch}{1.5}
\begin{tabular}{|l|l|c|c|}
\hline
Problem & $\faq$ formulation & Previous Algo. & {\bf Our Algo.}\\
\hline
\rowcolor{Red}
   $\sqcq$ & {\tiny $\displaystyle{\sum_{(x_1,\dots,x_f)}} 
        \textstyle{\bigoplus^{(f+1)}_{x_{f+1}}\cdots\bigoplus^{(n)}_{x_n}}
        \displaystyle{\prod_{S\in\calE}\psi_S(\mv x_S)}$}
        & No non-trivial algo & $\tilde
        O(N^{\faqw(\varphi)}+\repsize{\varphi})$\\
\rowcolor{Red}
& ~~~~~~~ where $\textstyle{\bigoplus^{(i)}}\in\{\max,\times\}$  & & \\
\rowcolor{Red}
   $\qcq$ &  {\tiny $\textstyle{\bigoplus^{(f+1)}_{x_{f+1}}\cdots\bigoplus^{(n)}_{x_n}}
           \displaystyle{\prod_{S\in\calE}\psi_S(\mv x_S)}$}
        & $\tilde
        O(N^{\problemname{PW}(\calH)}+\repsize{\varphi})$~\cite{DBLP:conf/lics/ChenD12}&$\tilde
        O(N^{\faqw(\varphi)}+\repsize{\varphi})$\\
\rowcolor{Red}
& ~~~~~~~ where $\textstyle{\bigoplus^{(i)}}\in\{\max,\times\}$ & & \\
\rowcolor{Red}
   $\scq$ & {\tiny $\displaystyle{\sum_{(x_1,\dots,x_f)} 
\max_{x_{f+1}}\cdots\max_{x_n}
        \prod_{S\in\calE}\psi_S(\mv x_S)}$}
        & $\tilde
        O(N^{\problemname{DM}(\calH)}+\repsize{\varphi})$~\cite{DBLP:conf/icdt/DurandM13}&
        $\tilde O(N^{\faqw(\varphi)}+\repsize{\varphi})$\\
\rowcolor{Red}
Joins  & $\bigcup_{\mv x}\bigcap_{S\in\calE} \psi_S(\mv x_S)$
        & $\tilde{O}\left(
        N^{\fhtw(\calH)}+\repsize{\varphi}\right)$~\cite{GM06} &
        $\tilde{O}\left( N^{\fhtw(\calH)}+\repsize{\varphi}\right)$\\
\rowcolor{Green}
   Marginal & $\displaystyle{\sum_{(x_{f+1},\dots,x_n)} \prod_{S\in\calE} \psi_S(\mv
   x_S)}$  & $\tilde O(N^{\htw(\varphi)}+\repsize{\varphi})$~\cite{KDLD05}
   &$\tilde O(N^{\faqw(\varphi)}+\repsize{\varphi})$\\
\rowcolor{Green}
   $\problemname{MAP}$ & $\displaystyle{\max_{(x_{f+1},\dots,x_n)} \prod_{S\in\calE}
\psi_S(\mv x_S)}$ &
   $\tilde O(N^{\htw(\varphi)}+\repsize{\varphi})$~\cite{KDLD05}
   & $\tilde O(N^{\faqw(\varphi)}+\repsize{\varphi})$\\
\rowcolor{Blue}
$\mcm$ & $\displaystyle{\sum_{x_2,\dots,x_{n}} \prod_{i=1}^{n}
\psi_{i,i+1}(x_i,x_{i+1})}$ & DP bound~\cite{MR2002e:68001}  & DP bound\\
\rowcolor{Blue}
   $\problemname{DFT}$ & {\tiny $\displaystyle{\sum_{\substack{(y_0,\dots,y_{m-1})\\\in\Z_p^m}}b_y\cdot \prod_{0\le
   j+k<m} e^{i2\pi\frac{x_j\cdot y_k}{p^{m-j-k}}}}$}& $O(N\log_p{N})$~\cite{fft}& $O(N\log_p{N})$\\
\hline
\end{tabular}
}
\caption{Runtimes of algorithms assuming optimal variable ordering is given.
    Problems shaded red are in CSPs and logic ($\D=\{0,1\}$  for CSP and $\D=\N$ for \#CSP), problems
shaded green fall under PGMs ($\D=\R_+$), and problems shaded blue 
fall under matrix operations ($\D=\mathbb{C}$). 
$N$ denotes the size of the largest factor (assuming they are represented with
the listing format; see Definition~\ref{def:listing}). 
$\htw(\varphi)$ is the notion of integral cover width defined in 
\cite{KDLD05} for $\pgm$.
$\problemname{PW}(\calH)$ is the {\em optimal width of a prefix graph} of 
$\calH$ from~\cite{DBLP:conf/lics/ChenD12} and 
$\problemname{DM}(\calH) = \poly(\Fss(\calH), \fhtw(\calH))$, where 
$\Fss(\calH)$ is the $[f]$-quantified star size~\cite{DBLP:conf/icdt/DurandM13}. 
$\repsize{\varphi}$ is the output size in listing representation.
Our width $\faqw(\varphi)$ is never worse than any of the three and
there are classes of queries where ours is unboundedly better than all three.
In DFT, $N=p^m$ is the length of the input vector.
$\tilde O$ hides a logarithmic factor in data
complexity and polynomial factor in query complexity.}
\label{tab:results}
}
\end{table*}

In light of the fact that many problems can be reduced
to $\faq$, Table~\ref{tab:results} presents a selected subset of corollaries
that our results imply.\footnote{The results listed in the table implicitly 
assumed the listing representation of input factors.} 
We list the results assuming the optimal variable ordering is already given.
(This holds true for both known results and our results.)
When the optimal variable ordering is not given, the exponents of $N$ in all
cases have to be changed to the best known approximating factors 
for the corresponding width.
In the $\faq$ case, that would be $O(\faqw^3(\varphi))$.

For each problem, the table lists the corresponding $\faq$ instance, 
the runtime of the previously best known algorithm, and the runtime of 
$\InsideOut$.
These are the problems that we would like to highlight, as they yield either
new results or an alternative interpretation of known results in the $\faq$
framework.

The results in Table~\ref{tab:results} roughly span three areas: 
(1) $\csp$s and Logic; 
(2) $\pgm$s and 
(3) Matrix operations. 
Except for joins, problems in area (1) need the full generality of our $\faq$ formulation, where
$\InsideOut$ either improves upon existing results or yields new results.
Problems in area (2) can already be reduced to $\faqcs$. Here, $\InsideOut$
improves upon known results since it takes advantage of Grohe and Marx's more
recent fractional hypertree width bounds. 
Finally, problems in area (3) of Table~\ref{tab:results} are classic.
$\InsideOut$ does not yield anything new here, but it is intriguing to be able
to explain the textbook dynamic programming algorithm for 
$\problemname{Matrix-Chain Multiplication}$~\cite{MR2002e:68001} 
as an algorithm to find a good
variable ordering for the corresponding $\faq$-instance.
The $\problemname{DFT}$ result is a re-writing of Aji and McEliece's
observation \cite{AM00}.\footnote{Note that we have further results on matrix 
vector multiplication for structured matrices (see
Appendix~\ref{sec:succ-rep}).}

It should be noted that the prior results on $\scq$ 
\cite{DBLP:conf/icdt/DurandM13} and 
$\qcq$ \cite{DBLP:conf/lics/ChenD12} focused on dichotomy theorems for
bounded-arity classes of input hypergraphs, not just on the best 
possible runtime one can get. Our $\faqw$ notion is a generalization
of fractional hypertree width, which steps into the unbounded arity 
world. See Marx~\cite{DBLP:journals/jacm/Marx13} for a more detailed 
discussion of known results on $\csp$ in the unbounded-arity case.

%
%

\section{Related work}
\label{sec:related}
Since $\faq$ encompasses so many areas, our related work discussion is 
necessarily incomplete. Appendix~\ref{app:sec:representation} 
discusses more related works on the factor/function representation issue.

\subsection{Problems on one semiring}

As was mentioned earlier, $\sumprod$ is 
the special case of $\faq$ when all variable aggregates are
semiring aggregate over {\em the same} semiring, and there is no free variable. This special case is powerful enough to capture a bunch of problems (e.g. it captures all CSPs). This problem was implicitly defined by Dechter \cite{DBLP:journals/ai/Dechter99}, who solved it
using variable elimination.

To the best of our knowledge, the $\faqcs$ problem was explicitly 
defined by Aji and McEliece~\cite{AM00} who called it the $\mpf$ problem (for
$\problemname{Marginalize the product function}$).\footnote{Though related problems had been defined before: see e.g.~\cite{faqcs-jacm}.} They presented a message
passing algorithm for $\faqcs$ and essentially showed that their algorithm meets
the treewidth bound. Their paper also lists a number of problems that are $\faqcs$
instances, including $\problemname{Matrix Chain Multiplication}$ (less specific
than our result, they just argue that essentially different variable orderings 
give rise to different ways of parenthesizing the matrix chain multiplication) and
$\problemname{Matrix}$ $\problemname{Vector}$ $\problemname{Multiplication}$. 
They showed that their general
algorithm contains $\FFT$ as a special case. We re-phrase their interpretation
of the $\FFT$ using $\InsideOut$. They also showed that many basic decoding 
problems in coding theory can be cast as $\faqcs$ instances.

Kohlas and Wilson~\cite{KW08} presented even more applications of the $\faqcs$
problem. The paper categorized various existing message passing algorithms
depending on what extra properties they need beyond $(\D,\oplus,\otimes)$ being
a commutative semiring. Their paper also explored algorithms for approximate
computations (while in this work we solely deal with exact computation). 
Approximate computations in $\pgm$s have been explored under the semiring 
framework~\cite{DBLP:conf/ijcai/RollonLD13}.

Most of the results in the $\pgm$ literature present algorithms that are shown 
to obtain the treewidth bound. To the best of our knowledge, the finest
hypergraph width parameter used to bound the performance of $\pgm$ inference 
algorithms is the 
integral hypertree width bound of~\cite{DBLP:journals/jcss/GottlobLS03}, which appeared 
in~\cite{KDLD05,DBLP:conf/ecai/DechterOM08}. 
See Sections~\ref{subsec:width} and~\ref{subsec:td} for a more 
detailed discussion on the various width parameters.

In the database literature, recently Koch 
\cite{DBLP:conf/pods/Koch10} described an algebraic query language called
$\problemname{AGCA}$ over `rings of databases' which is somewhat similar 
in spirit to $\faq$. This framework makes use of additive inverses to allow for
efficient view maintenance.

\subsection{Factorized databases}

Bakibayev et al.~\cite{DBLP:journals/pvldb/BakibayevKOZ13} and 
Olteanu and Z\'avodn\'y~\cite{OZ15} introduced the
notion of {\em factorized databases}, and showed how one can efficiently compute
join and aggregates over factorized databases. In hindsight there is much in
common between their approach and $\InsideOut$ applied to the single semiring
case of $\faqcs$.
Both approaches have the same runtime complexity, because
both are dynamic programming algorithms, $\InsideOut$ is bottom-up, and
factorized database computation is top-down (memoized).

The $\faq$ framework is more general in that it can handle multiple aggregate
types. Our contribution also involves the characterization of $\EVO$ and an 
approximation algorithm for $\faqw$. 
On the other hand, aspects of factorized database that $\faq$ does not handle 
include the evaluation of SQL queries and output size bounds
on the factorized representations.

\subsection{Width parameters}
\label{subsec:width}

Various notions of hypergraph `widths' have been developed over the years in
$\pgm$, $\csp$, and database theory.
In particular, two often-used properties of the input query are
{\em acyclicity} and {\em bounded width.}  

When the query is acyclic, the classic algorithm of
Yannakakis~\cite{dblp:conf/vldb/yannakakis81} for relational joins (and $\csp$s) 
runs in time linear in the input plus output size, modulo a log factor.
Similarly, Pearl's belief propagation algorithm~\cite{DBLP:conf/aaai/Pearl82}
works well for acyclic graphical models.
As we briefly touch upon in Appendix~\ref{app:subsec:quick:faqcs}, Yannakakis' algorithm is essentially
belief propagation on the Boolean semiring or set semiring. The algorithm can
also be reinterpreted using $\InsideOut$.

Subsequent works on databases and $\csp$s
further expand the classes of queries that can be
evaluated in polynomial time.  These works define progressively more
general width parameters for a query, which intuitively measure
how far a query is from being acyclic. Roughly, these results state
that if the corresponding width parameter is bounded by a
constant, then the query is `tractable,' i.e. there is a polynomial-time 
algorithm to evaluate it.  For example, Gyssens et
al. \cite{DBLP:journals/ai/GyssensJC94, DBLP:conf/adbt/GyssensP82}
showed that queries with bounded {\em degree of acyclicity} are
tractable.  Then came {\em query width} (qw) from Chekuri and
Rajaraman~\cite{DBLP:journals/tcs/ChekuriR00}, {\em hypertree width}
and {\em generalized hypertree width} (ghw) from Gottlob et al.
\cite{DBLP:journals/sigmod/Scarcello05,
DBLP:journals/jcss/GottlobLS03}, and {\em fractional hypertree width}
from Grohe and Marx \cite{DBLP:journals/talg/GroheM14}.
See~\cite{DBLP:conf/pods/GottlobGLS16} for a survey and ~\cite{DBLP:journals/corr/FischlGP16} for the latest in this line of work.
Marx developed stronger width parameters
called {\em adaptive width} and {\em submodular width}
\cite{DBLP:journals/mst/Marx11,DBLP:journals/jacm/Marx13},
which were recently extended to functional dependencies and degree bounds~\cite{DBLP:conf/pods/KhamisNS16,panda}.

In the $\pgm$ literature, the most common parameter is 
{\em treewidth} as the textbook variable elimination and message passing 
algorithms are often stated to run in time $O(N^{w+1})$ where $w$ is the tree 
width of the model \cite{MR2778120}. 
Freuder \cite{DBLP:conf/aaai/Freuder90} and Dechter and Pearl 
\cite{journals/ai/DechterP89} showed in late 1980s that $\csp$ instances with 
bounded treewidth are tractable.

In the logic/finite model theory literature, several width parameters were also
developed \cite{DBLP:journals/corr/abs-1203-3814, DBLP:conf/icdt/DurandM13,
DBLP:conf/lics/ChenD12}. We will describe them later in the relevant sections of
the paper.

\subsection{Finite model theory}
In \cite{pichler:2013}, Pichler and Skritek studied the $\scq$ problem in the special case where the query is acyclic. We refer to this special case as $\sacq$. In particular, they showed that $\sacq$ is tractable in data complexity (i.e. when the number of variables that we are counting over is a constant) and in query complexity (i.e. when all relations have constant sizes) but not in combined complexity where the problem turns out to be $\sharpP$-complete.

In \cite{DBLP:conf/icdt/DurandM13}, Durand and Mengel introduced a new parameter
for $\scq$ called the \emph{quantified star size}. It is basically a measure of
how free variables are connected in the query's hypergraph. Along with bounded
generalized hypertree width (or fractional hypertree width), bounded quantified
star size characterizes the classes of $\scq$ instances that are tractable in
the bounded arity (or bounded generalized hypertree width) case.

The quantified star size idea has been expanded later by applying it to the
\emph{core} of the query instead of the original query
\cite{DBLP:conf/pods/GrecoS14}. The core is a minimal subquery that is homomorphically equivalent to the 
original query. Because homomorphism does not preserve 
counts (i.e. it is not a one-to-one mapping), free variables have to be explicitly 
preserved by taking the core of the \emph{color query} of the original query. 
Further development along with new lower bounds appeared in 
\cite{chen_et_al:LIPIcs:2015:4980}.

$\qcq$ (second row of Table~\ref{tab:results}) has its own long line of research. An early and 
interesting result was the tractability of $\qcq$\ when the domain size, 
treewidth, and number of quantifier alternations are all constants 
\cite{DBLP:conf/ecai/Chen04}. More recently, Chen and Dalmau introduced a 
width parameter for $\qcq$ based on elimination orderings 
\cite{DBLP:conf/lics/ChenD12}. In particular, they take the minimum width 
over some variable orderings that are equivalent to the original query. 
They showed the tractability of $\qcq$ when their width is bounded. 

The runtime of $\InsideOut$ for $\cq$, $\scq$, and $\qcq$ are unboundedly 
better than the above results as shown in Table~\ref{tab:results}. 
Our result on $\sqcq$ is new: to the best of our knowledge no non-trivial efficient algorithms for $\sqcq$ were known prior to our work. Essentially, $\InsideOut$ is able
to unify the above results under the same umbrella. 

\section{Preliminaries}
\label{sec:prelim}

\subsection{Factors, their representations, and sizes}
\label{subsec:factors}

In relational database systems~\cite{DBLP:books/aw/AbiteboulHV95,
DBLP:books/cs/Maier83,DBLP:books/cs/Ullman89}, constraint satisfaction, 
and sparse matrix operations~\cite{Yuster:2005:FSM:1077464.1077466}, the
following representation of input and output factors is the most common:

\bdefn[Listing representation]
\label{def:listing}
In the {\em listing representation}, each factor is a table of all tuples of
the form $\inner{\mv x_S, \psi_S(\mv x_S)}$,
such that $\psi_S(\mv x_S)\neq \mv 0$.
In particular, entries not in the table are $\mv 0$-entries.
\label{defn:listing:rep}
\edefn

We will assume in most of this paper that all input and output factors are 
represented using the listing representation. Section~\ref{sec:representation}
discusses how our results still hold under other representations and the effect
they have on the computational landscape.

Recall that $\mv 0$ is the additive identity of the semiring(s) which also
annihilates any element of $\D$ under multiplication.

Let $W\subseteq [n]$ be some subset of variables, and 
$\mv y_W \in \prod_{i\in W} \Dom(X_i)$ be some given value tuple.
The {\em conditional factor} $\psi_S( \cdot \suchthat \mv y_W)$ is a function
from $\prod_{i\in S} \Dom(X_i)$ to $\mathbf D$ defined by 
\[ \psi_S( \mv x_S \suchthat \mv y_W)
    = \begin{cases}
        \mv 0 & \text{ if } S \cap W \neq \emptyset \text{ and } \mv x_{S\cap
    W} \neq \mv y_{S\cap W}\\
    \psi_S(\mv x_S) & \text{ otherwise.}
      \end{cases}
\]

For each factor $\psi_S$, define its {\em size} to be the number of non-zero
points under its domain:
\[ \repsize{\psi_S} := \left|\bigl\{ \mv x_S \suchthat 
                       \psi_S(\mv x_S) \neq \mv 0 
                     \bigr\}
               \right|. 
\]
(This is also the number of rows in the table representing $\psi_S$ in the list
representation.)
Let $T\subseteq S$ be arbitrary, then obviously we can write a factor size as
a sum of conditional factor sizes:
\begin{equation}\label{eqn:NS-T}
   \repsize{\psi_S} = \sum_{\mv y_T \in \prod_{i\in T}\Dom(X_i)} 
              \repsize{\psi_S( \cdot \suchthat \mv y_T)}.
\end{equation}
Throughout this paper, we use $N$ to denote the maximum over all input factor 
sizes.

\bdefn[Indicator projection]
For any two sets $S, T \subseteq[n]$ such that $S\cap T \neq \emptyset$, and a 
given factor $\psi_S$,
the function $\psi_{S/T} : \prod_{i \in S\cap T} \Dom(X_i) \to \mv D$
defined by
\[
   \psi_{S/T}(\mv x_{S \cap T}) :=
   \begin{cases}
      \mv 1 & \exists \mv x_{S-T} \text{ s.t. }
          \psi_S(\mv x_{S \cap T},\mv x_{S-T}) \neq \mv 0\\
      \mv 0 & \text{otherwise}
   \end{cases}
\]
is called the {\em indicator projection} of $\psi_S$ onto $T$.
\edefn

\subsection{$\agm$-bound and fractional cover numbers}
\label{subsec:agm:fractional}

Let $\calH=(\calV,\calE)$ be a hypergraph. Let $B\subseteq\calV$ be any subset
of vertices. 
An {\em integral edge cover} of $B$ using edges in $\calH$ is a feasible
solution $\vec\lambda =(\lambda_S)_{S\in\calE}$ to the following integer
program:
\begin{eqnarray*}
    \min && \sum_{S\in\calE} \lambda_S\\
    \text{s.t.}&& \sum_{S : v \in S} \lambda_S \geq 1, \ \ \forall v \in B\\
               && \lambda_S \in \{0,1\}, \ \ \forall S\in \calE,
\end{eqnarray*}
whose optimal objective value
is denoted by $\rho_\calH(B)$. The number $\rho_\calH(B)$ is called
the {\em integral edge cover number} of $B$.
Similarly, $\rho^*_\calH(B)$ is the optimal objective value of the relaxation
\begin{eqnarray}
    \min && \sum_{S\in\calE} \lambda_S\label{eqn:LP:rho-star}\\
    \text{s.t.}&& \sum_{S : v \in S} \lambda_S \geq 1, \ \ \forall v \in B\nonumber\\
               && \lambda_S \geq 0, \ \ \forall S\in \calE.\nonumber
\end{eqnarray}
Any feasible solution to the above linear program is called a {\em fractional
edge cover} of $B$ using edges in $\calH$.
Note that $\rho$ and $\rho^*$ are functions from $2^{\calV} \to \R_+$.

Fix some input $(\psi_S)_{S\in\calE}$ for our $\faq$ problem.
Let $\vec\lambda^* = (\lambda^*_S)_{S\in \calE}$ denote an optimal solution to
the linear program
\begin{eqnarray*}
    \min && \sum_{S\in\calE} \lambda_S \log_2 \repsize{\psi_S}\\
    \text{s.t.}&& \sum_{S : v \in S} \lambda_S \geq 1, \ \ \forall v \in B\\
               && \lambda_S \geq 0, \ \ \forall S\in \calE.
\end{eqnarray*}
Then, the quantity 
\begin{equation}
  \agm_\calH(B) := \prod_{S\in\calE} \repsize{\psi_S}^{\lambda^*_S}
  \label{eqn:agm}
\end{equation} 
is called the {\em $\agm$-bound} for $B$ using edges of $\calH$.
It is obvious that
\begin{equation}
\agm_\calH(B) \leq N^{\rho^*_\calH(B)}. 
\label{eqn:agm:rho*}
\end{equation}

We call the quantities $\rho_\calH(\calV), \rho^*_\calH(\calV)$, and 
$\agm_\calH(\calV)$ the integral cover number, fractional cover number, and
$\agm$-bound corresponding to $\calH$.
When $\calH$ is implicit from context, we shall drop the subscript $\calH$
and write $\rho(B), \rho^*(B)$, and $\agm(B)$ for the sake of
brevity.
Note that the $\agm$-bound is ``data-dependent'' in the sense that it is a
function of the input factors, while the cover numbers are only dependent on the
hypergraph. Thus, our notation $\agm(B)$ is under the implicit assumption that
the input factors are fixed.

\subsection{Tree decomposition, acyclicity, and width-parameters}
\label{subsec:td}

\bdefn[Tree decomposition]
Let $\calH = (\calV, \calE)$ be a hypergraph.
A {\em tree-decomposition} of $\calH$ is a pair $(T, \chi)$
where $T = (V(T), E(T))$ is a tree and $\chi : V(T) \to 2^{\calV}$ assigns to
each node of the tree $T$ a subset of vertices of $\calH$.
The sets $\chi(t)$, $t\in V(T)$, are called the {\em bags} of the 
tree-decomposition.  There are two properties the bags must satisfy
\bi
 \item[(a)] For any hyperedge $S \in \calE$, there is a bag $\chi(t)$, $t\in
     V(T)$, such that $S\subseteq \chi(t)$.
 \item[(b)] For any vertex $v \in \calV$, the set 
     $\{ t \suchthat t \in V(T), v \in \chi(t) \}$ is not empty and forms a 
 connected subtree of $T$.
\ei
\label{defn:TD}
\edefn

\bdefn[$\alpha$-acyclic]\label{defn:alpha-acyclic-td}
A hypergraph $\calH = (\calV, \calE)$ is $\alpha$-{\em acyclic} iff
there exists a tree decomposition 
$(T, \chi)$ in which every bag $\chi(t)$ is a hyperedge of $\calH$.
\edefn

When $\calH$ represents a join query, the tree $T$ in the above 
definition
is also called the {\em join tree} of the query. 
A query is $\alpha$-acyclic if and only if its hypergraph is $\alpha$-acyclic.  
While possessing many nice properties \cite{DBLP:journals/jacm/Fagin83}, 
the notion of $\alpha$-acyclicity is 
unsatisfying in some settings because we can turn any hypergraph into an $\alpha$-acyclic
hypergraph by adding a hyperedge covering all its vertices. This observation
motivates a second notion of acyclicity \cite{DBLP:journals/jacm/Fagin83}.

\bdefn[$\beta$-acyclicity]\label{defn:beta-acyclic}
A hypergraph $\calH$ is {\em $\beta$-acyclic} iff the hypergraph formed
by any subset of edges of $\calH$ is $\alpha$-acyclic.
\edefn

To define commonly used width parameters of hypergraphs, we follow the width 
function framework introduced by Adler \cite{adler:dissertation}.
Let $\calH=(\calV,\calE)$ be a hypergraph. 
Let $g : 2^\calV \to \mathbb R^+$ be a function that assigns a non-negative
real number to each subset of $\calV$.
Then, the {\em $g$-width} of a tree decomposition $(T, \chi)$ is 
$\max_{t\in V(T)} g(\chi(t))$.
The {\em $g$-width of $\calH$} is the {\em minimum} $g$-width
over all tree decompositions of $\calH$.
Note that the $g$-width of a hypergraph is a {\em Minimax} function.

\bdefn[Common width parameters]
Let $s$ be the following function:
$s(B) = |B|-1$, $\forall B \subseteq \calV$.
Then the {\em tree-width} of a hypergraph $\calH$, denoted by
$\tw(\calH)$, is exactly its $s$-width.
The {\em generalized hypertree width} of a hypergraph $\calH$, denoted by $\htw(\calH)$,
is the $\rho$-width of $\calH$, and
the {\em fractional hypertree width} of a hypergraph $\calH$,
denoted by $\fhtw(\calH)$, is the $\rho^*$-width of $\calH$.
\edefn

\subsection{Vertex/variable ordering and its equivalence to tree decomposition}
\label{subsec:VO}

Besides tree decompositions, there is another equivalent way to characterize 
and define ($\alpha$/$\beta$-) acyclicity and width parameters of hypergraphs
using a listing of vertices of a
hypergraph~\cite{MR1640205,bodlaender2006treewidth}.
The results in this section are probably well-known to researchers in this 
area, but we were not able to
track down a precise reference for some of our propositions below.
Their proofs are presented in Appendix~\ref{app:sec:td} for completeness.

\bdefn[Vertex/variable ordering]
A {\em vertex ordering} of a hypergraph $\calH=(\calV,\calE)$ is simply
a listing $\sigma = (v_1,\dots,v_n)$ of all vertices in $\calV$.
Because we use vertices (of $\calH$)
and variables more or less interchangeably in this paper,
the term {\em variable ordering} will also be used with the same semantic.
\edefn

In the literature ``elimination order'', ``elimination ordering'', or ``global
attribute order'' are also used in place of ``vertex
ordering''~\cite{bodlaender2006treewidth,MR1640205,nnrr,anrr}. 
However, we chose not to use ``elimination order'' here because a
vertex ordering is meant to be the {\em reverse} of a (GYO) elimination 
ordering, as we explain below.

\paragraph*{Elimination hypergraph sequence.}
Fix a vertex ordering $\sigma = (v_1,\dots,v_n)$ of $\calH$, for
$j=n,n-1,\dots,1$ we recursively define a sequence of $n$ hypergraphs 
$\calH^\sigma_n,\calH^\sigma_{n-1},\dots,\calH^\sigma_1$ as follows.
To avoid cumbersome super-scripting, we will denote the sequence as
$\calH_n,\calH_{n-1},\dots,\calH_1$ when the vertex ordering $\sigma$
is clear from context.
Define $\calH_n = (\calV_n,\calE_n) = (\calV,\calE) = \calH$. 
Let $\partial(v_n)$ be the set of hyperedges of $\calH_n$ incident to $v_n$,
and $U_n$ be the union of edges in $\partial(v_n)$:
\begin{eqnarray}
   \partial(v_n) &=& \bigl\{ S \in \calE_n \suchthat v_n \in S
   \bigr\},\label{eqn:partial:vn}\\
     U_n &=& \bigcup_{S \in \partial(v_n)} S. \label{eqn:Un}
\end{eqnarray}
For each $j=n-1, n-2, \dots, 1$, define the hypergraph
$\calH_j = (\calV_j, \calE_j)$ as follows.
\begin{eqnarray*}
    \calV_j &=& \left\{v_1,\dots,v_j\right\}\\
    \calE_j &=& \left(\calE_{j+1} - \partial(v_{j+1})\right)
    \cup \bigl\{ U_{j+1}  - \{v_{j+1}\} \bigr\}\\
 \partial(v_j)&=&\left\{ S \in \calE_j \suchthat v_j \in S\right\}\\
     U_j &=& \bigcup_{S\in\partial(v_j)} S.
\end{eqnarray*}
Again, strictly speaking the sets 
$\calV_j, \calE_j, \partial(v_j)$, and $U_j$ 
should have been denoted by
$\calV^\sigma_j, \calE^\sigma_j, \partial^\sigma(v_j)$, and $U^\sigma_j$.
But we drop the superscript as $\sigma$ is implicitly understood. 

\bdefn
The above sequence of hypergraphs is called the 
{\em elimination hypergraph sequence} 
associated with the vertex ordering $\sigma$.
There is an intimate relationship between tree decompositions and vertex
orderings, which can be proved by making use of the above 
elimination hypergraph sequence.
\label{defn:elimination:hypergraph:sequence}
\edefn

\bprop[$\alpha$-Acyclicity]
A hypergraph $\calH$ is $\alpha$-acyclic if and only if there is a 
vertex ordering $\sigma=(v_1,\dots,v_n)$ such that $U^\sigma_k \in \partial(v_k)$ for all 
$k\in [n]$.
\label{prop:alpha:acyclicity}
\eprop

\bprop[$\beta$-Acyclicity]
A hypergraph $\calH$ is $\beta$-acyclic if and only if there is a 
vertex ordering $\sigma=(v_1,\dots,v_n)$ such that the collection of
hyperedges in $\partial(v_k)$ form a nested inclusion chain, for all $k \in
[n]$. Furthermore, $\beta$-acyclicity can be verified in polynomial-time.
\label{prop:beta:acyclicity}
\eprop

\bdefn[Induced $g$-width]\label{defn:inducedFECWidth}
Let $\calH=(\calV,\calE)$ be a hypergraph.
Let $g: 2^{\calV} \to \mathbb R^+$ be a function and $\sigma=(v_1,\dots,v_n)$ be a 
vertex ordering
of $\calH$. Then, the {\em induced $g$-width} of $\sigma$ is the quantity
$\max_{k\in [n]} g(U^\sigma_k).$
When $g(B) = |B|-1$, this is called the {\em induced width} of $\sigma$.
When $g(B) = \rho_\calH(B)$, this is called the {\em induced 
integral edge cover width} of $\sigma$.
When $g(B) = \rho^*_\calH(B)$, this is called the {\em induced fractional
edge cover width} of $\sigma$.
\edefn

We next characterize three width parameters of a hypergraph using
vertex ordering. 
A function $g: 2^{\calV} \to \mathbb R^+$ is said to be {\em monotone}
if $g(A) \leq g(B)$ whenever $A\subseteq B$. We prove a generic lemma.

\blmm[$g$-width]
Let $g: 2^\calV \to \mathbb R^+$ be a monotone function.
A hypergraph $\calH = (\calV, \calE)$ has $g$-width at most $w$ if and only if
there exists a vertex ordering $\sigma$ of all vertices of 
$\calH$ such that the induced $g$-width of $\sigma$ is at most $w$.
\label{lmm:g-width}
\elmm
Because the functions $s(U_k) = |U_k|-1$, $\rho_\calH(U_k)$ and
$\rho^*_\calH(U_k)$ are all monotone, the following results follow immediately.

\bcor
A hypergraph $\calH = (\calV, \calE)$ has 
treewidth (respectively, generalized hypertree width, fractional hypertree width)
at most $w$ if and only if
there exists a vertex ordering $\sigma = (v_1,\dots, v_n)$ of 
$\calH$ such that for every $k \in [n]$ we have 
$|U^\sigma_k| \leq w+1$
(respectively, $\rho_\calH(U_k) \leq w$, $\rho^*_\calH(U_k) \leq w$).
\label{prop:vo-fhtw}
\ecor

Note that the above corollary is actually three corollaries.
The one regarding tree-width alone
is well-known in the probabilistic graphical model literature
\cite{journals/ai/DechterP89, MR985145}. 
The other two are probably folklore, but we were not able to find
them explicitly stated anywhere.

\section{The $\InsideOut$ Algorithm}
\label{sec:insideout}

\subsection{Algorithmic Warm-ups}
\label{subsec:warmup}

We first present a simple solution to the $\faqcs$ problem. Recall
$\faqcs$ denotes the special case of $\faq$ when all variable aggregates are
the same:
i.e. $\oplus^{(i)}=\oplus, \forall i>f$, and $(\D, \oplus, \otimes)$ is a 
commutative semiring. 
Our aim is to gently introduce the reader to the duality of backtracking-search
and dynamic programming, and to the main idea of variable elimination.

\subsubsection{Backtracking search}
\label{subsubsec:OI}

Consider the $\sumprod$ form of the $\faq$ expression~\eqref{eqn:gen:faq} when there is
no free variable and all aggregates are the same semiring aggregate. In this case,
we write the expression as 
\[
    \varphi = \bigoplus_{\mv x} \bigotimes_{S\in\calE}\psi_S(\mv x_S)
    = \bigoplus_{x_1 \in \Dom(X_1)} 
       \left( \bigoplus_{\mv x_{[n]-\{1\}}} 
       \bigotimes_{S\in\calE}\psi_S(\mv x_S \suchthat x_1) \right).
\]
We can evaluate this expression by going through each value of $x_1$ and 
computing the inner expression `conditioned' on this $x_1$.
The na\"ive implementation of this strategy wastes time
if there is any $x_1$ for which {\em some} conditional factor 
$\psi_S( \cdot \suchthat x_1)$ is identically $\mv 0$.
Thus, the obvious idea is to first compute the set $I_1$ of values $x_1$ for which
$\psi_S( \cdot  \suchthat x_1) \not\equiv \mv 0$ {\em for all} factors
$\psi_S$. Then, recursively compute the expression
\[
    \varphi = \bigoplus_{x_1 \in I_1}
       \left( \bigoplus_{\mv x_{[n]-\{1\}}} 
       \bigotimes_{S\in\calE}\psi_S(\mv x_S \suchthat x_1) \right).
\]
Given that the input factors are represented using the listing format 
(i.e.\ only non-$\mv 0$ entries are listed), computing the above expression 
recursively is a join algorithm in disguise, and any of the algorithms
from \cite{NPRR12, leapfrog,skew,anrr} works.
We call this the $\OI$ algorithm, as it evaluates the expression from the
outer-most aggregate to the inner-most. It will serve as 
the algorithmic building block of $\InsideOut$.
In fact, the $\OI$ algorithm works even if there were free variables.
The following is almost immediate.

\bthm
Let $\varphi$ be an $\faqcs$-query whose 
hypergraph $\calH=(\calV,\calE)$ has $n$ vertices and $m$ hyperedges.
Algorithm $\OI$ computes $\varphi$ in time
$O(n\cdot m \cdot \log N \cdot \agm_\calH(\calV))$,
where $N=\max_{S\in\calE}\repsize{\psi_S}$.
\label{thm:OI}
\ethm

\bp
Consider an $\faqcs$ query $\varphi$ with free variables $\mv X_{[f]}$.
As discussed, $\OI$ is basically {\sf LeapFrog Triejoin}~\cite{leapfrog}: (1) it 
finds -- by backtracking search -- all tuples
$\mv x_{[n]}$ for which $\psi_S(\mv x_S) \neq \mv 0$ for all $S \in \calE$.
For each found tuple, the algorithm adds the product
$\bigotimes_{S\in\calE}\psi_S(\mv x_S)$ to the entry $\varphi(\mv x_{[f]})$.
Hence, the overall runtime is dominated by the backtracking search itself, which
can be bounded by $O(n\cdot m\cdot \log N \cdot \agm_{\calH}(\calV))$ (see~\cite{skew}).
\ep

$\OI$ is backtracking search
\cite{DBLP:journals/cacm/DavisLL62,DBLP:dblp_journals/jacm/GolombB65} which was
known 50 years ago in the AI and constraint programming world.
In the $\pgm$ literature, the method of conditioning search is
similar, but the main theoretical objective is so that the conditioning graph is
acyclic \cite{Pearl:1986:FPS:9075.9076}.
The main advantage of backtracking search is that it requires very little extra
space. The main disadvantage is that it might have to resolve the same subproblem
multiple times. The duality between
backtracking search and dynamic programming is well-known in the constraint
programming literature \cite{Rossi:2006:HCP:1207782}.
The next section explores the other side of this duality: the dynamic
programming side.

\subsubsection{Dynamic programming with variable elimination}
\label{subsec:var:elim}

To solve the $\faqcs$ problem using variable elimination 
\cite{DBLP:journals/ai/Dechter99,MR1426261,zhangpoole94}, the idea is to 
``fold'' common factors, exploiting the distributive law:
\begin{eqnarray*}
   \varphi(\mv x_{[f]})   &=& \bigoplus_{x_{f+1}}\cdots
   \bigoplus_{x_n}
   \bigotimes_{S\in\calE} \psi_S(\mv x_S)\\
   &=&\bigoplus_{x_{f+1}}\cdots
      \bigoplus_{x_{n-1}}
      \bigotimes_{S\in\calE-\partial(n)} \psi_S(\mv x_S)
      \otimes \underbrace{
         \left( \bigoplus_{x_n} \bigotimes_{S\in\partial(n)}
\psi_S(\mv x_S)\right)}_{\text{new factor } \psi'_{U_n-\{n\}}},
\end{eqnarray*}
where the equality follows from the fact that $\otimes$ distributes over
$\oplus$, and recall from~\eqref{eqn:partial:vn} and~\eqref{eqn:Un}
that $\partial(n)$ denotes all edges incident to $n$ in $\calH$ 
and $U_n=\cup_{S\in\partial(n)} S$. 
Note that the problem of computing the intermediate factor $\psi'_{U_n-\{n\}}$
is {\em exactly} an $\faqcs$ instance, where there is only one bound variable
$X_n$, and $|U_n|-1$ free variables.
Assume for the moment that we can somehow efficiently compute $\psi'_{U_n-\{n\}}$.

After computing $\psi'_{U_n-\{n\}}$, the resulting problem is another instance of 
$\faqcs$ on a modified multi-hypergraph $\calH_{n-1}$, 
constructed from $\calH_n:=\calH$ by removing vertex $n$ along with
all edges in $\partial(n)$, and {\em adding back} a new hyperedge $U_n-\{n\}$.
Recursively, we continue this process until all variables $X_n,\dots,X_{f+1}$
are eliminated. Textbook treewidth-based results for $\pgm$ inference
are obtained this way \cite{MR2778120}.
In the database context (i.e.\ given an $\faq$-query over the Boolean semiring), the 
intermediate result $\psi'_{U_n-\{n\}}$ is essentially an intermediate
materialized relation of a query plan.

\subsection{The $\InsideOut$ Algorithm}
\label{subsec:IO}

\subsubsection{Introducing the indicator projections}
\label{subsubsec:IO:indic-proj}
While correct, basic variable elimination as described in
Section~\ref{subsec:var:elim} is potentially 
not very efficient for sparse input factors, i.e. factors whose sizes are
smaller than the product of the domain sizes. 
The main reason is that the product
that was factored out (i.e. $\otimes_{S\in\calE-\partial(n)} \psi_S(\mv x_S)$) might annihilate many entries of the intermediate result
$\psi'_{U_n-\{n\}}$, while we have spent so much time computing $\psi'_{U_n-\{n\}}$.
For example, for an $S\notin\partial(n)$ and tuple $\mv y_S$ such that $S\subseteq U_n$ and
$\psi_S(\mv y_S)=\mv 0$, we do not need to compute the entries
$\psi'_{U_n-\{n\}}(\mv x_{U_n-\{n\}})$ for which 
$\mv y_{S} = \mv x_{S}$: those entries will be killed later 
anyhow. 
The idea is then to only compute those 
$\psi'_{U_n-\{n\}}(\mv x_{U_n-\{n\}})$ values that will `survive' the other 
factors. One simple way to achieve this is to allow for the factors that were
factored out of the scope of $X_n$ to still participate in computing
$\psi'_{U_n-\{n\}}(\mv x_{U_n-\{n\}})$: 
\begin{equation}\label{eqn:maineqn}
    \psi'_{U_n-\{n\}}(\mv x_{U_n-\{n\}}) =
    \bigoplus_{x_n} \left[ \left( \bigotimes_{S\in\partial(n)} \psi_S(\mv x_S) \right)
        \otimes \left( 
            \bigotimes_{\substack{S\notin \partial(n),\\S\cap
            U_n\neq \emptyset}}
            \underbrace{\psi_{S/U_n}(\mv x_{S\cap U_n})}_{\text{indicator projection}}
\right)\right]
\end{equation}
For some $S \in \calE - \partial(n)$ with $S\cap U_n \neq \emptyset$, 
the participation of a factor $\psi_S$ in computing
$\psi'_{U_n-\{n\}}$ is only to ``confirm'' that entries computed are
not wasteful; thus, only their indicator projections participate and not the
real factors $\psi_S$ themselves.
In database terms, one can also think of the definition of $\psi'_{U_n-\{n\}}$
in~\eqref{eqn:maineqn} as a simultaneous semijoin reduction of
the main product $\bigotimes_{S\in\partial(n)}\psi_S$ with all of  
the ``tables'' $\psi_S$ for which $S\cap U_n \neq
\emptyset$.

The problem defined in~\eqref{eqn:maineqn} is an $\faqcs$ instance
with $|U_n|-1$ free variables, which we
can solve using the $\OI$ algorithm as described in Section~\ref{subsubsec:OI}.
Note the important point that, algorithmically we do 
not perform the ``semijoins'' individually; rather, we compute the multiway-join
using a worst-case optimal join algorithm.

\subsubsection{The general case}
\label{subsubsec:IO:gen-case}
We finally describe how $\InsideOut$ deals with a general $\faq$-query defined
by the expression~\eqref{eqn:gen:faq}.
Recall that each operator $\oplus^{(i)}$ for $f<i\leq n$ either forms a 
semiring with $\otimes$ or is $\otimes$ itself.
Let us see how the last variable can be eliminated in this general scenario.
The elimination depends on two cases.

\noindent
{\bf Case 1:} $(\D, \oplus^{(n)}, \otimes)$ forms a semiring. In this case
     we apply the same strategy as before: we compute the intermediate factor
     $\psi_{U_n-\{n\}}$ defined in~\eqref{eqn:maineqn} using $\OI$.
     After that, we get another $\faq$ instance on a hypergraph $\calH_{n-1}$
     constructed from $\calH_n:=\calH$ by removing vertex $n$ along with
     all edges in $\partial(n)$, and adding back a new hyperedge $U_n-\{n\}$.

\noindent
{\bf Case 2:} $\oplus^{(n)} = \otimes$. In this case, we rewrite
the expression as follows. 
\begin{eqnarray}
    \varphi(\mv x_{[f]})
    &=& \textstyle{\bigoplus^{(f+1)}_{x_{f+1}} \cdots
        \bigoplus^{(n)}_{x_{n}}
        \bigotimes_{S\in\calE}\psi_S(\mv x_S)}\nonumber\\
    &=& \textstyle{\bigoplus^{(f+1)}_{x_{f+1}} \cdots
        \bigoplus^{(n-1)}_{x_{n-1}}
        \left( \bigoplus^{(n)}_{x_{n}}
    \bigotimes_{S\in\calE}\psi_S(\mv x_S) \right)}\nonumber\\
    &=& \textstyle{\bigoplus^{(f+1)}_{x_{f+1}} \cdots
        \bigoplus^{(n-1)}_{x_{n-1}}
        \left( \bigotimes_{x_{n}}
    \bigotimes_{S\in\calE}\psi_S(\mv x_S) \right)}\nonumber\\
    &=& \textstyle{\bigoplus^{(f+1)}_{x_{f+1}} \cdots
    \bigoplus^{(n-1)}_{x_{n-1}}
    \left(
    \bigotimes_{S\in\calE}\bigotimes_{x_{n}}\psi_S(\mv x_S) \right)}\nonumber\\
    &=& \textstyle{\bigoplus^{(f+1)}_{x_{f+1}} \cdots
        \bigoplus^{(n-1)}_{x_{n}}
        \bigotimes_{S\notin\partial(n)} 
         \underbrace{\bigl(\psi_S(\mv x_S)\bigr)^{|\Dom(X_n)|}}_{\psi'_S}
        \otimes\bigl(
        \bigotimes_{S\in\partial(n)}
        \underbrace{\textstyle{\bigotimes_{x_n}} \psi_S(\mv x_S)}_{\psi'_{S-\{n\}}}
        \bigr)
        }\label{eqn:last:exp}
\end{eqnarray}
Thus, in Case 2, the new $\faq$ instance has as its input the factors
$\psi'_S(\mv x_S) := \psi_S(\mv x_S)^{|\Dom(X_n)|}$ for $S\not\in\partial(n)$,
and the factors
$\psi'_{S-\{n\}}(\mv x_{S-\{n\}}) := \otimes_{x_n} \psi_S(\mv x_S)$ for $S \in\partial(n)$. Also, the new instance has a hypergraph $\calH_{n-1}$ that is obtained from $\calH_n:=\calH=(\calV,\calE)$ by removing vertex $n$ from $\calV$ and from all hyperedges in $\calE$.

By repeatedly applying the above process on the resulting $\faq$ instance with hypergraph $\calH_{n-1}$, we eliminate the remaining bound variables $X_{n-1},\ldots, X_{f+1}$ and obtain a sequence of $\faq$ instances with hypergraphs $\calH_{n-2}, \ldots, \calH_{f}$.
We formally define below how to construct this sequence of hypergraphs.
The construction here differs from the one given earlier in Definition~\ref{defn:elimination:hypergraph:sequence} in how we eliminate vertices $k$ corresponding to product variables, i.e.~$\oplus^{(k)} = \otimes$.
(Also while we only described hypergraphs $\calH_n,\ldots,\calH_f$ in our discussion above,
for later convenience we will define below the additional hypergraphs $\calH_{f-1},\ldots,\calH_1$.)

\bdefn[Elimination hypergraph sequence]\label{defn:hyperSeq}
Let $\varphi $ be an $\faq$-query with hypergraph $\calH=(\calV=[n],\calE)$.
We define a sequence of hypergraphs
$\calH_k = (\calV_k=[k], \calE_k)$, for $k=n,n-1,\dots,1$, called the {\em elimination hypergraph sequence} of $\varphi$ (with respect to the input variable ordering $\sigma=(1,\ldots,n)$) as follows.
Let $\calH_n=(\calV_n=[n],\calE_n) := \calH$.
For $k=n,n-1,\dots,2$, construct
$\calH_{k-1}=(\calV_{k-1},\calE_{k-1})$ from $\calH_{k}=(\calV_{k},\calE_{k})$ as follows. Define
\begin{align*}
   \partial(k):= \{ S \in \calE_{k} \suchthat k \in S\},
   &&    U_{k} := \bigcup_{S\in\partial(k)} S,
   &&\calV_{k-1}:= \calV_{k}-\{k\}.
\end{align*}
\bi
\item If $k> f$ and $\oplus^{(k)} = \otimes$, then $\calE_{k-1}$ is obtained from 
$\calE_{k}$ by removing $k$ from all edges in $\calE_{k}$, i.e.
\[\calE_{k-1} := (\calE_k \setminus \partial(k)) \cup \{ S \setminus \{k\} \suchthat S \in \partial(k)\}.\]
\item Otherwise, $\calE_{k-1}$ is obtained from 
$\calE_{k}$ as follows:
\[ \calE_{k-1} := \left(\calE_{k} - \partial(k)\right)
\cup \bigl\{ U_{k}  - \{k\} \bigr\}.
\]
\ei
\edefn

We now discuss how to compute the new factors in Eq.~\eqref{eqn:last:exp}, namely $\psi'_S$ for $S\notin\partial(n)$ and $\psi'_{S-\{n\}}$ for $S\in\partial(n)$.

First, consider the factors $\psi'_S$ for $S \notin \partial(n)$.
When ``passing through'' a product aggregate, 
these factors are powered up, point-wise, by a power of $|\Dom(X_n)|$.
By repeated squaring, the number of multiplications needed is within
\[ \sum_{S\in\calE-\partial(n)} 2\cdot \repsize{\psi_S} \cdot \lceil\log_2 |\Dom(X_n)|\rceil. \]
So these factors are generally changed to new factors with the same size.

There is one case in which we do not have to power them up:
when $\psi_S(\mv x_S)$ is an idempotent element under the product aggregate
$\otimes$. In particular, if we knew that $\psi_S(\mv x_S) \in \{\mv 0, \mv 1\}$
for all $\mv x_S$, 
then $\psi_S(\mv x_S)^{|\Dom(X_n)|} = \psi_S(\mv x_S)$ and we can factor out
$\psi_S$ as in the semiring case.
This is indeed the case for the instances of $\faq$ that were reduced from
$\qcq$ and $\sqcq$ as shown in Example~\ref{ex:qcq} and Example~\ref{ex:sqcq}.
Motivated by those two examples, we define the following concept.

\bdefn[Idempotent product aggregate]\label{defn:idempotent product agg}
An aggregate $\oplus^{(k)}$ is called
an {\em idempotent product aggregate} (with respect to the input variable
ordering $\sigma$) if it is a product aggregate in which
for all $S \in \calE_k \setminus \partial(k)$, the (intermediate)
factor $\psi_S$ has as its range the idempotent elements of the operator
$\otimes$.
In particular, if $\psi_S(\mv x_S) \in \{\mv 0, \mv 1\}$ for all 
$S\in\calE_k \setminus \partial(k)$, then $\oplus^{(k)}$ is an idempotent
product aggregate whenever $\oplus^{(k)} = \otimes$.
\edefn

Note again that for the $\faq$ instances which are constructed from the
reductions from $\qcq$ and $\sqcq$, {\em all} product aggregates are
idempotent. In such a case, we can rewrite~\eqref{eqn:last:exp} as 
\[ \varphi(\mv x_{[f]}) = 
    \textstyle{\bigoplus^{(f+1)}_{x_{f+1}} \cdots
        \bigoplus^{(n-1)}_{x_{n}}
        \bigotimes_{S\notin\partial(n)} \psi_S(\mv x_S)
        \otimes\bigl(
        \bigotimes_{S\in\partial(n)}
        \underbrace{\textstyle{\bigotimes_{x_n}} \psi_S(\mv x_S)}_{\psi'_{S-\{n\}}}
        \bigr).
        }
\]

Now, consider the factors $\psi'_{S-\{n\}}$ for $S\in\partial(n)$ in~\eqref{eqn:last:exp}.
Each factor $\psi'_{S-\{n\}}$ can be thought of as a ``product marginalization''
of the factor $\psi_S$, where we ``marginalize out'' the $X_n$ variable.

To summarize, in Case 2 we can compute each one of the new factors $\psi'_{S}$ and $\psi'_{S-\{n\}}$ in~\eqref{eqn:last:exp} individually.
Therefore we do not have to solve the costly
intermediate $\faqcs$ instance $\psi'_{U_n-\{n\}}$ given by~\eqref{eqn:maineqn} as was done in Case 1.

\subsubsection{Output representation}
\label{subsubsec:IO:output}
Let $\calH_f = (\calV_f = [f], \calE_f)$ denote the hypergraph of the $\faq$
instance resulting from eliminating all bound variables $X_n, \ldots, X_{f+1}$, as described in Definition~\ref{defn:hyperSeq}.
With some abuse of notation, let $\psi_S, S\in\calE_f$ denote the set of input
factors to the $\calH_f$ instance.
The output to $\faq$ is now the expression
\begin{equation}
\label{eqn:output-listing}
\varphi(\mv x_{[f]}) =
        \textstyle{\bigotimes}_{S\in\calE_f} \psi_S(\mv x_S),
\end{equation}
which is an $\faq$ instance with {\em all} free variables.
We can compute the output directly by running $\OI$ on \eqref{eqn:output-listing}.
However, we can achieve a better runtime if we first compute the output in the {\em factorized representation} (see Olteanu and Z\'{a}vodn\'{y}~\cite{OZ15}) and then report it (i.e.~convert it to the listing representation).

In particular, to achieve the better runtime, we start with eliminating the free variables $X_f, X_{f-1},\ldots, X_1$.
To that end, we define the following $\mv{01}$-OR operator over the domain of $\{\mv 0, \mv 1\}$.

\bdefn[$\mv{01}$-OR]
Given $a, b\in\{\mv0, \mv 1\}$, define the {\em $\mv{01}$-OR} of $a$ and $b$, denoted by $a \myor b$, as follows.
\[ a \myor b := \begin{cases} \mv 0 & \text{ if } a=b=\mv 0\\
\mv 1 & \text{ otherwise}.
\end{cases}
\]
\label{defn:01-OR}
\edefn
Clearly $(\{\mv 0,\mv 1\}, \myor,\otimes)$ is a commutative semiring.
Using this semiring, we define the following
$\faqcs$ query that doesn't have any free variables and has a value that belongs to $\{\mv 0, \mv 1\}$:
\[ \overline\varphi() = \bigmyor_{\mv x_{[f]}} \bigotimes_{S\in\calE_f} 
   \psi_{S/S}(\mv x_S).
\]
Now, we run $\InsideOut$ on the above query eliminating the variables $X_f, X_{f-1}, \dots, X_1$ while computing the intermediate factors
$\psi_{U_k}, \psi_{U_k-\{k\}}$ for $k=f,f-1,\dots,1$, defined as follows:
\begin{eqnarray}
   \psi_{U_k}(\mv x_{U_k}) &=& \bigotimes_{\substack{S\in\calE_k\\ S\cap U_k
   \neq \emptyset}} \psi_{S/U_k}(\mv x_{S\cap U_k})\label{eqn:psi:Uk}\\
\psi_{U_k-\{k\}}(\mv x_{U_k-\{k\}}) &=&
\bigmyor_{x_k}\psi_{U_k}(\mv x_{U_k}).
\end{eqnarray}
For each $k$, both factors $\psi_{U_k}(\mv x_{U_k})$
and $\psi_{U_k-\{k\}}(\mv x_{U_k-\{k\}})$
can be computed by a single run to $\OI$.
Note that $U_1-\{1\}=\emptyset$, and $\psi_{U_1-\{1\}} =
\overline\varphi$. 

Finally, we report the output by running $\OI$ on the following expression instead of 
\eqref{eqn:output-listing}
\begin{equation}
\label{eqn:output-factorized}
\varphi(\mv x_{[f]}) =
        \left(\bigotimes_{S\in\calE_f} \psi_S(\mv x_S)\right) \otimes
        \left(\bigotimes_{k\in[f]} \psi_{U_k}(\mv x_{U_k})\right).
\end{equation}
The fact that~\eqref{eqn:output-factorized} defines exactly the same function 
as~\eqref{eqn:output-listing} follows from the following trivial observation:
for every fixed tuple $\mv x_{[f]}$, 
$\psi_S(\mv x_S) \neq \mv 0$ for all $S\in\calE_f$ implies
$\psi_{S/S}(\mv x_S) = \mv 1$ for all $S\in\calE_f$, which in turns implies
$\psi_{U_k}(\mv x_{U_k}) = \mv 1$, for all $k \in [f]$.

Eliminating free variables and then recovering them back 
by~\eqref{eqn:output-factorized} are equivalent to 
the two phases of the Yannakakis 
algorithm~\cite{dblp:conf/vldb/yannakakis81}.
Our implementation of $\InsideOut$ in LogicBlox follows the 
elimination/recovery method. 
The reason $\OI$ is faster on~\eqref{eqn:output-factorized} than
on~\eqref{eqn:output-listing} is that the factors $\psi_{U_k}$ help filter
out potential tuples $\mv x_{[f]}$ for which $\varphi(\mv x_{[f]})=\mv 0$.
We will see in the analysis below the effect of this idea in the overall runtime.
The entire algorithm is sketched out in Algorithm~\ref{algo:IO}.

\begin{algorithm}[th]
   \caption{$\InsideOut$ for $\faq$}
   \label{algo:IO}
   \begin{algorithmic}[1]
      \Require{Hypergraph $\calH=(\calV=[n],\calE)$, factors $\psi_S$, $S\in\calE$, set $F=[f]$ of free variables}
      \Require{$\faq$ query $\varphi$ in the form~\eqref{eqn:gen:faq}}
      \Statex
      \State $\calE_n \gets \calE$
      \For {($k \gets n$ {\bf downto} $1$)}
      \State $\partial(k) \gets \left\{S \in \calE_k \suchthat k
      \in S\right\}$
      \If {($k> f$ and $(\mv D,\oplus^{(k)}, \otimes)$ is a semiring) or ($k\leq f$)}
      \State $\displaystyle{U_k \gets \bigcup_{S\in\partial(k)} S}$
      \If {$k> f$}
      \State $\displaystyle{\psi_{U_k-\{k\}(\mv x_{U_k-\{k\}})} =
         \mathop{\textstyle{\bigoplus^{(k)}}}_{x_k} \left(
         \bigotimes_{S\in\partial(k)} \psi_S( \mv x_S ) \right)
         \otimes \left( \bigotimes_{\substack{S\in \calE_k - \partial(k),\\
               S\cap U_k\neq \emptyset}} \psi_{S/U_k} ( \mv x_{S\cap
            U_k})
         \right)}$ \label{ln:intermediate1}
      \Else
      \State $\psi_{U_k}(\mv x_{U_k}) =
      \displaystyle{
         \bigotimes_{\substack{S\in\calE_k,\\ S\cap
               U_k\neq \emptyset}}
         \psi_{S/U_k}(\mv x_{S\cap U_k})}$
      \State $\psi_{U_k-\{k\}}(\mv x_{U_k-\{k\}}) =
      \displaystyle{\bigmyor_{x_k} \psi_{U_k}(\mv
         x_{U_k})}$
      \EndIf
      \State $\calE_{k-1} \gets (\calE_k \setminus \partial(k)) \cup \{ U_k-\{k\}\}$ 
      \Else
      \For {each $S\in\partial(k)$}
      \State Compute the product marginalization factors $\psi_{S-\{k\}}(\mv x_{S-\{k\}}) = \bigotimes_{x_k} \psi_S(\mv x_S)$\label{ln:intermediate2}
      \EndFor
      \For {each $S\in\calE_k-\partial(k)$}  
      \If {Range of $\psi_S$ is {\bf not} idempotent wrt $\otimes$}
      \State $\psi_{S}(\mv x_{S}) = \psi_S(\mv x_S)^{|\Dom(X_k)|}$, for all $\mv x_S$ where
      $\psi_S(\mv x_S)$ is not $\otimes$-idempotent \label{ln:intermediate2-not-in-S}
      \EndIf
      \EndFor
      \State $\calE_{k-1} \gets (\calE_k \setminus \partial(k)) \cup \{ S \setminus \{k\} \suchthat S \in \partial(k)\}$
      \EndIf
      \EndFor
      \State Output $\varphi$ by running $\OI$ on $\faq$-expression~\eqref{eqn:output-factorized}\label{ln:output-phase}
   \end{algorithmic}
\end{algorithm}

\subsection{Analysis}
\label{subsec:analysis}
From notations introduced in Definition~\ref{defn:hyperSeq}, and recalling the $\agm$-bound definition~\eqref{eqn:agm}, 
the following is the main
theorem of this section.

\bthm
Suppose $\varphi$ is an $\faq$ query whose hypergraph $\calH=(\calV,\calE)$ has 
$n$ vertices and $m$ hyperedges.
Let $N:=\max_{S\in\calE}\repsize{\psi_S}$.
Define 
\begin{equation}
   K := [f] \cup \left\{ k\in [n]\setminus[f] \suchthat \oplus^{(k)} \neq \otimes \right\}.
   \label{eqn:the:set:K}
\end{equation}
For each $k\in [n] - K$, $S \in \calE_k\setminus\partial(k)$, define
\[ \Idem_k(\psi_S) = 
\begin{cases}
   0 & \text{ if the range of $\psi_S$ is idempotent w.r.t. $\otimes$}\\
   2\lceil \log_2|\Dom(X_k)|\rceil & \text{ otherwise.}
\end{cases}
\]
Then, the $\InsideOut$ algorithm (Algorithm~\ref{algo:IO}) 
computes $\varphi$ in time 
\begin{multline}
    O
   \left(n^2\cdot m+
   \sum_{k\in K} |U_k|^2 \cdot |\{ S\in \calE_k : S\cap U_k \neq \emptyset\}|
   \cdot \log N \cdot \agm_{\calH_k}(U_k) \right.
    \\+
\left.    \sum_{\substack{k\in [n]-K\\S\in\partial(k)}} |S| \cdot \repsize{\psi_S}
+
    \sum_{\substack{k\in [n]-K\\S\in\calE_k \setminus \partial(k)}}
    |S| \cdot\Idem_k(\psi_S)\cdot \repsize{\psi_S}
    + f^2\cdot(f+m)\cdot\log N\cdot\repsize{\varphi}
    \right).
\label{eqn:IO:runtime}
\end{multline}
\label{thm:IO}
\ethm

\bp
The runtime of $\InsideOut$ is the sum of the runtimes of $n$ variable
elimination steps plus the time needed to report the output at the end. 
For each $k = n, ..., 1$, we need $O(n\cdot m)$ time to compute $\partial(k)$ and $U_k$ (when $k\in K$) in a naive way, thus we have a data-independent term of $O(n^2\cdot m)$ in the runtime expression.
Moreover, for each $k = n, ..., 1$, the (data-dependent) cost of the $k$th-elimination step depends on whether the $k$th variable aggregate is a product aggregate or not:
\bi
 \item If $k>f$ and $\oplus^{(k)}=\otimes$ (i.e.~if $k\in[n]-K$), then the 
    runtime is the total time needed to compute the intermediate 
    factors shown in~\eqref{eqn:last:exp}.
    Based on the discussion of Case 2 in Section~\ref{subsubsec:IO:gen-case}, each factor $\psi'_S$ for $S\in\calE_k-\partial(k)$ can be computed in time 
    $O(|S|\cdot\Idem_k(\psi_S)\cdot \repsize{\psi_S})$, whereas each factor $\psi'_{S-\{k\}}$ for $S\in\partial(k)$ can be computed in time
    $O(|S| \cdot \repsize{\psi_S})$.
 \item If $k\in[K]$, then the runtime is dominated by the $\OI$ algorithm's 
    runtime to compute the intermediate factor $\psi_{U_k-\{k\}}$. From Theorem~\ref{thm:OI}, this runtime is 
    bounded by
    \[O \left(
   |U_k| \cdot |\{ S\in \calE_k : S\cap U_k \neq \emptyset\}|\cdot \log N_k\cdot\agm_{\calH_k}(U_k)  \right),\]
where $N_k:=\max\left\{\repsize{\psi_{S/U_k}} \suchthat S\in\calE_k, S\cap U_k \neq \emptyset\right\}$.
We can naively bound $N_k$ by $N^{|U_k|}$.
\ei
The final invocation of $\OI$ on \eqref{eqn:output-factorized} reports the 
output (in listing representation) in linear time (modulo a $\log$ factor) in
$\repsize{\varphi}$, just like the output phase of Yannakakis algorithm~\cite{dblp:conf/vldb/yannakakis81}. 
This is because the participations of the factors $\psi_{U_k}$,
$k\in[f]$ in
the formula ensures that every binding of the backtracking search algorithm is
part of an output tuple.
In particular, this final invocation of $\OI$ involves $f$ variables and at most $f+m$ factors (specifically $f$ factors $\psi_{U_k}$ for $k\in[f]$ and at most $m$ factors $\psi_S$ for $S\in\calE_f$).
Therefore, it runs in time $O(f\cdot(f+m)\cdot(\log \repsize{\varphi})\cdot\repsize{\varphi})$.
We can naively bound $\log \repsize{\varphi}$ by $f\log N$.
\ep

In the above discussion and analysis of $\InsideOut$, we eliminated the variables 
$X_n,X_{n-1},\dots,X_{1}$ in the order given by the input
$\faq$-expression~\eqref{eqn:gen:faq}. However, as Example~\ref{ex:var:order:effect}
shows, there is no reason to force $\InsideOut$ to follow this particular order. 
In particular, there might be a different variable ordering for which the
overall runtime of $\InsideOut$ is {\em a lot} smaller and the algorithm still
works correctly on that ordering.

\begin{ex}[Effect of variable ordering on runtime]\label{ex:var:order:effect}
We illustrate $\InsideOut$ and the effect of different variable orderings on the
runtime of $\InsideOut$ with an example.
Consider the following $\faq$ query (without free variables).
\[ \varphi = \max_{x_1} \max_{x_2} \prod_{x_3} \sum_{x_4} \max_{x_5} \max_{x_6}
             \psi_{\{1,5\}}\psi_{\{2,5\}}
             \psi_{\{1,3,4\}}
             \psi_{\{2,3,6\}}.
\]
Here, the support of each factor determines the variables so we do not write
down the parameters of each factor for the sake of brevity.
Also, in this example we consider $\D = \R_+$ and every input factor has range
$\D$.
Now, a straightforward
run of the $\InsideOut$ algorithm (Algorithm~\ref{algo:IO}) using the variable
ordering $(X_1,\dots,X_6)$ evaluates the above expression as follows.
\begin{eqnarray*}
 \varphi &=& 
 \max_{x_1} \max_{x_2} \prod_{x_3} \sum_{x_4} \max_{x_5} \max_{x_6}
             \psi_{\{1,5\}}\psi_{\{2,5\}}
             \psi_{\{1,3,4\}}
             \psi_{\{2,3,6\}}\\
 \text{(re-write)} &=&
 \max_{x_1} \max_{x_2} \prod_{x_3} \sum_{x_4} \max_{x_5} 
             \psi_{\{1,5\}}\psi_{\{2,5\}}
             \psi_{\{1,3,4\}}
             \max_{x_6}
             \psi_{\{2,3,6\}}\\
 \text{(spent $O(N)$-time)} &=&
 \max_{x_1} \max_{x_2} \prod_{x_3} \sum_{x_4} \max_{x_5} 
             \psi_{\{1,5\}}\psi_{\{2,5\}}
             \psi_{\{1,3,4\}}
             \psi_{\{2,3\}}\\
 \text{(re-write)} &=&
 \max_{x_1} \max_{x_2} \prod_{x_3} \sum_{x_4} 
             \psi_{\{1,3,4\}}
             \psi_{\{2,3\}}
             \max_{x_5} 
             \psi_{\{1,5\}}\psi_{\{2,5\}}\\
 \text{(spent $O(N^2)$-time)} &=&
 \max_{x_1} \max_{x_2} \prod_{x_3} \sum_{x_4} 
             \psi_{\{1,3,4\}}
             \psi_{\{2,3\}}
             \psi_{\{1,2\}}\\
 \text{(re-write)} &=&
 \max_{x_1} \max_{x_2} \prod_{x_3} 
             \psi_{\{2,3\}}
             \psi_{\{1,2\}}
             \sum_{x_4} 
             \psi_{\{1,3,4\}}\\
 \text{(spent $O(N)$-time)} &=&
 \max_{x_1} \max_{x_2} \prod_{x_3} 
             \psi_{\{2,3\}}
             \psi_{\{1,2\}}
             \psi_{\{1,3\}}\\
 \text{(re-write)} &=&
 \max_{x_1} \max_{x_2} 
             \left(\prod_{x_3} \psi_{\{2,3\}} \right)
             \left(\prod_{x_3} \psi_{\{1,2\}} \right)
             \left(\prod_{x_3} \psi_{\{1,3\}} \right)\\
 \text{(spent $\tilde O(N^2)$-time)} &=&
 \max_{x_1} \max_{x_2} 
             \psi_{\{2\}} 
             \bar \psi_{\{1,2\}}
             \psi_{\{1\}}\\
 \text{(re-write)} &=&
 \max_{x_1} \psi_{\{1\}}
             \max_{x_2} 
             \psi_{\{2\}} 
             \bar \psi_{\{1,2\}}\\
 \text{(spent $O(N^2)$-time)} &=&
 \max_{x_1} \psi_{\{1\}}
            \bar \psi_{\{1\}}\\
 \text{(spent $O(N)$-time)} &=&
 \psi_{\emptyset}
\end{eqnarray*}

Note that $\bar\psi_{\{1,2\}}(x_1,x_2) = \psi_{\{1,2\}}(x_1,x_2)^{|\Dom(X_3)|}$, for 
every $(x_1,x_2)$.
Because $\psi_{\{1,2\}}:=\max_{x_5}\psi_{\{1,5\}}\psi_{\{2,5\}}$, 
there can be up to $N^2$ non-zero entries in $\psi_{\{1,2\}}$.
Each one of those entries can be raised to a power of $|\Dom(X_3)|$ by repeated squaring using $O(\log N)$ multiplications for a total time of $O(N^2\log N)=\tilde O(N^2)$.
The overall runtime of the algorithm is thus $\tilde O(N^2)$.

Next, let us see how a slight change in the assumption of the input factors
allows us to conclude that the aggregate on $X_3$ is an idempotent product
aggregate, and that allows for a {\em different variable ordering} to be
equivalent to the original variable ordering, and that helps reduce the overall
runtime of $\InsideOut$.
Suppose we knew that all input factors have range $\{0,1\}$.
Then, we can evaluate the query as follows.
\begin{eqnarray*}
 \varphi &=& 
 \max_{x_1} \max_{x_2} \prod_{x_3} \sum_{x_4} \max_{x_5} \max_{x_6}
             \psi_{\{1,5\}}\psi_{\{2,5\}}
             \psi_{\{1,3,4\}}
             \psi_{\{2,3,6\}}\\
 \text{(re-write)} &=&
 \max_{x_1} \max_{x_2} \prod_{x_3} \sum_{x_4} 
             \max_{x_5} \psi_{\{1,5\}}\psi_{\{2,5\}}
             \psi_{\{1,3,4\}}
             \max_{x_6}
             \psi_{\{2,3,6\}}\\
 \text{(spent $O(N)$-time)} &=&
 \max_{x_1} \max_{x_2} \prod_{x_3} \sum_{x_4} 
             \max_{x_5} \psi_{\{1,5\}}\psi_{\{2,5\}}
             \psi_{\{1,3,4\}}
             \psi_{\{2,3\}}\\
 \text{(re-write)} &=&
 \max_{x_1} \max_{x_2} \prod_{x_3} \sum_{x_4} 
             \psi_{\{1,3,4\}}
             \psi_{\{2,3\}}
             (\max_{x_5} \psi_{\{1,5\}}\psi_{\{2,5\}})\\
 \text{(re-write)} &=&
 \max_{x_1} \max_{x_2} \prod_{x_3} 
             \psi_{\{2,3\}}
             (\max_{x_5} \psi_{\{1,5\}}\psi_{\{2,5\}})
             \sum_{x_4} 
             \psi_{\{1,3,4\}}\\
 \text{(spent $O(N)$-time)} &=&
 \max_{x_1} \max_{x_2} \prod_{x_3} 
             \psi_{\{2,3\}}
             (\max_{x_5} \psi_{\{1,5\}}\psi_{\{2,5\}})
             \psi_{\{1,3\}}\\
 \text{(re-write)} &=&
 \max_{x_1} \max_{x_2} 
             \left(\prod_{x_3} \psi_{\{2,3\}} \right)
             \left(\prod_{x_3} (\max_{x_5} \psi_{\{1,5\}}\psi_{\{2,5\}}) \right)
             \left(\prod_{x_3} \psi_{\{1,3\}} \right)\\
 \text{($\times$ is acting idempotently)} &=&
 \max_{x_1} \max_{x_2} 
             \left(\prod_{x_3} \psi_{\{2,3\}} \right)
             \left(\max_{x_5} \psi_{\{1,5\}}\psi_{\{2,5\}} \right)
             \left(\prod_{x_3} \psi_{\{1,3\}} \right)\\
 \text{(spent $O(N)$-time)} &=&
 \max_{x_1} \max_{x_2} 
             \psi_{\{2\}} 
             \left(\max_{x_5} \psi_{\{1,5\}}\psi_{\{2,5\}} \right)
             \psi_{\{1\}}\\
 \text{(re-arrange)} &=&
 \max_{x_1} \max_{x_2} \max_{x_5}
             \psi_{\{2\}} 
             \psi_{\{1,5\}}\psi_{\{2,5\}}
             \psi_{\{1\}}\\
             \text{(a {\bf crucial re-arrangement})} &=&
 \max_{x_5} \max_{x_1} \max_{x_2}
             \psi_{\{2\}} 
             \psi_{\{1,5\}}\psi_{\{2,5\}}
             \psi_{\{1\}}\\
 \text{(re-arrange)} &=&
 \max_{x_5} \max_{x_1} 
             \psi_{\{1,5\}}
             \psi_{\{1\}}
             \max_{x_2}
             \psi_{\{2\}} 
             \psi_{\{2,5\}}\\
 \text{(spent $O(N)$-time)} &=&
 \max_{x_5} \max_{x_1} 
             \psi_{\{1,5\}}
             \psi_{\{1\}}
             \psi_{\{5\}}\\
 \text{(re-arrange)} &=&
 \max_{x_5} \psi_{\{5\}}
 \max_{x_1} \psi_{\{1,5\}} \psi_{\{1\}}\\
 \text{(spent $O(N)$-time)} &=&
 \max_{x_5} \psi_{\{5\}} \bar\psi_{\{5\}}\\
 \text{(spent $O(N)$-time)} &=&
 \psi_\emptyset
\end{eqnarray*}

So in this case the fact that all factors inside the scope of
$\times$ have ranges which are $\{0,1\}$, the idempotent elements of $\times$,
we are able to reduce the runtime down to $O(N)$.
The key is, if we ran $\InsideOut$ with the variable ordering 
$(X_5, X_1, X_2, X_3, X_4, X_6)$ then we would have achieved a runtime of
$O(N)$.
\end{ex}

The main technical contributions of this paper lie in answering the following two questions:
\be 
\item How do we know which variable orderings are equivalent to the original
$\faq$-query expression? 
\item How do we find the ``best'' variable ordering among all
equivalent variable orderings in an efficient way (better than brute-force search)? 
\ee
The next sections formalize the above two questions.

\subsection{Equivalent variable orderings}
\label{subsec:equivalent:VO}

We first formalize the concept of ``equivalent variable ordering'':

\bdefn[$\EVO(\varphi)$]
Let $\varphi$ be an $\faq$-query written in the form \eqref{eqn:gen:faq} of Section~\ref{sec:faq-pbm}.
A {\em $\varphi$-equivalent variable ordering} is a vertex ordering 
$\sigma = (v_1,\dots,v_n)$ of the hypergraph $\calH$ satisfying the following
conditions:
\bi
 \item[(a)] The set $\{v_1,\dots,v_f\}$ is exactly $F = [f]$. In other words, in
 the $\varphi$-equivalent variable ordering, the free variables come first (in any order).
 \item[(b)] The function $\varphi'$ defined by
 \[ \varphi'(\mv x_F) :=
    \textstyle{\bigoplus^{(v_{f+1})}_{x_{v_{f+1}}} \cdots
    \bigoplus^{(v_n)}_{x_{v_n}}}
    \bigotimes_{S\in\calE}\psi_S(\mv x_S)
 \]
 is identical to the function $\varphi$ no matter what the input factors
 are.\footnote{Here, we assume that the variable domains, the range $\D$, and the 
 aggregates are fixed and known in advance, but the input factors are not.}
\ei
Let $\EVO(\varphi)$ denote the set of all $\varphi$-equivalent variable
orderings.

\label{eqn:phi-equiv-order}
\edefn

In many applications, we know a specific class of input factors that are allowed in the corresponding $\faq$ problem. For example, in logic
applications the input factors are often $\{\mv 0, \mv 1\}$-valued functions.
This motivates a stronger definition of $\EVO$. 

\bdefn[$\EVO(\varphi, \calF)$]
\label{defn:EVO-phi-F}
Let $\calF$ be a class of functions with range $\D$. 
Let $\EVO(\varphi, \calF)$ denote the set of all
$\varphi$-equivalent variable orderings under the promise that all input
factors come from $\calF$. In other words, the definition of $\EVO(\varphi,
\calF)$ is the same as that in Definition~\ref{eqn:phi-equiv-order}, except 
that we only require $\varphi'$ to be the same as $\varphi$ for all input
factors belonging to $\calF$.
\edefn

\subsection{$\faq$-width of a variable ordering}
\label{subsec:faq-width}

We now address the second question of what it means to be the ``best'' variable
ordering. With respect to the analysis of $\InsideOut$ in Theorem~\ref{thm:IO},
naturally the ``best'' variable ordering is a variable ordering in
$\EVO(\varphi)$ which minimizes the runtime~\eqref{eqn:IO:runtime}.
The $\EVO$ constraint makes it trickier to find an optimal
variable ordering for the general $\faq$ problem, as opposed to 
the typical variable elimination in the graphical model, matrix computation,
and constraint satisfaction domains where all variable 
orderings are in $\EVO(\varphi)$. 
To illustrate this point, Section~\ref{app:sec:quick:applications}
in the appendix presents
two examples (namely $\mcm$ and $\dft$) where minimizing 
\eqref{eqn:IO:runtime} is easy. These examples explain two entries in the
summary Table~\ref{tab:results}.

In general, however, without knowing a bit more about the structure of the 
problem it is
hard(er) to derive a general result on how to find a variable ordering
to minimize \eqref{eqn:IO:runtime}. In particular, in these examples we have
been lucky that every permutation of the variable aggregates yields an
expression that is equivalent to the original $\faq$ expression 
\eqref{eqn:gen:faq}.
This is because in these examples there is only {\em one} variable
aggregate which is the $+$ operator.
The second thing that helps with these examples is that we know
(a good estimate of) the sizes of the indicator projections; so we can plug them in
expression \eqref{eqn:IO:runtime}. In general, these sizes are highly
instance-dependent. 

Minimizing~\eqref{eqn:IO:runtime} is a bit unwieldy, thus we slightly relax
the runtime analysis of $\InsideOut$ to obtain a better behaved expression than
that in~\eqref{eqn:IO:runtime}.
\bprop
Following notations defined in Theorem~\ref{thm:IO} and the notion of $\rho^*_{\calH}$ from Section~\ref{subsec:agm:fractional},
the $\InsideOut$ algorithm 
(Algorithm~\ref{algo:IO}) computes $\varphi$ in time
\begin{equation}
   O\left(n^3\cdot m\cdot\log N\cdot N^{\max_{j\in K}\rho^*_\calH(U_j)}
    + f^2\cdot(f+m)\cdot\log N\cdot\repsize{\varphi}
    \right)
\label{eqn:simplified:IO:runtime}
\end{equation}
\label{prop:IO:runtime}
\eprop
\bp
For the sake of brevity,
define $w = \max_{j\in K}\rho^*_\calH(U_j)$.
We use the same analysis as that in the proof of Theorem~\ref{thm:IO}.
However, we further bound $\repsize{\psi_S}$ by $N^w$ and
also replace all the $\agm_{\calH_k}(U_k)$ terms by $N^w$.
Note that every factor $\psi_S$ is either an input factor ($S\in \calE$) or an
intermediate factor ($S = U_k-\{k\}$ for some $k$). 
\bi
\item If $\psi_S$ was an input factor then $\repsize{\psi_S}\leq N \leq
N^w$ because $w \geq 1$. If $\psi_S$ is an intermediate factor ($S=U_k-\{k\}$), 
then with exactly the same reasoning as in the proof of Theorem~\ref{thm:OI},  
$\repsize{\psi_S}$ can be bounded by $\agm_\calH(U_k)$
which is upperbounded
by $N^w$.
\item Next, for $k \in K$, we show that the computation time of $\psi_{U_k-\{k\}}$
using $\OI$ can be bounded by $O(n^2 \cdot m\cdot\log N\cdot\agm_\calH(U_k))$
instead of $O(n^2 \cdot m\cdot\log N\cdot\agm_{\calH_k}(U_k))$.\footnote{Note that the 
   bound $\agm_{\calH_k}(U_k)$ is computed from the best fractional edge 
   cover of $U_k$ using hyperedges $\calE_k$, which -- compared to $\calE$ 
   -- has additional intermediate hyperedges/relations and also lack
   hyperedges/relations from $\calE$ which belong to $\partial(j)$ for $j>k$.}
Since $\agm_\calH(U_k) \leq N^w$, we would be done.
This simple but subtle fact is best explained using the language of
database join. For each $k \in [n]$ and $S \in \calE_k$, define a relation
\[ R(\psi_S) := \{ \mv x_{S} \suchthat \psi_{S}(\mv x_{S}) \neq \mv 0\}. 
\]
The runtime of $\OI$ computing $\psi_{U_k-\{k\}}$ is precisely the same
runtime when we use a worst-case optimal join algorithm to compute the
natural join 
\[ Q = \ \Join_{\substack{S \in \calE_k\\ S \cap U_k \neq \emptyset}}
R(\psi_{S/U_k}).
\]
The runtime bound $\agm_{\calH_k}(U_k)$ is a bound on the maximum number of 
possible output tuples of $Q$. Now, consider the query
\[ Q' = \ \Join_{\substack{S \in \calE\\ S \cap U_k \neq \emptyset}}
R(\psi_{S/U_k}).
\]
Then, it is easy to see that $Q \subseteq Q'$, and thus the maximum number
of output tuples of $Q$ is bounded by the maximum number of output 
tuples of $Q'$, which is $\agm_{\calH}(U_k)$.
\ei
\ep

In addition to the term $\repsize{\varphi}$ required to report the output, 
the key parameter that controls the complexity of the algorithm is the
quantity $\max_{j\in K} \left\{ \rho^*_\calH(U_j)\right\}.$
This quantity is a function of the variable ordering $X_1,\dots,X_n$ we
chose to write the input query $\varphi$ on. As aforementioned, 
there might be multiple ways of writing the same $\faq$ query, leading to 
wildly different runtimes for $\InsideOut$. We have seen this effect in 
Example~\ref{ex:var:order:effect}, and have formalized what an 
``equivalent ordering'' means in the previous section.
The following definition follows naturally:

\bdefn[Fractional $\faq$-width of a variable ordering]\label{defn:faqw}
Let $\sigma$ be a $\varphi$-equivalent variable ordering.
Define the sequence of hypergraphs $\calH^\sigma_k$ along with the sets
$U^\sigma_k$ as in Definition~\ref{defn:hyperSeq} (but with respect to $\sigma$). 
The {\em fractional $\faq$ width} of a variable ordering $\sigma$ is the
quantity
\begin{equation}\label{eqn:faqw}
    \faqw(\sigma) := \max_{j\in K} \left\{ \rho^*_\calH(U^\sigma_j)\right\}.
\end{equation}
\edefn
(Recall the Definition of $K$ from~\eqref{eqn:the:set:K}. Recall also from Section~\ref{sec:faq-pbm} that we are only interested in $\faq$ queries for which there is at least one semiring aggregate. Therefore $K$ in the above definition cannot be empty.)

From the above definition, we can interpret Proposition~\ref{prop:IO:runtime} as basically saying that $\InsideOut$ runs in time $\tilde O(N^{\faqw(\sigma)}+\repsize{\varphi})$, where $\tilde O$ hides a factor that is polynomial in query size and logarithmic in data size.

In the next few sections, we study the main problem of how to select a 
$\varphi$-equivalent variable 
ordering $\sigma$ with the minimum $\faqw(\sigma)$. 

\bdefn[$\faqw(\varphi)$]
The following quantity is called the {\em $\faq$-width} of an $\faq$-query
$\varphi$:
\[
    \faqw(\varphi) :=
    \min \left\{ \faqw(\sigma) \suchthat \sigma \in \EVO(\varphi) \right\} 
\]
\label{defn:faqw-phi}
\edefn

\bdefn[$\faqw(\varphi,\calF)$]
Given an $\faq$-query $\varphi$ and a class $\calF$ of functions with range $\D$,
let $\EVO(\varphi,\calF)$ be defined as in Definition~\ref{defn:EVO-phi-F}.
The $\faq$-width of $\varphi$ w.r.t. the class $\calF$ of input functions is defined as follows:
\[
\faqw(\varphi, \calF) :=
\min \left\{ \faqw(\sigma) \suchthat \sigma \in \EVO(\varphi, \calF) \right\} 
\]
\label{defn:faqw-phi-F}
\edefn

In some cases, $\EVO(\varphi)$ consists of all $n!$ permutations, making it
``easy'' to solve the above optimization problem. Appendix~\ref{app:subsec:quick:faqcs} 
presents several immediate consequences of this easy case. In particular,
$\faqw$ generalizes fractional hypertree widths ($\fhtw$,
see~\cite{DBLP:journals/talg/GroheM14}), because the following follows directly
from Corollary~\ref{prop:vo-fhtw}:

\bprop
Let $\varphi$ be an $\faq$ query with hypergraph $\calH$.
If $\EVO(\varphi)$ contains all $n!$ variable orderings, then 
$\faqw(\varphi) = \fhtw(\calH)$.
\label{prop:faqw=fhtw}
\eprop

In general, however,
the question of determining whether a given variable ordering
$\sigma$ belongs to $\EVO(\varphi)$ is a tricky
question to answer formally. In particular, the answer depends on what exactly
we meant by a variable aggregate, a factor aggregate, the variable domain sizes,
and the range $\D$. This difficulty is analogous to the situation in logic when
one wants to decide whether two (first-order, e.g.) formulas of specific forms
are logically equivalent \cite{MR1860010}\footnote{We thank Balder ten Cate for
pointing out to us the essence of the difficulty and the reference.}. 

Our approach to solving this problem was outlined in 
Figure~\ref{fig:contrib:summary}, and is summarized again as follows.
\bi 
 \item We define a class of variable orderings for a given input $\faq$-query 
       $\varphi$. This class will be precisely the set of linear extensions 
       $\LE(P)$ of a partially ordered set (poset) on variables called the 
       {\em precedence poset} $P$.
       The precedence poset is defined on a tree called the {\em expression tree} of
       the input query $\varphi$.  
       The expression tree can be constructed in polynomial time in query
       complexity.
  \item We show that every variable ordering in $\LE(P)$ is $\varphi$-equivalent. 
        This is the ``soundness'' of $\LE(P)$.
 \item We define a combinatorial notion called {\em component-wise
       equivalence}, which is a relation between pairs of variable orderings.
       We show that component-wise equivalence preserves $\EVO$-membership {\em
       and} preserves $\faqw$.
  \item We show that every variable ordering in $\EVO$ is component-wise
     equivalent to some ordering in $\LE(P)$. This is called 
     ``strong completeness'' of $\LE(P)$. In particular, by tracing 
     component-wise equivalent variable 
     orderings starting from $\LE(P)$, one can list {\em all} of $\EVO$
     in exponential time and find the ordering that minimizes $\faqw$.
     (Moreover, by directly applying the definition of component-wise equivalence, 
     we can check whether a given variable ordering is in $\EVO$ in polynomial time.)
  \item However, we can do better. We prove that $\LE(P)$
       variable orderings is all we need to consider,
       because every $\varphi$-equivalent variable ordering $\sigma$
       either belongs to $\LE(P)$ or $\faqw(\sigma) = \faqw(\pi)$ for some
       $\pi\in \LE(P)$. This shows the ``completeness'' of $\LE(P)$ as far as
       the width is concerned.
\ei

The completeness result rests on the assumption that different variable
aggregates do not commute. (Even this simple statement needs clarification, which
is done in Proposition~\ref{prop:commute}.)
Because the final result is a bit technically involved, 
we present our results incrementally in several stages,
by relaxing one assumption at a time.
Each time an assumption is relaxed, a couple of ideas are introduced to deal
with the relaxation. It should be noted, however, that in the end there is only
one theorem and one algorithm.

In Appendix~\ref{app:subsec:quick:faqcs} 
we describe the above steps when applied to $\faqcs$ without free 
variables (i.e. $\sumprod$), which is the case when determining $\EVO(\varphi)$ is trivial.
Appendix~\ref{app:subsec:2blocks} covers a simple but 
non-trivial case when $\varphi$ has two blocks of semiring aggregates.
The reader who would like to read at a slower pace can start with those two
sections in the appendix, where we also connect our results to known results in
$\pgm$, joins, Yannakakis algorithm, and $\scq$.

In Section~\ref{subsec:evo:Kblocks}, we present our solution for the case when there 
is an arbitrary number of semiring aggregates but no product aggregates. This 
is when the idea of an expression tree is introduced.
Finally, in Sections~\ref{subsec:evo:inner:faq} and~\ref{subsec:evo:general:faq}
we cover the most general cases.

\section{Characterizing equivalent variable orderings}
\label{sec:evo}

\subsection{{\large $\faq$} with only semiring aggregates}
\label{subsec:evo:Kblocks}

This section presents the characterization of $\EVO$ results for $\faq$ where every 
variable aggregate forms a semiring with the product aggregate. 
(Note that there can be an arbitrary number of different types of semirings.) 
In particular, we consider the $\faq$-query of the form
\begin{equation}\label{eqn:main:faqKblocks}
    \varphi(\mv x_{[f]}) =
    \textstyle{\bigoplus^{(f+1)}_{x_{f+1}} \cdots
    \bigoplus^{(n)}_{x_{n}}}
    \bigotimes_{S\in\calE}\psi_S(\mv x_S)
\end{equation}
where $(\D,\oplus^{(i)}, \otimes)$ is a commutative semiring for every $i>f$ (i.e. the
set $K$ as defined in~\eqref{eqn:the:set:K} is $[n]$).

The main aim in this section is to illustrate the
key technical  idea of the {\em expression tree}.
The expression tree defines the {\em precedence poset}.
One key component of our completeness results is the notion of \emph{component-wise equivalence}
between two variable orderings. Component-wise equivalence preserves $\faq$-width  and $\varphi$-equivalence.
We will show 
two important facts about the precedence poset:
 \bi
 \item (Soundness) Every linear extension of the precedence poset is a 
       $\varphi$-equivalent variable ordering.
 \item (Completeness) Every $\varphi$-equivalent variable ordering is component-wise equivalent to (hence has the same
     $\faq$-width as) some linear extension of the precedence poset.
     Therefore, $\EVO(\varphi)$ is \emph{completely} characterized using component-wise equivalence and the precedence poset.
     Moreover, to compute/approximate $\faqw(\varphi)$, we \emph{only} need to consider linear extensions of the precedence poset.
 \ei
In Section~\ref{sec:approx}, we will use the structure of the expression tree to compute a 
variable ordering to approximate $\faqw(\varphi)$, using an approximation
algorithm for $\fhtw$ as a blackbox.
Thus, the expression tree is crucial in guiding the construction of a good
variable ordering.

\subsubsection{Expression tree and precedence poset}

The expression tree is defined on a sequence of 
{\em tagged variables} along with a hypergraph.
In such a sequence, every vertex $i$ (or equivalently variable $X_i$)
is tagged with its corresponding operator $\oplus^{(i)}$; or, if the variable
is a free variable then its tag is {\em free}.
Given a sequence $\sigma$ of tagged variables, a {\em tag block} is a 
maximal subsequence of consecutive variables in $\sigma$ with the same tag.
The first tag block of a sequence $\sigma$ of tagged variables 
is the longest prefix of $\sigma$ consisting of variables of the same tag.

\bdefn[Expression tree]
\label{defn:expr-tree}
The {\em expression tree} for
$\varphi$ is a rooted tree $P$. Every node of the tree is a set of variables.
We construct the expression tree using two steps:
the {\em compartmentalization step} and the {\em compression step}.
In the compartmentalization step, we construct the expression tree based on the
connected component structures of the $\faq$-query relative to the hypergraph
structure $\calH$. In the compression step we collapse the tree to make it
shorter whenever possible.

{\bf Compartmentalization.}
In this step, initially we start off with the
sequence of variables with their corresponding tags exactly as written in
\eqref{eqn:main:faqKblocks}. In particular, the sequence starts with $f$ free variables
(whose tags are `free'), and then the $i$'th variable with tag $\oplus^{(i)}$
for $i=f+1,\dots,n$. 
For technical reasons, we add a dummy variable $X_0$ to the beginning
of the sequence with a free tag too. So the sequence we start off with is
the following
\[
   \sigma = 
    \Bigl\langle
        (X_0,\text{`free'}),  \dots, (X_f,
        \text{`free'}), \Bigr.\\
        \Bigl. (X_{f+1}, \oplus^{(f+1)}), \dots, (X_n, \oplus^{(n)})
    \Bigr\rangle.
\]
The vertex $X_0$ is an isolated vertex of the hypergraph $\calH$.
Now given a tagged variable sequence $\sigma$ and a hypergraph $\calH$, we 
build the tree by constructing a node $L$ containing all variables in the 
first tag block $L$ of $\sigma$, and node $L$  has a tag which is the same tag as that of its variables. This node $L$ will be the root of the
expression tree.
(The effect of the dummy variable $X_0$ is that, even if
the original query has no free variable, the first block $L$ still has $X_0$ 
in it.)
If $L$ contains all variables already, then naturally the tree has only one
node.
Otherwise, for each connected component 
$C = (\calV(C), \calE(C))$ of $\calH - L$ we construct
a sequence $\sigma_C$ of tagged variables by listing all variables 
in $\calV(C)$ in exactly the same relative order as they appeared in $\sigma$. 
From the sequence $\sigma_C$ and the hypergraph $C$, we recursively construct
the expression tree $P_C$. 
Finally, we connect all the roots of 
(sub)expression trees $P_C$ to the node $L$. 
This completes the compartmentalization step.
After everything is done, we also remove the dummy variable $X_0$ from the 
expression tree $P$.
If originally there was no free variable, the tree has an empty 
root node and the subtrees correspond to the connected components of $\calH$. 
(See Example~\ref{ex:intuition-semirings} and
Figure~\ref{fig:semirings:compart}.)

     {\bf Compression.}
Now, in the expression tree $P$ that resulted from the compartmentalization step,
as long as there is still a node $L$ whose tag is the same as a child node
$L'$ of the tree $P$, we merge the child into $L$; namely, we set $L := L \cup
L'$, remove $L'$, and connect all subtrees under $L'$ to become subtrees of $L$.
Repeat this step until no further merging is possible. (See Figure~\ref{fig:semirings:compress}.)
\edefn
Note that the compression step can make some nodes $L$ larger and the final 
tree $T$ shorter than the tree that resulted from the compartmentalization step 
alone. This step is crucial for getting the correct expression tree.
If $\varphi$ is an instance of $\faqcs$, 
then $P$ is a tree of depth $\leq 1$ where the root node contains all free 
variables (if any) and its children (if any) contain the rest of the variables.

\begin{ex}[Intuition behind expression tree]\label{ex:intuition-semirings}
Consider the following $\faq$ query that has two different semiring 
aggregates ($\sum$ and $\max$) and no free variables:
\[
\varphi = \sum_{x_1} \sum_{x_2} \max_{x_3} \sum_{x_4} \sum_{x_5} \max_{x_6} 
\max_{x_7} \psi_{12} \psi_{135} \psi_{14} \psi_{246} \psi_{27} \psi_{37}.
\]
($\psi_{12}$ above denotes a factor whose support is $\{X_1, X_2\}$, and so on.)
The hypergraph of $\varphi$ is depicted in Figure~\ref{fig:compart:a}. The 
compartmentalization step of the construction of the expression tree is depicted 
in Figures~\ref{fig:compart:b} through \ref{fig:compart:d}. 
Figures~\ref{fig:compress:a} and ~\ref{fig:compress:b} depict the compression 
step. The final expression tree appears on the right of Figure~\ref{fig:compress:b}.
\end{ex}

\pgfmathsetmacro{\subfigyscale}{.97}


\pgfkeys{/tikz,
vertex/.style={line width=1, radius=.25},
edge/.style={line width=1.25},
treenode/.style={line width=1.1, rounded corners, anchor=center},
hypergraph/.style={shift={(5, .6)}, scale=.8, 
	yscale=.85, every node/.style={transform shape}},
exprtree/.style={shift={(14, 2.8)},
	scale=.6, yscale=1, xscale=1}}

\newcommand{\drawtreenode}[1]{
	\draw[treenode]  (-1.45, -.5) rectangle (1.45, .5);
	\draw[treenode] (0, 0) node {#1};
}

\newcommand{\drawframe}{
	\draw[white] (1, 0) -- (17, 0) -- (17, 3.3) -- (1, 3.3) --cycle;
	\draw [gray, dashed] (9.2, .4) -- (9.2, 3);
}


\colorlet{exprColor}{black}
\colorlet{exprColor0}{red!75!black}
\colorlet{exprColor1}{blue}
\colorlet{exprColor00}{magenta!50!black}
\colorlet{exprColor01}{olive!60!black}
\colorlet{exprColor10}{teal!60!black}

\begin{figure}[htp!]
\begin{center}
\centering
\everymath{\displaystyle}

\subfloat[
	\color{exprColor}
	$\varphi=$
	$\sum_{x_1}$
	$\sum_{x_2}$
	$\max_{x_3}$
	$\sum_{x_4}$
	$\sum_{x_5}$
	$\max_{x_6}$
	$\max_{x_7}$
	$\psi_{12}$
	$\psi_{135}$
	$\psi_{14}$
	$\psi_{246}$
	$\psi_{27}$
	$\psi_{37}$
	{\color{white}$\Biggl)$}]
	{\label{fig:compart:a}\begin{tikzpicture}[yscale=\subfigyscale,scale=0.8]
		\draw[white] (1, 0) -- (17, 0) -- (17, 3.3) -- (1, 3.3) --cycle;
	\begin{scope}[hypergraph, shift={(5.5, 0)}]
		\draw[vertex] (0,0) circle node (v2) {$\mathbf{2}$};
		\draw[vertex] (0,3) circle node (v1) {$\mathbf{1}$};

		\draw[edge] (0, 1.5) ellipse (.5 and 2.25); 

		\draw[vertex] (-2,1.5) circle node (v3) {$\mathbf{3}$};

		\draw[vertex] (-4,3) circle node (v5) {$\mathbf{5}$};
		\draw[edge, rounded corners=7 mm]
							(-5, 3.35)--
							(1, 3.35)--
							(-2, .8)--cycle; 

		\draw[vertex] (-4,0) circle node (v7) {$\mathbf{7}$};
		\draw[edge] (-2, 0) ellipse (2.7 and .5); 
		\draw[edge, rotate around={36:(-3, .75)}] (-3, .75) ellipse (2 and .5); 

		\draw[vertex] (3,3) circle node (v4) {$\mathbf{4}$};
		\draw[edge] (1.5, 3) ellipse (2.25 and .5); 
			
		\draw[vertex] (3,0) circle node (v6) {$\mathbf{6}$};
		\draw[edge, rounded corners=4.5 mm]
						(-.9, -.4)--
						(3.45, -.4)--
						(3.45, 4)--cycle; 
	\end{scope}
	\end{tikzpicture}
}

\subfloat[
		$\varphi=$
	{\color{exprColor}
		$\sum_{x_1}$
		$\sum_{x_2}$
		$\psi_{12}$}
	{\color{exprColor0}
		$\Biggl($
		$\max_{x_3}$
		$\sum_{x_5}$
		$\max_{x_7}$
		$\psi_{135}$
		$\psi_{27}$
		$\psi_{37}$
		$\Biggr)$}
	{\color{exprColor1}
		$\Biggl($
		$\sum_{x_4}$
		$\max_{x_6}$
		$\psi_{14}$
		$\psi_{246}$
		$\Biggr)$}]
	{\label{fig:compart:b}\begin{tikzpicture}[yscale=\subfigyscale,scale=0.8]
		\drawframe
	\begin{scope}[hypergraph]
	{\color{exprColor}
		\draw[vertex] (0,0) circle node (v2) {$\mathbf{2}$};
		\draw[vertex] (0,3) circle node (v1) {$\mathbf{1}$};

		\draw[edge] (0, 1.5) ellipse (.5 and 2.25); 
	}

	{\color{exprColor0}
		\draw[vertex] (-2,1.5) circle node (v3) {$\mathbf{3}$};

		\draw[vertex] (-4,3) circle node (v5) {$\mathbf{5}$};
		\draw[edge, rounded corners=7 mm]
							(-5, 3.35)--
							(1, 3.35)--
							(-2, .8)--cycle; 

		\draw[vertex] (-4,0) circle node (v7) {$\mathbf{7}$};
		\draw[edge] (-2, 0) ellipse (2.7 and .5); 
		\draw[edge, rotate around={36:(-3, .75)}] (-3, .75) ellipse (2 and .5); 
	}
	{\color{exprColor1}
		\draw[vertex] (3,3) circle node (v4) {$\mathbf{4}$};
		\draw[edge] (1.5, 3) ellipse (2.25 and .5); 
			
		\draw[vertex] (3,0) circle node (v6) {$\mathbf{6}$};
		\draw[edge, rounded corners=4.5 mm]
						(-.9, -.4)--
						(3.45, -.4)--
						(3.45, 4)--cycle; 
	}
	\end{scope}
	
	\begin{scope}[exprtree]
		{\color{exprColor}
			\drawtreenode{$\mathbf{1, 2}_{~\sum}$};
		}

		\begin{scope}[shift={(-3, -2)}]
			{\color{exprColor0}
				\drawtreenode{$\mathbf{3, 5, 7}$};
				\draw[treenode] (.5, .5)--(2.5, 1.5);
			}
		\end{scope}

		\begin{scope}[shift={(3, -2)}]
			{\color{exprColor1}
				\drawtreenode{$\mathbf{4, 6}$};
				\draw[treenode] (-.5, .5)--(-2.5, 1.5);
			}
		\end{scope}

	\end{scope}
	\end{tikzpicture}
}

\subfloat[
		$\varphi=$
	{\color{exprColor}
		$\sum_{x_1}$
		$\sum_{x_2}$
		$\psi_{12}$}
	{\color{exprColor0}$\Biggl($}
	{\color{exprColor0}$\max_{x_3}$}
	{\color{exprColor00}
		$\sum_{x_5}$
		$\psi_{135}$}
	{\color{exprColor01}
		$\max_{x_7}$
		$\psi_{27}$
		$\psi_{37}$}
	{\color{exprColor0}$\Biggr)$}
	{\color{exprColor1}
		$\Biggl($
		$\sum_{x_4}$
		$\max_{x_6}$
		$\psi_{14}$
		$\psi_{246}$
		$\Biggr)$}]
	{\label{fig:compart:c}\begin{tikzpicture}[yscale=\subfigyscale,scale=0.8]
		\drawframe
	\begin{scope}[hypergraph]
	{\color{exprColor}
		\draw[vertex] (0,0) circle node (v2) {$\mathbf{2}$};
		\draw[vertex] (0,3) circle node (v1) {$\mathbf{1}$};

		\draw[edge] (0, 1.5) ellipse (.5 and 2.25); 
	}

	{\color{exprColor0}
		\draw[vertex] (-2,1.5) circle node (v3) {$\mathbf{3}$};
		{\color{exprColor00}
			\draw[vertex] (-4,3) circle node (v5) {$\mathbf{5}$};
			\draw[edge, rounded corners=7 mm]
							(-5, 3.35)--
							(1, 3.35)--
							(-2, .8)--cycle; 
		}
		{\color{exprColor01}
			\draw[vertex] (-4,0) circle node (v7) {$\mathbf{7}$};
			\draw[edge] (-2, 0) ellipse (2.7 and .5); 
			\draw[edge, rotate around={36:(-3, .75)}] (-3, .75) ellipse (2 and .5); 
		}
	}
	{\color{exprColor1}
		\draw[vertex] (3,3) circle node (v4) {$\mathbf{4}$};
		\draw[edge] (1.5, 3) ellipse (2.25 and .5); 
			
		\draw[vertex] (3,0) circle node (v6) {$\mathbf{6}$};
		\draw[edge, rounded corners=4.5 mm]
						(-.9, -.4)--
						(3.45, -.4)--
						(3.45, 4)--cycle; 
	}
	\end{scope}
	
	\begin{scope}[exprtree]
		{\color{exprColor}
			\drawtreenode{$\mathbf{1, 2}_{~\sum}$};
		}

		\begin{scope}[shift={(-3, -2)}]
			{\color{exprColor0}
				\drawtreenode{$\mathbf{3}_{~\max}$};
				\draw[treenode] (.5, .5)--(2.5, 1.5);
			}
			
			\begin{scope}[shift={(-2.5, -2)}]
				{\color{exprColor00}
					\drawtreenode{$\mathbf{5}_{~\sum}$};
					\draw[treenode] (.5, .5)--(2, 1.5);
				}
			\end{scope}
			\begin{scope}[shift={(2.5, -2)}]
				{\color{exprColor01}
					\drawtreenode{$\mathbf{7}_{~\max}$};
					\draw[treenode] (-.5, .5)--(-2, 1.5);
				}
			\end{scope}
		\end{scope}

		\begin{scope}[shift={(3, -2)}]
			{\color{exprColor1}
				\drawtreenode{$\mathbf{4, 6}$};
				\draw[treenode] (-.5, .5)--(-2.5, 1.5);
			}
		\end{scope}

	\end{scope}
	\end{tikzpicture}
}

\subfloat[
		$\varphi=$
	{\color{exprColor}
		$\sum_{x_1}$
		$\sum_{x_2}$
		$\psi_{12}$}
	{\color{exprColor0}$\Biggl($}
	{\color{exprColor0}$\max_{x_3}$}
	{\color{exprColor00}
		$\sum_{x_5}$
		$\psi_{135}$}
	{\color{exprColor01}
		$\max_{x_7}$
		$\psi_{27}$
		$\psi_{37}$}
	{\color{exprColor0}$\Biggr)$}
	{\color{exprColor1!}$\Biggl($}
	{\color{exprColor1!}
		$\sum_{x_4}$
		$\psi_{14}$}
	{\color{exprColor10}
		$\max_{x_6}$
		$\psi_{246}$}
	{\color{exprColor1}$\Biggr)$}]
	{\label{fig:compart:d}\begin{tikzpicture}[yscale=\subfigyscale,scale=0.8]
		\drawframe
	\begin{scope}[hypergraph]
	{\color{exprColor}
		\draw[vertex] (0,0) circle node (v2) {$\mathbf{2}$};
		\draw[vertex] (0,3) circle node (v1) {$\mathbf{1}$};

		\draw[edge] (0, 1.5) ellipse (.5 and 2.25); 
	}

	{\color{exprColor0}
		\draw[vertex] (-2,1.5) circle node (v3) {$\mathbf{3}$};
		{\color{exprColor00}
			\draw[vertex] (-4,3) circle node (v5) {$\mathbf{5}$};
			\draw[edge, rounded corners=7 mm]
							(-5, 3.35)--
							(1, 3.35)--
							(-2, .8)--cycle; 
		}
		{\color{exprColor01}
			\draw[vertex] (-4,0) circle node (v7) {$\mathbf{7}$};
			\draw[edge] (-2, 0) ellipse (2.7 and .5); 
			\draw[edge, rotate around={36:(-3, .75)}] (-3, .75) ellipse (2 and .5); 
		}
	}
	{\color{exprColor1}
		\draw[vertex] (3,3) circle node (v4) {$\mathbf{4}$};
		\draw[edge] (1.5, 3) ellipse (2.25 and .5); 
	}
	{\color{exprColor10}
		\draw[vertex] (3,0) circle node (v6) {$\mathbf{6}$};
		\draw[edge, rounded corners=4.5 mm]
						(-.9, -.4)--
						(3.45, -.4)--
						(3.45, 4)--cycle; 
	}
	\end{scope}
	
	\begin{scope}[exprtree]
		{\color{exprColor}
			\drawtreenode{$\mathbf{1, 2}_{~\sum}$};
		}

		\begin{scope}[shift={(-3, -2)}]
			{\color{exprColor0}
				\drawtreenode{$\mathbf{3}_{~\max}$};
				\draw[treenode] (.5, .5)--(2.5, 1.5);
			}
			
			\begin{scope}[shift={(-2.5, -2)}]
				{\color{exprColor00}
					\drawtreenode{$\mathbf{5}_{~\sum}$};
					\draw[treenode] (.5, .5)--(2, 1.5);
				}
			\end{scope}
			\begin{scope}[shift={(2.5, -2)}]
				{\color{exprColor01}
					\drawtreenode{$\mathbf{7}_{~\max}$};
					\draw[treenode] (-.5, .5)--(-2, 1.5);
				}
			\end{scope}
		\end{scope}

		\begin{scope}[shift={(3, -2)}]
			{\color{exprColor1}
				\drawtreenode{$\mathbf{4}_{~\sum}$};
				\draw[treenode] (-.5, .5)--(-2.5, 1.5);
			}
			\begin{scope}[shift={(0, -2)}]
			{\color{exprColor10!}
				\drawtreenode{$\mathbf{6}_{~\max}$};
				\draw[treenode] (0, .5)--(0, 1.5);
			}
		\end{scope}
		\end{scope}

	\end{scope}
	\end{tikzpicture}
}

\caption{The compartmentalization step of the expression tree from Example~\ref{ex:intuition-semirings}, depicted using colors. (\ref{fig:compart:a}) depicts the hypergraph of the $\faq$ query: $\varphi = \sum_{x_1} \sum_{x_2} \max_{x_3} \sum_{x_4} \sum_{x_5} \max_{x_6} \max_{x_7} \psi_{12} \psi_{135} \psi_{14} \psi_{246} \psi_{27} \psi_{37}$. For simplicity, the dummy free variable $X_0$ is ignored in this example. (\ref{fig:compart:b}) shows the first part of the compartmentalization step, where the first tag block is $L=\{1, 2\}$. After removing $L$, the query breaks into two connected components: the {\color{DarkRed}red} and the {\color{DarkBlue}blue}. The expression tree at this point appears on the right. Each color is used to denote correspondence between parts of the query expression, hypergraph, and expression tree having that color. (\ref{fig:compart:c}) shows how to apply compartmentalization recursively on the red component, while (\ref{fig:compart:d}) shows the blue component compartmentalization.}
\Description{The compartmentalization step of the expression tree from Example~\ref{ex:intuition-semirings}, depicted using colors.}
\label{fig:semirings:compart}
\end{center}
\end{figure}


\begin{figure}[htp!]
\begin{center}
\centering
\everymath{\displaystyle}

\subfloat[
		$\varphi=$
	{\color{exprColor}
		$\sum_{x_1}$
		$\sum_{x_2}$
		$\psi_{12}$}
	{\color{exprColor0}$\Biggl($}
	{\color{exprColor0}$\max_{x_3}$}
	{\color{exprColor0}
		$\max_{x_7}$
		$\psi_{27}$
		$\psi_{37}$}
	{\color{exprColor00}
		$\sum_{x_5}$
		$\psi_{135}$}
	{\color{exprColor0}$\Biggr)$}
	{\color{exprColor1}$\Biggl($}
	{\color{exprColor1}
		$\sum_{x_4}$
		$\psi_{14}$}
	{\color{exprColor10}
		$\max_{x_6}$
		$\psi_{246}$}
	{\color{exprColor1}$\Biggr)$}]
	{\label{fig:compress:a}\begin{tikzpicture}[yscale=\subfigyscale,scale=0.8]
		\drawframe
	\begin{scope}[hypergraph]
	{\color{exprColor}
		\draw[vertex] (0,0) circle node (v2) {$\mathbf{2}$};
		\draw[vertex] (0,3) circle node (v1) {$\mathbf{1}$};

		\draw[edge] (0, 1.5) ellipse (.5 and 2.25); 
	}

	{\color{exprColor0}
		\draw[vertex] (-2,1.5) circle node (v3) {$\mathbf{3}$};
		{\color{exprColor00}
			\draw[vertex] (-4,3) circle node (v5) {$\mathbf{5}$};
			\draw[edge, rounded corners=7 mm]
							(-5, 3.35)--
							(1, 3.35)--
							(-2, .8)--cycle; 
		}
		{\color{exprColor0}
			\draw[vertex] (-4,0) circle node (v7) {$\mathbf{7}$};
			\draw[edge] (-2, 0) ellipse (2.7 and .5); 
			\draw[edge, rotate around={36:(-3, .75)}] (-3, .75) ellipse (2 and .5); 
		}
	}
	{\color{exprColor1}
		\draw[vertex] (3,3) circle node (v4) {$\mathbf{4}$};
		\draw[edge] (1.5, 3) ellipse (2.25 and .5); 
	}
	{\color{exprColor10}
		\draw[vertex] (3,0) circle node (v6) {$\mathbf{6}$};
		\draw[edge, rounded corners=4.5 mm]
						(-.9, -.4)--
						(3.45, -.4)--
						(3.45, 4)--cycle; 
	}
	\end{scope}
	
	\begin{scope}[exprtree]
		{\color{exprColor}
			\drawtreenode{$\mathbf{1, 2}_{~\sum}$};
		}

		\begin{scope}[shift={(-3, -2)}]
			{\color{exprColor0}
				\drawtreenode{$\mathbf{3, 7}_{~\max}$};
				\draw[treenode] (.5, .5)--(2.5, 1.5);
			}
			
			\begin{scope}[shift={(0, -2)}]
				{\color{exprColor00}
					\drawtreenode{$\mathbf{5}_{~\sum}$};
					\draw[treenode] (0, .5)--(0, 1.5);
				}
			\end{scope}
		\end{scope}

		\begin{scope}[shift={(3, -2)}]
			{\color{exprColor1}
				\drawtreenode{$\mathbf{4}_{~\sum}$};
				\draw[treenode] (-.5, .5)--(-2.5, 1.5);
			}
			\begin{scope}[shift={(0, -2)}]
			{\color{exprColor10!}
				\drawtreenode{$\mathbf{6}_{~\max}$};
				\draw[treenode] (0, .5)--(0, 1.5);
			}
		\end{scope}
		\end{scope}

	\end{scope}
	\end{tikzpicture}
}

\subfloat[
	$\varphi=$
	{\color{exprColor}
		$\sum_{x_1}$
		$\sum_{x_2}$}
	{\color{exprColor}
		$\sum_{x_4}$}
	{\color{exprColor}
		$\psi_{12}$}
	{\color{exprColor}
		$\psi_{14}$}
	{\color{exprColor0}$\Biggl($}
	{\color{exprColor0}$\max_{x_3}$}
	{\color{exprColor0}
		$\max_{x_7}$
		$\psi_{27}$
		$\psi_{37}$}
	{\color{exprColor00!}
		$\sum_{x_5}$
		$\psi_{135}$}
	{\color{exprColor0!}$\Biggr)$}
	{\color{exprColor10}$\Biggl($}
	{\color{exprColor10}
		$\max_{x_6}$
		$\psi_{246}$}
	{\color{exprColor10}$\Biggr)$}]
	{\label{fig:compress:b}\begin{tikzpicture}[yscale=\subfigyscale,scale=0.8]
		\drawframe
	\begin{scope}[hypergraph]
	{\color{exprColor}
		\draw[vertex] (0,0) circle node (v2) {$\mathbf{2}$};
		\draw[vertex] (0,3) circle node (v1) {$\mathbf{1}$};

		\draw[edge] (0, 1.5) ellipse (.5 and 2.25); 
	}

	{\color{exprColor0}
		\draw[vertex] (-2,1.5) circle node (v3) {$\mathbf{3}$};
		{\color{exprColor00}
			\draw[vertex] (-4,3) circle node (v5) {$\mathbf{5}$};
			\draw[edge, rounded corners=7 mm]
							(-5, 3.35)--
							(1, 3.35)--
							(-2, .8)--cycle; 
		}
		{\color{exprColor0}
			\draw[vertex] (-4,0) circle node (v7) {$\mathbf{7}$};
			\draw[edge] (-2, 0) ellipse (2.7 and .5); 
			\draw[edge, rotate around={36:(-3, .75)}] (-3, .75) ellipse (2 and .5); 
		}
	}
	{\color{exprColor}
		\draw[vertex] (3,3) circle node (v4) {$\mathbf{4}$};
		\draw[edge] (1.5, 3) ellipse (2.25 and .5); 
	}
	{\color{exprColor10}
		\draw[vertex] (3,0) circle node (v6) {$\mathbf{6}$};
		\draw[edge, rounded corners=4.5 mm]
						(-.9, -.4)--
						(3.45, -.4)--
						(3.45, 4)--cycle; 
	}
	\end{scope}
	
	\begin{scope}[exprtree]
		{\color{exprColor}
			\drawtreenode{$\mathbf{1, 2, 4}_{~\sum}$};
		}

		\begin{scope}[shift={(-3, -2)}]
			{\color{exprColor0}
				\drawtreenode{$\mathbf{3, 7}_{~\max}$};
				\draw[treenode] (.5, .5)--(2.5, 1.5);
			}
			
			\begin{scope}[shift={(0, -2)}]
				{\color{exprColor00}
					\drawtreenode{$\mathbf{5}_{~\sum}$};
					\draw[treenode] (0, .5)--(0, 1.5);
				}
			\end{scope}
		\end{scope}

		\begin{scope}[shift={(3, -2)}]
			{\color{exprColor10}
				\drawtreenode{$\mathbf{6}_{~\max}$};
				\draw[treenode] (-.5, .5)--(-2.5, 1.5);
			}
		\end{scope}

	\end{scope}
	\end{tikzpicture}
}

\caption{The compression step of the expression tree from Example~\ref{ex:intuition-semirings}. Recall that Figure~(\ref{fig:compart:d}) (right) depicted the expression tree at the end of the compartmentalization step. (\ref{fig:compress:a}) shows a compression where node $\{7\}$ is merged into its parent $\{3\}$ (since they both have the same tag ``$\max$''). (\ref{fig:compress:b}) shows another compression where $\{4\}$ is merged into $\{1, 2\}$. Since no further compression is possible, (\ref{fig:compress:b}) (right) depicts the final expression tree.}
\Description{The compression step of the expression tree from Example~\ref{ex:intuition-semirings}.}
\label{fig:semirings:compress}
\end{center}
\end{figure}

\bdefn[Precedence poset]\label{defn:precedencePoset}
The expression tree defines a partial order on the variables. 
Abusing notation we will also use $P$ to denote the partial order $([n],
\preceq)$.
In this poset, for any pair $u, v\in [n]$, we write $u \preceq_P v$ if $u=v$ or $u$ belongs
to a strict ancestor of $v$ in the expression poset $P$.
In particular, different variables in the same node of
the expression tree are not comparable in this partial order.
We call this partial order the {\em precedence poset}.
Let $\LE(P)$ denote the set of all linear extensions of the poset.
\edefn

\subsubsection{What makes two aggregates different?}
\label{subsec:differentaggregates}

Before proving soundness and completeness, we need a small
technical detour.
Recall that aggregates are simply binary operators under $\D$.

\bdefn[Different aggregates]
Two aggregates $\oplus$ and $\bar \oplus$ are {\em
different} if there is a pair $a,b\in \D$ such that
\[ a \oplus b \neq a \bar \oplus b. \]
Otherwise they are {\em identical}.
\edefn

\bdefn[Mutually commutative aggregates]\label{defn:commutativity}
Two aggregates $\oplus$ and $\bar \oplus$ are said to be {\em mutually commutative}
if for every $a,b,c,d\in \D$, we have
$(a \oplus b) \bar \oplus (c \oplus d) = (a \bar \oplus c) \oplus (b \bar \oplus
d).$
\label{defn:commute}
\edefn

Recall that in $\faq$, all semirings share the same $\mv 0$ (since one `$\mv 0$' must
annihilate the rest). Thus, if we select
$a = d = \mv 0$ in the above equality, then we obtain
$b \bar \oplus c =  b \oplus c$, for every $b, c\in \D$.
This means 

\bprop\label{prop:commute}
Any two aggregates are mutually commutative if and only if they are identical.
\eprop

(Note that it is possible for semantically different aggregates to be 
identical under $\D$ by accident. For example, in the $\{0,1\}$ domain $\min$
and $\times$ are identical.)
In this paper, we assume that two different aggregates in the input
$\faq$-expression are not functionally identical. 
Recall that we also assumed $|\Dom(X_i)|\geq 2$ for every $i\in [n]$ 
(otherwise the aggregate on
that $X_i$ is trivial and can be ignored). 

\bprop
Suppose $\oplus$ and $\bar\oplus$ are different binary operators
(under the domain $\D$), then for every $i, j \in [n]$, there is a
function $\phi_{ij}: \Dom(X_i)\times\Dom(X_j) \to \D$ for which
\begin{equation}
   \mathop{\textstyle{\bigoplus}}_{x_i\in\Dom(X_i)} 
   \mathop{\textstyle{\bar\bigoplus}}_{x_j\in\Dom(X_j)} \phi_{ij}(x_i, x_j)
    \neq
    \mathop{\textstyle{\bar\bigoplus}}_{x_j\in\Dom(X_j)}
    \mathop{\textstyle{\bigoplus}}_{x_i\in\Dom(X_i)} 
   \phi_{ij}(x_i, x_j).
\label{eqn:non-commutative-ineq}
\end{equation}
\label{prop:non-commutative}
\eprop
\bp
From the analysis above, the two operators do not commute. Hence, there are four
members $a,b,c,d\in\D$ so that
$(a \oplus b) \bar \oplus (c \oplus d) \neq (a \bar \oplus c) \oplus (b \bar \oplus d).$
Fix arbitrary elements $x^1_i \neq x^2_i\in \Dom(X_i)$ and
$x^1_j \neq x^2_j \in \Dom(X_j)$. Define
\begin{equation}
    \phi_{ij}(x_i,x_j) :=
    \begin{cases}
        a & \text{ if } (x_i,x_j) = (x^1_i,x^1_j)\\
        b & \text{ if } (x_i,x_j) = (x^2_i,x^1_j)\\
        c & \text{ if } (x_i,x_j) = (x^1_i,x^2_j)\\
        d & \text{ if } (x_i,x_j) = (x^2_i,x^2_j)\\
        \mv 0 & \text{ otherwise}.
    \end{cases}
    \label{eqn:non-commutative-defn}
\end{equation}
Then,
$   \mathop{\textstyle{\bigoplus}}_{x_i}
   \mathop{\textstyle{\bar\bigoplus}}_{x_j}\phi_{ij}(x_i,x_j) = 
       (a \bar \oplus c) \oplus (b \bar \oplus d)
   \neq (a \oplus b) \bar \oplus (c \oplus d)
   = 
   \mathop{\textstyle{\bar \bigoplus}}_{x_j}
   \mathop{\textstyle{\bigoplus_{x_i}}}\phi_{ij}(x_i,x_j).$
\ep

Given $i, j \in [n]$, $x^1_i \neq x^2_i\in \Dom(X_i)$, $x^1_j \neq x^2_j \in \Dom(X_j)$, we define an `identity' function 
$\phi^{(\mv I)}_{ij}: \Dom(X_i)\times\Dom(X_j) \to \D$ as follows.
\begin{equation}
    \phi_{ij}^{(\mv I)}(x_i,x_j) :=
    \begin{cases}
        \mv 1 & \text{ if } (x_i,x_j) = (x^1_i,x^1_j) \text{ or } (x_i,x_j) = (x^2_i,x^2_j)\\
        \mv 0 & \text{ otherwise}.
    \end{cases}
    \label{eqn:identity-defn}
\end{equation}
We will use both $\phi_{ij}$ and $\phi_{ij}^{(\mv I)}$ in the proofs below.

\subsubsection{Soundness and completeness}

We are now fully equipped to show that $\LE(P)$ is sound.

\bthm[$\LE(P) \subseteq \EVO(\varphi)$]\label{thm:LE subseteq EVO K blocks case}
Every linear extension of the precedence poset $P$ is $\varphi$-equivalent.
\ethm
\bp
Let $P$ be the expression tree constructed using only the compartmentalization
step. This expression tree already defines a poset on variables. We will first
show that every linear extension of this {\em compartmentalization poset} 
is $\varphi$-equivalent.
We prove this claim by induction on the number of tag blocks of the input
sequence.
Let $\sigma$ denote the input sequence of tagged variables with input hypergraph
$\calH$.

In the base case $\sigma$ has only one tag block.
All variables in the sequence belong to 
the same node of the compartmentalization expression tree. 
This means every permutation of variables is a linear extension of the poset, 
which is what we expect because every permutation is $\varphi$-equivalent.

In the inductive step, suppose $\sigma$ has at least two tag blocks with the
first block being the set $L$ of variables.
Then, each sub-sequence $\sigma_C$ for each connected component of $\calH-L$
defines an $\faq$-expression $\varphi_C$ on the set of
conditional factors 
$\bigl\{ \psi_S( \cdot \suchthat \mv x_L) \suchthat S\in\calE \wedge
S\cap \calV(C) \neq \emptyset \bigr\}$.
When we condition on the first $L$ variables, the
expression $\varphi(\cdot \suchthat \mv x_L)$ completely factorizes 
into a product of the $\faq$-expressions $\varphi_C$. 
(Another way to put this is that $\varphi$ can be written as a series of
aggregates on variables in $L$, with a product of $\varphi_C$ inside.)
By induction, every linear extension of the
compartmentalization poset for $\sigma_C$ is $\varphi_C$-equivalent. Those
linear extensions can be put together in an arbitrary interleaving way to form
$\varphi(\cdot \suchthat \mv x_L)$. 
This observation completes the proof of the claim, because every linear
extension of the expression poset for $\varphi$ consists of variables in $L$,
followed by arbitrary interleavings of linear extensions of the expression
posets for the $\varphi_C$.

Next, we consider the expression tree after the compression step. We show
that every linear extension of the final precedence poset is
$\varphi$-equivalent by induction on the number of merges of a child node $L'$
to a parent node $L$ (which both must have the same tag). 
To see this, we can take a linear extension $\sigma$
of the expression tree {\em before} the merge where all
variables in $L$ and in $L'$ are consecutive in $\sigma$.
Then, because all variables in $L \cup L'$ have the same tag, we can permute
them in any way and still obtain a $\varphi$-equivalent variable ordering.
\ep

It would have been nice if every $\varphi$-equivalent variable ordering is a
linear extension of $P$. Unfortunately this is not true. Consider the following
$\faq$-query
\[ \varphi = \sum_{x_1}\sum_{x_2}\max_{x_3}\max_{x_4}\sum_{x_5}
    \psi_{15}\psi_{25}\psi_{13}\psi_{24},
\]
where all factors have range $\R_+$ so that all variables are semiring
variables. In this case the expression tree has four nodes: one empty root, a
node containing $\{1,2,5\}$ and two children $\{3\}$ and $\{4\}$. The linear
extensions will enforce that $X_1,X_2,X_5$ come before $X_3$ and $X_4$.
Notice that the original variable ordering $\sigma=(1,2,3,4,5)$ is \emph{not} a linear extension of the precedence poset $P$. Hence, this example already shows that $\LE(P)$ is a \emph{proper} subset of $\EVO(\varphi)$, i.e. $\LE(P)\subsetneq\EVO(\varphi)$.
Moreover, this example gives us some hints about what needs to be done with $\LE(P)$ in order to obtain all of $\EVO(\varphi)$, and at the same time shows that those variable orderings that are missing from $\LE(P)$ are irrelevant to the computation of $\faqw(\varphi)$, as we explain next.
In particular,
it is easy to see that we can rewrite $\varphi$ as follows.
\begin{eqnarray*}
    \varphi &=& \sum_{x_1}\sum_{x_2}\max_{x_3}\max_{x_4}\sum_{x_5}
    \psi_{15}\psi_{25}\psi_{13}\psi_{24}\\
    &=& \sum_{x_5} \sum_{x_1}\sum_{x_2}\max_{x_3}\max_{x_4}
    \psi_{15}\psi_{25}\psi_{13}\psi_{24}\\
    &=& \sum_{x_5} \left( \sum_{x_1}
    \max_{x_3}
    \psi_{15}
    \psi_{13}
\right) \cdot
\left(
    \sum_{x_2}
    \max_{x_4}
    \psi_{25}
\psi_{24}\right).
\end{eqnarray*}
From the above, we know that when conditioned on $X_5$, the expression factorizes and we can interleave
the two factors together in any way we want allowing for $4$ to come before $1$, or 
for $3$ to come before $2$; namely,
\begin{eqnarray*}
    \varphi 
    &=& \sum_{x_5} \sum_{x_2}\max_{x_4}\sum_{x_1}\max_{x_3}
    \psi_{15}\psi_{25}\psi_{13}\psi_{24}\\
    &=& \sum_{x_5} 
    \sum_{x_1}\max_{x_3}
    \sum_{x_2}\max_{x_4}
    \psi_{15}\psi_{25}\psi_{13}\psi_{24}.
\end{eqnarray*}
However, it can be verified that 
$$\faqw(5, 2, 4, 1, 3) = \faqw(5, 1, 3, 2, 4) = \faqw(5, 1, 3, 2, 4)
= \faqw(\sigma)
$$ for any $\sigma \in \LE(P)$ where $5$ comes first in $\sigma$,
and where $P$ is the expression tree
of the query. Note that we can take the linear extensions of the
factorized components $\{1, 3\}$ and $\{2, 4\}$
and interleave them in any way, as long as we still
respect their relative order within each component.
However, these interleavings  do not 
add anything of value as far as the $\faqw$ is concerned.

Another way to think about the above example is that we could have arbitrarily selected 
 one variable in the first tag block of $\varphi$, then constructed a
compartmentalization expression tree with that variable as the root. (One
variable at a time instead of one tag block at a time.) Then, by the same
reasoning we used in the proof of 
Theorem~\ref{thm:LE subseteq EVO K blocks case}, every linear extension of this
`variable-wise' poset is $\varphi$-equivalent. However, this idea alone also
does not capture all of $\EVO(\varphi)$ because it will forbid the selection of $X_5$ as the first
variable in the above example. Thus, it is crucial that we construct
the compressed expression tree first, to determine 
which variable {\em can} come first in a
$\varphi$-equivalent variable ordering.
Ultimately, the set $\LE(P)$ gives us a canonical way of listing 
the variable orderings that really matter in evaluating $\varphi$.

In what follows, we implement the above informal discussion and intuition
by showing that every $\varphi$-equivalent variable ordering 
has the same width as some ordering in $\LE(\varphi)$.
The following lemma says that the expression tree indeed gives us a 
complete list of variables that can occur first (after the free variables) in 
any $\varphi$-equivalent  ordering.

\blmm
For every variable ordering $\pi=(u_1, \ldots, u_n)\in\EVO(\varphi)$, the
variable $u_{f+1}$ must belong to a child node of the root of the expression
tree.\footnote{Recall that $\{u_1, \ldots, u_f\}$ are the free variables, which
are located in the root of the expression tree. And, if $f=0$ then the root of
the expression tree is empty.}
\label{lmm:u_f+1 in L}
\elmm
\bp
Let $\varphi^\pi$ denote the function defined by the $\faq$-query with
$\pi$ as the variable ordering (over the same input factors as $\varphi$).
Our aim is to show that if the conclusion of the lemma does not hold then there 
exist input factors $\psi_S$ for which $\varphi^\pi \not\equiv \varphi$.

Suppose for the sake of contradiction that 
$u_{f+1}$ belongs to a node $L$ whose parent is $L_p$, and $L_p$ is not the root
of the expression tree $P$.
Let $\overline L_a$ denote union of all the (strict) ancestors of $L_p$ in the 
expression tree. 
From the construction of the expression tree, the
vertices in the set $\bigl\{ u_{f+1} \bigr\} \cup L_p$ belong 
to the same connected component of the 
graph $\calH- \overline L_a$.
Let $i_0:=u_{f+1}, i_1, \ldots, i_k \in L_p$ be the shortest path in the 
Gaifman graph\footnote{Given a hypergraph $\calH=(\calV,\calE)$, the \emph{Gaifman graph} of $\calH$ is a graph $G=(V,E)$ where $V:=\calV$ and $E:=\{\{i, j\} \suchthat \exists S \in \calE\text{ where } \{i, j\}\subseteq S\}.$\label{fnote:gaifman}} 
of $\calH - \overline L_a$ from $u_{f+1}$ to $L_p$.
Then, the vertices $i_1,\dots,i_{k-1}$ do not belong to 
$L_p \cup \overline L_a$; and, there are distinct hyperedges 
$S_1,\dots,S_k$ of $\calH$ such that 
$\{i_{j-1},i_{j}\} \subseteq S_{j}$ for all $j \in [k]$.

For each variable $\ell \in \{i_0, \ldots, i_k\}$, we fix two arbitrary 
values $x_\ell^1\neq x_\ell^2\in \Dom(X_\ell)$ and for each 
$\ell'\in[n]\setminus\{i_0, \ldots, i_k\}$, we fix one arbitrary value 
$e_{\ell'}\in\Dom(X_{\ell'})$. 
For the sake of brevity, denote $\textstyle{\bigoplus = \bigoplus^{(i_0)}}$ and 
$\textstyle{\bar\bigoplus = \bigoplus^{(i_k)}}$. 

Now, we construct an input set of factors $\psi_S$, $S\in\calE$, for which
$\varphi^\pi \not\equiv \varphi$.
\bi
 \item Define the factor $\psi_{S_k}$ by
\begin{equation}
\psi_{S_k}(\mv x_{S_k}):=
\begin{cases}
\phi_{i_{k-1} i_k}(x_{i_{k-1}}, x_{i_k}) & \text{if $x_{\ell'} = e_{\ell'}$ for all $\ell'\in S_k\setminus \{i_0, \ldots, i_k\}$}\\
\mv 0 & \text{otherwise},
\end{cases}
\label{eq:construction:psi_S_k}
\end{equation}
where $\phi_{i_{k-1}i_k}$ is the function defined in \eqref{eqn:non-commutative-defn}.
 \item For every $j\in[k-1]$, define 
\begin{equation}
    \psi_{S_{j}}(\mv x_{S_{j}}) :=
    \begin{cases}
        \phi_{i_{j-1}i_j}^{(\mv I)}(x_{i_{j-1}}, x_{i_j}) & \text{if $x_{\ell'} = e_{\ell'}$ for all $\ell'\in S_j \setminus \{i_0, \ldots, i_k\}$}\\
        \mv 0 & \text{otherwise},
    \end{cases}
\label{eq:construction:psi_S_j}
\end{equation}
where $\phi_{i_{j-1}i_j}^{(\mv I)}$ is the `identity' function defined in \eqref{eqn:identity-defn}.
(Think of these factors as little $2 \times 2$ identity matrices.)
 \item Finally, for every $S'\in \calE \setminus \{S_1, \ldots, S_k\}$, define 
\[ 
    \psi_{S'}(\mv x_{S'}) :=
    \begin{cases}
        \mv 1 & \text{if $x_{\ell'} = e_{\ell'}$ for all $\ell' \in S' \setminus \{i_0, \ldots, i_k\}$}\\
        \mv 0 & \text{otherwise}.
    \end{cases}
\]
\ei
Because $i_0$ is the first in $\pi$ after the free variables, 
$\varphi^\pi(e_1,\dots,e_f)$ will evaluate to the left 
hand side of \eqref{eqn:non-commutative-ineq}. 
(Imagine running the $\InsideOut$
algorithm to evaluate $\varphi^\pi$.) 
Next, to get a contradiction, we pick an ordering 
$\sigma\in\LE(P)$ such that $i_0$ precedes all variables that are at 
the same or lower level in the expression tree. 
By Theorem~\ref{thm:LE subseteq EVO K blocks case}, 
we know that $\sigma\in\EVO(\varphi)$. 
In $\sigma$, $i_k$ precedes $i_0$ which in turn 
precedes all of $\{i_1, \ldots, i_{k-1}\}$. Hence, when we compute
$\varphi(e_1,\dots,e_f)$ 
using the ordering $\sigma$ (and the $\InsideOut$ algorithm) we get
the right hand side of \eqref{eqn:non-commutative-ineq}.
Thus, $\varphi^\pi \not\equiv \varphi$ as desired.
\ep

The next definition realizes the intuition that if we construct the precedence
tree using the one-variable-at-a-time strategy (as opposed to the
one-tag-block-at-a-time strategy), then we can interleave the linear extensions
of connected components arbitrarily and still get a variable ordering 
which is $\varphi$-equivalent with the same $\faq$-width.
Since the interleaving can happen at any level, the definition is inductive.

\bdefn[Component-wise equivalence]
\label{defn:CWE}
Let $\varphi$ be an \faq-query where all variable aggregates are semiring
aggregates. Let
$\sigma=(v_1, \ldots, v_n)\in\EVO(\varphi)$ be a variable ordering.
Let $\pi=(u_1, \ldots, u_n)$ be another variable ordering with
$\{u_1,\dots,u_f\} = F$.
Then, $\pi$ is said to be \emph{component-wise equivalent} 
(or $\CW$-equivalent) to $\sigma$ if and only if:
\bi
\item either $n=1$,
\item or $\calH$ has at least two connected components, and for each connected
    component $C=(\calV(C), \calE(C))$ of $\calH$, $\pi_C$ is $\CW$-equivalent
    to $\sigma_C$, where $\sigma_C$ (respectively, $\pi_C$) is the 
    variable ordering of $\calV(C)$ that
    is consistent with $\sigma$ (respectively, $\pi$),
\item or $u_{1}=v_{1}$, and for each connected
    component $C=(\calV(C), \calE(C))$ of $\calH-\{v_{1}\}$,
    $\pi_C$ is $\CW$-equivalent to $\sigma_C$, where $\sigma_C$ 
    (respectively, $\pi_C$) is the
    ordering of $\calV(C)$ that is consistent with $\sigma$ 
    (respectively, $\pi$).
\ei
Given a set of variable orderings $\Lambda \subseteq \EVO(\varphi)$, we use 
$\CWE(\Lambda)$ to denote the set of all variable orderings that are 
$\CW$-equivalent to some variable ordering in $\Lambda$.
\edefn

\bprop\label{prop:CWE preserves faqw}
Let $\pi$ be a variable ordering that is $\CW$-equivalent to 
$\sigma \in \EVO(\varphi)$. Then, we have
$\pi\in\EVO(\varphi)$ and
$\faqw(\sigma)=\faqw(\pi)$.
\eprop
\bp
We can prove that $\pi\in\EVO(\varphi)$ by induction in a very similar way to the proof of Theorem~\ref{thm:LE subseteq EVO K blocks case} (the compartmentalization poset). The proof is thus omitted.

To prove that $\faqw(\sigma)=\faqw(\pi)$, we will need to prove a stronger induction hypothesis.
In particular, we inductively prove that $\sigma$ and $\pi$ induce identical rooted tree decompositions (or sets of tree decompositions if the hypergraph is not connected) (see Proposition~\ref{prop:Uk-equal}).

The base case of $n=1$ holds trivially.
The inductive step where $\calH$ has at least two connected components (i.e. the second case in Definition~\ref{defn:CWE}) also holds trivially.
Consider the inductive step where $u_1=v_1$ (i.e. the third case in Definition~\ref{defn:CWE}).
By induction for each connected component $C=(\calV(C),\calE(C))$ of $\calH-\{v_1\}$,
$\sigma_C$ and $\pi_C$ induce the same rooted tree decomposition $(T_C, \chi_C)$.
For each hyperedge $S_C\in \calE(C)$ corresponding to a hyperedge $(S:=S_C\cup\{v_1\})\in \calE$, let $t$ be the node of $T_C$ that is closest to the root of $T_C$ and satisfies $S_C \subseteq\chi_C(t)$.
(Because $(T_C,\chi_C)$ is induced by a variable ordering $\sigma_C$, $t$ must be unique for a fixed $S_C$, since $\chi_C(t)$ must correspond to $U^{\sigma_C}_{j}$ for the unique $v_j\in\calV(C)$ where $S_C\in\partial^{\sigma_C}(v_j)$. See Sections~\ref{subsec:VO} and~\ref{app:sec:td} and Proposition~\ref{prop:Uk-equal} for more details.)
We add $\{v_1\}$ to $\chi_C(t)$ and to $\chi_C(t')$ for all ancestors $t'$ of $t$ in $T_C$.
We construct a tree decomposition $(T, \chi)$ of $\calH$ by creating a root node $t$ with $\chi(t)=\{v_1\}$ and connecting all the tree decompositions $(T_C,\chi_C)$ to become subtrees of $t$.
Now the tree decomposition $(T,\chi)$ is induced by both $\sigma$ and $\pi$.
\ep

The following theorem shows the completeness part. 

\bthm[$\EVO(\varphi)=\CWE(\LE(P))$]
A variable ordering $\sigma$ is $\varphi$-equivalent if and only if it is \CW-equivalent to some linear extension of the precedence poset $P$.
\label{thm:kblocks:completeness}
\ethm
\bp
We only need to show that $\EVO(\varphi)$ $\subseteq$ $\CWE(\LE(P))$ because the
reverse containment follows from Proposition~\ref{prop:CWE preserves faqw} and 
Theorem~\ref{thm:LE subseteq EVO K blocks case}.
Also, without loss of generality we can assume that the root $F$ of the expression 
tree $P$ is empty, and it has one child node $L$.
(If there were different connected components, they can interleave arbitrarily
and we prove each of them separately.)

Fix an arbitrary $\sigma=(v_1, \ldots, v_n)\in\EVO(\varphi)$. 
Then $v_1\in L$ by Lemma~\ref{lmm:u_f+1 in L}.
For each connected component $C$ of $\calH-\{v_1\}$, define a 
sub-query $\varphi_{C}$ with variable ordering $\sigma_C$ 
on the conditional factors 
$\left\{
    \psi_S(\cdot\suchthat x_{v_1})
    \suchthat S \in\calE \wedge S \cap \calV(C) \neq \emptyset
\right\},
$
where $\sigma_{C}$ is the subsequence of $\sigma$ obtained by
picking out vertices in $\calV(C)$. 
Let $P_C$ be the precedence poset of the expression tree for 
$\varphi_{C}$. By induction on the number of variables, we know 
$\sigma_{C}\in \EVO(\varphi_C) \subseteq \CWE(\LE(P_C)).$ 
Hence, there exists 
$\pi_{C}\in\LE(P_C)$ that is \CW-equivalent to $\sigma_{C}$.

The expression tree $P_C$ for $\varphi_{C}$ consists of an empty root 
(with `free' tag). This root has only one child node $L_C$.
The subtree rooted at $L_C$ is called an $L_C$-subtree.
The expression tree $P$ can be constructed from all the $L_C$-subtrees as follows.
We create a root $R$ containing a single node $v_1$ and we add an empty parent to $R$.
Now we attach all the $L_C$ subtrees to become subtrees of $R$ (such that each node $L_C$ is now a child node of $R$.)
Then, we merge into $R$ every child node $L_C$ that has the same tag as $R$ (which is the tag of $v_1$). The resulting tree is exactly the expression tree $P$,
and the resulting $R=L$ is the union of $\{v_1\}$ with all nodes $L_C$ that have the same tag as $v_1$.

From this observation,
we can pick a variable ordering $\pi$ that is consistent with all $\pi_{C}$ 
such that $\pi$ starts with $v_1$, followed by variables in $L-\{v_1\}$.
(Recall that because each $\pi_C\in \LE(P_C)$, each $\pi_C$ must start with the variables in $L_C$.)
It follows that $\pi\in\LE(P)$ and $\pi$ is \CW-equivalent to $\sigma$.
\ep

Now, we give an example of component-wise equivalence 
(Definition~\ref{defn:CWE}) and the role it plays in completeness (Theorem~\ref{thm:kblocks:completeness}).
\label{app:completeness-example}
\begin{ex}[Component-wise equivalence and completeness]
Consider the following $\faq$ query with two different semiring aggregates ($\sum$ and $\max$) and three variables, all are bound.
\[\varphi=\sum_{x_1}\max_{x_2}\sum_{x_3}\psi_{12}\psi_{13}.\]
($\psi_{ij}$ above denotes an input factor whose support is $\{X_i, X_j\}$.)
For this query,  
$$\EVO(\varphi)=\{(1, 2, 3), (1, 3, 2), (3, 1, 2)\}.$$
Ignoring the dummy free variable $X_0$, the expression tree of $\varphi$ consists of a root with the tag `$\sum$' containing the variables $\{X_1, X_3\}$ and a single child node with tag `$\max$' containing $\{X_2\}$.
Therefore, $\LE(P)=\{(1, 3, 2), (3, 1, 2)\}\subseteq \EVO(\varphi),$ as suggested by Theorem~\ref{thm:LE subseteq EVO K blocks case}. Note that the original ordering $(1, 2, 3)\notin\LE(P)$. However, $(1, 2, 3)$ is component-wise equivalent to $(1, 3, 2)$ (See Definition~\ref{defn:CWE}). Therefore,
\[\CWE(\LE(P))=\{(1, 3, 2), (1, 2, 3), (3, 1, 2)\}=\EVO(\varphi),\]
just as predicted by Theorem~\ref{thm:kblocks:completeness}.
Moreover, by Proposition~\ref{prop:CWE preserves faqw}, $\faqw((1, 2, 3))=\faqw((1, 3, 2))=1$. Therefore, when searching for the variable ordering with the best $\faq$-width, the ordering $(1, 2, 3)$ is redundant, hence it is sufficient to consider $\LE(P)$, just as suggested by Corollary~\ref{cor:faqw-char}.
In particular, removing those redundant variable ordering is {\em precisely} what makes it possible to {\em efficiently} search for and find (an approximation of) the best variable ordering , as we will see later in Section~\ref{sec:approx}.
\end{ex}

Proposition~\ref{prop:CWE preserves faqw} and 
Theorem~\ref{thm:kblocks:completeness} imply the following result, which is 
precisely what we need to approximate $\faqw(\varphi)$ in Section~\ref{subsec:approx:Kblocks}.

\bcor
$\faqw(\varphi) =
    \min \left\{ \faqw(\sigma) \suchthat \sigma \in \LE(P) \right\}.$
\label{cor:faqw-char}
\ecor

\subsection{$\faq$ with an inner $\faq$-formula closed under idempotent elements}
\label{subsec:evo:inner:faq}
Now, we generalize the results of Section~\ref{subsec:evo:Kblocks}
to $\faq$-expressions that have ``idempotent" product aggregates (in addition to semiring aggregates).
In particular, this section considers $\faq$-expressions of the following special form (that is still more
general than that of Section~\ref{subsec:evo:Kblocks}). 

\subsubsection{Problem specification}

Before defining the special case of $\faq$ that we solve in this section,
we need to define and clarify a couple of concepts.
Let $\D_I \subseteq \D$ be a set of idempotent elements of $\otimes$
under the domain $\D$.
A variable aggregate $\oplus$ is said to be {\em closed under $\D_I$} if
$a \oplus b \in \D_I$ whenever both $a$ and $b$ belong to $\D_I$. (Recall that
$\oplus$ may or may not be the same as $\otimes$).
In the $\faq$ context, the two elements $\mv 0$ and $\mv 1$ are 
idempotent elements of $\otimes$.
Hence, the canonical example is $\D_I = \{\mv 0, \mv 1\}$:
$\max$ and $\times$ are closed under $\{0,1\}$,
$\vee$ and $\wedge$ are closed under $\{\true,\false\}$,
$\cup$ and $\cap$ are closed under $\{2^U,\emptyset\}$, etc.
If $\D$ is a matrix domain then there might be more than two idempotent elements
under matrix multiplication.

Note that if a semiring aggregate $\oplus$ is closed under $\D_I$, then it
is also closed under $\D_I \cup \{\mv 0\}$; however, it is not necessarily
closed under $\D_I \cup \{\mv 1\}$.
On the other hand, a product aggregate is always closed under $\D_I \cup \{\mv
0, \mv 1\}$.

\bdefn[Identical aggregates under a subset $\D'\subseteq \D$]
Given a set $\D'\subseteq \D$ and two aggregates $\oplus, \bar\oplus$ 
(that are not necessarily closed under $\D'$), $\oplus, \bar\oplus$ are said 
to be \emph{identical under $\D'$} if for all $a, b\in \D'$, we have
\[a\oplus b=a\bar\oplus b.\]
Note that two aggregates might be identical under $\D'$ but not under $\D$
(by accident).
\edefn

\bdefn[Problem Formulation]
\label{defn:inner-faq}
In this section, we consider an $\faq$-query $\varphi$ of the following
form:
\begin{equation}\label{eqn:faqIdempotent}
    \varphi(\mv x_{[f]}) =
    \textstyle{
        \bigoplus^{(f+1)}_{x_{f+1}} \cdots \bigoplus_{x_{f+\ell}}^{(f+\ell)}
        \bigoplus^{(f+\ell+1)}_{x_{f+\ell+1}} \cdots 
        \bigoplus^{(n)}_{x_{n}}
    }
    \bigotimes_{S\in\calE}\psi_S(\mv x_S),
\end{equation}
where 
\be
 \item All input factors have range $\D_I$, such that $\D_I$ is a set of
     idempotent elements of $\otimes$, $\{ \mv 0, \mv 1 \} \subseteq \D_I$,
     and $\otimes$ is closed under $\D_I$.\footnote{If $\D_I$ is the set of {\em
         all} idempotent elements of $\otimes$, then $\otimes$ is closed under
         $\D_I$, because for any two elements $a,b\in\D_I$, we have
         $(a \otimes b) \otimes (a \otimes b) = a \otimes (b \otimes b) \otimes
         a = a \otimes b \otimes a = (a \otimes a) \otimes b = a \otimes b.$}
 \item The integer $\ell$ satisfies $0 \leq \ell \leq n-f$.
 \item For $f+1\leq i \leq f+\ell$, $\textstyle{\bigoplus^{(i)}}$ is a semiring aggregate that is not closed under $\D_I$.
 In particular, there exist $g_1,g_2\in\D_I$ where $g_1 \textstyle{\bigoplus^{(i)}} g_2 \neq \left(g_1\textstyle{\bigoplus^{(i)}} g_2\right)^e$ for any integer $e \geq 2$ (which implies $g_1 \textstyle{\bigoplus^{(i)}} g_2 \notin \D_I$).
 \label{itm:inner:not-closed-semiring}
 \item For $f+\ell+1 \leq i \leq n$, $\textstyle{\bigoplus^{(i)}}$ 
     is closed under $\D_I$. (It
     could be either a product or semiring aggregate.)
  \item For every $f+1\leq i, j \leq n$ such that $\textstyle{\bigoplus^{(i)}}$ and $\textstyle{\bigoplus^{(j)}}$ are semiring aggregates, if $\textstyle{\bigoplus^{(i)}}$ and $\textstyle{\bigoplus^{(j)}}$ are not identical under $\D$, then they are not identical under $\D_I$.
  \label{itm:inner:faq:non-identical}
\ee
A compact way to write the above $\faq$ expression is
\begin{equation} \varphi(\mv x_F) = 
    \textstyle{
        \bigoplus^{(f+1)}_{x_{f+1}} \cdots \bigoplus_{x_{f+\ell}}^{(f+\ell)}
        \varphi'
    }
 \label{eq:varphi:phi'}
\end{equation}
where $\varphi'$ is a general $\faq$ expression whose aggregates are closed
under $\D_I$.
This special case of $\faq$ captures the $\qcq$ and $\sqcq$ instances
from Example~\ref{ex:qcq} and Example~\ref{ex:sqcq}, both of which can be
written in the above form.
\edefn
\begin{ex}[$\sqcq$-revisited]
In $\sqcq$, $\varphi'$ is basically a $\qcq$ formula.
In particular, although the input factors have range $\{0, 1\}$, the output is 
computed under the range $\D=\N$. Hence by choosing $\D_I=\{0, 1\}$, we can 
compute the ``$\qcq$-part'' of $\sqcq$ over $\D_I$ exploiting the fact 
that all product aggregates are idempotent (as in 
Definition~\ref{defn:idempotent product agg}), and the ``$\#$-part'' of 
$\sqcq$ over the full domain $\D$ while enjoying the fact that there are 
no product aggregates.

More specifically, in $\sqcq$ we have two aggregates ($\max, \times$) that are closed under $\D_I$ (those are equivalent to the logical $\vee, \wedge$), and one aggregate $+$ which is not (e.g. $1+1\notin\{0, 1\}$). Notice that because of this, $+$ cannot be identical under $\D_I$ to either $\max$ or $\times$ (even if we use any different arithmetic interpretation of the logical $\vee, \wedge$ other than $\max, \times$). Hence, $\sqcq$ satisfies the last condition above in the problem formulation.
\end{ex}

In this special case of $\faq$, as long as the variable ordering lists 
$\mv X_{[f+\ell]}$ first (in any order), then all product aggregates are
idempotent (see Definition~\ref{defn:idempotent product agg}).
Our aim is to find a good variable ordering for $\varphi$ in the set
$\EVO(\varphi, \calF(\D_I))$, where $\calF(\D_I)$ is the set of factors whose 
range is $\D_I$ (See Definition~\ref{defn:EVO-phi-F}).
 Notice that in this context, 
$\EVO(\varphi, \calF(\D_I))$ is a stronger notion and harder to deal with than 
$\EVO(\varphi)$. In particular, a variable ordering $\sigma$ can still be in
$\EVO(\varphi, \calF(\D_I))$ even if the output evaluates differently under 
$\sigma$ for some input factors whose range is $\D$, as long as those 
factors don't have range $\D_I\subseteq \D$.

\subsubsection{Non-mutually-commutative aggregates vs. non-identical aggregates}
\label{subsec:differentaggregates:prod}
From the discussion of Section~\ref{subsec:differentaggregates}, a necessary 
condition for two variable aggregates (possibly products) to be mutually commutative 
under $\D_I$ is that for all $a, b, c, d\in \D_I$, we have
\[
(a \oplus b) \bar \oplus (c \oplus d) =
   (a \bar \oplus c) \oplus (b \bar \oplus d).
   \label{eq:commute}
\]
(Notice that $\oplus$, $\bar\oplus$ are not necessarily closed under $\D_I$, and that is why the above necessary condition is not sufficient for the two aggregates to be mutually commutative under $\D_I$). We recognize two cases:
\bi
\item If both $\oplus$ and $\bar\oplus$ are semiring aggregates, then by selecting $a=d=\mv 0$ (since $\mv 0\in\D_I$), we obtain $b \bar\oplus c=b\oplus c$ for every $b, c\in\D_I$. Hence, no two semiring aggregates commute under $\D_I$ unless they are identical under $\D_I$. However, our problem formulation requires any two semiring aggregates that are identical under $\D_I$ to be identical under $\D$ as well.
\item If exactly one of the two aggregates is a product, say $\bar\oplus=\otimes$, then by selecting $a=d=\mv 0$ and $b=c=\mv 1$, we get a contradiction ($\mv 1=\mv 0$). Hence, $\oplus, \otimes$ do not commute under $\D_I$.
\item If exactly one of the two aggregates is a product, say $\bar\oplus=\otimes$, while the other ($\oplus$) is a semiring aggregate that is not closed under $\D_I$, then by our problem formulation above there is an alternative choice of $a,b,c,d$
that shows that $\bar\oplus$ and $\oplus$ do not commute. In particular, since there exist $g_1,g_2\in\D_I$ where $g_1\oplus g_2 \neq (g_1 \oplus g_2)^2$, we can select $a=c=g_1$ and $b=d=g_2$ and get a contradiction. This alternative choice will be useful later.
\ei
We infer the following variant of Proposition~\ref{prop:non-commutative}: (Notice that the range of $\phi_{ij}$ is now $\D_I$ instead of $\D$.)
\bprop
Suppose that $\oplus$ and $\bar\oplus$ are either semiring aggregates that are not identical under $\D$ or one of them is a product aggregate (while the other is not). Then for every $i, j \in [n]$, there is a
function $\phi_{ij}: \Dom(X_i)\times\Dom(X_j) \to \D_I$ for which
\begin{equation}
    \textstyle{\bigoplus_{x_i\in\Dom(X_i)} \bar\bigoplus_{x_j\in\Dom(X_j)} \phi_{ij}(x_i, x_j)
    \neq
    \bar\bigoplus_{x_j\in\Dom(X_j)}
   \bigoplus_{x_i\in\Dom(X_i)} 
   \phi_{ij}}(x_i, x_j).
\end{equation}
\label{prop:non-commutative-var}
\eprop
\bp
If both $\oplus$ and $\bar\oplus$ are semiring aggregates, then $\phi_{ij}$ is defined exactly the same as in the proof of Proposition~\ref{prop:non-commutative}.
If exactly one aggregate is a product, say $\bar\oplus=\otimes$, 
then $\phi_{ij}$ is defined as follows.
Fix arbitrarily elements $x^1_i \neq x^2_i\in \Dom(X_i)$, $x^1_j\in \Dom(X_j)$ such that $x^1_j$ is not the last in $\Dom(X_j)$ (i.e. there is $x^2_j\in \Dom(X_j)$ such that $x^2_j>x^1_j$). Define
\begin{equation}
    \phi_{ij}(x_i,x_j) :=
    \begin{cases}
        \mv 1 & \text{ if } \left(x_i= x^1_i \wedge x_j\leq x^1_j \right) \text{ or } \left(x_i= x^2_i \wedge x_j> x^1_j\right) \\
        \mv 0 & \text{ otherwise}.
    \end{cases}
    \label{eqn:non-commutative-defn-prod}
\end{equation}
Then,
\[
    \textstyle{\bigoplus_{x_i}\bar\bigoplus_{x_j}\phi_{ij}(x_i,x_j) = 
       \mv 0
   \neq \mv 1
   = \bar \bigoplus_{x_j}\bigoplus_{x_i}\phi_{ij}(x_i,x_j).}
\]
\ep
Note that in the above proof if exactly one aggregate is a product, say $\bar\oplus=\otimes$, while the other ($\oplus$) is a semiring aggregate that is {\em not} closed under $\D_I$, then we can utilize our problem formulation in Definition~\ref{defn:inner-faq} and define $\phi_{ij}(x_i,x_j)$ alternatively as follows:
Let $g_1, g_2\in \D_I$ satisfy $g_1 \oplus g_2\neq (g_1\oplus g_2)^e$ for all integers $e\geq 2$ (as guaranteed by Definition~\ref{defn:inner-faq}).
Fix arbitrary elements $x^1_i, x^2_i\in \D_I$.
\begin{equation}
\phi_{ij}(x_i,x_j) := \phi'_i(x_i) :=
\begin{cases}
g_1 & \text{ if } x_i= x^1_i\\
g_2 & \text{ if } x_i= x^2_i\\
\mv 0 & \text{ otherwise}.
\end{cases}
\label{eqn:non-commutative-defn-prod-non-closed}
\end{equation}
Then,
\begin{equation}
\textstyle{\bigoplus_{x_i}\bar\bigoplus_{x_j}\phi_{ij}(x_i,x_j) = 
   g_1 \oplus g_2
   \neq (g_1 \oplus g_2)^{|\Dom(X_j)|}
   = \bar \bigoplus_{x_j}\bigoplus_{x_i}\phi_{ij}(x_i,x_j).}
\label{eq:no-swap-product-semiring}
\end{equation}
(Recall from Section~\ref{sec:faq-pbm} that $|\Dom(X_j)|\geq 2$.)
The alternate function $\phi_{ij}(x_i,x_j)$ given by \eqref{eqn:non-commutative-defn-prod-non-closed}
depends only on $x_i$.
This nice property will be useful later.

\subsubsection{Expression tree and precedence poset}
\label{subsubsec:inner:faq:expr-tree}

The approach in this section and the next mirrors that of 
Section~\ref{subsec:evo:Kblocks}.
However, dealing with the product aggregates (under the $\varphi'$ part of the
query, as in~\eqref{eq:varphi:phi'}) requires extra care. In particular, the corresponding variables do not
play a role in determining the connected components when we construct the
expression tree.
And they do not contribute directly to the width $\faqw(\sigma)$, because
$\faqw(\sigma)$ is defined only over $U_k$ for $k\in K$ (where $K$ was defined by \eqref{eqn:the:set:K}).
While the discussion in this section is meant for $\faq$ of the form \eqref{eqn:faqIdempotent}, it might be helpful if the reader
uses $\sqcq$ as the running example for this section.

\bdefn[Extended components and dangling set of $(\calH, W)$]
\label{defn:extended-component}
Given a hypergraph  $\calH=(\calV,\calE)$ and a set $W\subseteq \calV$,
the extended components of $(\calH, W)$ are defined as follows:
For each connected component $C=(\calV(C),\calE(C))$ of $\calH-W$,
we construct a hypergraph $(\calV'(C),\calE'(C))$ as follows:
\begin{eqnarray*}
   \calV'(C) &:=& \calV(C) \ \cup \ \bigl\{ w \in W \suchthat
   \exists S \in \calE \text{ where }
   S \cap \calV(C) \neq\emptyset \text{ and } w \in S \bigr\}\\
   \calE'(C) &:=& \bigl\{ S \in\calE \suchthat S\cap\calV(C) \neq \emptyset\bigr\}
\end{eqnarray*}
The hypergraphs $(\calV'(C),\calE'(C))$ are called the {\em extended components} of $(\calH, W)$.

Moreover the set
\[
D := \bigcup_{\substack{S\in \calE\\ S \subseteq W}} S
\]
is called the {\em dangling set} of $(\calH, W)$.
\edefn

\bdefn[Expression tree]
\label{defn:expr-tree-prod}
The {\em expression tree} for the query $\varphi$ is a rooted tree $P$. 
Every node of the tree is a set of variables, and the tree is constructed
via a {\em query extension} step, a {\em compartmentalization} step and a {\em compression} step.
The query extension step is completely new compared to Section~\ref{subsec:evo:Kblocks},
the compartmentalization step is a bit trickier than that of Section~\ref{subsec:evo:Kblocks},
while the compression step is identical.

{\bf Query extension.}
Given the input $\faq$-query $\varphi$ of the form~\eqref{eqn:faqIdempotent},
we extend $\varphi$ as follows:
For every $i$ satisfying $f+1\leq i \leq f+\ell$ (hence $\textstyle{\bigoplus^{(i)}}$ is a semiring aggregate that is not closed under $\D_I$) and for every $j$ where $\textstyle{\bigoplus^{(j)}=\bigotimes}$ (hence $f+\ell+1 \leq j \leq n$), if
there is no hyperedge $S$ satisfying $\{i, j\}\subseteq S$, then we add a new hyperedge $\{i,j\}$ and a corresponding input factor $\psi_{\{i,j\}}(x_i, x_j)$ to the input $\faq$ query $\varphi$.
(The new factor $\psi_{\{i,j\}}(x_i, x_j)$ can be defined to be identically $\mv 1$ so that it does not change the value of $\varphi$.)

{\bf Compartmentalization.}
In this step, initially we start off with the
sequence of variables with their corresponding tags exactly as they appear in the $\faq$ query that resulted from the query extension step. We also apply the same trick of adding a dummy free
variable $X_0$ as was done in Section~\ref{subsec:evo:Kblocks}, i.e. we start off
with the sequence
\[ \sigma = 
    \left\langle
        (X_0,\text{``free''}), (X_1,\text{``free''}), \dots, (X_f, \text{``free''}),
        (X_{f+1}, \oplus^{(f+1)}), \dots, (X_n, \oplus^{(n)})
    \right\rangle
\]
and with the hypergraph $\calH$ (of the extended query) which has an extra isolated 
vertex $X_0$ marked with a ``free'' tag.

Now given a tagged variable sequence $\sigma$ and a hypergraph $\calH$, we 
build the tree by constructing a node $L$ containing the first tag block.
Then, we do the following.
\bi 
 \item Let $W$ be the set of all product variables in $\sigma$ 
       that do not belong to $L$.
 \item We construct the extended components $(\calV'(C), \calE'(C))$ of $(\calH-L, W)$ and we also construct the dangling set $D$ of $(\calH-L, W)$, as in Definition~\ref{defn:extended-component}.
 \item Now, for each extended component $(\calV'(C),\calE'(C))$, 
     we construct a sequence $\sigma_C$ (of tagged variables) by listing all 
     variables in $\calV'(C)$ in exactly the same relative order they 
     appeared in $\sigma$. From the sequence $\sigma_C$ and the hypergraph 
     $(\calV'(C), \calE'(C))$ we recursively construct the expression tree 
     $P_C$. 
 \item Finally, we connect all subtrees $P_C$ to the node $L$. {\bf And}, we
     create a node that contains the dangling set $D$ (if $D$ is not empty)
     and connect it to $L$ as well. 
 \item The dummy variable $X_0$ is removed after everything is done.
      This completes the compartmentalization step.
Note that after compartmentalization all variables in the same node of $P$ 
have the same tag.
\ei

{\bf Compression.}
Now, in the expression tree $P$ that resulted from the compartmentalization step,
as long as there is still a node $L$ whose tag is the same as a child node
$L'$ of the tree $P$, we merge the child into $L$.
Repeat this step until no further merging is possible.
\edefn

Before proceeding with the proof, we present an example to illustrate the computation of the expression tree in the presence of product aggregates.

\begin{ex}[Intuition behind expression tree]\label{ex:intuition2}
   Consider the following $\faq$ query that has product aggregates ($\prod$) in addition to two different semiring aggregates ($\sum$ and $\max$):
   \[\varphi=\sum_{x_1}\max_{x_2}\prod_{x_3}\max_{x_4}\max_{x_5}\prod_{x_6}\psi_{123}\psi_{125}\psi_{14}\psi_{34}\psi_{6}.\]
   ($\psi_{123}$ above denotes an input factor $\psi_{123}(x_1,x_2,x_3)$, and so on\ldots)
   Suppose all input factors $\psi_{123},\psi_{125},\ldots$ have range $\D_I=\{0, 1\}$.
   Let $\D$ be $\N$.
   Both $\max$ and $\times$ are closed under $\{0, 1\}$, while $+$ is not. Hence, by comparison to the format in \eqref{eqn:faqIdempotent}, $f=0$ and $\ell=1$. Notice that because $\max$ is closed under $\D_I$ while $+$ is not, $\max$ and $+$ are not identical under $\D_I$. Hence, the last condition in Definition~\ref{defn:inner-faq} is satisfied.
   The hypergraph of the above query is depicted in Figure~\ref{fig:prod:init:a}.
   The query extension step of the construction of the expression tree is depicted in
   Figure~\ref{fig:prod:init:b}.
   The compartmentalization step is depicted in Figure~\ref{fig:prod:compart}, and the compression step is depicted in Figure~\ref{fig:prod:compress}.
   The final expression tree appears on the right of Figure~\ref{fig:prod:compress:a}.
\end{ex}


\colorlet{exprColor}{black}
\colorlet{exprColor0}{red!75!black}
\colorlet{exprColor1}{blue}
\colorlet{exprColor00}{olive!60!black}
\colorlet{exprColor01}{magenta!50!black}
\colorlet{exprColor10}{teal!60!black}

\begin{figure}[htp!]
   \captionsetup[subfigure]{justification=centering}
\begin{center}
\centering
\everymath{\displaystyle}

\subfloat[
	$\varphi=$
	$\sum_{x_1}$
	$\max_{x_2}$
	$\prod_{x_3}$
	$\max_{x_4}$
	$\max_{x_5}$
   $\prod_{x_6}$
	$\psi_{123}$
	$\psi_{125}$
	$\psi_{14}$
	$\psi_{34}$
   $\psi_{6}$]
	{\label{fig:prod:init:a}\begin{tikzpicture}[yscale=\subfigyscale,scale=.9]
      \color{exprColor}
		\draw[white] (1, 0) -- (17, 0) -- (17, 3.3) -- (1, 3.3) --cycle;
	\begin{scope}[hypergraph, shift={(5.5, 0)}]
		\draw[vertex] (0,3) circle node (v1) {$\mathbf{1}$};
      \draw[vertex] (0,0) circle node (v3) {$\mathbf{3}$};
      \draw[vertex] (-3,3) circle node (v2) {$\mathbf{2}$};
      \draw[vertex] (-3,0) circle node (v5) {$\mathbf{5}$};
      \draw[vertex] (2.5,1.5) circle node (v4) {$\mathbf{4}$};
      
      \draw[edge, rotate around={31:(2.5/2, 1.5/2)}]
         (2.5/2, 1.5/2) ellipse ( 2.1 and .5); 
      \draw[edge, rotate around={-31:(2.5/2, 3-1.5/2)}]
         (2.5/2, 3-1.5/2) ellipse ( 2.1 and .5); 
      
      \draw[edge, rounded corners=3.5 mm]
         (.35, -.9)--
         (.35, 3.33)--
         (-3.8, 3.33)--cycle; 
      
      \draw[edge, rounded corners=6 mm]
         (-3.45, -1.1)--
         (-3.45, 3.45)--
         (1, 3.45)--cycle; 
      
      \draw[vertex] (2.5,3) circle node (v6) {$\mathbf{6}$};
      \draw[edge] (v6) circle (.35); 
	\end{scope}
	\end{tikzpicture}
}

\subfloat[
$\varphi=$
$\sum_{x_1}$
$\max_{x_2}$
$\prod_{x_3}$
$\max_{x_4}$
$\max_{x_5}$
$\prod_{x_6}$
$\psi_{123}$
$\psi_{125}$
$\psi_{14}$
$\psi_{34}$
$\psi_{6}$
$\psi_{16}$]
{\label{fig:prod:init:b}\begin{tikzpicture}[yscale=\subfigyscale,scale=.9]
   \color{exprColor}
   \draw[white] (1, 0) -- (17, 0) -- (17, 3.3) -- (1, 3.3) --cycle;
   \begin{scope}[hypergraph, shift={(5.5, 0)}]
   \draw[vertex] (0,3) circle node (v1) {$\mathbf{1}$};
   \draw[vertex] (0,0) circle node (v3) {$\mathbf{3}$};
   \draw[vertex] (-3,3) circle node (v2) {$\mathbf{2}$};
   \draw[vertex] (-3,0) circle node (v5) {$\mathbf{5}$};
   \draw[vertex] (2.5,1.5) circle node (v4) {$\mathbf{4}$};
   
   \draw[edge, rotate around={31:(2.5/2, 1.5/2)}]
   (2.5/2, 1.5/2) ellipse ( 2.1 and .5); 
   \draw[edge, rotate around={-31:(2.5/2, 3-1.5/2)}]
   (2.5/2, 3-1.5/2) ellipse ( 2.1 and .5); 
   
   \draw[edge, rounded corners=3.5 mm]
   (.35, -.9)--
   (.35, 3.33)--
   (-3.8, 3.33)--cycle; 
   
   \draw[edge, rounded corners=6 mm]
   (-3.45, -1.1)--
   (-3.45, 3.45)--
   (1, 3.45)--cycle; 
   
   \draw[vertex] (2.5,3) circle node (v6) {$\mathbf{6}$};
   \draw[edge] (v6) circle (.35); 
   \draw[edge] (1.25, 3) ellipse (1.9 and .45); 
   \end{scope}
   \end{tikzpicture}
}

\caption{The query extension step of the construction of expression tree from Example~\ref{ex:intuition2}. Fig.~\ref{fig:prod:init:a} shows the hypergraph of original query $\displaystyle\varphi=\sum_{x_1}\max_{x_2}\prod_{x_3}\max_{x_4}\max_{x_5}\prod_{x_6}\psi_{123}\psi_{125}\psi_{14}\psi_{34}\psi_{6}$.
Fig.~\ref{fig:prod:init:b} shows the query after adding the new factor $\psi_{16}$
(which had to be added because $\sum_{x_1}$ is a semiring aggregate that is not closed under $\D_I$, $\prod_{x_6}$ is a product aggregate, and there is no hyperedge already containing $\{1, 6\}$).}
\Description{The query extension step of the construction of expression tree from Example~\ref{ex:intuition2}.}
\label{fig:prod:init}
\end{center}
\end{figure}


\colorlet{exprColor}{black}
\colorlet{exprColor0}{red!75!black}
\colorlet{exprColor1}{blue}
\colorlet{exprColor2}{brown}
\colorlet{exprColor00}{olive!60!black}
\colorlet{exprColor01}{magenta!50!black}
\colorlet{exprColor10}{teal!60!black}

\begin{figure}[htp!]
   \captionsetup[subfigure]{justification=centering}
\begin{center}
\centering
\everymath{\displaystyle}

\subfloat[]
      [$\varphi=$
      {\color{exprColor}
         $\sum_{x_1}$}
      $\max_{x_2}$
      $\prod_{x_3}$
      $\Biggl($
      {\color{exprColor0}
         $\biggl($
         $\max_{x_5}$
         $\psi_{123}$
         $\psi_{125}$
         $\biggr)$}
      {\color{exprColor1}
         $\biggl($
         $\max_{x_4}$
         $\psi_{14}$
         $\psi_{34}$
         $\biggr)$}
      {\color{exprColor2}
         $\biggl($
         $\prod_{x6}$
         $\psi_{6}$
         $\psi_{16}$
         $\biggr)$}
      $\Biggr)$\\
      $=$
      {\color{exprColor}
         $\sum_{x_1}$}
      $\max_{x_2}$
      {\color{exprColor0}
         $\biggl($
         $\prod_{x_3}$
         $\max_{x_5}$
         $\psi_{123}$
         $\psi_{125}$
         $\biggr)$}
      {\color{exprColor1}
         $\biggl($
         $\prod_{x_3}$
         $\max_{x_4}$
         $\psi_{14}$
         $\psi_{34}$
         $\biggr)$}
      {\color{exprColor2}
         $\biggl($
         $\prod_{x3}$
         $\prod_{x6}$
         $\psi_{6}$
         $\psi_{16}$
         $\biggr)$
      }\\
      $=$
      {\color{exprColor}
         $\sum_{x_1}$}
      {\color{exprColor0}
         $\biggl($
         $\max_{x_2}$
         $\prod_{x_3}$
         $\max_{x_5}$
         $\psi_{123}$
         $\psi_{125}$
         $\biggr)$}
      {\color{exprColor1}
         $\biggl($
         $\prod_{x_3}$
         $\max_{x_4}$
         $\psi_{14}$
         $\psi_{34}$
         $\biggr)$}
      {\color{exprColor2}
         $\biggl($
         $\prod_{x6}$
         $\psi_{6}$
         $\psi_{16}$
         $\biggr)$
      }]
	{\label{fig:prod:compart:a}\begin{tikzpicture}[yscale=\subfigyscale,scale=.9]
   \color{exprColor}
   \drawframe
   \begin{scope}[hypergraph]
   \draw[exprColor, vertex] (0,3) circle node (v1) {$\mathbf{1}$};
   \draw[exprColor0, vertex] (-.5,0) circle node {$\mathbf{3}$};
   \draw[exprColor1, vertex] (+.5,0) circle node {$\mathbf{3}$};
   \draw[exprColor0, vertex] (-3,3) circle node (v2) {$\mathbf{2}$};
   \draw[exprColor0, vertex] (-3,0) circle node (v5) {$\mathbf{5}$};
   \draw[exprColor1, vertex] (2.5,1.5) circle node (v4) {$\mathbf{4}$};
   
   \draw[exprColor1, edge, rotate around={37:(1.5, 1.5/2)}]
      (1.5, 1.5/2) ellipse (1.8 and .5); 
   \draw[exprColor1, edge, rotate around={-31:(2.5/2, 3-1.5/2)}]
      (2.5/2, 3-1.5/2) ellipse ( 2.1 and .5); 
   
   \draw[exprColor0, edge, rounded corners=3.5 mm]
      (-.25, -.9)--
      (.35, 3.33)--
      (-3.8, 3.33)--cycle; 
   
   \draw[exprColor0, edge, rounded corners=6 mm]
      (-3.45, -1.1)--
      (-3.45, 3.45)--
      (1, 3.45)--cycle; 
      
      \draw[exprColor2, vertex] (2.5,3) circle node (v6) {$\mathbf{6}$};
      \draw[exprColor2, edge] (v6) circle (.35); 
      \draw[exprColor2, edge] (1.25, 3) ellipse (1.9 and .45); 
   \end{scope}
   
   \begin{scope}[exprtree,scale=.75, shift={(-3, 0)}]
      \drawtreenode{$\mathbf{1}~_{\sum}$};
      
      \begin{scope}[shift={(-3, -2)}]
         {\color{exprColor0}
            \drawtreenode {$\mathbf{2, 3, 5}$};
            \draw[treenode] (.5, .5)--(2.5, 1.5);
         }
      \end{scope}
      
      \begin{scope}[shift={(3, -2)}]
         {\color{exprColor1}
            \drawtreenode{$\mathbf{3, 4}$};
            \draw[treenode] (-.5, .5)--(-2.5, 1.5);
         }
      \end{scope}
      
      \begin{scope}[shift={(7, -2)}]
      {\color{exprColor2}
         \drawtreenode{$\mathbf{6}_{~\prod}$};
         \draw[treenode] (-.9, .5)--(-6, 1.5);
      }
      \end{scope}
   \end{scope}
   \end{tikzpicture}
}

\subfloat[]
[$\varphi=$
{\color{exprColor}
   $\sum_{x_1}$}
{\color{exprColor0}
   $\biggl($
   $\max_{x_2}$
{\color{exprColor00}
   $\max_{x_5}$
   $\psi_{125}$}
{\color{exprColor01}
   $\prod_{x_3}$
   $\psi_{123}$}
   $\biggr)$}
{\color{exprColor1}
   $\biggl($
   $\prod_{x_3}$
{\color{exprColor10}
   $\max_{x_4}$
   $\psi_{14}$
   $\psi_{34}$}
   $\biggr)$}
{\color{exprColor2}
   $\biggl($
   $\prod_{x6}$
   $\psi_{6}$
   $\psi_{16}$
   $\biggr)$
}]
{\label{fig:prod:compart:b}\begin{tikzpicture}[yscale=\subfigyscale,scale=.9]
   \color{exprColor}
   \drawframe
   \begin{scope}[hypergraph]
   \draw[exprColor, vertex] (0,3) circle node (v1) {$\mathbf{1}$};
   \draw[exprColor01, vertex] (-.5,0) circle node {$\mathbf{3}$};
   \draw[exprColor1, vertex] (+.5,0) circle node {$\mathbf{3}$};
   \draw[exprColor0, vertex] (-3,3) circle node (v2) {$\mathbf{2}$};
   \draw[exprColor00, vertex] (-3,0) circle node (v5) {$\mathbf{5}$};
   \draw[exprColor10, vertex] (2.5,1.5) circle node (v4) {$\mathbf{4}$};
   
   \draw[exprColor10, edge, rotate around={37:(1.5, 1.5/2)}]
   (1.5, 1.5/2) ellipse (1.8 and .5); 
   \draw[exprColor10, edge, rotate around={-31:(2.5/2, 3-1.5/2)}]
   (2.5/2, 3-1.5/2) ellipse ( 2.1 and .5); 
   
   \draw[exprColor01, edge, rounded corners=3.5 mm]
   (-.25, -.9)--
   (.35, 3.33)--
   (-3.8, 3.33)--cycle; 
   
   \draw[exprColor00, edge, rounded corners=6 mm]
   (-3.45, -1.1)--
   (-3.45, 3.45)--
   (1, 3.45)--cycle; 
   
   \draw[exprColor2, vertex] (2.5,3) circle node (v6) {$\mathbf{6}$};
   \draw[exprColor2, edge] (v6) circle (.35); 
   \draw[exprColor2, edge] (1.25, 3) ellipse (1.9 and .45); 
   \end{scope}
   
   \begin{scope}[exprtree,scale=.75, shift={(-3,0)}]
   \drawtreenode{$\mathbf{1}~_{\sum}$};
   
   \begin{scope}[shift={(-3, -2)}]
   {\color{exprColor0}
      \drawtreenode {$\mathbf{2}~_{\max}$};
      \draw[treenode] (.5, .5)--(2.5, 1.5);
      \begin{scope}[shift={(-2.5, -2)}]
      {\color{exprColor00}
         \drawtreenode{$\mathbf{5}_{~\max}$};
         \draw[treenode] (.5, .5)--(2, 1.5);
      }
      \end{scope}
      \begin{scope}[shift={(2.5, -2)}]
      {\color{exprColor01}
         \drawtreenode{$\mathbf{3}_{~\prod}$};
         \draw[treenode] (-.5, .5)--(-2, 1.5);
      }
      \end{scope}
   }
   \end{scope}
   
   \begin{scope}[shift={(3, -2)}]
   {\color{exprColor1}
      \drawtreenode{$\mathbf{3}~_{\prod}$};
      \draw[treenode] (-.5, .5)--(-2.5, 1.5);
   }
      \begin{scope}[shift={(0, -2)}]
         {\color{exprColor10}
            \drawtreenode{$\mathbf{4}_{~\max}$};
            \draw[treenode] (0, .5)--(0, 1.5);
         }
      \end{scope}
   \end{scope}
   
         \begin{scope}[shift={(7, -2)}]
   {\color{exprColor2}
      \drawtreenode{$\mathbf{6}_{~\prod}$};
      \draw[treenode] (-.9, .5)--(-6, 1.5);
   }
   \end{scope}
   \end{scope}
   \end{tikzpicture}
}

\caption{The compartmentalization step of the construction of the expression tree from Example~\ref{ex:intuition2}, depicted using colors.
For simplicity, the dummy free variable $X_0$ is ignored in this example.
Fig.~\ref{fig:prod:compart:a} shows the first part of the compartmentalization step, where the first tag block $L=\{1\}$ and the set of product variables $W=\{3, 6\}$.
After removing both $L$ and $W$, we get two connected components: $\color{exprColor0}\{2,5\}$ and $\color{exprColor1}\{4\}$ (in addition to an empty component $\color{exprColor2}\{\}$).
By adding back the product variables that were removed from hyperedges of each component, we get the two \emph{extended components}: $\color{exprColor0}\{2,3,5\}$ and $\color{exprColor1}\{3,4\}$.
We also get the \emph{dangling set} $\color{exprColor2}\{6\}$.
We can see that for each fixed value $x_1$, the conditional query $\varphi(\cdot|x_1)$ can indeed be factored into a product of three sub-queries corresponding to the two extended components $\color{exprColor0}\{2,3,5\}$ and $\color{exprColor1}\{3,4\}$ along with the dangling set $\color{exprColor2}\{6\}$.
(Note that closure and idempotence of $\times$ under $\D_I$ are both used to prove that
$\color{exprColor2}\prod_{x_3}\prod_{x_6}\psi_6\psi_{16}=\prod_{x_6}\psi_6\psi_{16}$.)
Fig.~\ref{fig:prod:compart:b} show the compartmentalization of each one of the resulting two extended components.}
\Description{The compartmentalization step of the construction of the expression tree from Example~\ref{ex:intuition2}, depicted using colors.}
\label{fig:prod:compart}
\end{center}
\end{figure}


\begin{figure}[htp!]
\begin{center}
\centering
\everymath{\displaystyle}

\subfloat[]
[$\varphi=$
{\color{exprColor}
   $\sum_{x_1}$}
{\color{exprColor0}
   $\biggl($
   $\max_{x_2}$
   {\color{exprColor0}
      $\max_{x_5}$
      $\psi_{125}$}
   {\color{exprColor01}
      $\prod_{x_3}$
      $\psi_{123}$}
   $\biggr)$}
{\color{exprColor1}
   $\biggl($
   $\prod_{x_3}$
   {\color{exprColor10}
      $\max_{x_4}$
      $\psi_{14}$
      $\psi_{34}$}
   $\biggr)$}
{\color{exprColor2}
   $\biggl($
   $\prod_{x6}$
   $\psi_{6}$
   $\psi_{16}$
   $\biggr)$
}]
{\label{fig:prod:compress:a}\begin{tikzpicture}[yscale=\subfigyscale,scale=.9]
   \color{exprColor}
   \drawframe
   \begin{scope}[hypergraph]
   \draw[exprColor, vertex] (0,3) circle node (v1) {$\mathbf{1}$};
   \draw[exprColor01, vertex] (-.5,0) circle node {$\mathbf{3}$};
   \draw[exprColor1, vertex] (+.5,0) circle node {$\mathbf{3}$};
   \draw[exprColor0, vertex] (-3,3) circle node (v2) {$\mathbf{2}$};
   \draw[exprColor0, vertex] (-3,0) circle node (v5) {$\mathbf{5}$};
   \draw[exprColor10, vertex] (2.5,1.5) circle node (v4) {$\mathbf{4}$};
   
   \draw[exprColor10, edge, rotate around={37:(1.5, 1.5/2)}]
   (1.5, 1.5/2) ellipse (1.8 and .5); 
   \draw[exprColor10, edge, rotate around={-31:(2.5/2, 3-1.5/2)}]
   (2.5/2, 3-1.5/2) ellipse ( 2.1 and .5); 
   
   \draw[exprColor01, edge, rounded corners=3.5 mm]
   (-.25, -.9)--
   (.35, 3.33)--
   (-3.8, 3.33)--cycle; 
   
   \draw[exprColor0, edge, rounded corners=6 mm]
   (-3.45, -1.1)--
   (-3.45, 3.45)--
   (1, 3.45)--cycle; 
   
   \draw[exprColor2, vertex] (2.5,3) circle node (v6) {$\mathbf{6}$};
   \draw[exprColor2, edge] (v6) circle (.35); 
   \draw[exprColor2, edge] (1.25, 3) ellipse (1.9 and .45); 
   \end{scope}
   
   \begin{scope}[exprtree,scale=.75, shift={(-3, 0)}]
   \drawtreenode{$\mathbf{1}~_{\sum}$};
   
   \begin{scope}[shift={(-3, -2)}]
   {\color{exprColor0}
      \drawtreenode {$\mathbf{2,5}~_{\max}$};
      \draw[treenode] (.5, .5)--(2.5, 1.5);
      \begin{scope}[shift={(0, -2)}]
      {\color{exprColor01}
         \drawtreenode{$\mathbf{3}_{~\prod}$};
         \draw[treenode] (0, .5)--(0, 1.5);
      }
      \end{scope}
   }
   \end{scope}
   
   \begin{scope}[shift={(3, -2)}]
   {\color{exprColor1}
      \drawtreenode{$\mathbf{3}~_{\prod}$};
      \draw[treenode] (-.5, .5)--(-2.5, 1.5);
   }
   \begin{scope}[shift={(0, -2)}]
   {\color{exprColor10}
      \drawtreenode{$\mathbf{4}_{~\max}$};
      \draw[treenode] (0, .5)--(0, 1.5);
   }
   \end{scope}
   \end{scope}
    \begin{scope}[shift={(7, -2)}]
      {\color{exprColor2}
         \drawtreenode{$\mathbf{6}_{~\prod}$};
         \draw[treenode] (-.9, .5)--(-6, 1.5);
      }
   \end{scope}
   \end{scope}
   \end{tikzpicture}
}

\caption{The compression step of the expression tree from Example~\ref{ex:intuition2}.
In the expression tree in Fig.~\ref{fig:prod:compart:b} that resulted from the compartmentalization step, we can merge node $\{5\}$ into its parent $\{2\}$ resulting in the above tree.}
\Description{The compression step of the expression tree from Example~\ref{ex:intuition2}.}
\label{fig:prod:compress}
\end{center}
\end{figure}

The following proposition follows directly from Definition~\ref{defn:extended-component}. In particular, since vertices in $W$ are not taken into account while determining the connectivity of extended components of $(\calH, W)$ removing a subset $U\subseteq W$ cannot break a single extended component into multiple ones.
\bprop\label{prop:one-extended-component}
Given a hypergraph $\calH=(\calV,\calE)$ and a set $W\subseteq \calV$,
if $(\calH, W)$ has only one extended component (and an empty dangling set), then for any $U\subseteq W$,
$(\calH-U, W-U)$ has only one extended component (and an empty dangling set) as well.
\eprop

\bcor\label{cor:single-product-child}
The construction of the expression tree from Definition~\ref{defn:expr-tree-prod} satisfies the following properties:
\bi
\item[(a)] In the compartmentalization step, whenever the tag of $L$ is $\otimes$, $(\calH-L, W)$ cannot have more than one extended component (and the dangling set must be empty).
\item[(b)] In the compression step, there cannot be a node $L$ and a child node $L'$ that both have the tag $\otimes$.
\item[(c)] In the final expression tree, each node whose tag is $\otimes$ must have at most one child, and the tag of that child (if it exists) is not $\otimes$.
\ei
\ecor
\bp
First, we prove part (a). Initially because we are adding a dummy
free variable $X_0$, the first tag block $L=F$ has the tag ``free''.
To construct the expression tree, we recursively construct an expression tree for each extended
component $(\calV'(C), \calE'(C))$ of $(\calH-L, W)$ (where $W$ is the set of product variables in $\calH$ that are {\em not} in $L$). For each one of those extended components, we recursively construct an expression tree for each extended component, and so on\ldots Hence apart from the time when the first tag block $L$ was $F$, we can assume that in the compartmentalization step, $(\calH, \overline W)$ has only one extended component (and an empty dangling set) where $\overline W$ is the set of {\em all} product variables in $\calH$.

Now, suppose $(\calH, \overline W)$ has a single extended component (and an empty dangling set).
Let $L$, the first tag block of $\calH$, have the tag $\otimes$.
By Proposition~\ref{prop:one-extended-component}, $(\calH-L, W:=\overline W-L)$ has only one extended component (and an empty dangling set), which proves part (a).
Moreover, the first tag block for this extended component cannot have the tag $\otimes$ because otherwise this tag block should have been part of $L$ itself (by definition of tag block).
This proves part (b).
Part (c) follows from (a) and (b).
\ep

Unlike free and semiring variables, it is possible for the same product variable to appear multiple times in the expression tree. In Example~\ref{ex:intuition2} above, $X_3$ occurs twice in the expression tree, depicted in Fig.~\ref{fig:prod:compress:a}. 
We can think of those two occurrences of $X_3$ as two different variables $X_3$ and $X'_3$ 
with identical domains $\Dom(X_3) = \Dom(X'_3)$ and rewrite $\varphi$ as:
\begin{eqnarray*}
\varphi=
{\color{exprColor}
   \sum_{x_1}}
{\color{exprColor0}
   \biggl(
   \max_{x_2}
   {\color{exprColor0}
      \max_{x_5}
      \psi_{125}}
   {\color{exprColor01}
      \prod_{x_3}
      \psi_{123}}
   \biggr)}
{\color{exprColor1}
   \biggl(
   \prod_{x_3'}
   {\color{exprColor10}
      \max_{x_4}
      \psi_{14}
      \psi_{34}}
   \biggr)}
{\color{exprColor2}
   \biggl(
   \prod_{x6}
   \psi_{6}
   \psi_{16}
   \biggr)
}
\end{eqnarray*}

Luckily, the distribution of product variables (with their multiple copies) in
the expression tree exhibits some nice properties, some of which do not even
hold for semiring variables. Those nice properties enable us to take care of
multiple copies of product variables.

In particular, given integers $i,j\in[n]-[f], i<j$, where
$\oplus^{(i)}=\oplus^{(j)}$ is a semiring aggregate, it is possible for $X_j$ to
be a strict ancestor of $X_i$ in the expression tree, due to the compression
step. ($X_j$ is a strict ancestor of $X_i$ if $X_j \in L$, $X_i \in L'$, and $L$
is a strict ancestor of $L'$ in the expression tree.)
For example, consider the following $\faq$-query.
\[\varphi=
\textstyle{\bigoplus_{x_1}\bar\bigoplus_{x_2}\bigoplus_{x_3}\bigoplus_{x_4}
\psi_{12}\psi_{23}\psi_{14}}.
\]
The expression tree roughly corresponds to the following re-writing of the
query:
\[\varphi=
\textstyle{\bigoplus_{x_1}\bigoplus_{x_4}\left[\bar\bigoplus_{x_2}\left(\bigoplus_{x_3}
\psi_{12}\psi_{23}\psi_{14}\right)\right]}.
\]
Although $X_3$ comes before $X_4$ in the original expression, $X_4$ is now a
strict ancestor of $X_3$ in the expression tree.
However, the following lemma says that the above scenario is impossible if
$\oplus^{(i)}=\oplus^{(j)}$ is a product aggregate. The lemma also allows us to
show that the comparison relation formed by the expression tree is indeed a
partial order. 

\blmm\label{lmm:no circular products}
The expression tree satisfies the following properties.
\bi
\item[(a)] For any $i\in[n]-[f]$ such that $\oplus^{(i)}=\otimes$, no copy of
     $X_i$ is a strict ancestor of another copy of $X_i$.
\item[(b)] For any $i,j\in[n]-[f]$ such that $i < j$ and $\oplus^{(i)}=\oplus^{(j)}=\otimes$, no copy of $X_j$ is a strict ancestor of any copy of $X_i$.
\ei
\elmm
\bp
First, we prove that the expression tree before the compression step 
(after performing only the query extension and compartmentalization steps)
satisfies the above properties.
The proof is by induction on the number of tag blocks. In the base case where there is only one tag block, the expression tree has only one node and the lemma follows trivially.

In the inductive step, suppose the tagged sequence has at least two tag blocks with $L$
being the first tag block. 
The expression tree $P$ can be constructed by constructing the expression trees $P_C$ of
the extended components $(\calV'(C), \calE'(C))$ and then connecting them to the root node $L$.
No $P_C$ contains any variable in $L$.
Assuming the properties hold for each $P_C$, they will hold for $P$ as well.

Finally, we prove that the final expression step (after the compression step)
satisfies the above properties.
The proof is by induction on the number of merges of a child node $L'$ into a parent node $L$ (which both have the same tag).
Thanks to part (b) of Corollary~\ref{cor:single-product-child}, we only need to consider the case where the tag of $L$ (and $L'$) is not $\otimes$.
After merging, only $L'$ can become an ancestor of some new nodes (that were not descendants of $L'$ before merging).
But since the tag of $L'$ is not $\otimes$, the induction holds.
\ep

Now we are ready to show that the expression tree defines a partial order on the
variables.
Let $P$ be the expression tree of $\varphi$ as defined in
Definition~\ref{defn:expr-tree-prod}. Define a binary relation 
$\preceq_P \subseteq \calV \times \calV$ on the variables $\calV$ as follows.
For any pair $u, v\in \calV$, we write $u \preceq_P v$ if $u=v$ or $u$ belongs
to a strict ancestor of $v$ in the expression poset $P$.
(Note that the same variable $u$ might occur several times in the expression
tree, hence it is not immediately obvious that $\preceq_P$ is indeed a partial
order.)

\bcor\label{cor:indeed a poset}
The binary relation $\preceq_P$ defines a partially ordered set.
\ecor
\bp
Reflexivity and transitivity hold trivially.
We check the antisymmetry property of $\preceq_P$.
Suppose $u \preceq_P v$, $v \preceq_P u$, but $u \neq v$. It cannot be the case
that both $u$ and $v$ are semiring variables, because each semiring variable
occurs only once in $P$. If $u$ is a semiring variable and $v$ is not, then $v$
is both a strict ancestor and a strict descendant of $u$. This means $v$ is a
strict ancestor of a copy of itself, violating part (a) of
Lemma~\ref{lmm:no circular products}. So we are left with the case when both $u$
and $v$ are product variables. But in this case, one of them (say, $u$)
comes before the other (say, $v$) in the original expression $\varphi$, and 
thus by part (b) of Lemma~\ref{lmm:no circular products} $v$ cannot be an
ancestor of $u$.
\ep

The above corollary justifies the correctness of the precedence poset
definition.

\bdefn[Precedence poset]
The {\em precedence poset} is the partially ordered set $P = (\calV, \preceq_P)$.
(We abuse notation and use $P$ to denote the poset also.)
Also, as in Definition~\ref{defn:precedencePoset} we let
$\LE(P)$ denote the set of linear extensions of $P$.
\label{defn:precedencePoset:prod}
\edefn

We next prove the soundness of $\LE(P)$. For that we need a new lemma.

\blmm
\label{lmm:non-closed-semiring}
The precedence poset $P$ satisfies the following property: For every $f+1\leq i \leq f+\ell$
(hence $\textstyle{\bigoplus^{(i)}}$ is a semiring aggregate that is not closed under $\D_I$) and for every $j$ where $\textstyle{\bigoplus^{(j)}=\bigotimes}$, we have $i \preceq_P j$.
\elmm
\bp
The lemma holds thanks to the query extension step of the construction of the expression tree (Definition~\ref{defn:expr-tree-prod}).
First, we prove that the expression tree after the query extension and compartmentalization steps (and before the compression step) satisfies the following property:
Every variable $X_i$ for $f+1\leq i \leq f+\ell$ is a strict ancestor of every product variable $X_j$ in the expression tree.
The proof is by induction on the number of tag blocks.
The base case of one tag block holds trivially.
In the inductive step, suppose the tagged sequence has at least two tag blocks with $L$ being the first tag block.
We recognize two cases:

\noindent
{\bf Case 1.}
If the tag of $L$ is $\otimes$, then there cannot be any variable $X_i$ for $f+1\leq i \leq f+\ell$ since $X_i$ should have appeared in an earlier tag block by Definition~\ref{defn:inner-faq}. Hence, the property holds trivially.

\noindent
{\bf Case 2.}
If the tag of $L$ is not $\otimes$, then each extended component $(\calV'(C),\calE'(C))$ that contains any variable $X_i$ for $f+1\leq i \leq f+\ell$
must also contain every product variable $X_j$ thanks to the factors $\psi_{\{i,j\}}$ that were added in the query extension step. Assuming the property holds for the expression tree for each extended component $(\calV'(C),\calE'(C))$, it will hold for the combined expression tree as well.

Finally, we prove that the expression tree after the compression step satisfies the above property as well. We do that by induction on the number of merges of a child node $L'$ into a parent node $L$ (which both have the same tag).
When we merge, the only constraints $i\preceq_P j$ that are removed from $\preceq_P$
are the ones where $i\in L$, $j\in L'$. But since for those constraints we have $\oplus^{(i)}=\oplus^{(j)}$, the induction holds.
\ep

\bthm[$\LE(P) \subseteq \EVO(\varphi, \calF(\D_I))$]\label{thm:LE=EVO logic case}
Every linear extension of the precedence poset is $\varphi$-equivalent.
\label{thm:sound with prod}
\ethm
\bp
Similar to the proof of Theorem~\ref{thm:LE subseteq EVO K blocks case}, we first
consider the expression tree constructed using only the query extension and compartmentalization steps, without the compression step. (In Example~\ref{ex:intuition2}, that would be the poset defined by the tree on the right of Figure~\ref{fig:prod:compart:b}.)
From the proof of Lemma~\ref{lmm:no circular products} and then repeating the same proof of Corollary~\ref{cor:indeed a poset}, this expression tree also defines a partially ordered set, which we call the {\em compartmentalization poset}.
We first prove soundness for the compartmentalization poset.
A slightly tricky issue in this case compared to that of Theorem~\ref{thm:LE subseteq EVO K blocks case} is the multiple occurrences of some product 
variables. 
Let $\sigma=(1,\ldots,n)$ be the variable ordering used to write $\varphi$.
We prove the soundness of the compartmentalization poset by induction on the number of tag blocks of the input sequence. 

The base of one single tag block holds trivially.
In the inductive step, suppose $\sigma$ has at least two tag blocks with $L$
being the first tag block.
Let $W$ be the set of product variables not in $L$.
For each extended component $(\calV'(C), \calE'(C))$
of $(\calH-L, W)$, we form a sub-sequence $\sigma_C$, which defines an
$\faq$-expression $\varphi_C$. This expression $\varphi_C$ is on the 
conditional factors $\psi_S( \cdot \suchthat \mv x_L)$ for each $S\in\calE$ 
such that $S\cap \calV(C) \neq \emptyset$. (Recall that $\calV(C)$ is different 
from $\calV'(C)$ because $\calV'(C)$ also contains some product variables from $W$.)
We also construct the expression $\varphi_D$ for the dangling set $D$ of $(\calH-L, W)$.
This expression is on the conditional factors
$\psi_S( \cdot \suchthat \mv x_L)$ for each $S\in\calE$ 
such that $S\setminus L \subseteq W$.

{\bf Claim 1.}
The conditional factor $\varphi(\cdot|\mv x_L)$ can be factorized into a product of the $\faq$-expressions $\varphi_C$ and $\varphi_D$ (if $D$ is not empty).

To prove Claim 1, first notice that $\varphi(\cdot|\mv x_L)$ can be written as
\begin{eqnarray*}
   \varphi(\mv x_L|\mv x_L) &=& 
   \bigoplus_{\mv x_{[n]\setminus L}}
   \bigotimes_{S\in \calE}\psi_{S}(\mv x_S|\mv x_L)\\
   &=&
   \bigoplus_{\mv x_{[n]\setminus L}}
   \bigotimes_{C}
   \underbrace{\left(\bigotimes_{S:S\cap \calV(C)\neq \emptyset}
   \psi_{S}(\mv x_S|\mv x_L)\right)}_{=:\overline\varphi_C(\mv x_{\calV'(C)\cup L}|\mv x_L)}
   \otimes
   \underbrace{\left(\bigotimes_{S:S\setminus L\subseteq W}
   \psi_{S}(\mv x_S|\mv x_L)\right)}_{=:\overline\varphi_D(\mv x_{D\cup L}|\mv x_L)}
\end{eqnarray*}
For simplicity, suppose that the dangling set $D$ is empty, i.e. there is no hyperedge $S$ where $S\setminus L \subseteq W$.
Depending on the aggregate $\oplus^{(n)}$, we recognize two cases:
\begin{itemize}
   \item If $\oplus^{(n)}$ is a semiring aggregate, let $C_n$ be the (unique) component where $n\in\calV(C_n)$.
   Then, $\varphi(\cdot|\mv x_L)$ can be written as
   \begin{eqnarray*}
   \varphi(\mv x_L|\mv x_L)&=&
   \bigoplus_{\mv x_{[n-1]\setminus L}}
   \bigoplus_{x_{n}}^{(n)}
   \bigotimes_{C}
   \overline\varphi_C(\mv x_{\calV'(C)\cup L}|\mv x_L)
   \\
   &=&\bigoplus_{\mv x_{[n-1]\setminus L}}
   \left[\left(
   \bigoplus_{x_{n}}^{(n)}
   \overline\varphi_{C_n}(\mv x_{\calV'(C_n)\cup L}|\mv x_L)
   \right)
   \otimes
   \bigotimes_{C:n\notin\calV(C)}
   \overline\varphi_C(\mv x_{\calV'(C)\cup L}|\mv x_L)\right]
   \end{eqnarray*}
   \item If $\oplus^{(n)}$ is a product aggregate, then there can be multiple components $C$ where $n \in \calV'(C)$ and we can rewrite $\varphi(\cdot|\mv x_L)$ as
   \begin{eqnarray}
   \varphi(\mv x_L|\mv x_L)&=&
   \bigoplus_{\mv x_{[n-1]\setminus L}}
   \bigotimes_{x_{n}}
   \bigotimes_{C}
   \overline\varphi_C(\mv x_{\calV'(C)\cup L}|\mv x_L)
   \nonumber\\
   &=&\bigoplus_{\mv x_{[n-1]\setminus L}}
   \left[\bigotimes_{C:n\in\calV'(C)}
   \left(\bigotimes_{x_{n}}
   \overline\varphi_C(\mv x_{\calV'(C)\cup L}|\mv x_L)\right)
   \otimes
   \bigotimes_{C:n\notin\calV'(C)}
   \overline\varphi_C(\mv x_{\calV'(C)\cup L}|\mv x_L)^{|\Dom(X_n)|}\right]
   \nonumber\\
   &=&\bigoplus_{\mv x_{[n-1]\setminus L}}
   \left[\bigotimes_{C:n\in\calV'(C)}
   \left(\bigotimes_{x_{n}}
   \overline\varphi_C(\mv x_{\calV'(C)\cup L}|\mv x_L)\right)
   \otimes
   \bigotimes_{C:n\notin\calV'(C)}
   \overline\varphi_C(\mv x_{\calV'(C)\cup L}|\mv x_L)\right]
   \label{eqn:soundness:prod:decompose}
   \end{eqnarray}
   The last step above relies on the idempotence of $\otimes$ under $\D_I$.
\end{itemize}
By repeatedly applying the above process on the aggregates $\oplus^{(n-1)},\ldots,\oplus^{(|L|+1)}$, we can factorize $\varphi(\cdot|\mv x_L)$ into
\begin{equation}
\varphi(\mv x_L|\mv x_L)=
\bigotimes_{C}
\underbrace{\left(\bigoplus_{\mv x_{\calV'(C)}}
\overline\varphi_C(\mv x_{\calV'(C)\cup L}|\mv x_L)\right)}_{=:\varphi_C(\mv x_L|\mv x_L)}
\label{eqn:soundness:factor}
\end{equation}
Note that by Definition~\ref{defn:inner-faq} for every product aggregate $\oplus^{(i)}=\otimes$, all subsequent aggregates $\oplus^{(i+1)},\ldots,\oplus^{(n)}$ are closed under $\D_I$.
Hence, we can repeatedly use idempotence in the same way as above.
Note also that if the dangling set $D$ was not empty,
then eq.~\eqref{eqn:soundness:factor} would have an additional factor
$\varphi_D(\mv x_L|\mv x_L):=\bigotimes_{\mv x_D}\overline\varphi_D(\mv x_{D\cup L}|\mv x_L)$.

Now, let $\pi=(u_1,\ldots,u_n)$ be an arbitrary linear extension of the compartmentalization
poset for $\varphi$ (hence $\{u_1,\ldots,u_{|L|}\}=L$). 
For each extended component $(\calV'(C), \calE'(C))$, 
let $\pi_C$ denote the subsequence of $\pi$ obtained by
picking out variables in $\calV'(C)$. Then, $\pi_C$ is a linear extension of the
(compartmentalization) poset for $\varphi_C$, and thus $\pi_C \in
\EVO(\varphi_C, \calF(\D_I))$ by induction.
Let $k:=u_{|L|+1}$.
We recognize two cases depending on the aggregate $\oplus^{(k)}$.
\begin{itemize}
   \item If $\oplus^{(k)}$ is a semiring aggregate, let $C_k$ be the (unique) component where $k\in\calV(C_k)$ (hence $\pi_{C_k}$ must begin with the variable $k$).
   Then, we can rewrite $\varphi(\cdot|\mv x_L)$ as
   \begin{eqnarray*}
   \varphi(\mv x_L|\mv x_L)
   &=&
   \left[\bigoplus^{(k)}_{x_{k}}
   \left(\bigoplus_{\mv x_{\calV'(C_k)\setminus \{k\}}}
   \overline\varphi_{C_k}(\mv x_{\calV'(C_k)\cup L}|\mv x_L)\right)\right]
   \otimes
   \bigotimes_{C:k\notin \calV(C)}
   \left(\bigoplus_{\mv x_{\calV'(C)}}
   \overline\varphi_C(\mv x_{\calV'(C)\cup L}|\mv x_L)\right)
   \\
   &=&
   \bigoplus^{(k)}_{x_{k}}
   \left[
   \left(\bigoplus_{\mv x_{\calV'(C_k)\setminus \{k\}}}
   \overline\varphi_{C_k}(\mv x_{\calV'(C_k)\cup L}|\mv x_L)\right)
   \otimes
   \bigotimes_{C:k\notin \calV(C)}
   \left(\bigoplus_{\mv x_{\calV'(C)}}
      \overline\varphi_C(\mv x_{\calV'(C)\cup L}|\mv x_L)\right)
   \right]
   \end{eqnarray*}
   \item If $\oplus^{(k)}$ is a product aggregate, then there can be multiple components $C$ for which $k\in\calV'(C)$, and for each such component $C$, we have $\pi_{C}$ begins with the variable $k$.
   Moreover by Lemma~\ref{lmm:non-closed-semiring}, all subsequent aggregates in $\pi$,
   $\oplus^{(u_{|L|+2})},\ldots,\oplus^{(u_{n})}$ are closed under $\D_I$.
   Therefore, we can rewrite $\varphi(\cdot|\mv x_L)$ as
   \begin{eqnarray}
      &&\varphi(\mv x_L|\mv x_L)\nonumber\\
      &=&
   \left[\bigotimes_{x_k}
   \bigotimes_{C:k\in\calV'(C)}
   \left(\bigoplus_{\mv x_{\calV'(C)\setminus\{k\}}}
      \overline\varphi_C(\mv x_{\calV'(C)\cup L}|\mv x_L)\right)
   \right]
      \otimes
      \bigotimes_{C:k\notin\calV'(C)}
      \left(\bigoplus_{\mv x_{\calV'(C)}}
      \overline\varphi_C(\mv x_{\calV'(C)\cup L}|\mv x_L)\right)
      \nonumber\\&=&
      \bigotimes_{x_k}
      \left[\bigotimes_{C:k\in\calV'(C)}
      \left(\bigoplus_{\mv x_{\calV'(C)\setminus\{k\}}}
      \overline\varphi_C(\mv x_{\calV'(C)\cup L}|\mv x_L)\right)
      \otimes
      \bigotimes_{C:k\notin\calV'(C)}
      \left(\bigoplus_{\mv x_{\calV'(C)}}
      \overline\varphi_C(\mv x_{\calV'(C)\cup L}|\mv x_L)\right)
      \right]
      \label{eqn:soundness:prod:interleave}
   \end{eqnarray}
   The last step above relies on $\otimes$ and all aggregates $\oplus_{\mv x_{\calV'(C)}}$ being closed under $\D_I$ (as guaranteed by Lemma~\ref{lmm:non-closed-semiring}) and on $\otimes$ being idempotent under $\D_I$.
\end{itemize}

By repeatedly applying the above process on the aggregates $\oplus^{(u_{|L|+2})},\ldots,\oplus^{(u_{n})}$,
we can show that the $\faq$-expression $\varphi^\pi$
defined by the variable ordering $\pi$ is identical to $\varphi$.
This proves soundness for the compartmentalization poset.

Finally, we prove soundness for the final poset after the compression step. Recall from part (b) of Corollary~\ref{cor:single-product-child} that product variables play no role in the compression step.
Hence, this part of the proof is identical to the corresponding part of the proof of Theorem~\ref{thm:LE subseteq EVO K blocks case}.
\ep

To prove completeness of $\LE(P)$, we will need the following lemma which basically says that in any equivalent variable ordering, product variables must appear after semiring variables that are not closed under $\D_I$.

\blmm
Every variable ordering 
$\pi=(u_1, \ldots, u_n)\in\EVO(\varphi, \calF(\D_I))$ satisfies the following property:
for every variable $u_i$ where $\oplus^{(u_i)}$ is a semiring aggregate that is not closed under $\D_I$ and for every variable $u_j$ where $\oplus^{(u_j)}=\otimes$, we have $i < j$.
\label{lmm:non-closed-semiring-completeness}
\elmm
\bp
Suppose otherwise that there is a variable $u_i$ where $\oplus^{(u_i)}$ is a semiring aggregate that is not closed under $\D_I$ and another variable $u_j$ where $\oplus^{(u_j)}=\otimes$ and $i > j$.
Note that by Definition~\ref{defn:inner-faq}, $u_i$ must appear before $u_j$ in the original $\faq$ expression $\varphi$, i.e. $u_i < u_j$.
Let $\varphi^{\pi}$ be the expression $\varphi$ with the aggregates permuted using the variable ordering $\pi$.
We define input factors $\psi_S$ for which $\varphi^{\pi}\not\equiv\varphi$ as follows.
For every variable $\ell\in[n]\setminus\{u_i, u_j\}$, we fix one arbitrary value 
$e_{\ell}\in\Dom(X_{\ell})$. 
For the sake of brevity, denote $\oplus = \oplus^{(u_i)}$ and 
$\bar\oplus = \oplus^{(u_j)}$.
Pick an arbitrary input factor $\psi_{S_i}$ where $u_i \in S_i$ and define $\psi_{S_i}$ as follows (where $\phi'_{u_i}(x_{u_i})$ was defined by~\eqref{eqn:non-commutative-defn-prod-non-closed}):
\[
\psi_{S_i}(\mv x_{S_i}) :=
\begin{cases}
\phi'_{u_i}(x_{u_i}) & \text{ if $x_{l}=e_{l}$ for all $l\in S_i\setminus\{u_i, u_j\}$}\\
\mv 0 &\text{ if  there is $l\in S_i\setminus\{u_i, u_j\}$ where $x_{l}\neq e_{l}$ and $l$ is a semiring variable}\\
\mv 1 &\text{ otherwise}
\end{cases}
\]
Every other input factor $\psi_{S}$ for $S\in\calE\setminus\{S_i\}$ is defined as follows:
\[
\psi_{S}(\mv x_S) :=
\begin{cases}
\mv 1 & \text{ if $x_{l}=e_{l}$ for all $l\in S\setminus\{u_i, u_j\}$}\\
\mv 0 &\text{ if  there is $l\in S\setminus\{u_i, u_j\}$ where $x_{l}\neq e_{l}$ and $l$ is a semiring variable}\\
\mv 1 &\text{ otherwise}
\end{cases}
\]
Note that $\varphi$ evaluates to the left-hand side of \eqref{eq:no-swap-product-semiring} while $\varphi^{\pi}$ evaluates to the right-hand side (where $i$ and $j$ in \eqref{eq:no-swap-product-semiring} are replaced here by $u_i$ and $u_j$ respectively).
\ep

We now follow the script of Section~\ref{subsec:evo:Kblocks} to prove the
completeness of $\LE(P)$.

\blmm
For every variable ordering 
$\pi=(u_1, \ldots, u_n)\in\EVO(\varphi, \calF(\D_I))$, the
variable $u_{f+1}$ must belong to a child node of the root of the expression 
tree.
\label{lmm:u_f+1 in L with prod}
\elmm
\bp[Proof Sketch]
The proof is similar to that of Lemma~\ref{lmm:u_f+1 in L}. The main difference is that instead of considering a connected component, we will be considering an extended component. Hence in the path $(i_0:=u_{f+1}, i_1, \ldots, i_k \in L_p)$, we can assume that none of the intermediate vertices $i_1, \ldots, i_{k-1}$ is a product variable. In addition, instead of using the $\phi_{i_{k-1}i_k}$ defined in Proposition~\ref{prop:non-commutative}, we will be using the one defined in Proposition~\ref{prop:non-commutative-var} (whose range is $\D_I$). Also for all $S\in\calE$, instead of $\psi_S(\mv x_S)$ being $\mv 0$ whenever there is a variable $\ell' \in S \setminus \{i_0, \ldots, i_k\}$ for which $x_{\ell'}\neq e_{\ell'}$, the value of $\psi_S(\mv x_S)$ could now be $\mv 0$ or $\mv 1$ depending on whether $\bigoplus^{(\ell')}$ is a semiring or product aggregate. In particular, if there is $\ell' \in S \setminus \{i_0, \ldots, i_k\}$ for which $x_{\ell'}\neq e_{\ell'}$ and $\bigoplus^{(\ell')}$ is a semiring aggregate, then $\psi_S(\mv x_S)=\mv 0$. Otherwise, if there is $\ell' \in S \setminus \{i_0, \ldots, i_k\}$ for which $x_{\ell'}\neq e_{\ell'}$ and $\bigoplus^{(\ell')}$ is a product aggregate, then $\psi_S(\mv x_S)=\mv 1$. Otherwise, $\psi_S(\mv x_S)$ is defined as in the proof of Lemma~\ref{lmm:u_f+1 in L}.

Note the subtle issue that if some factor $\psi_S$ that we used in the above construction was a factor that was added to $\varphi$ in the query extension step of the construction of the expression tree (Definition~\ref{defn:expr-tree-prod}), then the above construction does not work (because the original $\faq$ query $\varphi$ does not have such a factor).
Therefore, we need to verify that non of the factors $\psi_{S_1}, \ldots, \psi_{S_k}$ defined by equations~\eqref{eq:construction:psi_S_k} and \eqref{eq:construction:psi_S_j}
corresponds to a factor that was added in the query extension step (and not to an original input factor).
\bi
\item None of the vertices $i_1, \ldots, i_{k-1}$ is a product variable, and therefore none of the factors $\psi_{S_2}, \ldots, \psi_{S_{k-1}}$ could have been added in the query extension step (since each factor that is added in the query extension step has one product variable and one semiring variable that is not closed under $\D_I$).
\item It is not possible for $i_k$ to be a product variable and $i_{k-1}$ to be a semiring variable that is not closed under $\D_I$ at the same time.
This is because $i_k\in L_p$, $i_{k-1}\notin L_p \cup \overline L_a$ (hence $i_k$ is a strict ancestor of $i_{k-1}$ in the expression tree $P$) however by Lemma~\ref{lmm:non-closed-semiring} we should have had $i_{k-1}\preceq_P i_k$.
Therefore, $\psi_{S_k}$ in \eqref{eq:construction:psi_S_k} could not have been added in the query extension step.
\item If $i_0=u_{f+1}$ was a product variable and $i_1$ was a semiring variable that is not closed under $\D_I$, then $\psi_{S_1}$ could have been added in the query extension step. However in this particular case, we could have used Lemma~\ref{lmm:non-closed-semiring-completeness} directly to show that $\pi\notin\EVO(\varphi, \calF(\D_I))$.
\ei
\ep

The definition of componentwise-equivalence remains the same as that of
Section~\ref{subsec:evo:Kblocks} except that instead of taking connected components, we will be taking extended components (See Definition~\ref{defn:extended-component}).
\bdefn[Componentwise-equivalence]\label{defn:CWE with prod}
Let $\varphi$ be an \faq-query of the form \eqref{eqn:faqIdempotent},
$\calH$ be the hypergraph of $\varphi$,
$W$ be the set of all product variables in $\varphi$, $\sigma=(v_1, \ldots, v_n)\in\EVO(\varphi, \calF(\D_I))$ be a variable ordering.
Let $\pi=(u_1, \ldots, u_n)$ be another variable ordering with
$\{u_1,\dots,u_f\} = F$.
Then, $\pi$ is said to be \emph{componentwise-equivalent} 
(or shortly $\CW$-equivalent) to $\sigma$ if and only if:
\bi
\item either $n=1$,
\item or $(\calH, W)$ has at least two extended components, and for each extended 
    component $(\calV'(C), \calE'(C))$ of $(\calH, W)$, $\pi_C$ is $\CW$-equivalent
    to $\sigma_C$, where $\sigma_C$ (respectively, $\pi_C$) is the 
    variable ordering of $\calV'(C)$ that
    is consistent with $\sigma$ (respectively, $\pi$),
\item or $u_{1}=v_{1}$, and for each extended 
    component $(\calV'(C), \calE'(C))$ of $(\calH-\{v_{1}\}, W-\{v_1\})$,
    $\pi_C$ is $\CW$-equivalent to $\sigma_C$, where $\sigma_C$ 
    (respectively, $\pi_C$) is the
    ordering of $\calV'(C)$ that is consistent with $\sigma$ 
    (respectively, $\pi$).
\ei
Given a set of variable orderings $\Lambda \subseteq \EVO(\varphi, \calF(\D_I))$, we use 
$\CWE(\Lambda)$ to denote the set of all variable orderings that are 
$\CW$-equivalent to some variable ordering in $\Lambda$.
\edefn
Notice that variables in the dangling set can be placed anywhere later in the ordering, and hence they are ignored in the above definition.

The following proposition can be proved in exactly the same way as Proposition~\ref{prop:CWE preserves faqw}, hence the proof is omitted.
\bprop\label{prop:CWE preserves faqw with prod}
Let $\pi$ be a variable ordering that is $\CW$-equivalent to 
$\sigma \in \EVO(\varphi, \calF(\D_I))$, we have
$\pi\in\EVO(\varphi, \calF(\D_I))$ and
$\faqw(\sigma)=\faqw(\pi)$.
\eprop

\bthm[$\EVO(\varphi, \calF(\D_I))=\CWE(\LE(P))$]
Let $\varphi$ be an $\faq$-expression of the form \eqref{eqn:faqIdempotent}.
A variable ordering $\sigma$ is $\varphi$-equivalent if and only if it is
\CW-equivalent to some ordering $\pi$ which is a linear extension of the
precedence poset $P$.
\label{thm:completness with prod}
\ethm
\bp
Similar to the proof of Theorem~\ref{thm:kblocks:completeness}.
The containment $\CWE(\LE(P))\subseteq\EVO(\varphi, \calF(\D_I))$ follows from Proposition~\ref{prop:CWE preserves faqw with prod} and 
Theorem~\ref{thm:sound with prod}. Hence, we only need to prove the other 
direction. 

We prove the containment
$\EVO(\varphi, \calF(\D_I)) \subseteq \CWE(\LE(P))$ by induction with 
a slightly stronger induction hypothesis.
We show that, if $\sigma$ is $\varphi$-equivalent, then there is a linear 
extension $\pi \in \LE(P)$ that is \CW-equivalent to $\sigma$ and all product variables in $\pi$ have exactly the same relative order 
as that in $\sigma$.

Without loss of generality we can assume that the root of the expression tree 
$F$ is empty, and it has only one child node $L$.
Let $\calH$ be the hypergraph of $\varphi$, and $W$ be the set of all product variables in $\varphi$.
Fix an arbitrary $\sigma=(v_1, \ldots, v_n)\in\EVO(\varphi)$. By Lemma~\ref{lmm:u_f+1 in L with prod}, $v_1\in L$.
We recognize two cases:
\bi 
\item If the tag of $L$ is not $\otimes$, then for each extended component $(\calV'(C),\calE'(C))$  of $(\calH-\{v_1\}, W)$, define a 
sub-query $\varphi_{C}$ with variable ordering $\sigma_C$ 
on the conditional factors 
\[ \left\{
    \psi_S(\cdot\suchthat x_{v_1})
    \suchthat S \in\calE \wedge S \cap \calV(C) \neq \emptyset
\right\},
\]
where $\sigma_{C}$ is the subsequence of $\sigma$ obtained by
picking out vertices in $\calV'(C)$. 
Let $P_C$ be the precedence poset of the expression tree for 
$\varphi_{C}$. By induction on the number of variables, we know that
there is a variable ordering
$\pi_{C}\in\LE(P_C)$ that is \CW-equivalent to $\sigma_{C}$ where product variables (if any) in $\pi_{C}$ maintain their relative order in $\sigma_C$.
Although the same product variable(s) might appear in different extended components,
we can pick a variable ordering $\pi$ that is consistent with all $\pi_{C}$ 
such that $\pi$ starts with $v_1$, followed by variables in $L-\{v_1\}$.
(Recall that the expression tree $P$ can be constructed by creating a root node $R$ containing $\{v_1\}$ with subtrees which are the expression trees $P_C$ and then mering into $R$ the children $L_C$ that have the same tag as $v_1$. This tag is not a product in here.)
In particular, to construct $\pi$, we start off with a sequence $\pi$ which is a subsequence of $\sigma$ consisting of only product variables. Then for each $\pi_C$, we insert the semiring variables of $\pi_C$ into $\pi$ while maintaining the relative order of $\pi_C$.
Moreover, for every $\pi_C$ where the root $L_C$ of $P_C$ has the same (semiring) tag as $v_1$, 
we make sure that the (semiring) variables in $L_C$ are inserted to (and remain at) the beginning of $\pi$.
Finally, we insert $v_1$ to the beginning of $\pi$.
We can verify that the final $\pi$ starts with $v_1$ followed by variables in $L-\{v_1\}$, $\pi$ is consistent with all $\pi_C$, $\pi\in\LE(P)$, $\pi$ is \CW-equivalent to $\sigma$, and product variables in $\pi$ maintain their relative order in $\sigma$.
\item If the tag of $L$ is $\otimes$, then $(\calH-\{v_1\}, W-\{v_1\})$ cannot have more than on extended component $(\calV'(C),\calE'(C))=\calH-\{v_1\}$, by Proposition~\ref{prop:one-extended-component}.
Let $\varphi_C$, $\sigma_C$, and $P_C$ be defined as before.
Then, by induction there is a variable ordering $\pi_{C}\in\LE(P_C)$ that is \CW-equivalent to $\sigma_{C}$ where product variables (if any) in $\pi_{C}$ maintain their relative order in $\sigma_C$.
Let $\pi$ start with $v_1$ followed by $\pi_C$.
It follows that $\pi\in\LE(P)$, $\pi$ is \CW-equivalent to $\sigma$, and product variables in $\pi$ maintain their relative order in $\sigma$.
\ei
\ep
Recall Definition~\ref{defn:faqw-phi-F} of $\faqw(\varphi, \calF)$.
\bcor We have
$$\faqw(\varphi, \D_I) =
    \min \left\{ \faqw(\sigma) \suchthat \sigma \in \LE(P) \right\}.$$ 
\label{cor:faqw-char-with-prod}
\ecor

\subsection{General $\faq$ query}
\label{subsec:evo:general:faq}

The only case of $\faq$ that does not fall under the previous section is the case where we have non-idempotent product aggregates. This general case is not very natural as there aren't many practical examples that can only be represented using this form. For completeness, we describe how to handle it in this section.

Definition~\ref{defn:idempotent product agg} defines idempotence from an algorithmic point of view: A product aggregate is called idempotent at the time it is about to be eliminated in $\InsideOut$ if the ranges of all factors containing it at that time are idempotent w.r.t. to $\otimes$. This definition not only depends on the specific elimination ordering being used, but also on the input factors (i.e. data-dependent). While this definition makes sense from an algorithmic point of view, it does not capture idempotence at semantic level.

From a semantic point of view, we can only reason about whether the product is idempotent under the entire domain $\D$ (or at least a closed subset of the domain $\D_I$, as we did in Section~\ref{subsec:evo:inner:faq}). More specifically in this section, instead of thinking of idempotence as a property of each product aggregate, we will think of it as a property of the product operator $\otimes$ under the domain $\D$. A product operator $\otimes$ is \emph{idempotent under $\D$} if and only if $a\otimes a=a$ for all $a\in \D$. Here, we are interested in the case where $\otimes$ is \emph{not} idempotent under $\D$.

We will explain the basic ideas using a simple example.
\begin{ex}[$\faq$ with non-idempotent $\otimes$ under $\D$]
\label{ex:faq:non-idem}
Consider the following $\faq$.
\[\varphi=\sum_{x_1}\prod_{x_2}\sum_{x_3}\psi_{13}(x_1, x_3)\psi_{2}(x_2).\]
Notice that $\varphi$ consists of two connected components: $\{1, 3\}$ and $\{2\}$.
(And using Definition~\ref{defn:extended-component} with $W=\{2\}$ being the set of product variables, $(\calH, W)$ consists of one extended component $\calV'(C)=\{1, 3\}$ in addition to a dangling set $D=\{2\}$). However, assuming the product is non-idempotent under $\D$, $\varphi$ is written as
\[\varphi=\left[\sum_{x_1}\left(\sum_{x_3}\psi_{13}(x_1, x_3)\right)^{|\Dom(X_2)|}\right]\left[\prod_{x_2}\psi_{2}(x_2)\right].\]
Although $x_2$ is connected to neither $x_1$ nor $x_3$, it imposes an order on them.
However if we construct the expression tree using Definition~\ref{defn:expr-tree}
or Definition~\ref{defn:expr-tree-prod} (without the query extension step, which was described only in the context of $\faq$ queries of the form given by Definition~\ref{defn:inner-faq}),
then the expression tree is going to be equivalent to the following expression:
\[\varphi=\left[\sum_{x_1, x_3}\psi_{13}(x_1, x_3)\right]\left[\prod_{x_2}\psi_{2}(x_2)\right].\]
The above expression fails to capture the fact that $x_1$ must come before $x_3$.
(Notice that this situation does not necessarily happen between sibling nodes in the expression tree. It could have happened between arbitrary nodes that are incomparable under the precedence poset of the expression tree.)

To solve this issue, we will use a different (and more general) query extension step than the one in Definition~\ref{defn:expr-tree-prod}, then we will apply the same compartmentalization and compression steps from Definition~\ref{defn:expr-tree-prod} on the extended query.
In particular, the query extension step that we will use here consists of extending all input factors in the original $\faq$ query by adding all product variables to each one of them. In this example, we extend $\psi_{13}$ into $\psi'_{123}$ which is defined for all $(x_1, x_2, x_3)\in\Dom(X_1)\times\Dom(X_2)\times\Dom(X_3)$ as follows.
\[\psi'_{123}(x_1, x_2, x_3)=\psi_{13}(x_1, x_3).\]
The extended $\varphi$ (let's call it $\varphi'$) will be
\[\varphi'=\sum_{x_1}\prod_{x_2}\sum_{x_3}\psi'_{123}(x_1, x_2, x_3)\psi_{2}(x_2).\]
The expression tree of the extended query $\varphi'$ (constructed using the compartmentalization and compression steps of Definition~\ref{defn:expr-tree-prod}) is going to be equivalent to the following expression.
\[\varphi'=\left[\sum_{x_1}\prod_{x_2}\sum_{x_3}\psi'_{123}(x_1, x_2, x_3)\right]\left[\prod_{x_2}\psi_{2}(x_2)\right].\]
Now, this new expression forces $x_1$ to come before $x_3$ in the order. Notice that the above extension takes place only at semantic level (i.e. during the construction of the expression tree). At algorithmic level, we will still be working over the original $\varphi$. For example after eliminating $x_3$ above in $\InsideOut$, $x_2$ is going to be eliminated from $\sum_{x_3}\psi'_{123}(x_1, x_2, x_3)$ by raising the original $\sum_{x_3}\psi_{13}(x_1, x_3)$ to the power of $|\Dom(X_2)|$ (using repeated squaring).
\end{ex}

\bdefn[Expression tree and precedence poset]
Given an $\faq$ query $\varphi$ (as defined in Section~\ref{sec:faq-pbm}) with hypergraph $\calH$ and domain $\D$ where $\otimes$ is not idempotent under $\D$, let $K$ be the set of semiring aggregates and free variables (as defined by \eqref{eqn:the:set:K}) and $\bar K:=[n]-K$ be the set of product aggregates. We define an extended $\faq$ query $\varphi'$ with domain $\D$ and hypergraph $\calH'=([n], \calE')$ where
\[\calE':=\left\{S\cup\bar K\suchthat S\in\calE\right\},\]
and $\varphi': \prod_{i\in [f]} \Dom(X_i) \to \D$ is defined as
\[
\varphi'(\mv x_{[f]}) :=
    \textstyle{\bigoplus^{(f+1)}_{x_{f+1}} \cdots
    \bigoplus^{(n)}_{x_{n}}}
    \bigotimes_{S'\in\calE'}\psi'_{S'}(\mv x_{S'}),
\]
where for each $S\cup\bar K\in \calE'$, $\psi'_{S\cup \bar K}: \prod_{i\in S\cup \bar K} \Dom(X_i) \to \D$ is defined as
\[\psi'_{S\cup \bar K}(\mv x_{S\cup \bar K}):=\psi_S(\mv x_S).\]
The expression tree of $\varphi$ is constructed by applying
the compartmentalization and compression steps of Definition~\ref{defn:expr-tree-prod} on the extended query $\varphi'$.
The precedence poset is constructed by applying Definition~\ref{defn:precedencePoset:prod} on the expression tree of $\varphi$.
\label{defn:expr-tree:non-idem}
\edefn

Notice that because every factor in $\varphi'$ contains all product variables $\bar K$, it doesn't really matter anymore whether product is idempotent or not. In particular, the situation of having to raise to a power of $|\Dom(X_i)|$ won't be happening anymore in $\varphi'$ (at semantic level). Hence, we can run the exact same analysis from Section~\ref{subsec:evo:inner:faq} on $\varphi'$ in order to prove that:
\bthm[$\LE(P) \subseteq \EVO(\varphi)$]
Every linear extension of the precedence poset $P$ is $\varphi$-equivalent.
\label{thm:soundness:general}
\ethm
\bp
The proof is the same as that of Theorem~\ref{thm:sound with prod} except for two differences.
In particular, in~\eqref{eqn:soundness:prod:decompose} there won't be any component $C$ for which $n\notin\calV'(C)$ (because $\oplus^{(n)}$ is a product aggregate and Definition~\ref{defn:expr-tree:non-idem} adds $n$ to every $\calV'(C)$).
Because of that, we won't need the idempotence assumption anymore in ~\eqref{eqn:soundness:prod:decompose}.
Similarly, in~\eqref{eqn:soundness:prod:interleave} there won't be any component $C$ for which $k\notin\calV'(C)$, hence we won't need idempotence or Lemma~\ref{lmm:non-closed-semiring} (which we don't have here).
\ep

Notice again that we are merely using $\varphi'$ to semantically determine which orderings $\sigma$ are in $\EVO(\varphi)$. The algorithm (e.g. \InsideOut) will still be running on $\varphi$.

While the above definition of the expression tree captures completeness at an intuitive level (as we argued for in Example~\ref{ex:faq:non-idem}), achieving completeness in a rigorous way requires long definitions and unnatural assumptions. Because this class of $\faq$ is not very well-motivated by practical examples, we skip the rigorous completeness definitions/proofs for this section and leave them as an open problem.

\bopm
\label{opm:completeness:general}
Prove that $\EVO(\varphi)=\CWE(\LE(P))$ for $\faq$ queries of the general form given in Section~\ref{sec:faq-pbm}.
\eopm

\section{Approximating the $\faq$-width}
\label{sec:approx}

Given a hypergraph $\calH$ and a rational number $w\geq 1$ as input, it is $\np$-hard to decide whether $\fhtw(\calH)\leq w$~\cite{Marx:2010:AFH:1721837.1721845}.
Recently, it has been shown that even for a fixed $w$, and even for $w=2$, the problem is still $\np$-hard~\cite{DBLP:journals/corr/FischlGP16}.
By extension, the problem of computing a tree decomposition with the (optimal) fractional hypertree width is also $\np$-hard 
(since having that tree decomposition enables computing $\fhtw$ in polynomial time).
As was shown in Proposition~\ref{prop:faqw=fhtw}, our $\faq$-width ($\faqw$) is an exact generalization of the fractional hypertree width: it coincides with $\fhtw$ for
$\sumprod$ queries and for queries where all variables are free.
By extension, computing $\faqw$ and finding a tree decomposition with the optimal $\faqw$ are both $\np$-hard.

Marx~\cite{Marx:2010:AFH:1721837.1721845} had suggested an approximation algorithm that, given a hypergraph $\calH$, outputs a tree decomposition of $\calH$ whose fractional hypertree width is $O(\fhtw^3(\calH))$ in time $O(|\calH|^{O(\fhtw^3(\calH))})$.
Hence Marx's algorithm runs in polynomial time in $|\calH|$ when $\fhtw(\calH)$ is a constant.
In this section, we aim to design a polynomial-time approximation algorithm for the $\faq$-width.
Our algorithm is going to use any approximation algorithm for $\fhtw$
(such as Marx's) as a blackbox, 
and our approximation guarantee is going to be in terms of the approximation guarantee of the blackbox algorithm.
Our approximation algorithm is also going to rely on the expression tree constructed in Section~\ref{sec:evo}.

\subsection{$\faq$ with only semiring aggregates}
\label{subsec:approx:Kblocks}

Recall the definition of the $\faq$-width (Definition~\ref{defn:faqw}):
\[ \faqw(\sigma) := \max_{k\in K} \left\{ \rho^*_\calH(U_k)\right\}. \]
When there is no product aggregate, $K$ (as defined in~\eqref{eqn:the:set:K})
is {\em exactly} $[n]$.
Let $P$ be the expression tree constructed from the query $\varphi$ in the form
\eqref{eqn:main:faqKblocks}.
We define notation that will be used throughout this section.
Let $C$ be a node of the expression tree $P$. (This means $C$ is a set of
variables of $\varphi$, and all variables in $C$ have the same tag.)
Let $L$ be the parent node of $C$ (if any) in the expression tree $P$. 
We define the following sets:
\begin{eqnarray}
    \bar \calE(C) &:=& \bigl\{
        S\in\calE \suchthat S \cap C' \neq \emptyset \nonumber\\
        & &\quad\quad\quad \text{ for some $C'$ node in the subtree of $P$ rooted at $C$}.
    \bigr\} \label{eqn:barcalEKblock}\\
    S_{L,C} &:=& L \cap \left( \bigcup_{S \in \bar\calE(C)} S \right)\label{eqn:S_{L,C}}\\
    U(C)  &:=& \bigcup_{L' \text{ an ancestor of $C$}} \left(L' \cap
\bigcup_{S\in \bar\calE(C)} S\right).\label{eqn:U(C)}
\end{eqnarray}
If $C$ has no parent $L$, then $U(C) = \emptyset$ by default and $S_{L,C}$ is undefined.
We think of the set $S_{L,C}$ as the contribution of all the nodes in the
$C$-branch to $L$, and the set $U(C)$ as the contribution of all the nodes in
the $C$-branch to all the (strict) ancestors of $C$. 
Next, for every node $L$ in the expression tree $P$,
define the hypergraph $\calH_L$ as follows.
\bi
 \item If $L$ is a leaf node of $P$, 
     then $\calH_L = \calH[L]$, where $\calH[L]$ denotes the subgraph of
     $\calH$ induced by $L$.
     \footnote{Given $\calH=(\calV,\calE)$ and $L\subseteq \calV$, the hypergraph $\calH[L]:=(L,\calE[L])$ is defined by $\calE[L]:=\{S\cap L\suchthat S\in\calE\}$.\label{fnote:induced:graph}}
 \item If $L$ is not a leaf node of $P$, then $\calH_L = (L, \calE_L)$, where
 \begin{eqnarray*}
    \calE_L &:=& \bigl\{ S \cap L \suchthat 
             (S\in \calE) \wedge 
             (S \cap L  \neq \emptyset) \wedge 
      (S \cap C = \emptyset, \forall \text{ descendant $C$ of } L) \bigr\} \\
         && \cup \bigl\{ S_{L,C} \suchthat C \text{ a child of } L \bigr\}.
      \end{eqnarray*}
     In other words, $\calE_L$ is the set of all projections of $S$ down to $L$ for
     all hyperedges $S$ for which $S$ intersects $L$ but not any descendant of
     $L$ in the expression tree; and for each child $C$ of $L$, $\calE_L$ also contains the projection onto $L$ of the union of all hyperedges $S$ that intersect (some descendant 
     of) $C$.
\ei

We next prove a simple lowerbound for $\faqw(\varphi)$ that leads to an
approximation algorithm for computing $\faqw(\varphi)$ using an approximation
algorithm for $\fhtw$ as a blackbox.

\blmm\label{lmm:main:lowerboundKblocks}
For any node $L$  in the expression tree,
\begin{eqnarray*}
   \faqw(\varphi) &\geq& \fhtw(\calH_L)\\
    \faqw(\varphi) &\geq& \rho^*_\calH(U(L)).
\end{eqnarray*}
\elmm
\bp
To show the first inequality, by Corollary~\ref{cor:faqw-char}, it is sufficient to 
prove that $\faqw(\sigma) \geq \fhtw(\calH_L)$
for any variable ordering $\sigma=(v_1,\dots,v_n) \in \LE(P)$ and
for any node $L$ in the expression tree. 

If $L$ is a leaf node of the expression tree, then 
$\faqw(\sigma) \geq \fhtw(\calH) \geq \fhtw(\calH_L)$ 
because $\calH_L$ is an induced subgraph of $\calH$.
Now, suppose $L$ is not a leaf node. 
For any child $C$ of $L$, let $k$ be the smallest integer 
such that $v_k$ belongs to some node in the subtree rooted at $C$.
Then, due to the fact that $\sigma\in\LE(P)$, 
if $v_j \in L$ then $j<k$.
The set $U^\sigma_k$ is {\em precisely} equal to $U(C) \cup \{v_k\}$.
This is because each time we eliminate a vertex belonging to 
any node in the subtree rooted at $C$, we insert back a
hyperedge interconnecting all its neighbors (to the next hypergraph in the
hypergraph sequence). And so by the time we reach $U^\sigma_k$ all of the nodes
in $U(C)$ belong to the same hyperedge of $\calH^\sigma_k$.
It follows that $(U^\sigma_k -\{v_k\}) \cap L = U(C) \cap L = S_{L,C}$ (since $L\cap L'=\emptyset$ for any ancestor $L'$ of $L$).
From this observation we obtain:
\begin{eqnarray*}
    \faqw(\varphi) &=&
\min_\sigma\left(\max 
   \left\{ 
     \rho^*_\calH(U^\sigma_k) \suchthat k \in [n]
   \right\}\right)\\
   &\geq&
   \min_\sigma\left(\max 
   \left\{ 
     \rho^*_\calH(U^\sigma_j) \suchthat j \in L
   \right\}\right)\\
   (\text{$\rho^*$ is monotone})   
   &\geq&
   \min_\sigma\left(\max 
   \left\{ 
     \rho^*_\calH(U^\sigma_j \cap L) \suchthat j \in L
   \right\}\right)\\
   &\geq&
   \min_\tau \left(\max 
   \left\{ 
     \rho^*_\calH(U^\tau_j) \suchthat j \in L
   \right\}\right)\\
   &=& \fhtw(\calH_L).
\end{eqnarray*}
In the above, $\tau$ is taken only over all variable orderings of $L$
instead of the entire set $[n]$.

We next prove the second inequality that, for any 
$\sigma = (v_1,\dots,v_n) \in\LE(\varphi)$,
and any node $C$ in the expression tree, we have
$\faqw(\sigma) \geq \rho^*_\calH(U(C))$. 
As we have observed above, let $k$ be the smallest integer such that
$v_k \in C$, then $U_k^\sigma = U(C) \cup \{v_k\}$. Hence, because $\rho^*$ is
monotone,
\[
    \faqw(\sigma) = \max_{j\in [n]} \rho^*_\calH(U^\sigma_j)
    \geq \rho^*_\calH(U^\sigma_k)
    \geq \rho^*_\calH(U(C)).
\]
\ep

\bthm\label{thm:faqKblocks}
Let $\varphi$ be any $\faq$ query whose hypergraph is $\calH$  and all variable aggregates are
semiring aggregates.
Suppose there is an approximation algorithm that,
given any hypergraph $\calH'$, outputs a tree decomposition 
of $\calH'$ with fractional hypertree width at
most $g(\fhtw(\calH'))$ in time $t(|\calH'|, \fhtw(\calH'))$ for some non-decreasing functions $g, t$. 
Then, we can in time $|\calH|\cdot t(|\calH|, \faqw(\varphi))$
compute a $\varphi$-equivalent vertex ordering $\sigma$ such that
\[ \faqw(\sigma) \leq \faqw(\varphi) + g(\faqw(\varphi)). \]
\ethm
\bp
We use the blackbox approximation algorithm for $\fhtw$ to construct a tree 
decomposition $(T_L, \chi_L)$ for every hypergraph $\calH_L$ where $L$ is a 
node in the expression tree. 
Then, from each of those tree decompositions, 
we construct a variable ordering $\sigma_L$ for variables in the set $L$ in the standard way.
Finally, we construct the variable ordering $\sigma$ for $[n]$ by concatenating
all the $\sigma_L$ together in any way that respects the
precedence partial order.

Suppose $\sigma = (v_1,\dots,v_n)$ is the resulting variable ordering.
Consider an arbitrary  vertex $v_k$. 
Let $L$ be the node of the precedence tree
that contains $v_k$. Let $B$ be the bag in $(T_L, \chi_L)$ that $v_k$ belonged to
when it was eliminated to construct $\sigma_L$. Then, using the same argument as
in the proof of Lemma~\ref{lmm:main:lowerboundKblocks}, we can show that
\begin{equation}\label{eqn:crucial}
U^\sigma_k \subseteq B \cup U(L).
\end{equation}
To see this, first consider the simpler case when $L$ is a leaf node of the
expression tree. Then, when we eliminate $v_k$ the set $U_k$ is the union of the
sets $S\in\partial(v_k)$. The part $U_k \cap L$ is covered by $B$ because within
$L$ the elimination algorithm works on $\calH_L$. The part $U_k \setminus L$ is
covered by the maximum {\em residue} left over from eliminating all vertices in
$L$. The residue is precisely the set $U(L)$, because every time we eliminate
a vertex we collect all its neighbors together into a hyperedge. 
By the time the last vertex from $L$
is eliminated, the entire set $U(L)$ becomes a hyperedge.
Now, if $L$ is not a leaf node, the situation is exactly the same except for the
fact that we work on the graph $\calH_L$ which is not necessarily the same as
$\calH[L]$. The hypergraph $\calH_L$ contains the restrictions on $L$ of all
the residues of the subtrees under $L$.

Next, from Lemma~\ref{lmm:main:lowerboundKblocks} and from the fact that $\rho^*$ is
subadditive, relation \eqref{eqn:crucial} implies
\begin{eqnarray*}
   \rho^*_\calH(U^\sigma_k) 
   &\leq& \rho^*(B) + \rho^*(U(L))\\
   &\leq& g(\fhtw(\calH_L)) + \faqw(\varphi)\\
   &\leq& g(\faqw(\varphi)) + \faqw(\varphi).
 \end{eqnarray*}
Finally,
\[ \faqw(\sigma) = \max_{k\in [n]} \rho^*_\calH(U^\sigma_k)
    \leq g(\faqw(\varphi)) + \faqw(\varphi).
\]
\ep

\subsection{$\faq$ with an inner $\faq$-formula closed
under idempotent elements}
\label{subsec:approx:inner:faq}

The key 
difference between 
this section and Section~\ref{subsec:approx:Kblocks} is that $K$ (defined by~\eqref{eqn:the:set:K}) is not necessarily
equal to $[n]$: instead it is now the union of $F=[f]$ and the set of semiring variables.
Moreover instead of using $\faqw(\varphi)$ given by Definition~\ref{defn:faqw-phi},
we will now use $\faqw(\varphi, \D_I)$ given by Definition~\ref{defn:faqw-phi-F}.
We will still follow the same strategy as that of
Section~\ref{subsec:approx:Kblocks}, though the definition of the sets
$S_{L,C}$ for each node $L$ and a child $C$ of $L$
is a bit more delicate.

\bdefn[Semiring node and product node]
A node in the expression tree $P$ is called a {\em semiring node} if its 
variables
have a tag forming a semiring with $\otimes$. Otherwise, the node is called a
{\em product node}.
\edefn

Let $C$ be a node of the expression tree $P$. 
Let $L$ be the parent node of $C$ (if any) in the expression tree $P$. 
We define $\bar \calE(C)$ differently from the previous section as follows:
\begin{eqnarray}
    \bar \calE(C) &:=& \bigl\{
        S\in\calE \suchthat S \cap C' \neq \emptyset \text{ for some {\em semiring} node $C'$}\nonumber\\
        & &\quad\quad\quad\quad\quad\quad\quad\quad
        \text{ in the subtree of $P$ rooted 
               at $C$}.
    \bigr\}\label{eqn:barcalEinnerfaq}
\end{eqnarray}
(Compare the above definition with \eqref{eqn:barcalEKblock}.)
We define $S_{L,C}$ and $U(C)$ using \eqref{eqn:S_{L,C}} and \eqref{eqn:U(C)} 
(where the $\bar\calE(C)$ referred to in \eqref{eqn:S_{L,C}} and \eqref{eqn:U(C)} is now the one defined by \eqref{eqn:barcalEinnerfaq}).
Note in particular that if a node does not have any semiring descendant then
$U(C)$ is empty. 
We think of $S_{L,C}$ as the {\em residue} imposed on $S$ 
from the process of eliminating all
semiring variables under the $C$-subtree. 

We next define the hypergraphs $\calH_L$, similar to what was defined in
Section~\ref{subsec:approx:Kblocks}, with one small difference:
\bi
 \item If $L$ is a leaf node of $P$, then $\calH_L = \calH[L]$, the subgraph of
       $\calH$ induced by $L$.
 \item If $L$ is not a leaf node of $P$, then $\calH_L = (L, \calE_L)$, where
\begin{eqnarray*} \calE_L &:=& \bigl\{ S \cap L \suchthat 
             (S\in \calE) \wedge 
             (S \cap L  \neq \emptyset) \wedge
             (S \cap C = \emptyset, \forall \text{ {\em semiring} descendant $C$ of } L)
         \bigr\}\\
         & &\cup \bigl\{ S_{L,C} \suchthat C \text{ a child of } L \bigr\}.
\end{eqnarray*}
     Here is where $\calE_L$ is different from the corresponding definition of
     $\calE_L$ in Section~\ref{subsec:approx:Kblocks}. We only take the projection of
     the hyperedges $S$ for which $S$ does not intersect any semiring 
     descendant of $L$. The key point to notice is that, if $S$ intersects some
     semiring descendant, then its contribution to the node $L$ is summarized
     already in some $S_{L,C}$.
\ei

We now can follow the script of Section~\ref{subsec:approx:Kblocks}.
Note that the bound only holds for the semiring nodes.

\blmm\label{lmm:lowerboundLogic}
For any semiring node $L$ in the expression tree, we have
\begin{eqnarray*}
    \faqw(\varphi, \D_I) &\geq& \fhtw(\calH_L)\\
    \faqw(\varphi, \D_I) &\geq& \rho^*_\calH(U(L)).
\end{eqnarray*}
\elmm
\bp
If the semiring node $L$ is a leaf node of the expression tree, then
\begin{eqnarray*}
   \faqw(\varphi, \D_I) &=&
   \min_{\sigma\in\LE(P)}\left(\max 
   \left\{ 
   \rho^*_\calH(U^\sigma_k) \suchthat k \in K
   \right\}\right)\\
   &\geq&
   \min_{\sigma\in\LE(P)}\left(\max 
   \left\{ 
   \rho^*_\calH(U^\sigma_j) \suchthat j \in L
   \right\}\right)\\
   (\text{$\rho^*$ is monotone})   
   &\geq&
   \min_{\sigma\in\LE(P)}\left(\max 
   \left\{ 
   \rho^*_\calH(U^\sigma_j \cap L) \suchthat j \in L
   \right\}\right)\\
   &\geq&
   \min_\tau \left(\max 
   \left\{ 
   \rho^*_\calH(U^\tau_j) \suchthat j \in L
   \right\}\right)\\
   &=& \fhtw(\calH_L),
\end{eqnarray*}
where $\tau$ above ranges over all variable orderings of $L$.
If the semiring node $L$ is not a leaf node, then the rest of the proof is the same as that of Lemma~\ref{subsec:approx:Kblocks} except for a couple of minor differences:
We will only consider \emph{semiring} nodes of the expression tree
and \emph{semiring} vertices in the hypergraph sequence.
Also we will consider $k \in K$ instead of $k \in [n]$.
\ep

We can now design the approximation algorithm.

\bthm\label{thm:faqLogic}
Let $\varphi$ be any $\faq$ query of the form \eqref{eqn:faqIdempotent} whose hypergraph is $\calH$.
Suppose there is an approximation algorithm that,
given any hypergraph $\calH'$, outputs a tree decomposition 
of $\calH'$ with fractional hypertree width at
most $g(\fhtw(\calH'))$ in time $t(|\calH'|, \fhtw(\calH'))$ for some non-decreasing functions $g, t$. 
Then, we can in time $|\calH|\cdot t(|\calH|, \faqw(\varphi, \D_I))$
compute a vertex ordering $\sigma\in\EVO(\varphi,\D_I)$ such that
\[ \faqw(\sigma) \leq \faqw(\varphi, \D_I) + g(\faqw(\varphi, \D_I)). \]
\ethm
Recall that in the proof of the corresponding Theorem~\ref{thm:faqKblocks} from the previous Section~\ref{subsec:approx:Kblocks}, we applied the blackbox approximation algorithm on each node $L$ of the expression tree to obtain an ordering $\sigma_L$ of variables in $L$, and then we obtained an ordering $\sigma$ of all variables by concatenating together $\sigma_L$ for all nodes $L$ in an ascending order by the depth of $L$ in the expression tree, i.e. starting from the root node and going down the expression tree.

In contrast while trying to prove Theorem~\ref{thm:faqLogic} in the same way as above, we will encounter two key differences. First, for every product node $L$, we don't need to use the blackbox approximation algorithm to obtain an ordering $\sigma_L$ because product variables play no role in the definition of the $\faq$-width, hence we can just order $L$ arbitrarily.
Second, even if we obtain an ordering $\sigma_L$ for every semiring node $L$, it is not very clear how to obtain the overall variable ordering $\sigma$ because product variables can occur multiple times in the expression tree.
In particular, if we try to obtain $\sigma$ in the same way as above by going through every node $L$ of the expression tree ordered by depth and listing $\sigma_L$ when the node $L$ is a semiring node and an arbitrary order of $L$ when $L$ is a product node, then the resulting $\sigma$ might list the same product variable multiple times in multiple contradictory locations, hence it might not be a valid variable ordering!
\bp[Proof of Theorem~\ref{thm:faqLogic}]
Define a set $\overline\calV$ whose items are either product variables or nodes $L$ where $L$ is a semiring node in the expression tree, i.e.
\[\overline\calV:=\left\{v \in \calV \suchthat \text{$v$ is a product variable}\right\}\cup\left\{L \suchthat \text{$L$ is a semiring node in $P$}\right\}\]
Define a binary relation $\preceq_{\overline P}\subseteq \overline \calV\times \overline \calV$ as follows.
For every pair $u, v\in\overline\calV$, we have $u\preceq_{\overline P} v$ iff
\bi
\item either $u=v$,
\item or there are two nodes $L, L'$ in the expression tree $P$ where $L$ is a strict ancestor of $L'$ and
\[(u = L \vee u \in L)\wedge (v = L' \vee v \in L')\]
\ei
By a very similar proof to that of Corollary~\ref{cor:indeed a poset}, we can prove that $\preceq_{\overline P}$ indeed defines a partial order on $\overline\calV$.
Let $\overline \sigma$ be an ordering of $\overline\calV$ which is an arbitrary linear extension of this partial order $\preceq_{\overline P}$.
We can obtain an ordering $\sigma$ of the original set of variables $\calV$ using the ordering $\overline\sigma$ (of $\overline\calV$) as follows.
For every $L$ appearing in the ordering $\overline\sigma$ where $L$ is a semiring node of $P$, we use the 
blackbox approximation algorithm to construct a tree decomposition 
$(T_L, \chi_L)$ for the hypergraph $\calH_L$.
Then, from this tree decomposition, 
we construct a variable ordering $\sigma_L$ for variables in the set $L$.
Finally, we replace the occurrence of $L$ in $\overline \sigma$ by 
the variable ordering $\sigma_L$.
After repeating these steps for every $L$, we call the final variable ordering $\sigma$. Then $\sigma$ is a linear extension of the
precedence poset $P$.

Suppose $\sigma = (v_1,\dots,v_n)$ is the resulting variable ordering.
Consider an arbitrary vertex $v_k$ in a semiring node $L$.
Following the argument in the proof of Theorem~\ref{thm:faqKblocks}, we 
can see that $U^\sigma_k \subseteq B \cup U(L)$.
From Lemma~\ref{lmm:lowerboundLogic} and from the fact that $\rho^*$ is
subadditive, it follows that
\[ \rho^*_\calH(U^\sigma_k) \leq \rho^*(B) + \rho^*(U(L))
    \leq g(\fhtw(\calH_L)) + \faqw(\varphi, \D_I)
    \leq g(\faqw(\varphi, \D_I)) + \faqw(\varphi, \D_I).
\]
The rest of the proof is identical to that of Theorem~\ref{thm:faqKblocks}.
\ep

Note that requiring different copies of the same variable to be consecutive is
not necessary for $\InsideOut$ to work. The algorithm works even if we consider
them to be different variables on the same domain.
However, the collapsing of different copies back to the original copy is needed
for the rigor of the definition of $\EVO(\varphi, \calF(\D_I))$.
Note also that Theorem~\ref{thm:faqLogic} implies 
Theorem~\ref{thm:faq2blocks}
and Theorem~\ref{thm:faqKblocks}.

Given a hypergraph $\calH$ with fractional hypertree width of $\fhtw(\calH)$,
Marx's approximation algorithm~\cite{Marx:2010:AFH:1721837.1721845}
computes a tree decomposition of $\calH$ whose fractional hypertree width is $O(\fhtw^3(\calH))$ in time $O(|\calH|^{O(\fhtw^3(\calH))})$.
By plugging this algorithm into Theorem~\ref{thm:faqLogic}, we obtain an approximation algorithm for $\faqw(\varphi, \D_I)$ that, given an $\faq$ query $\varphi$, computes a variable ordering $\sigma\in\EVO(\varphi,\D_I)$ with $\faqw(\sigma)=O(\faqw^3(\varphi,\D_I))$
in time $O(|\calH|^{O(\faqw^3(\varphi, \D_I))})$.
If $\faqw(\varphi, \D_I)$ is bounded by a constant, then the runtime becomes $\poly(|\calH|)=\tilde O(1)$. (Recall from Table~\ref{tab:results} that in this paper we use $\tilde O$ to hide a factor that is polynomial in the query size.)
From this discussion, we obtain the following corollary.

\bcor\label{cor:faqLogic}
Let $\varphi$ be any $\faq$ query of the form defined in
\eqref{eqn:faqIdempotent}.
Suppose $\faqw(\varphi, \D_I)\leq c$ for some constant c.
Then, the query can be answered in time
\[\tilde O\left(N^{O(\faqw^3(\varphi, \D_I))} +\repsize{\varphi}\right).\]
\ecor

\subsubsection{Corollaries in Logic}

Using the reduction in Example~\ref{ex:qcq} from $\qcq$ to $\faq$, we obtain

\bcor
$\qcq$ is tractable for the class of quantified conjunctive queries $\varphi$
where $\faqw$ is bounded.
\ecor

To determine which classes of $\qcq$ formulas are tractable,
Chen and Dalmau \cite{DBLP:conf/lics/ChenD12} defined the notion of 
a {\em prefixed graph} and its width. 
The definition is in the beginning of Section III of their paper. 
%
The prefix graph's width corresponds exactly to to the quantity
$\max_{\sigma} \max_{k\in K} \{ |U_k| \}$.
Since $\rho^*_\calH(U_k) \leq |U_k|$, $\faqw(\varphi, \D_I)$ is a stronger notion and
our result implies the main positive result in their paper (the first part of
Theorem 3.1).
Furthermore, we can construct families of $\qcq$ instances for which
$\faqw(\varphi, \D_I)$ is bounded but the prefix graph's width is unbounded.
For example, consider the following quantified conjunctive query

\[ \Phi = \forall X_1 \cdots \forall X_n \exists X_{n+1} 
    \left( S(X_1,\dots,X_n) \wedge \bigwedge_{i\in [n]} R(X_i, X_{n+1})
    \right).
\]

Chen-Dalmau's prefix graph's width is $n+1$ in this case, but 
$\faqw$ is $2$. Consequently, our result is strictly stronger.
The next corollary resolves an open question posed at the end of 
Durand and Mengel's paper \cite{DBLP:conf/icdt/DurandM13}.

\bcor
$\sqcq$ is tractable for the class of quantified conjunctive queries $\varphi$
whose $\faqw$ is bounded.
\ecor

\section{Input and Output Representation}
\label{sec:representation}

Thus far we stated our results for the case when the input and output factors 
are in the listing representation (Definition~\ref{def:listing}). Tracing back
at the key steps of $\InsideOut$ (and $\OI$) algorithm, it is easy to see
that the algorithm works for a more general setting than just the listing 
representation. Section~\ref{subsec:oracle} presents the two conditions that we 
need from the input factor representation for all our previous arguments to go 
through. In Sections~\ref{subsec:compressed-repr} and~\ref{subsec:sat-repr}, we 
consider more compressed forms of input factor representations than the listing 
representation. In particular, Section~\ref{subsec:compressed-repr} considers 
other representations that can be reduced to the listing representation (at the 
cost of having to compute a modified $\faq$), which means we can still use 
$\InsideOut$ as is. In Section~\ref{subsec:sat-repr} we consider even more 
succinct representations where the reductions in 
Section~\ref{subsec:compressed-repr} are too costly. In particular, for the 
special case of $\sat$ and $\ssat$, we show that we can replace $\OI$ with a more 
specific algorithm, which allows us to recover known results on polynomial-time 
solvability of certain classes of $\sat$ and $\ssat$ formulas. While these 
results only re-prove existing results, we consider these as an illustration of 
the power of the $\faq$ (and variable elimination) framework. We then switch 
to considering situations where one might be interested in output representations 
other than the listing representation in Section~\ref{sec:output-rep}. Finally 
in Section~\ref{sec:composition}, we consider the case when the output and 
input representations match, which naturally leads to the notion of composing 
$\faq$s. $\faq$ composition is also needed in Section~\ref{subsec:compressed-repr}.
For the sake of clarity, we will almost exclusively focus on $\faqcs$ in this section.

\subsection{The factor oracle}
\label{subsec:oracle}

In many application areas of $\faq$, there are {\em many} ways the input factors
can be represented.
Appendix~\ref{app:sec:representation} lists some of the main
representations collected from areas such as logic, $\pgm$, and matrix
computation. 
It should be obvious that the computational complexity of 
$\faq$ is highly dependent on how the inputs are specified. To generalize our results,
 we advocate for the following oracle model, which is
 sufficiently general to capture existing models of
input and output representations.
We note that our assumptions here are met by
all of the representations discussed here and in 
Appendix~\ref{app:sec:representation} -- the only difference is 
in the price the oracle pays to answer the query.

\begin{assumption}[\Assumption]\label{assumption:conQuery}
   We assume that $\Dom(X_i)$ are totally ordered.\footnote{We order the
domain arbitrarily if there is no natural total order; e.g. $\false < \true$ 
for a Boolean problem, or blue $<$ green $<$ red in a coloring problem.}
Each factor oracle for $\psi_S$ is capable of answering the following query
called the {\em \conQuery}.
Let $0 \leq k \leq n-1$ be any integer such that $k+1\in S$.
Let $\mv x_{[k]} = (x_1,\dots,x_k)$ be a vector such that
     $\psi_S(\cdot \suchthat \mv x_{[k]}) \not\equiv \mv 0$.
Let $y \in \Dom(X_{k+1})$ be arbitrary.
Then, the factor oracle for $\psi_S$ can return a {\em minimum} value 
$x_{k+1} > y$ for which $\psi_S( \cdot \suchthat \mv x_{[k+1]}) \not\equiv \mv 0$,
       i.e.
       \[ x^S_{k+1} = \min
           \left\{ x_{k+1} \suchthat (x_{k+1} > y) \wedge
               \bigl(\psi_S( \cdot \suchthat \mv x_{k+1}) \not\equiv \mv 0\bigr)
           \right\}.
       \]
If no such $x^S_{k+1}$ exists, then $+\infty$ (just a symbol) is returned.
\end{assumption}
A {\em value query} returns the value $\psi_S(\mv x_S)$ for some given $\mv x_S$. For simplicity we
will also call those value queries \conQueries.

The second assumption about the input oracles is the following.

\begin{assumption}[\ProdMar]\label{assumption:ProdMar}
    Let $\psi_S$ be an input oracle and $i \in S$.
    The input oracle for $\psi_S$ can return another oracle on $S-\{i\}$,
    denoted by $\psi_{S-\{i\}}$, defined by
    \[ \psi_{S-\{i\}}(\mv x_{S-\{i\}}) = 
        \bigotimes_{x_i \in \Dom(X_i)} \psi_S(\mv x_S).
    \]
    The number of $\otimes$ operations performed to compute the
    factor $\psi_{S-\{i\}}$ is bounded by $\repsize{\psi_S}$.

This assumption is reasonable because as soon as 
$\psi_S(\mv x_S) = \mv 0$ we can infer that 
$\psi_{S-\{i\}}(\mv x_{S-\{i\}}) = \mv 0$.
Note also that if we use the listing representation for $\psi_{S-\{i\}}$,
then we can compute entries of
   $\psi_{S-\{i\}}$ using at most $\repsize{\psi_S}$ \conQueries.
\end{assumption}

Appendix~\ref{app:subsec:oracle} explains why the listing representation
satisfies both of the above assumptions.
By re-examining the previous runtime analysis of $\InsideOut$, one can check 
that in the more general factor oracle model, one can prove the following 
generalization of Theorem~\ref{thm:IO}:
\bthm
Suppose the \Assumption and \ProdMar are satisfied.
Using notations established in Theorem~\ref{thm:IO},
the $\InsideOut$ algorithm applied to $\varphi$
(Algorithm~\ref{algo:IO}) satisfies the following properties.
\bi
 \item[(i)] The number of \conQueries to the factor oracles is at most
\begin{multline}
    \sum_{k\in K} |U_k| \cdot |\{S\in\calE : S\cap U_k \neq \emptyset\}|
    \cdot \agm(U_k)
    +
    \sum_{\substack{k\notin K\\S\in\partial(k)}} \repsize{\psi_S}\\
    +
    \sum_{\substack{k\notin K\\S\in\calE_k-\partial(k)}}
    \repsize{\psi_S}\cdot\Idem_k(\psi_S)
    + f(f+m)\repsize{\varphi}.
    \label{eqn:P4}
\end{multline}
 \item[(ii)] the total number of $\oplus^{(k)}$ operations performed is at most
    $\sum_{k\in K} |U_k| \cdot \agm(U_k)$.
 \item[(iii)] the number of $\otimes$ operations performed is at most
\begin{multline}
    \sum_{\substack{k\in K}}
    \left(|\{S\in\calE : S\cap U_k \neq \emptyset\}| -1\right)
    \cdot
    \agm(U_k)
    +
    \sum_{\substack{k\notin K\\S\in\partial(k)}} \repsize{\psi_S}\\
    +
    \sum_{\substack{k\notin K\\S\in\calE_k-\partial(k)}}
    \repsize{\psi_S}\cdot\Idem_k(\psi_S)
    +f(f+m)\repsize{\varphi}.
    \label{eqn:P6}
\end{multline}
\ei
\label{thm:IO-gen}
\ethm

\subsection{Input representations parsimoniously reducible to $\faq$}
\label{subsec:compressed-repr}

We briefly mention other representation choices for input factors. More details 
(including formal definitions) can be found in Appendix~\ref{app:sec:representation}.

We begin with a representation that is more wasteful than the listing 
representation: the ``truth table'' representation, which is the norm in dense 
matrix computation as well as inference in probabilistic graphical models. The 
difference from listing representation is that tuples that result in a 
$\mv{0}$ are also explicitly listed. It is trivial to convert this representation 
into the listing representations.

Beyond the truth tables, other known representations are more succinct than the 
listing representation. In the $\csp$ literature, these include generalized 
disjunctive normal form (GDNF) that was first considered by Chen and
Grohe~\cite{DBLP:journals/jcss/ChenG10a} and decision
diagram representation of Chen and Grohe~\cite{DBLP:journals/jcss/ChenG10a},
which is a generalization of the well-studied ordered binary decision
diagrams or OBDDs. In the database literature, 
Olteanu and Z\'{a}vodn\'{y}~\cite{OZ15} considered factorized  representations 
for conjunctive queries. Fast matrix vector multiplication has many examples of 
more succinct representations than the listing representation (e.g. the $\dft$). 
In the $\pgm$ world, Algebraic Decision Diagrams (or ADDs) were introduced in 
Bahar et al.~\cite{DBLP:journals/fmsd/BaharFGHMPS97}.

A common property of all of the above succinct representations is that they can 
all be reduced to the listing representation at the expense of having to solve 
a more complicated $\faq$ problem. In particular, this leads to the problem of 
composing $\faq$ problems, which is discussed in Section~\ref{sec:composition}. 

\subsection{Input representations not parsimoniously reducible to $\faq$}
\label{subsec:sat-repr}

Thus far we have analyzed the runtime of the $\InsideOut$ algorithm in terms of
the sizes of the input factors, which are the number of non-zero points of each
input factor. If the input factors were represented using the listing
representation, then $O( |S| \cdot \repsize{\psi_S} )$ is precisely the input size for 
the factor $\psi_S$ (assuming each functional value is of size $O(1)$). 
However, in some applications the input factors are much more
compactly represented, and the representation of $\psi_S$ may be exponentially 
smaller than $\repsize{\psi_S}$. 
The canonical example is the $\sat$ problem, where a CNF clause $\psi_S$
has size $\repsize{\psi_S} = 2^{|S|} - 1$, yet represented using only $|S|$ bits. Note that in this case we could reduce the input to the listing representation but this would suffer an exponential blowup in size, which is too expensive. In the rest of the section, we outline how we can modify our framework to handle this succinct representation of $\sat$ and $\ssat$ problems.

The conditional-search $\OI$ algorithm  is no longer a
good choice for such compact input representation, because its runtime depends 
heavily on the number of non-zero elements in each input factor.
Also, we cannot expect the same sort of results from
Section~\ref{sec:insideout} to hold for $\sat$ or $\ssat$, because even if the input
query was $\alpha$-acyclic the problem is still $\np$-hard or $\sharpP$-hard,
respectively. (Recall that $\alpha$-acyclic hypergraphs have $\fhtw = 1$.)
Recently, however, there were a couple of interesting results showing that
$\sat$ and $\ssat$ are tractable for $\beta$-acyclic queries
\cite{ordyniak_et_al:LIPIcs:2010:2855,
braultbaron_et_al:LIPIcs:2015:4910}.

In the relational join problem, there is also a phenomenon discovered recently
where the computational difficulty we face comes from precisely the same problem of
compact input factor representations. In \cite{nnrr}, modulo a technical assumption,
we showed that the minimum number of comparisons a comparison-based join
algorithm performs is in the same order as the minimum number of rectangular
boxes of a certain format needed to cover the entire output space. We designed
an algorithm called $\ms$ whose core functionality is to determine whether or
not a set of rectangular boxes covers the entire output space. 
In \cite{anrr}, the $\tetris$ algorithm relaxed the assumption about the input 
boxes, and formally defined this $\problemname{box cover problem}$ (or $\bcp$). 
In both cases, $\bcp$'s difficulty lies precisely in the fact that the
boxes represent points {\em not} in the output, and thus a
conditional-search-style algorithm such as $\OI$ does not
work well. 
In $\ms$ and $\tetris$, the collection of supports of the rectangular boxes
forms a hypergraph.
Similar to the $\sat$ and $\ssat$ case, if this hypergraph is only
$\alpha$-acyclic, then the box cover problem is $\np$-hard. But, if it is
$\beta$-acyclic then there is a linear time algorithm solving the box cover
problem.

All of the four algorithms can be explained using the $\InsideOut$/$\OI$ duality
framework. The $\InsideOut$ algorithm remains the same -- it is just variable
elimination -- but we have to tailor the $\OI$ algorithm to suit the compact 
encoding. In particular, $\bcp$, $\sat$, and $\ssat$ can all be reduced to
$\faq$ instances in which each of the input factors $\psi_S$ is of a special
form called the {\em box factor}.

\bdefn[Box factor]
\label{def:box}
A {\em box} with support $S=\{i_1,\dots,i_s\}$ is a tuple 
$\mv B = \inner{I_{i_1},\dots,I_{i_s}}$ where $I_{i_j}\subseteq \Dom(X_{i_j})$ is an
interval in the domain $\Dom(X_{i_j})$. The box is a set of points $\mv x \in
\prod_{i=1}^n \Dom(X_i)$ for which $x_{i_j} \in I_{i_j}$ for all $j\in [s]$.
A factor $\psi_S: \prod_{i_j\in S} \Dom(X_{i_j}) \to \D$ is called a {\em box
factor} if there is a box $\mv B$ and a $c \in \D$ for which
\[ \psi_S(\mv x) = 
   \begin{cases} 
      c & \text{ if } \mv x \in \mv B\\
      \mv 1 & \text{ if } \mv x \notin \mv B.
   \end{cases} 
\]
\edefn
In $\sat$ and $\ssat$, each CNF clause is a box factor. For example, in the
clause $(X_1 \vee \bar X_2 \vee X_3)$, the box is 
$\inner{\{\false\}, \{\true\}, \{\false\}}$ (with $c=\false$).
For $\sat$ the domain $\D$ is $\{\true,\false\}$.
For $\ssat$ the domain is $\R_+$.
Each rectangular box in $\bcp$ is clearly a box factor with Boolean domain, and 
looking for a point not covered by all the boxes is the same as solving the
corresponding $\faq$ instance.

The first key to the algorithm in \cite{ordyniak_et_al:LIPIcs:2010:2855} for $\sat$,
the algorithm in \cite{braultbaron_et_al:LIPIcs:2015:4910} for $\ssat$, 
$\ms$ in \cite{nnrr}, and $\tetris$ in \cite{anrr} is that we can still run
variable elimination, but each of the intermediate factors $\psi_{U_k-\{k\}}$ is
computed and represented as a product of box factors.
The second key observation is that, for $\beta$-acyclic hypergraphs, there is a
variable ordering $\sigma=(v_1,\dots,v_n)$ called the {\em nested elimination
order} (NEO) such that, for every $k\in [n]$, the collection of sets in
$\partial(v_k)$ forms a chain (see Proposition~\ref{prop:beta:acyclicity} and
its proof in Appendix~\ref{app:sec:td}). 
This nesting
allows us to keep the box factor representation of the intermediate factors
compact, i.e. each intermediate factor $\psi_{U_k-\{v_k\}}$ is a product of a
``small'' number of box factors.

Since $\ms$ and $\tetris$ take a bit of work to set up properly, in the
following section we explain the algorithms in
\cite{ordyniak_et_al:LIPIcs:2010:2855} and
\cite{braultbaron_et_al:LIPIcs:2015:4910} using this idea.

\subsubsection{$\InsideOut$ for $\sat$}

Consider the $\sat$ problem, where we want to know if there exists one
satisfying assignment. What does ``eliminating a variable'' mean in this case?
We have a set of clauses that contain the variable $X_n$. In each such clause,
$X_n$ occurs either as a positive literal $X_n$, or a negated variable
$\bar X_n$. 
Let $\partial_P(X_n)$ be the set of clauses containing $X_n$,
and $\partial_N(X_n)$ be the set of clauses containing $\bar X_n$.
Then, $\partial(X_n) = \partial_P(X_n) \cup \partial_N(X_n)$.
We want to construct a new boolean factor $\psi_{U_n-\{X_n\}}$ where
\begin{eqnarray*}
 \psi_{U_n-\{X_n\}} &=&
   \left( \bigwedge_{C\in\partial(X_n)} C|_{x_n=\true} \right) \vee
   \left( \bigwedge_{C\in\partial(X_n)} C|_{x_n=\false} \right)\\
   &=&
   \left( \bigwedge_{C_N\in\partial_N(X_n)} C_N|_{x_n=\true} \right) \vee
   \left( \bigwedge_{C_P\in\partial_P(X_n)} C_P|_{x_n=\false} \right)\\
   &=&
   \bigwedge_{C_N\in\partial_N(X_n), C_P\in\partial_P(X_n)} 
     \left( C_N|_{x_n=\true} \vee C_P|_{x_n=\false} \right)
\end{eqnarray*}

But $C_N|_{x_n=\true}$ and $C_P|_{x_n=\false}$ are simply the original clauses
$C_N$ and $C_P$ with $X_n$ eliminated completely.
Hence, the rule is: for every clause $C_i$ containing $X_n$
and every clause $C_j$ containing $\bar X_n$,
create a new clause $C_{ij} = C_i \vee C_j$ {\em without} $X_n$ in
it.\footnote{In proof complexity, the clause $C_{ij}$ is called a resolvent of
the two clauses $C_i$ and $C_j$.} 
And we have a new instance of the $\sat$ problem.
This procedure is known as the {\em Davis-Putnam} procedure for 
$\sat$-solving.
Under this procedure, we do not need to enumerate all $2^{|U_n|}$ truth
assignments, so it might be faster depending on the input.
On the other hand, we will run into a combinatorial explosion problem
in the number of clauses.
For example, after one step we might have up to $\Omega(m^2)$ clauses.
Dechter and Rish \cite{DBLP:conf/kr/DechterR94} showed 
that this algorithm runs in time at most $O(n^39^w)$, where $w$ is the 
appropriate notion of width for the $\sat$ formula.
Of course this is no better than the rough runtime of $O(mn^22^w)$, 
but the advantage is that it creates a {\em theory} for the proof system.
We will not delve any further on this point.

What is important to note here is that the representation of the intermediate
factor $\psi_{U_n-\{X_n\}}$ is a product of box factors. 
One interesting class of input CNF formulas is the class of $\beta$-acyclic 
CNF formulas. Under this class, if the vertex ordering is a nested elimination
order, then we don't have the clause (or box factor) explosion problem. 
This is due to the fact that the set $\partial(X_k)$ forms a chain,
so a variable set in $\partial(X_k)$ is a subset of another
variable set in $\partial(X_k)$.
When the variables in $C_i$ are a subset of variables in $C_j$,
the new clause $C_{ij}$ is either a tautology (which can be removed), or 
is equal to $C_j$.
(When the resolvent is a subset of one of the two resolvers, the
resolution is called {\em subsumption resolution}.)
Hence, every time we eliminate a variable $X_n$, we do not increase the
number of clauses.
So we have a polynomial-time algorithm for solving 
$\beta$-acyclic CNF formulas, a result that was only proved recently
\cite{ordyniak_et_al:LIPIcs:2010:2855}.

\bthm[Ordyniak, Paulusma, and 
Szeider \cite{ordyniak_et_al:LIPIcs:2010:2855}]
$\sat$ is polynomial-time solvable for the class of $\beta$-acyclic CNF
formulas.
\ethm

\subsubsection{$\InsideOut$ for $\ssat$}

For the $\ssat$ problem, each clause corresponds to 
a function to $\{0,1\}$. In particular, $\ssat$ can be viewed as an $\faqcs$ over 
the semiring $(\R_+, +, \times)$ where each clause $C$ (whose variables are 
$\vars(C)$) corresponds to a factor $\psi_{\vars(C)}$ defined as
\[\psi_{\vars(C)}(\mv x_{\vars(C)}):=
\begin{cases}
1 & \text{ if } \mv x_{\vars(C)} \text{ satisfies }C,\\
0 & \text{ otherwise}.
\end{cases}\]
We will be working with a slightly more general version of $\ssat$, called $\wssat$, where each clause $C$ has an associated weight ($\weight(C)\in\R_+$) such that the corresponding factor $\psi_{\vars(C)}$ becomes
\[\psi_{\vars(C)}(\mv x_{\vars(C)}):=
\begin{cases}
1 & \text{ if } \mv x_{\vars(C)} \text{ satisfies }C,\\
\weight(C) & \text{ otherwise}.
\end{cases}\]

(Obviously, if each clause has weight $0$, then $\wssat$ reduces back to
$\ssat$. Further, note that this is a re-statement of the box factor formulation from Definition~\ref{def:box}
specialized to this problem.) We chose to work with $\wssat$ because eliminating
a variable from an $n$-variable $\wssat$ instance results in an $(n-1)$-variable $\wssat$ instance, as we will see shortly.

Eliminating a variable $X_n$ in $\wssat$ means defining a set $\calC'_n$ of weighted clauses over the variables $U_n-\{X_n\}$ such that for all $\mv x$
\begin{equation}
\label{eq:ssat:U_n-X_n}
\underbrace{\prod_{C'\in\calC'_n}\psi_{\vars(C')}(\mv x_{\vars(C')})}_{\psi_{U_n-\{X_n\}}(\mv x_{U_n-\{X_n\}})}
=\sum_{x_n}\prod_{C\in\partial(X_n)}\psi_{\vars(C)}(\mv x_{\vars(C)}\suchthat x_n).
\end{equation}

A clause $C'$ is called \emph{monochromatic} with respect to a set of (weighted) clauses $\calC$ if for every clause $C\in\calC$, either $(C \implies C')$ or $(C \vee C'\equiv \true)$. The color of $C'$ with respect to $\calC$, denoted by $\clr_{\calC}(C')$, is defined as:
\[\clr_{\calC}(C'):=\prod_{C\in\calC \suchthat C\implies C'}\weight(C).\]

Given a clause $C$ and a variable $X\in\vars(C)$, we will be using $[C]_{-X}$ to denote the clause that results from $C$ by dropping either the literal $X$ or $\bar X$ (whichever occurs in $C$).
 Let $\calC_n$ be the set of clauses that contained $X_n$ but with $X_n$ now removed from them, i.e.
\[\calC_n:=\left\{\left[C\right]_{-X_n} \suchthat C\in\partial(X_n)\right\}.\]
Let $\pi_n=\left(X_{i_1}, X_{i_2}, \ldots, X_{i_\ell}\right)$ be some arbitrarily-fixed order of $U_n-\{X_n\}$. We can define $\calC'_n$ to be the set of clauses that are ``minimal'' w.r.t $\pi_n$ and monochromatic w.r.t $\calC_n$. Formally,

\begin{equation}
\calC'_n:=\left\{C' \text{ is a clause } \Big|
\begin{array}{l}
\vars(C')=\left\{X_{i_1}, X_{i_2}, \ldots, X_{i_{\ell'}}\right\}\text{ for some $\ell'\in[\ell]$, and}\\
\text{$C'$ is monochromatic w.r.t $\calC_n$ while $\left[C'\right]_{-X_{i_{\ell'}}}$ is not.}
\label{eq:beta:calC'_n}
\end{array}
\right\}.
\end{equation}
Notice that for each $C'\in\calC'_n$, both $C'\vee X_n$ and $C'\vee \bar X_n$ are monochromatic w.r.t $\partial(X_n)$.
Let 
\begin{eqnarray*}
\weight(C')&=&\clr_{\partial(X_n)}(C'\vee X_n)+\clr_{\partial(X_n)}(C'\vee \bar X_n)\\
&=&\clr_{\partial_P(X_n)}(C'\vee X_n)+\clr_{\partial_N(X_n)}(C'\vee \bar X_n).
\end{eqnarray*}
It is not hard to verify that the above definition of $\calC'_n$ indeed satisfies \eqref{eq:ssat:U_n-X_n}. However, the size of $\calC'_n$ could be larger than $|\partial(X_n)|$, and $|\calC'_i|$ for $i<n$ could be even much larger.

\emph{$\beta$-acyclic case:}
When the CNF formula is $\beta$-acyclic, it is convenient to work with a nested
elimination order. Let $X_n$ be the last variable in such an order. Thanks to
Proposition~\ref{prop:beta:acyclicity}, we can choose $\pi_n=\left(X_{i_1}, X_{i_2}, \ldots, X_{i_\ell}\right)$ such that for each clause $C\in\partial(X_n)$,
\[\vars(C)=\left\{X_n\right\}\cup\left\{X_{i_1}, X_{i_2}, \ldots, X_{i_{k}}\right\}\text{ for some $k\in[\ell]$}.\]
While the above looks very convenient, if we define $\calC'_n$ as in \eqref{eq:beta:calC'_n}, we might end up with a clause $C'\in\calC'_n$ for which $\vars(C')\neq \vars(C)-\left\{X_n\right\}$ for every $C\in\partial(X_n)$. After eliminating $X_n$, the remaining hypergraph might not be $\beta$-acyclic anymore because of the new hyperedge $\vars(C')$.

To remedy the situation, let $\left(C_1, \ldots, C_{|\partial(X_n)|}\right)$ be the clauses of $\partial(X_n)$ sorted in ascending order by $|\vars(C_i)|$. In particular, for each $i<j \in\left\{1, \ldots, |\partial(X_n)|\right\}$, we have
\[|\vars(C_i)|\leq |\vars(C_j)|\iff\vars(C_i)\subseteq \vars(C_j).\]
(Ties can be broken arbitrarily. The above equivalence follows from
Proposition~\ref{prop:beta:acyclicity} along with the fact that $X_n$ is the last in NEO.)
Define
\begin{eqnarray*}
\partial_P^{\leq i}(X_n)&:=&\partial_P(X_n)\cap\left\{C_j\right\}_{j\leq i},\\
\partial_N^{\leq i}(X_n)&:=&\partial_N(X_n)\cap\left\{C_j\right\}_{j\leq i},\\
\partial_P^{< i}(X_n)&:=&\partial_P(X_n)\cap\left\{C_j\right\}_{j< i},\\
\partial_N^{< i}(X_n)&:=&\partial_N(X_n)\cap\left\{C_j\right\}_{j< i}.
\end{eqnarray*}
We will choose $\calC'_n:=\left\{C'_0, C'_1, \ldots, C'_{|\partial(X_n)|}\right\}$, where $C'_0$ is an empty clause whose weight is 2, and for each $i\in\left\{1, \ldots, |\partial(X_n)|\right\}$, $C'_i:=\left[C_i\right]_{-X_n}$ has weight
\[\weight(C'_i):=
\begin{cases}
0 \text{ if } \clr_{\partial_P^{< i}(X_n)}(C'_i\vee X_n)+\clr_{\partial_N^{< i}(X_n)}(C'_i\vee \bar X_n)=0,\\
\frac{\clr_{\partial_P^{\leq i}(X_n)}(C'_i\vee X_n)+\clr_{\partial_N^{\leq
i}(X_n)}(C'_i\vee \bar X_n)}{\clr_{\partial_P^{< i}(X_n)}(C'_i\vee
X_n)+\clr_{\partial_N^{< i}(X_n)}(C'_i\vee \bar X_n)}
\text{ otherwise.}
\end{cases}
\]
While the above definition of $\calC'_n$ still satisfies \eqref{eq:ssat:U_n-X_n}, we now have for each $C'_i\in\calC'_n$, either $\vars(C'_i)=\emptyset$ or $\vars(C'_i)=\vars(C_i)-\{X_n\}$ for $C_i\in\partial(X_n)$. Hence after eliminating $X_n$, the remaining hypergraph is still $\beta$-acyclic. Moreover, $|\calC'_n|=|\partial(X_n)|$, which guarantees that after eliminating $X_n$, the resulting $\wssat$ instance has the same size as the original. We conclude that:
\bthm[Brault-Baron,  Capelli, and Mengel 
\cite{braultbaron_et_al:LIPIcs:2015:4910}]
$\ssat$ is polynomial-time solvable for the class of $\beta$-acyclic CNF
formulas.
\ethm

\subsection{Output representation}
\label{sec:output-rep}

Our framework, in addition to being able to handle multiple input 
representations, is also able to handle different output representations. In 
particular, we describe how one can modify $\InsideOut$ to obtain different 
output representations. 

Inspired by the factorized database ideas developed
by Olteanu and Z\'{a}vodn\'{y}~\cite{OZ15}, the extension to aggregates
in Bakibayev et al. \cite{DBLP:journals/pvldb/BakibayevKOZ13}, and
inspired by the discussion in Section~\ref{subsec:compressed-repr} 
regarding using an 
$\faq$-instance as an input factor, we observe that an $\faq$-expression not 
only defines a new function, 
but also stores the computation needed to compute that function.
There is a spectrum of tradeoffs one can explore in terms of the time it takes to
query into a function stored in a ``compressed'' form and the time it takes 
to decompress it.

In order to explain this idea more cleanly, and to simplify the presentation 
a bit, we will only consider $\faqcs$ instances, i.e.
\[    \varphi(\mv x_F) =
    \bigoplus_{\mv x_{[n]-F}}
    \bigotimes_{S\in\calE}\psi_S(\mv x_S),
\]
where $F=[f]$ for some $f\in[n]$.
Furthermore, we will assume that $\D=\{0,1\}$.

We now illustrate how we can modify $\InsideOut$ so that we can represent the 
output in three different output representations (in addition to the default listing representation). We will compare these output representations across the following three axes:
\begin{enumerate}
\item {\em Output pre-processing time:} This is the time needed to compute the output in the required form;
\item {\em Value query time:} This is the time needed to check if $\varphi(\mv y)=1$ given the query $\mv y\in\prod_{i=1}^f\Dom(X_i)$; and
\item {\em Enumeration delay:} Given the output representation, how much time it takes between reporting two consecutive output tuples, given that ultimately we want to list all output tuples.
\end{enumerate}

Recall from Section~\ref{subsubsec:IO:output} that 
$\InsideOut$ runs until it has eliminated the variables $X_{n},\dots,X_{f+1}$ 
and then runs $\OI$ on the resulting $\faqcs$ instance $\varphi_f$ corresponding to $\calH_f$. 
In two of the options below, we will see how we can change this last step. 
Further, we will only talk about the runtime of the algorithm {\em after} it 
has eliminated the variables $X_{n},\dots,X_{f+1}$.

\paragraph{Listing representation.} This is the default behavior of 
$\InsideOut$. After spending $\tilde O(N^{\faqw(\sigma)})$ time in eliminating variables $X_n,\ldots,X_{f+1}$, it runs $\OI$ on the resulting query \eqref{eqn:output-listing} taking time $\tilde O(\agm(F))$ to report the output $\varphi$
(where $\repsize{\varphi} \leq \agm(F)$). Here, the output pre-processing time is $\tilde O(\agm(F))$, the value query time is $\tilde O(1)$, and we have a constant enumeration delay.

Next, we consider two options where we have a smaller output pre-processing 
time while we can still handle the value query and enumeration operations 
(at potentially a larger cost than above).

\paragraph{$\faq$ representation.} Another option is to just not do any output
pre-processing {\em nor} eliminate the free variables. Note that the $\faq$ instance $\varphi_f$ constructed when $\InsideOut$ eliminates $X_{f+1}$ is a valid representation of the output. Further, the value query time is still $\tilde O(1)$ (by checking if for every $T\in \calE_f$, $\psi_T(\mv y_T)=1$). 
However, there is no way to construct an enumeration algorithm with a constant delay from this representation. Finally, this representation has the advantage that if one considers the input as an $\faq$ instance (see Section~\ref{sec:succ-rep}), then there is a nice symmetry in that the output here is also an $\faq$ instance. Given the discussion in Section~\ref{sec:succ-rep}, this generalizes the framework of 
Olteanu and Z\'{a}vodn\'{y}~\cite{OZ15} 
from factorized representation to $\faq$ representation.

\paragraph{$\tilde O(1)$-delay enumeration representation.} 
Another option is to eliminate free variables but avoid running the $\OI$ algorithm at the last step (i.e. drop line~\ref{ln:output-phase} of Algorithm~\ref{algo:IO}). In this case,
we already have the output stored in a factorized representation like that
from the factorized database framework.
If the output was only required to be in this form, then essentially our 
runtime is $\tilde O(N^{\faqw(\sigma)})$ (without the $+\repsize{\varphi}$ term).

In this output representation, the value query time is $\tilde O(1)$
(since one can check if for every 
$i\in [f]$, $\psi_{U_i}(\mv y_{U_i})=1$). Perhaps more importantly, 
one can design a $\tilde O(1)$-delay enumeration function. Assume that the 
$\psi_{U_i}$ are stored as a BTree/trie (with sort order $X_1,X_2,\dots,X_f$). 
Start with the `first' $x_1$ such that $\psi_{U_1}(x_1)=1$ and then figure 
out the `first' $x_2$ such that $\psi_{U_2}(x_1,x_2)=1$ and so on. It is 
easy to check that there is $\tilde O(1)$ delay in outputting every two
consecutive output tuples.

\paragraph{$\faq$ composition and message passing}

Being able to efficiently enumerate entries of a function defined by an $\faq$ 
expression might have applications, say in answering a conjunctive query in a
database. Thus, the $\tilde O(1)$-delay enumeration representation is a decent
choice. 
However, as we mentioned in Section~\ref{subsec:compressed-repr} there are cases where we want to use the output
of one $\faq$-query to feed into the input of another $\faq$-query. 
Thus, the output representation should allow us to efficiently answer
\conQueries and product marginalization queries (e.g. universal quantifiers in logic applications). 

The problem is, we do not know in advance in which variable ordering the future
\conQueries and product marginalization queries will be posed. If they were
posed in the same order as $X_1,\dots,X_f$, then the $\tilde O(1)$-delay
representation is sufficient. We can easily answer a \conQuery using the
intermediate factors $\psi_{U_i}$, $i\in [f]$, efficiently in $\tilde
O(1)$ time.

This is where we can use the playbook of the graphical model literature. The
key reason that message passing (or belief propagation) is advantageous over
variable elimination is that it prepares the graphical model for future
(unknown) queries.
The message passing algorithm is essentially variable elimination run in all
directions at once. We compute the tree decomposition of the graph $\calH_f$
(after using $\InsideOut$ to eliminate $X_i$, $i>f$.) Then, we run the message
passing algorithm on the bags of the tree decomposition. The overall complexity
is still $\tilde O( N^{\fhtw(\calH_f)})$, but after convergence all
the bags are in calibrated state, and they allow for answering \conQueries
along any GYO elimination order of the tree decomposition. 
We leave the tradeoffs involved in realizing such an idea for a future work; but
we would like to point out that Olteanu and Z\'{a}vodn\'{y}~\cite{OZ15} has
taken a step toward this direction in the database domain.

\subsection{Composition of $\faq$ instances} 
\label{sec:composition}

We consider how the fractional hypertree width changes when we compose $\faq$ instances. More precisely we consider the following problem:
\begin{quote} 
Let $\calH^0=(\calV,\calE^0)$ be a hypergraph. For every $e\in\calE^0$, let $\calH^1_e=(e,\calE^1_e)$ be a hypergraph. Let $\calH^1$ denote the collection of hypergraphs $\{\calH^1_e\}_{e\in\calE^0}$. Now consider the following {\em composed} hypergraph $\calH^0\circ\calH^1$ with $\calV$ as the set of nodes and the edge set $\calE^{01}=\cup_{e\in \calE^0} \calE^1_e$. How does $\fhtw(\calH^0\circ\calH^1)$ behave with respect to $\fhtw(\calH^0)$ and $\fhtw(\calH^1_e)$ (or in terms of other widths of these hypergraphs) for $e\in\calE^0$?
\end{quote}

We begin with a simple observation:
\bprop
\label{prop:fhtw-comp-agm}
For every hypergraph $\calH^0$ and a corresponding collection of hypergraphs $\calH^1$,
\[\fhtw(\calH^0\circ\calH^1)\le \fhtw(\calH^0)\cdot\max_{e\in\calE^0} \rho^*(\calH^1_e).\]
\eprop
\bp
Let $(T,\chi)$ be a tree decomposition for $\calH^0$ with $\rho^*$-width of $w=\fhtw(\calH^0)$. It is easy to check that $(T,\chi)$ is also a valid tree decomposition for $\calH^0\circ\calH^1$. Further, we claim that for every bag $B$ in $T$, $\rho^*_{\calH^0\circ\calH^1}(B)\le w \cdot\max_{e\in\calE^0} \rho^*(\calH^1_e),$ which would complete the proof.

Finally, we argue the claim. Fix any bag $B$ and let $x_e$ for every $e\in\calE^0$ such that $e\cap B\neq\emptyset$ be the optimal edge cover for $B$ (i.e. $\sum_{e\in\calE^0: e\cap B\neq\emptyset} x_e\le w$). For every edge $e\in \calE^0$, let $y^e_{e'}$ for every $e'\in\calE^1_e$ be an optimal fractional edge cover for $\calH^1_e$. Then it is easy to check that the following is a valid edge cover for $B$ using edges from $\calH^0\circ\calH^1$: for every $e'\in\calE^{01}$ such that $e'\cap B\neq\emptyset$, define 
\[z_{e'}=\sum_{e\in\calE^0: e'\cap e\cap B\neq \emptyset, e'\calE^1_e} x_{e}\cdot y^e_{e'}.\]
The claim follows by noting that
\[\sum_{e'\in\calE^{01}: e'\cap B\neq \emptyset} z_{e'}\le \left(\sum_{e\in\calE^0: e\cap B\neq \emptyset} x_{e}\right)\cdot \left(\max_{e\in\calE^0: B\cap e\neq \emptyset} \sum_{e'\in \calE^1_e:e'\cap e\cap B\neq\emptyset} y^e_{e'}\right),\]
which completes the proof.
\ep

\begin{rmk}
We note that the proof above gives a better bound than that stated in Proposition~\ref{prop:fhtw-comp-agm} for specific hypergraphs. However, we chose to present a uniform bound for all hypergraphs for its elegance.
\end{rmk}

It is natural to wonder if one can improve upon the bound of Proposition~\ref{prop:fhtw-comp-agm} in the worst-case. In particular, it is natural to wonder if we can prove a bound of the form
\[\fhtw(\calH^0\circ\calH^1)\le O\left(\fhtw(\calH^0)\cdot\max_{e\in\calE^0} \fhtw(\calH^1_e)\right).\]
We argue next that such a bound is not achievable.
\blmm
\label{lem:fthw^2-counter-ex}
There exists hypergraphs $\calH^0$ and a family of corresponding hyerpgraphs $\calH^1$ such that there is an unbounded gap of $\Omega(|\calV|)$ between $\fhtw(\calH^0\circ\calH^1)$ and $\fhtw(\calH^0)\cdot\max_{e\in\calE^0} \fhtw(\calH^1_e)$
\elmm
\bp Let $\calV=\{a_1,\dots,a_n,b_1,\dots,b_n\}$. The hypergraph $\calH^0$ has the following $n$ hyperedges: $e_i= \{a_1,\dots,a_n,b_i\}$. Note that this is essentially a star graph with $\fhtw(\calH^0)=1$. Further, for every $i\in [n]$, the graph $\calH^1_{e_1}$ is the star graph with $a_i$ as the center and $a_1,\dots,a_{i-1},a_{i+1},\dots,a_n,b_i$ as the leaves. Note that for every $i\in [n]$, $\fhtw(\calH^1_{e_i})=1$.

However, note that $\calH^0\circ\calH^1$ has a $K_n$ as a subgraph (in particular the subgraph on $\{a_1,\dots,a_n\}$ forms a clique). Further the only other edges are the $n$ ``spokes''-- $(a_i,b_i)$ for $i\in [n]$. It can be verified that $\fhtw(\calH^0\circ\calH^1)\ge n$, which completes the proof.
\ep

We now come back to the question of proving an upper bound on $\fhtw(\calH^0\circ\calH^1)$. We first note that the argument in proof of Proposition~\ref{prop:fhtw-comp-agm} is wasteful since it does not make use of any tree decomposition representation of the hypergraphs in $\calH^1$. Next we describe a simple algorithm that tries to take advantage of this choice.

We begin with an optimal $\rho^*$-width tree decomposition $(T,\chi)$ of $\calH^0$ as in the proof of Proposition~\ref{prop:fhtw-comp-agm}. Consider an arbitrary bag $B$ in this tree decomposition. For every edge $e\in\calE^0$ such that $e\cap B\neq\emptyset$, let $(T_e,\chi_e)$ be an optimal $\rho^*$-width tree decomposition for $\calH^1_e$. Root this tree at an arbitrary bag $r_e$. 
 Build  new tree $T'$ as follows: for each bag $B$ we create a new bag $B'$ as follows. For every edge $e\in\calE^0$ such that $e\cap B\neq\emptyset$, add $r_e$ to $B'$ and hang the rest of tree $T_e$ from $B'$. Using the argument in the proof of Proposition~\ref{prop:fhtw-comp-agm}, one can argue that for every bag $B'$, $\rho^*_{\calH^0\circ\calH^1}(B')\le \fhtw(\calH^0)\cdot\max_{e\in\calE^0} \fhtw(\calH^1_e)$. Define $\chi'$ as follows: for every $t\in V(T)$, if $B=\chi(t)$, then $\chi'(t)=B'$. For every other vertex $t$ (that comes from some $T_e$), $\chi'(t)=\chi_e(t)$. However, $(T',\chi')$ is {\em not} a valid tree decomposition because it might not satisfy the running intersection property. We fix this with obvious greedy `patchup' phase. The final tree $T''=T'$ but we will modify the bags by defining a new map $\chi''$. For every vertex $v\in\calV$, consider the sub-forest of $T$ with vertices whose bags contain $v$. Add in the set of vertices $t'$ to this forest such that the resulting sub-graph is a tree and for each such $t'$, $\chi''(t')\gets \chi''(t)\cup \{t'\}$ (for every $t$, $\chi''(t)$ is initialized to $\chi'(t)$). Note that when the patchup phase is done $(T'',\chi'')$ is indeed a valid tree decomposition for $\calH^0\circ\calH^1$. It is not too hard to verify that this tree decomposition gives the following bound:

\blmm
\label{lem:gen-algo-compose-bound}
The following holds
\begin{eqnarray*}
   \fhtw(\calH^0\circ\calH^1) &\le& \max_{t\in V(T'')}
   \rho^*_{\calH^0\circ\calH^1}(\chi''(t))\\ & \le&
   \fhtw(\calH^0)\cdot\max_{e\in\calE^0} \fhtw(\calH^1_e) + \max_{t\in V(T'')}
\rho^*_{\calH^0\circ\calH^1}(\chi''(t)-\chi'(t)).\end{eqnarray*}
\elmm

We finish with two remarks:
\begin{enumerate}
\item In the worst-case the algorithm will result in $(T'',\chi'')=(T,\chi)$ (after reducing $(T'',\chi'')$), in which case are back in the proof of Proposition~\ref{prop:fhtw-comp-agm}.
\item Lemma~\ref{lem:gen-algo-compose-bound} for the hard instance in proof of Lemma~\ref{lem:fthw^2-counter-ex} translates into a bound of $n+1$, which is essentially tight.
\end{enumerate}

\section{Concluding Remarks}

The algorithms and ideas developed in this paper are not just on paper.
Implementations of join algorithms based on fractional hypertree width 
for counting graph patterns have shown that the theory predicts what happens
in practice very well: as they are faster
than existing commercial systems by at least an order of 
magnitude~\cite{DBLP:conf/sigmod/NguyenABKNRR14} for selected 
queries. Far beyond graph patterns, we have implemented
$\InsideOut$ within the commerical {\sf LogicBlox} database system~\cite{LB} 
with great performance results.
Moreover, learning from the beautiful work of Olteanu and
Schleich~\cite{DBLP:journals/pvldb/OlteanuS16, DBLP:conf/sigmod/SchleichOC16},
we realized~\cite{indblearn} that $\InsideOut$
can be used to train a large class of machine learning models {\em inside}
the database. Our implementation showed orders of magnitude speedup over the
traditional data modeler route
of exporting the data and running it through \texttt{R} or \texttt{Python}.

Real-world implementation of $\InsideOut$ faces two additional hurdles.
The first problem is that we do not just have materialized
predicates as inputs,
we also have predicates such as $a<b$, $a+b=c$, negations and so on.
These predicates do not have a ``size''.
To solve this problem, one solution is to set the ``size'' of those
predicates to be $\infty$
while computing the $\agm$-bound. For instance, if we have a sub-query of the
form
$Q \leftarrow R(a,b),S(b,c),a+b=c$, where $R$ and $S$ are input
materialized predicates of size $N$, then by setting the size of $a+b=c$
to be infinite, $\agm(Q) = N^2$. This solution does
not work for two reasons. (1) If we knew $a+b=c$, then it is easy to infer
that $|Q| \leq N$ and also to compute $Q$ in time $\tilde O(N)$:
scan over tuples in $R$, use $a+b=c$ to compute $c$, and see if $(b,c) \in S$.
In other words, the $\agm$-bound is no longer tight.
(2) The solution may give an $\infty$-bound when the output size is clearly
bounded. Consider, for example, the query $Q \leftarrow R(a),S(b),a+b=c$; in
this case, $\{a,b,c\}$ is the only hyperedge covering vertex $c$ in the
fractional edge cover.
Our implementation at {\sf LogicBlox} makes use of generalizations of $\agm$ to
queries with functional dependencies and immaterialized predicates (such as
$a+b=c$). These new bounds are based on a linear program
whose variables are marginal entropies 
\cite{DBLP:conf/pods/KhamisNS16,panda}.

The second problem is to select a good variable ordering to run $\InsideOut$
on. In principle, one does not have to use the $\agm$-bound or the
bounds from~\cite{DBLP:conf/pods/KhamisNS16,panda} to estimate the cost of
an $\faq$ subquery. If one were to implement $\InsideOut$ inside any RDBMS, one
could poll that RDBMS's optimizer to figure out the cost of a
given variable ordering. However, there are $n!$ variable orderings, and
optimizer's cost estimation is time-consuming. Furthermore, some subqueries
have inputs which are intermediate results.
Hence, it is much faster to compute a variable ordering minimizing the
$\faqw$ of the query, defined on the bounds
in~\cite{DBLP:conf/pods/KhamisNS16,panda}. As the problem is $\np$-hard,
either an approximation algorithm (from Section~\ref{sec:approx}) or a greedy heuristic suffices
in our experience.

\bibliographystyle{acm}
\bibliography{main}

\def\cprime{$'$}
\begin{thebibliography}{10}

\bibitem{DBLP:books/aw/AbiteboulHV95}
{\sc Abiteboul, S., Hull, R., and Vianu, V.}
\newblock {\em Foundations of Databases}.
\newblock Addison-Wesley, 1995.

\bibitem{anrr}
{\sc {Abo Khamis}, M., Ngo, H.~Q., R{\'{e}}, C., and Rudra, A.}
\newblock Joins via geometric resolutions: Worst-case and beyond.
\newblock In {\em Proceedings of the 34th {ACM} Symposium on Principles of
  Database Systems, {PODS} 2015, Melbourne, Victoria, Australia, May 31 - June
  4, 2015\/} (2015), T.~Milo and D.~Calvanese, Eds., {ACM}, pp.~213--228.

\bibitem{faq-pods16}
{\sc Abo~Khamis, M., Ngo, H.~Q., and Rudra, A.}
\newblock {FAQ}: {Q}uestions {A}sked {F}requently.
\newblock In {\em Proceedings of the 35th ACM SIGMOD-SIGACT-SIGAI Symposium on
  Principles of Database Systems\/} (New York, NY, USA, 2016), PODS '16, ACM,
  pp.~13--28.

\bibitem{DBLP:conf/pods/KhamisNS16}
{\sc Abo~Khamis, M., Ngo, H.~Q., and Suciu, D.}
\newblock Computing join queries with functional dependencies.
\newblock In {\em Proceedings of the 35th ACM SIGMOD-SIGACT-SIGAI Symposium on
  Principles of Database Systems\/} (New York, NY, USA, 2016), PODS '16, ACM,
  pp.~327--342.

\bibitem{panda}
{\sc {Abo Khamis}, M., Ngo, H.~Q., and Suciu, D.}
\newblock What do {S}hannon-type inequalities, submodular width, and
  disjunctive datalog have to do with one another?
\newblock {\em CoRR abs/1612.02503\/} (2016).

\bibitem{adler:dissertation}
{\sc Adler, I.}
\newblock {\em Width functions for hypertree decompositions}.
\newblock 2006.
\newblock Ph.D. Dissertation, Albert-Ludwigs-Universit\"at Freiburg. 2006.

\bibitem{DBLP:journals/corr/abs-1203-3814}
{\sc Adler, I., and Weyer, M.}
\newblock Tree-width for first order formulae.
\newblock {\em Logical Methods in Computer Science 8}, 1 (2012).

\bibitem{aji-thesis}
{\sc Aji, S.~M.}
\newblock {\em Graphical models and iterative decoding}.
\newblock PhD thesis, California Institute of Technology, 2000.

\bibitem{AM00}
{\sc Aji, S.~M., and McEliece, R.~J.}
\newblock The generalized distributive law.
\newblock {\em {IEEE} Transactions on Information Theory 46}, 2 (2000),
  325--343.

\bibitem{LB}
{\sc Aref, M., ten Cate, B., Green, T.~J., Kimelfeld, B., Olteanu, D., Pasalic,
  E., Veldhuizen, T.~L., and Washburn, G.}
\newblock Design and implementation of the {L}ogic{B}lox system.
\newblock In {\em Proceedings of the 2015 {ACM} {SIGMOD} International
  Conference on Management of Data, Melbourne, Victoria, Australia, May 31 -
  June 4, 2015\/} (2015), T.~Sellis, S.~B. Davidson, and Z.~G. Ives, Eds.,
  {ACM}, pp.~1371--1382.

\bibitem{MR985145}
{\sc Arnborg, S., and Proskurowski, A.}
\newblock Linear time algorithms for {NP}-hard problems restricted to partial
  {$k$}-trees.
\newblock {\em Discrete Appl. Math. 23}, 1 (1989), 11--24.

\bibitem{AGM08}
{\sc Atserias, A., Grohe, M., and Marx, D.}
\newblock Size bounds and query plans for relational joins.
\newblock In {\em FOCS\/} (2008), IEEE Computer Society, pp.~739--748.

\bibitem{DBLP:journals/fmsd/BaharFGHMPS97}
{\sc Bahar, R.~I., Frohm, E.~A., Gaona, C.~M., Hachtel, G.~D., Macii, E.,
  Pardo, A., and Somenzi, F.}
\newblock Algebraic decision diagrams and their applications.
\newblock {\em Formal Methods in System Design 10}, 2/3 (1997), 171--206.

\bibitem{DBLP:journals/pvldb/BakibayevKOZ13}
{\sc Bakibayev, N., Kocisk{\'{y}}, T., Olteanu, D., and Zavodny, J.}
\newblock Aggregation and ordering in factorised databases.
\newblock {\em {PVLDB} 6}, 14 (2013), 1990--2001.

\bibitem{faqcs-jacm}
{\sc Bistarelli, S., Montanari, U., and Rossi, F.}
\newblock Semiring-based constraint satisfaction and optimization.
\newblock {\em J. ACM 44}, 2 (Mar. 1997), 201--236.

\bibitem{MR1640205}
{\sc Bodlaender, H.~L.}
\newblock Treewidth: algorithmic techniques and results.
\newblock In {\em Mathematical foundations of computer science 1997
  ({B}ratislava)}, vol.~1295 of {\em Lecture Notes in Comput. Sci.} Springer,
  Berlin, 1997, pp.~19--36.

\bibitem{bodlaender2006treewidth}
{\sc Bodlaender, H.~L.}
\newblock Treewidth: characterizations, applications, and computations.
\newblock In {\em International Workshop on Graph-Theoretic Concepts in
  Computer Science\/} (2006), Springer, pp.~1--14.

\bibitem{MR1860010}
{\sc B{\"o}rger, E., Gr{\"a}del, E., and Gurevich, Y.}
\newblock {\em The classical decision problem}.
\newblock Universitext. Springer-Verlag, Berlin, 2001.
\newblock Reprint of the 1997 original.

\bibitem{braultbaron_et_al:LIPIcs:2015:4910}
{\sc Brault-Baron, J., Capelli, F., and Mengel, S.}
\newblock {Understanding Model Counting for beta-acyclic CNF-formulas}.
\newblock In {\em 32nd International Symposium on Theoretical Aspects of
  Computer Science (STACS 2015)\/} (Dagstuhl, Germany, 2015), E.~W. Mayr and
  N.~Ollinger, Eds., vol.~30 of {\em Leibniz International Proceedings in
  Informatics (LIPIcs)}, Schloss Dagstuhl--Leibniz-Zentrum fuer Informatik,
  pp.~143--156.

\bibitem{brouwer-kolen-1980}
{\sc Brouwer, A., and Kolen, A.}
\newblock A super-balanced hypergraph has a nest point.
\newblock Tech. Report.

\bibitem{DBLP:journals/siamcomp/CaiCL13}
{\sc Cai, J., Chen, X., and Lu, P.}
\newblock Graph homomorphisms with complex values: {A} dichotomy theorem.
\newblock {\em {SIAM} J. Comput. 42}, 3 (2013), 924--1029.

\bibitem{MR3105918}
{\sc Cai, J.-Y., Lu, P., and Xia, M.}
\newblock The complexity of complex weighted {B}oolean \#{CSP}.
\newblock {\em J. Comput. System Sci. 80}, 1 (2014), 217--236.

\bibitem{DBLP:journals/tcs/ChekuriR00}
{\sc Chekuri, C., and Rajaraman, A.}
\newblock Conjunctive query containment revisited.
\newblock {\em Theor. Comput. Sci. 239}, 2 (2000), 211--229.

\bibitem{DBLP:conf/ecai/Chen04}
{\sc Chen, H.}
\newblock Quantified constraint satisfaction and bounded treewidth.
\newblock In {\em Proceedings of the 16th Eureopean Conference on Artificial
  Intelligence, ECAI'2004, including Prestigious Applicants of Intelligent
  Systems, {PAIS} 2004, Valencia, Spain, August 22-27, 2004\/} (2004),
  pp.~161--165.

\bibitem{DBLP:conf/lics/ChenD12}
{\sc Chen, H., and Dalmau, V.}
\newblock Decomposing quantified conjunctive (or disjunctive) formulas.
\newblock In {\em Proceedings of the 27th Annual {IEEE} Symposium on Logic in
  Computer Science, {LICS} 2012, Dubrovnik, Croatia, June 25-28, 2012\/}
  (2012), {IEEE} Computer Society, pp.~205--214.

\bibitem{DBLP:journals/jcss/ChenG10a}
{\sc Chen, H., and Grohe, M.}
\newblock Constraint satisfaction with succinctly specified relations.
\newblock {\em J. Comput. Syst. Sci. 76}, 8 (2010), 847--860.

\bibitem{chen_et_al:LIPIcs:2015:4980}
{\sc Chen, H., and Mengel, S.}
\newblock {A Trichotomy in the Complexity of Counting Answers to Conjunctive
  Queries}.
\newblock In {\em 18th International Conference on Database Theory (ICDT
  2015)\/} (Dagstuhl, Germany, 2015), M.~Arenas and M.~Ugarte, Eds., vol.~31 of
  {\em Leibniz International Proceedings in Informatics (LIPIcs)}, Schloss
  Dagstuhl--Leibniz-Zentrum fuer Informatik, pp.~110--126.

\bibitem{fft}
{\sc Cooley, J.~W., and Tukey, J.~W.}
\newblock {An algorithm for the machine calculation of complex Fourier series}.
\newblock {\em Mathematics of Computation 19\/} (1965), 297--301.

\bibitem{MR2002e:68001}
{\sc Cormen, T.~H., Leiserson, C.~E., Rivest, R.~L., and Stein, C.}
\newblock {\em Introduction to algorithms}, second~ed.
\newblock MIT Press, Cambridge, MA, 2001.

\bibitem{DBLP:journals/cacm/DavisLL62}
{\sc Davis, M., Logemann, G., and Loveland, D.~W.}
\newblock A machine program for theorem-proving.
\newblock {\em Commun. {ACM} 5}, 7 (1962), 394--397.

\bibitem{DBLP:journals/ai/Dechter99}
{\sc Dechter, R.}
\newblock Bucket elimination: {A} unifying framework for reasoning.
\newblock {\em Artif. Intell. 113}, 1-2 (1999), 41--85.

\bibitem{DBLP:conf/ecai/DechterOM08}
{\sc Dechter, R., Otten, L., and Marinescu, R.}
\newblock On the practical significance of hypertree vs. treewidth.
\newblock In {\em {ECAI} 2008 - 18th European Conference on Artificial
  Intelligence, Patras, Greece, July 21-25, 2008, Proceedings\/} (2008),
  M.~Ghallab, C.~D. Spyropoulos, N.~Fakotakis, and N.~M. Avouris, Eds.,
  vol.~178 of {\em Frontiers in Artificial Intelligence and Applications},
  {IOS} Press, pp.~913--914.

\bibitem{journals/ai/DechterP89}
{\sc Dechter, R., and Pearl, J.}
\newblock Tree clustering for constraint networks.
\newblock {\em Artificial Intelligence 38}, 3 (1989), 353--366.

\bibitem{DBLP:conf/kr/DechterR94}
{\sc Dechter, R., and Rish, I.}
\newblock Directional resolution: The davis-putnam procedure, revisited.
\newblock In {\em Proceedings of the 4th International Conference on Principles
  of Knowledge Representation and Reasoning (KR'94). Bonn, Germany, May 24-27,
  1994.\/} (1994), J.~Doyle, E.~Sandewall, and P.~Torasso, Eds., Morgan
  Kaufmann, pp.~134--145.

\bibitem{DBLP:conf/icdt/DurandM13}
{\sc Durand, A., and Mengel, S.}
\newblock Structural tractability of counting of solutions to conjunctive
  queries.
\newblock In {\em Joint 2013 {EDBT/ICDT} Conferences, {ICDT} '13 Proceedings,
  Genoa, Italy, March 18-22, 2013\/} (2013), W.~Tan, G.~Guerrini, B.~Catania,
  and A.~Gounaris, Eds., {ACM}, pp.~81--92.

\bibitem{DBLP:journals/jcss/DurandM14}
{\sc Durand, A., and Mengel, S.}
\newblock The complexity of weighted counting for acyclic conjunctive queries.
\newblock {\em J. Comput. Syst. Sci. 80}, 1 (2014), 277--296.

\bibitem{DBLP:journals/jacm/Fagin83}
{\sc Fagin, R.}
\newblock Degrees of acyclicity for hypergraphs and relational database
  schemes.
\newblock {\em J. ACM 30}, 3 (1983), 514--550.

\bibitem{DBLP:journals/corr/FischlGP16}
{\sc Fischl, W., Gottlob, G., and Pichler, R.}
\newblock General and fractional hypertree decompositions: Hard and easy cases.
\newblock In {\em ACM PODS 2018, to appear\/} (2018).

\bibitem{DBLP:conf/aaai/Freuder90}
{\sc Freuder, E.~C.}
\newblock Complexity of k-tree structured constraint satisfaction problems.
\newblock In {\em Proceedings of the 8th National Conference on Artificial
  Intelligence. Boston, Massachusetts, July 29 - August 3, 1990, 2 Volumes.\/}
  (1990), H.~E. Shrobe, T.~G. Dietterich, and W.~R. Swartout, Eds., {AAAI}
  Press / The {MIT} Press, pp.~4--9.

\bibitem{DBLP:conf/uai/GogateD13}
{\sc Gogate, V., and Domingos, P.}
\newblock Structured message passing.
\newblock In {\em Proceedings of the Twenty-Ninth Conference on Uncertainty in
  Artificial Intelligence, Bellevue, WA, USA, August 11-15, 2013\/} (2013),
  {AUAI} Press, Corvallis, Oregon.

\bibitem{Goldberg:2010:CDP:1958055.1958080}
{\sc Goldberg, L.~A., Grohe, M., Jerrum, M., and Thurley, M.}
\newblock A complexity dichotomy for partition functions with mixed signs.
\newblock {\em SIAM J. Comput. 39}, 7 (Aug. 2010), 3336--3402.

\bibitem{DBLP:dblp_journals/jacm/GolombB65}
{\sc Golomb, S.~W., and Baumert, L.~D.}
\newblock Backtrack programming.
\newblock pp.~516--524.

\bibitem{DBLP:conf/pods/GottlobGLS16}
{\sc Gottlob, G., Greco, G., Leone, N., and Scarcello, F.}
\newblock Hypertree decompositions: Questions and answers.
\newblock In {\em Proceedings of the 35th {ACM} {SIGMOD-SIGACT-SIGAI} Symposium
  on Principles of Database Systems, {PODS} 2016, San Francisco, CA, USA, June
  26 - July 01, 2016\/} (2016), T.~Milo and W.~Tan, Eds., {ACM}, pp.~57--74.

\bibitem{DBLP:journals/jcss/GottlobLS03}
{\sc Gottlob, G., Leone, N., and Scarcello, F.}
\newblock Robbers, marshals, and guards: game theoretic and logical
  characterizations of hypertree width.
\newblock {\em J. Comput. Syst. Sci. 66}, 4 (2003), 775--808.

\bibitem{ontheuniversalrelation}
{\sc Graham, M.~H.}
\newblock On the universal relation, 1980.
\newblock Tech. Report.

\bibitem{DBLP:conf/pods/GrecoS14}
{\sc Greco, G., and Scarcello, F.}
\newblock Counting solutions to conjunctive queries: structural and hybrid
  tractability.
\newblock In Hull and Grohe \cite{DBLP:conf/pods/2014}, pp.~132--143.

\bibitem{GM06}
{\sc Grohe, M., and Marx, D.}
\newblock Constraint solving via fractional edge covers.
\newblock In {\em SODA\/} (2006), ACM Press, pp.~289--298.

\bibitem{DBLP:journals/talg/GroheM14}
{\sc Grohe, M., and Marx, D.}
\newblock Constraint solving via fractional edge covers.
\newblock {\em {ACM} Transactions on Algorithms 11}, 1 (2014), 4.

\bibitem{DBLP:dblp_journals/fttcs/Guruswami06}
{\sc Guruswami, V.}
\newblock Algorithmic results in list decoding.

\bibitem{MR2590822}
{\sc Guruswami, V., Umans, C., and Vadhan, S.}
\newblock Unbalanced expanders and randomness extractors from
  {P}arvaresh-{V}ardy codes.
\newblock {\em J. ACM 56}, 4 (2009), Art. 20, 34.

\bibitem{DBLP:journals/ai/GyssensJC94}
{\sc Gyssens, M., Jeavons, P., and Cohen, D.~A.}
\newblock Decomposing constraint satisfaction problems using database
  techniques.
\newblock {\em Artif. Intell. 66}, 1 (1994), 57--89.

\bibitem{DBLP:conf/adbt/GyssensP82}
{\sc Gyssens, M., and Paredaens, J.}
\newblock A decomposition methodology for cyclic databases.
\newblock In {\em Advances in Data Base Theory\/} (1982), pp.~85--122.

\bibitem{DBLP:journals/ijar/Darwiche96}
{\sc Huang, C., and Darwiche, A.}
\newblock Inference in belief networks: {A} procedural guide.
\newblock {\em Int. J. Approx. Reasoning 15}, 3 (1996), 225--263.

\bibitem{DBLP:conf/pods/2014}
{\sc Hull, R., and Grohe, M.}, Eds.
\newblock {\em Proceedings of the 33rd {ACM} {SIGMOD-SIGACT-SIGART} Symposium
  on Principles of Database Systems, PODS'14, Snowbird, UT, USA, June 22-27,
  2014\/} (2014), {ACM}.

\bibitem{KDLD05}
{\sc Kask, K., Dechter, R., Larrosa, J., and Dechter, A.}
\newblock Unifying tree decompositions for reasoning in graphical models.
\newblock {\em Artif. Intell. 166}, 1-2 (2005), 165--193.

\bibitem{khatri-rao}
{\sc Khatri, C.~G., and Rao, C.~R.}
\newblock Solutions to some functional equations and their applications to
  characterization of probability distributions.
\newblock {\em Sankhya: The Indian Journal of Statistics, Series A 30}, 2
  (1968).

\bibitem{DBLP:conf/pods/Koch10}
{\sc Koch, C.}
\newblock Incremental query evaluation in a ring of databases.
\newblock In {\em Proceedings of the Twenty-Ninth {ACM} {SIGMOD-SIGACT-SIGART}
  Symposium on Principles of Database Systems, {PODS} 2010, June 6-11, 2010,
  Indianapolis, Indiana, {USA}\/} (2010), J.~Paredaens and D.~V. Gucht, Eds.,
  {ACM}, pp.~87--98.

\bibitem{KW08}
{\sc Kohlas, J., and Wilson, N.}
\newblock Semiring induced valuation algebras: Exact and approximate local
  computation algorithms.
\newblock {\em Artif. Intell. 172}, 11 (2008), 1360--1399.

\bibitem{MR2778120}
{\sc Koller, D., and Friedman, N.}
\newblock {\em Probabilistic graphical models}.
\newblock Adaptive Computation and Machine Learning. MIT Press, Cambridge, MA,
  2009.
\newblock Principles and techniques.

\bibitem{DBLP:books/cs/Maier83}
{\sc Maier, D.}
\newblock {\em The Theory of Relational Databases}.
\newblock Computer Science Press, 1983.

\bibitem{Marx:2010:AFH:1721837.1721845}
{\sc Marx, D.}
\newblock Approximating fractional hypertree width.
\newblock {\em ACM Trans. Algorithms 6}, 2 (Apr. 2010), 29:1--29:17.

\bibitem{DBLP:journals/mst/Marx11}
{\sc Marx, D.}
\newblock Tractable structures for constraint satisfaction with truth tables.
\newblock {\em Theory Comput. Syst. 48}, 3 (2011), 444--464.

\bibitem{DBLP:journals/jacm/Marx13}
{\sc Marx, D.}
\newblock Tractable hypergraph properties for constraint satisfaction and
  conjunctive queries.
\newblock {\em J. {ACM} 60}, 6 (2013), 42.

\bibitem{MR2010377}
{\sc Newman, M. E.~J.}
\newblock The structure and function of complex networks.
\newblock {\em SIAM Rev. 45}, 2 (2003), 167--256 (electronic).

\bibitem{nnrr}
{\sc Ngo, H.~Q., Nguyen, D.~T., R{\'{e}}, C., and Rudra, A.}
\newblock Beyond worst-case analysis for joins with minesweeper.
\newblock In Hull and Grohe \cite{DBLP:conf/pods/2014}, pp.~234--245.

\bibitem{indblearn}
{\sc Ngo, H.~Q., Nguyen, X., Olteanu, D., and Schleich, M.}
\newblock In-database factorized learning, 2017.
\newblock Manuscript.

\bibitem{NPRR12}
{\sc Ngo, H.~Q., Porat, E., R{\'e}, C., and Rudra, A.}
\newblock Worst-case optimal join algorithms.
\newblock In {\em PODS\/} (2012), M.~Lenzerini and M.~Benedikt, Eds., ACM,
  pp.~37--48.

\bibitem{MR2874136}
{\sc Ngo, H.~Q., Porat, E., and Rudra, A.}
\newblock Efficiently decodable error-correcting list disjunct matrices and
  applications (extended abstract).
\newblock In {\em Automata, languages and programming. {P}art {I}}, vol.~6755
  of {\em Lecture Notes in Comput. Sci.} Springer, Heidelberg, 2011,
  pp.~557--568.

\bibitem{skew}
{\sc Ngo, H.~Q., R{\'e}, C., and Rudra, A.}
\newblock Skew strikes back: New developments in the theory of join algorithms.
\newblock In {\em SIGMOD RECORD\/} (2013), pp.~5--16.

\bibitem{DBLP:conf/sigmod/NguyenABKNRR14}
{\sc Nguyen, D.~T., Aref, M., Bravenboer, M., Kollias, G., Ngo, H.~Q.,
  R{\'{e}}, C., and Rudra, A.}
\newblock Join processing for graph patterns: An old dog with new tricks.
\newblock In {\em Proceedings of the Third International Workshop on Graph Data
  Management Experiences and Systems, {GRADES} 2015, Melbourne, VIC, Australia,
  May 31 - June 4, 2015\/} (2015), pp.~2:1--2:8.

\bibitem{OS00}
{\sc Olshevsky, V., and Shokrollahi, M.~A.}
\newblock Matrix-vector product for confluent cauchy-like matrices with
  application to confluent rational interpolation.
\newblock In {\em Proceedings of the Thirty-Second Annual {ACM} Symposium on
  Theory of Computing, May 21-23, 2000, Portland, OR, {USA}\/} (2000), F.~F.
  Yao and E.~M. Luks, Eds., {ACM}, pp.~573--581.

\bibitem{DBLP:journals/pvldb/OlteanuS16}
{\sc Olteanu, D., and Schleich, M.}
\newblock {F:} regression models over factorized views.
\newblock {\em {PVLDB} 9}, 13 (2016), 1573--1576.

\bibitem{OZ15}
{\sc Olteanu, D., and Z\'{a}vodn\'{y}, J.}
\newblock Size bounds for factorised representations of query results.
\newblock {\em ACM Trans. Datab. Syst. 40}, 1 (2015).

\bibitem{ordyniak_et_al:LIPIcs:2010:2855}
{\sc Ordyniak, S., Paulusma, D., and Szeider, S.}
\newblock {Satisfiability of Acyclic and Almost Acyclic CNF Formulas}.
\newblock In {\em {FSTTCS 2010}\/} (2010), vol.~8.

\bibitem{Pagh:2012:CTC:2133845.2133976}
{\sc Pagh, R., and Tsourakakis, C.~E.}
\newblock Colorful triangle counting and a mapreduce implementation.
\newblock {\em Inf. Process. Lett. 112}, 7 (Mar. 2012), 277--281.

\bibitem{P00}
{\sc Pan, V.~Y.}
\newblock Nearly optimal computations with structured matrices.
\newblock In {\em Proceedings of the Eleventh Annual {ACM-SIAM} Symposium on
  Discrete Algorithms, January 9-11, 2000, San Francisco, CA, {USA.}\/} (2000),
  D.~B. Shmoys, Ed., {ACM/SIAM}, pp.~953--962.

\bibitem{DBLP:conf/aaai/Pearl82}
{\sc Pearl, J.}
\newblock Reverend bayes on inference engines: {A} distributed hierarchical
  approach.
\newblock In {\em Proceedings of the National Conference on Artificial
  Intelligence. Pittsburgh, PA, August 18-20, 1982.\/} (1982), D.~L. Waltz,
  Ed., {AAAI} Press, pp.~133--136.

\bibitem{Pearl:1986:FPS:9075.9076}
{\sc Pearl, J.}
\newblock Fusion, propagation, and structuring in belief networks.
\newblock {\em Artif. Intell. 29}, 3 (Sept. 1986), 241--288.

\bibitem{pichler:2013}
{\sc Pichler, R., and Skritek, S.}
\newblock Tractable counting of the answers to conjunctive queries.
\newblock {\em J. Comput. Syst. Sci. 79}, 6 (Sept. 2013), 984--1001.

\bibitem{DBLP:conf/ijcai/RollonLD13}
{\sc Rollon, E., Larrosa, J., and Dechter, R.}
\newblock Semiring-based mini-bucket partitioning schemes.
\newblock In {\em {IJCAI} 2013, Proceedings of the 23rd International Joint
  Conference on Artificial Intelligence, Beijing, China, August 3-9, 2013\/}
  (2013), F.~Rossi, Ed., {IJCAI/AAAI}.

\bibitem{Rossi:2006:HCP:1207782}
{\sc Rossi, F., Beek, P.~v., and Walsh, T.}
\newblock {\em Handbook of Constraint Programming (Foundations of Artificial
  Intelligence)}.
\newblock Elsevier Science Inc., New York, NY, USA, 2006.

\bibitem{DBLP:journals/sigmod/Scarcello05}
{\sc Scarcello, F.}
\newblock Query answering exploiting structural properties.
\newblock {\em SIGMOD Record 34}, 3 (2005), 91--99.

\bibitem{DBLP:conf/sigmod/SchleichOC16}
{\sc Schleich, M., Olteanu, D., and Ciucanu, R.}
\newblock Learning linear regression models over factorized joins.
\newblock In {\em SIGMOD\/} (2016), pp.~3--18.

\bibitem{DBLP:conf/www/SuriV11}
{\sc Suri, S., and Vassilvitskii, S.}
\newblock Counting triangles and the curse of the last reducer.
\newblock In {\em WWW\/} (2011), pp.~607--614.

\bibitem{Tarjan:1984:SLA:1169.1179}
{\sc Tarjan, R.~E., and Yannakakis, M.}
\newblock Simple linear-time algorithms to test chordality of graphs, test
  acyclicity of hypergraphs, and selectively reduce acyclic hypergraphs.
\newblock {\em SIAM J. Comput. 13}, 3 (July 1984), 566--579.

\bibitem{tracy-singh}
{\sc Tracy, D.~S., and Singh, R.~P.}
\newblock A new matrix product and its applications in partitioned matrix
  differentiation.
\newblock {\em Statistica Neerlandica 26}, 4 (1972), 143--157.

\bibitem{DBLP:conf/icdm/Tsourakakis08}
{\sc Tsourakakis, C.~E.}
\newblock Fast counting of triangles in large real networks without counting:
  Algorithms and laws.
\newblock In {\em Proceedings of the 8th {IEEE} International Conference on
  Data Mining {(ICDM} 2008), December 15-19, 2008, Pisa, Italy\/} (2008),
  {IEEE} Computer Society, pp.~608--617.

\bibitem{DBLP:books/cs/Ullman89}
{\sc Ullman, J.~D.}
\newblock {\em Principles of Database and Knowledge-Base Systems, Volume II}.
\newblock Computer Science Press, 1989.

\bibitem{leapfrog}
{\sc Veldhuizen, T.~L.}
\newblock Triejoin: A simple, worst-case optimal join algorithm.
\newblock In {\em ICDT\/} (2014), pp.~96--106.

\bibitem{dblp:conf/vldb/yannakakis81}
{\sc Yannakakis, M.}
\newblock Algorithms for acyclic database schemes.
\newblock In {\em VLDB\/} (1981), pp.~82--94.

\bibitem{yu1984determining}
{\sc Yu, C., and Ozsoyoglu, M.}
\newblock On determining tree-query membership of a distributed query.
\newblock {\em Informatica 22}, 3 (1984), 261--282.

\bibitem{Yuster:2005:FSM:1077464.1077466}
{\sc Yuster, R., and Zwick, U.}
\newblock Fast sparse matrix multiplication.
\newblock {\em ACM Trans. Algorithms 1}, 1 (July 2005), 2--13.

\bibitem{zhangpoole94}
{\sc Zhang, N., and Poole, D.}
\newblock {A simple approach to Bayesian network computations}.
\newblock In {\em Proceedings of the Tenth Canadian Conference on Artificial
  Intelligence\/} (1994), pp.~171--178.

\bibitem{MR1426261}
{\sc Zhang, N.~L., and Poole, D.}
\newblock Exploiting causal independence in {B}ayesian network inference.
\newblock {\em J. Artificial Intelligence Res. 5\/} (1996), 301--328.

\end{thebibliography}

\newpage

\appendix
\section{More examples of problems reducible to $\faq$}
\label{app:sec:reductions}

This section presents many examples showing how $\faq$ captures a wide range of problems,
such as graphical model inference, matrix multiplication, constraint satisfaction,
quantified conjunctive query evaluation, etc. Some of the reductions to
$\faqcs$ are already discussed in the seminal work of 
Dechter~\cite{DBLP:journals/ai/Dechter99},
Aji and McEliece \cite{AM00}, and Kohlas and Wilson~\cite{KW08}; other
reductions to $\faqcs$ are new; all reductions to the general $\faq$ problem
(over multiple semirings) are new.
We first present the examples for $\faqcs$ by considering the following semirings.
\bi
 \item $(\{\true,\false\}, \vee, \wedge)$: the {\em Boolean semiring}
 \item $(\R, +, \times)$: the {\em sum-product semiring}
 \item $(\R_+, \max, \times)$: the {\em max-product semiring}, which is
     essentially equivalent to the $(\R, \min, +)$ semiring.\footnote{If the
         ranges of factors are non-negative, then we can take log of the factors
     to turn product into sum.}
 \item $(2^U, \cup, \cap)$: the {\em set semiring}, where
 $U$ is some set called the universe.
 In this case, $\emptyset$ is the additive identity, and $U$ itself is the
 multiplicative identity.
\ei

\subsection{The Boolean semiring}

\begin{ex}[{\sf Satisfiability}]
    Let $\varphi$ be a CNF formula over $n$ Boolean variables $X_1,\dots,X_n$.
    Let $\calH=(\calV,\calE)$ be the hypergraph of $\varphi$.
    Then, each clause of $\varphi$ is a factor $\psi_S$, and
    the question of whether $\varphi$ is satisfiable is the same as 
    evaluating the constant function
    \[ \varphi = \bigvee_{\mv x} \bigwedge_{S\in\calE} \psi_S(\mv x_S). \]
    Note that in this case each factor is very compactly represented.
    The size of each factor $\psi_S$ is $O(|S|)$, i.e. linear in the number of
    variables that the factor is on. We will have much more to say about such
    compact representation in Sections 
    \ref{sec:representation} and
\ref{app:sec:representation}.
\end{ex}

\begin{ex}[$\problemname{$k$-colorability}$]
    Let $G=(V,E)$ be a graph. 
    Define the following instance of $\faqcs$.
    Let $\calH=G$, and
    $\Dom(X_v) = [k]$ for every $v\in V$.
    Define a factor $\psi_{uv}$ for every edge $uv\in E$ to be the predicate
    $\psi_{uv}(c_1,c_2) = (c_1\neq c_2)$.
    The question of whether $G$ is $k$-colorable is
    equivalent to evaluating the following constant function:
    \[ \varphi = \bigvee_{\mv x} \bigwedge_{uv \in E} \psi_{uv}(x_u,x_v). \]
    Even for non-constant $k$, each factor in this example is also very
    compactly represented, amounting to an inequality.
\end{ex}

\begin{ex}[Boolean conjunctive query]
    The Boolean conjunctive query evaluation problem ($\bcq$) can be written
    as follows. The query $\Phi$ has the attribute set $\{X_1,\ldots, X_n\}$ and the input relation set $\atoms(\Phi)$.
    Each relation $R\in \atoms(\Phi)$ is on attribute set $\vars(R)\subseteq \{X_1,\ldots, X_n\}$.
    We want to know if there exists a tuple satisfying all relations:
    \[ \Phi = \exists_{X_1} \cdots \exists_{X_n} 
        \bigwedge_{R\in \atoms(\Phi)} R(\vars(R))
    \]
    Define a hypergraph $\calH=(\calV,\calE)$, where $\calV$ is the set of all
    attributes, and $\calE = \{ \vars(R) \suchthat R\in \atoms(\Phi)\}$.
    For each $S\in \calE$ corresponding to relation $R$, 
    there is a factor $\psi_S$ where
    $\psi_S(\mv x_S) = (\mv x_S \in R)$.
    Then, the problem is reduced to evaluating the constant function
    \[ \varphi = \bigvee_{\mv x} \bigwedge_{S\in\calE} \psi_S(\mv x_S). \]
    Note that in this case, the typical input relation encoding is to {\em list}
    all tuples that belong to a given input relation. Thus, the inputs are not
    as compact as the first two examples.
\end{ex}

\begin{ex}[Constraint satisfaction]
\label{ex:csp}
The \problemname{Constraint Satisfaction Problem}
($\csp$) is reducible to $\faqcs$ in the obvious way. Note that
$\sat$, $\problemname{$3$-colorability}$, $\bcq$ are all special 
cases of $\csp$.
\end{ex}

To summarize, the subtle issue of input encodings already shows up in the 
above examples. In $\sat$ each input factor $\psi_S$ is encoded using 
a clause of size $O(|S|)$. In $\bcq$ we use the {\em listing
encoding}, where each input factor is encoded with a table of entries whose 
functional values are $\true$.
The $\csp$ problem as defined above is underspecified.

\begin{ex}[Conjunctive query]
\label{ex:conj}
    In the general conjunctive query evaluation ($\cqe$) problem, 
    the query has existential quantifiers over a subset of variables
    in $[n]-F$. It should be obvious that conjunctive query evaluation is
    reduced to the following form of $\faqcs$.
    The problem is to compute the output function
    \[ \varphi(\mv x_F) = \bigvee_{\mv x_{[n]-F}} 
                          \bigwedge_{S\in\calE} \psi_S(\mv x_S). \]
\end{ex}

\begin{ex}[Natural join query]
    The natural join query is the $\cqe$ problem when all variables are free.
Essentially, a natural join query is a {\em quantifier-free conjunctive query}, 
after some pre-processing steps.
The reason we need pre-processing is because, strictly speaking in a
conjunctive query the same relation might appear several times, and the same
variable might appear more than once in the same atom.
This point is immaterial at this point of our discussion.
\end{ex}

\begin{ex}[Coding theory]
Here's a typical setting in coding theory. Let $q$ be a prime power, and $\F_q$
denote the field of order $q$. Let $n\geq k\geq 1$ be integers. Then, an
{\em $(n,k)_q$-code} $C$ is simply a subset of $\F_q^n$ of size $|\F_q^k|$. If
every pair of codewords (i.e. members of $C$) have Hamming distance at least
$d$, then we call the code an $(n,k,d)_q$-code. The 
$\problemname{List Recovery}$ problem
\cite{DBLP:dblp_journals/fttcs/Guruswami06}
in (modern) coding theory is the following problem. We are given, for each
position $i\in [n]$ of the code, a subset $S_i \subseteq \F_q$ of symbols. The
objective is to ``recover'' the set of all codewords $\mv c = (c_1,\dots,c_n)
\in C$ for which $c_i \in S_i$ for all $i$. (Technically, we also want to give
some lax where $c_i\in S_i$ for only a fraction of the positions $i$; but that
requirement is not important for our discussion here.) List recoverable codes
are codes for which, if the $S_i$ are ``small'' then the list of codewords to be
recovered (satisfying the above condition) is also ``small.'' Of course, we 
also would like the recovery process to be as fast as possible. 

The list recovery paradigm has found important applications. It should be noted
that when $|S_i|=1$ for all $i$ then we get back the list decoding problem.
Powerful {\em expander graphs} are constructed using the Parvaresh-Vardy family
of codes, precisely because they are list recoverable \cite{MR2590822}. 
In \cite{MR2874136}, we
used list recovery to construct very good group testing schemes.

In the {\sf list recovery} problem, we have $n+1$ factors: $\psi_{[n]}$, 
and $\psi_i$ for $i\in [n]$, where
\begin{eqnarray*}
    \psi_{[n]}(\mv c) &=& \true \text{ iff } \mv c \in C\\
    \psi_{i}(c) &=& \true \text{ iff } c \in S_i.
\end{eqnarray*}
The $\faqcs$ instance is
\[ \varphi(c_1,\dots,c_n) = \psi_{[n]}(c_1,\dots,c_n) \wedge 
                            \bigwedge_{i=1}^n \psi_i(c_i).
\]
\end{ex}

\subsection{The sum-product semiring}

\begin{ex}[Complex network analysis]
In complex network analysis we often want to count the number of
occurrences of a given small (induced or non-induced) subgraph $H$ inside of
a massive graph $G$. This problem is pervasive in social network and biological
network analysis, where each occurrence of the subgraph is a ``pattern'' that
one wants to mine from the network. 
A canonical example is the {\sf triangle counting} problem, where we want to
count the number of triangles in a given graph $G=(V,E)$.
The number of triangles is used to compute the clustering coefficients 
\cite{DBLP:conf/www/SuriV11,MR2010377} and transitivity            
ratio \cite{DBLP:conf/icdm/Tsourakakis08,
Pagh:2012:CTC:2133845.2133976}.
We reduce {\sf triangle counting} to $\faqcs$ as follows.
The $\faqcs$'s hypergraph is
$\calH = (\calV,\calE)$, where
\begin{eqnarray*}
    \calV &=& \{1,2,3\},\\
    \calE &=& \{\{1,2\},\{1,3\},\{2,3\}\}
\end{eqnarray*}
The domains are $\Dom(X_i) = V$, for $i\in[3]$.
The factors are all the same: $\psi_{12}=\psi_{13}=\psi_{23}=\psi$,
where 
$$\psi(u,v) = 
   \begin{cases}
       1 & \text{if } \{u, v\}\in E, u<v\\
       0 & \text{otherwise.}
   \end{cases}
$$
(In terms of input encoding, the factor $\psi$ can be represented with a 
data structure of size $O(|E|)$, for example.)
The problem is to compute the constant function
\[ \varphi = \sum_{x_1\in V}\sum_{x_2\in V} \sum_{x_3\in V} 
             \psi(x_1,x_2) \cdot \psi(x_1,x_3) \cdot \psi(x_2,x_3).
\]
The same strategy can be used to count the number of
occurrences of $H$ in $G$, where $H$ is a fixed graph.
For example, $H$ can be a $k$-clique.

\end{ex}

\brmk
If we'd like to list {\em induced subgraphs}, there should be other 
factors indicating the {\em non-existence} of edges in the subgraph pattern.
These factors correspond to inequalities, and they can be compactly represented.
\ermk

\begin{ex}[$\ssat$]
In $\sat$ we want a Boolean answer: ``is the formula satisfiable or not?'' 
In $\ssat$ we want a more specific piece of information: ``exactly how many
satisfying assignments are there?'' 
Let $\Phi$ be a CNF formula over $n$ Boolean variables $X_1,\dots,X_n$.
Let $\calH=(\calV,\calE)$ be the hypergraph of $\Phi$.
Then for each clause $C$ of $\Phi$, there is a factor $\psi_S$ where $\mv X_S$ is the set of variables occurring in $C$ and $\psi_S$ is defined
by
\[ \psi_S(\mv x_S) = 
     \begin{cases} 
         1 & \text{ if $\mv x_S$ satisfies $C$}\\
         0 &\text{ otherwise.}
     \end{cases}
\]
The problem of counting the number of satisfying assignments is
the same as evaluating the constant function
\[ \varphi = \sum_{\mv x} \prod_{S\in\calE} \psi_S(\mv x_S). \]
\end{ex}

\begin{ex}[$\problemname{Permanent}$]
$\ssat$ is $\sharpP$-complete.
Another canonical $\sharpP$-complete problem is {\sf Permanent}, which is the
problem of evaluating the {\em permanent} $\perm(\mv A)$ of a given {\em binary}
square matrix $\mv A$. 
Let $S_n$ denote the symmetric group on $[n]$, then
the permanent of $\mv A = (a_{ij})_{i,j=1}^n$ is defined as follows. 
\[ \perm(\mv A) := \sum_{\pi \in S_n} \prod_{i=1}^n a_{i\pi(i)}. \]
Note that $\perm(\mv A)$ has exactly the same form as $\det(\mv A)$ written 
using {\em Leibniz formula} except that the signs are all $1$. 

The {\sf Permanent} problem can be written in the sum-product form
as follows. 
Given the input matrix $\mv A = (a_{ij})$, 
each row index $i\in[n]$
corresponds to a vertex $i$ and has a corresponding singleton factor $\psi_i$ 
where $\psi_i(j) = a_{ij}$. 
Moreover, there are $\binom n 2$ factors $\psi_{jk}$ for $j\neq k \in [n]$,
where $\psi_{jk}(x,y)= 1$ if $x\neq y$ and $0$ if $x=y$.
The problem is to evaluate the constant function
\[ \varphi = \sum_{\mv x} \prod_{i\in [n]} \psi_i(x_i) \prod_{(j\neq k)}
    \psi_{jk}(x_j,x_k).
\]
\end{ex}
\begin{ex}[Probabilistic graphical models]
\label{ex:pgm}
We consider {\em probabilistic graphical model}s ($\pgm$s) on discrete finite 
domains. Without loss of generality, we restrict ourselves to 
{\em undirected graphical models} (also called {\em Markov Random Field}, 
{\em factor graph}, {\em Gibbs distribution}, etc.) 
If the input model is directed, we {\em moralize} it to get an undirected model.
The model can be represented by a hypergraph $\calH=(\calV,\calE)$, where
there are $n$ discrete random variables $X_1,\dots,X_n$ on finite domains 
$\Dom(X_1), \dots, \Dom(X_n)$ respectively, and $m$ {\em factors} 
(also called {\em potential functions}):
\[ \psi_S : \prod_{i \in S} \Dom(X_i) \to \R_+. \]
Typically, we want to {\em learn} the model and perform {\em inference} 
from the model.\footnote{From a Bayesian point of view, there is no difference
between the two. ``Learning'' refers to the task of estimating the
parameters of a model given the observed data. 
For example, we often estimate
the parameters using the {\em Maximum A Posteriori} (MAP) estimate 
$\hat\theta = \argmax_\theta p(\theta \suchthat \mv x_A) = 
\argmax_{\theta} \left\{\log p(\mv x_A \suchthat \theta) + 
\log p(\theta)\right\}$.
Here, $\mv x_A$ is the set of observed values.
(MAP is also called MPE which stands for {\em most probable explanation},
because it is the {\em mode} of the posterior distribution.)
``Inference'' refers to the task of computing $p(\mv x_B \suchthat \mv x_A,
\theta)$, where $\mv x_B$ are the {\em hidden variables} and $\theta$ are
known parameters.}
For example, we might want to
\bi
 \item Compute the marginal distribution of some set of variables
 \item Compute the conditional distribution $p(\mv x_A \suchthat \mv x_B)$
     of some set of variables $A$ given specific values to another set of 
     variables $\mv x_B$.
 \item Compute $\argmax_{\mv x_A} p(\mv x_A \suchthat \mv x_B)$ (for MAP
     queries, for example).
\ei
When we condition on some variables, we can restrict the factors to only
those entries that match the conditioned variables.
It is obvious that the first two questions above are special cases 
of the $\faqcs$ problem on the sum-product semiring.
(The third question is on the max-product semiring.)
These are well-known facts \cite{MR2778120}.
\end{ex}

Since matrix-vector multiplication is a special case of matrix-matrix 
multiplication, Example~\ref{ex:matrix mult} also shows that the matrix 
vector multiplication problem is a special case of the $\faqcs$ problem.
Given the above, it also follows that computing the Discrete Fourier 
Transform is a special case of $\faqcs$. Next, we present another 
interpretation of the DFT from Aji's thesis~\cite{aji-thesis}, which 
immediately shows that our algorithm implies the Fast Fourier Transform (FFT).

\begin{ex}[Discrete Fourier Transform] 
\label{ex:dft}
Recall that the discrete Fourier transform (DFT) is the matrix vector
multiplication where $A_{xy}=e^{i2\pi\frac{x\cdot y}{n}}$. For this example, we
will consider the case when $n=p^m$ for some prime $p$ and integer $m\ge 1$.
Recall that the DFT is defined as follows:
\[\varphi(x)=\sum_{y=0}^{n-1} b_y\cdot e^{i2\pi\frac{x\cdot y}{n}}.\]
Write $x=\sum_{i=0}^{m-1} x_i\cdot p^i$ and $y=\sum_{j=0}^{m-1} y_j\cdot p^j$ in their base-$p$ form. Then we can re-write the transform above as follows:
\[\varphi(x_0,x_1,\dots,x_{m-1})=\sum_{(y_0,\dots,y_{m-1})\in\F_p^m}b_y\cdot e^{i2\pi\frac{\sum_{0\le j+k\le 2m-2} x_j\cdot y_k\cdot p^{j+k}}{n}}.\]
Recalling that $n=p^m$, note that for any $j+k\ge m$, we have $e^{i2\pi\frac{x_j\cdot y_k\cdot p^{j+k}}{n}}=1$. Thus, the above is equivalent to
\[\varphi(x_0,x_1,\dots,x_{m-1})=\sum_{(y_0,\dots,y_{m-1})\in\F_p^m}b_y\cdot \prod_{0\le j+k<m} e^{i2\pi\frac{x_j\cdot y_k}{p^{m-j-k}}}.\]

The above immediately suggests the following reduction to $\faqcs$. Let
$\calH=(\calV,\calE)$ where
$\calV=\{X_0,X_1,\dots,X_{m-1},Y_0,Y_1,\dots,Y_{m-1}\}$ and $\calE$ has an edge
$(X_j,Y_k)$ for every $j,k\in \{0,1,\dots,m-1\}$ such that $j+k<m$. Further,
there is another edge $(Y_0,Y_1,\dots,Y_{m-1})$. The variable domains are
$\Dom(X_i)=\Dom(Y_i)=\{0,1,\dots,p-1\}$. For every $j,k\in\F_p$ such that
$j+k<m$, the corresponding factor $$\psi_{X_j,Y_k}:\F_p^2\to \D$$ is defined 
as 
$$\psi_{X_j,Y_k}(x,y)=e^{i2\pi\frac{x\cdot y}{p^{m-j-k}}}.
$$ 
Finally, the factor $\psi_{\mv Y}:\F_p^m\to\mathbf D$ is defined as 
$$\psi_{\mv Y}(y_0,y_1,\dots,y_{m-1})=b_{(y_0,y_1,\dots,y_{m-1})}.$$ 
Then the output is
\[\varphi(x_0,x_1,\dots,x_{m-1})=\sum_{(y_0,\dots,y_{m-1})\in\F_p^m}\psi_Y(y_0,y_1,\dots,y_{m-1})\cdot \prod_{0\le j+k<m} \psi_{X_j,Y_k}(x_j,y_k).\]
\end{ex}

\begin{ex}[Graph homomorphism function]
    Given a (non-hyper) graph $G$ and an $m \times m$ symmetric matrix 
    $\mv A = (a_{ij})$. The {\em graph homomorphism function} 
    is defined by
    \[ Z_{\mv A}(G) = \sum_{\zeta: V \to [m]} \prod_{uv\in E(G)} a_{\zeta(u),
        \zeta(v)}.
    \]
    (The reader is referred to the masterpieces by Cai et al.
        \cite{DBLP:journals/siamcomp/CaiCL13}, Goldberg et al.
        \cite{Goldberg:2010:CDP:1958055.1958080},
        references therein for a long and fascinating history of this problem.)
    The corresponding $\faq$ instance is
    \[ \varphi = \sum_{x_1}\cdots\sum_{x_n} \prod_{uv\in E(G)}
        \psi_{uv}(x_u,x_v),
    \]
    where $\Dom(X_i) = [m]$, and $\psi_{uv}: [m] \times [m] \to \D$ are
    identical factors, mapping $\psi_{uv}(x_u,x_v) = a_{x_u,x_v}$.
\end{ex}

\begin{ex}[Holant problem]
    The Holant problems can also be expressed using $\faqcs$.
    Essentially, in a Holant instance there is a function $f_v$ for each vertex
    of a graph $G$, where the function is on the set of incident edges. Each
    edge can be assigned a domain value. The problem is to compute the sum 
    over all edge assignments of the product of all vertex factors.
    See \cite{MR3105918} for more details.
\end{ex}

\subsection{The max-product semiring}

In addition to MAP queries in $\pgm$s which are well-known to be reducible to
$\faqcs$ on the max-product semiring, the following is another common instance
of the max-product semiring.

\begin{ex}[Maximum Likelihood Decoder for Linear Codes]
We present the instantiation of Maximum Likelihood Decoding (MLD) for linear codes as an instance of $\faqcs$ from Aji and McEliece~\cite{AM00}. We consider the decoding problem from a discrete memoryless channel. We assume that the alphabet is $\F_2$ and given $y,x\in\F_2$, $\psi_p(y,x)$ denotes the probability of the receiver receiving $y$ given that $x$ was transmitted over the channel. Let $C$ be a binary linear code with dimension $k$ and block length $n$. Then given a received word $\mv y\in\F_2^n$ and a codeword $\mv c\in C$, the probability that the receiver receives $\mv y$ when $\mv c$ was transmitted is given by
\[\mathrm{Pr}[\mv y|\mv c]=\prod_{i=1}^n \psi_p(y_i,c_i).\]
The maximum likelihood decoder, given $\mv y\in\F_2^n$ outputs the codeword $\mv c\in C$ that is most likely, i.e. it outputs:
\[\arg\max_{\mv c\in C} \mathrm{Pr}[\mv y|\mv c].\]
For simplicity, we will concentrate on the related problem of computing the most likely probability, i.e.
\begin{equation}
\label{eq:mld-org}
\max_{\mv c\in C} \mathrm{Pr}[\mv y|\mv c].
\end{equation}

Define the hypergraph $\calV=([n],\calE)$ as follows. There is a singleton edge $\{i\}$ for every $i\in [n]$ (with the corresponding factor $\psi_i(Y_i,X_i)=\psi_p(Y_i,X_i)$). Further, the linear code $C$ has an $(n-k)\times n$ parity check matrix $H$ such that $\mv c\in C$ if and only if $H\cdot \mv c^T=\mv 0$. Then for each row $H_j$, we have $\mathrm{supp}(H_j)\in\calE$. (The factor $\psi_{H_j}(\mv x_{\mathrm{supp}(H_j)})$ is defined to be $1$ if the corresponding parity check is satisfied by $\mv x$ and $0$ otherwise.)
It is easy to verify that the $\faqcs$ instance below is equivalent to the problem in~\eqref{eq:mld-org}:
\[\varphi=\max_{\mv \mv x\in\F_q^n} \prod_{i\in [n]} \psi_i(y_i,x_i)\cdot \prod_{j\in [n-k]} \psi_{H_j}(\mv x_{\mathrm{supp}(H
_j)}).\]
\label{ex:mld}
\end{ex}

\subsection{The set semiring}
\label{subsec:set-semiring}
\begin{ex}[Natural join query]
    \label{ex:natural join}
    Consider the natural join query
    \[ \Phi = \ \Join_{R\in \atoms(\Phi)} R. \]
    Let $\calH=(\calV,\calE)$ be the query's hypergraph. We define a factor
    $\psi_S$ for each $S\in\calE$ as follows. 
    The domain of the set semiring 
    is $\mv D = 2^U$, where $U = \prod_{i=1}^n\Dom(X_i)$.
    Let $R$ be the relation corresponding to $S$. Then, $\psi_S$
    \[ \psi_S(\mv x_S) =
        \begin{cases}
            \bigl\{ \mv t \suchthat \pi_S(\mv t) = \mv x_S \bigr\} & 
                \text{ if } \mv x_S \in R\\
            \emptyset & \text{ if } \mv x_S \notin R\\
        \end{cases}
    \]
    Computing the output of $\Phi$ becomes computing the constant set-valued
    function
    \[ \varphi = \bigcup_{\mv x} \bigcap_{S\in\calE} \psi_S(\mv x_S). \]
\end{ex}

\subsection{Applications of the $\faq$ problem}
\label{subsec:faq-applications}

The $\faqcs$ problem is quite general, as we have seen above. However, 
there are at least two classes of problems which are not captured by the basic 
$\faqcs$ formulation above. 

First, consider the problem of counting the number of answers to a conjunctive
query. This is called the $\scq$ problem \cite{pichler:2013}.
If the input query is quantifier-free, then we {\em can} formulate the problem
as $\faqcs$ on the sum-product semiring by mapping $\true$ to $1$ and $\false$ to
$0$. However, when the query does have existential quantifiers, the
straightforward mapping does not work.

Second, consider the general $\problemname{Quantified Conjunctive Query}$ 
($\qcq$) problem, which has both existential and universal quantifiers (and a 
conjunction of atoms inside). It is not
possible to reduce this problem to $\faqcs$.
The related problem of counting the number of solutions to such
formulas ($\sqcq$) is even harder.


\begin{ex}[$\scq$]\label{ex:scq}
    Let $\Phi$ be a conjunctive formula of the form
    \[ \Phi(X_1,\dots,X_f) = \exists X_{f+1} \dots X_n 
    \left( \bigwedge_{R\in\atoms(\Phi)} R \right), \]
    and the problem is to {\em count} the number of assignments
    $(x_1,\dots,x_f)$ such that $\Phi(x_1,\dots,x_f)$ is satisfied.
    Then, we can reduce this problem to $\faq$ with the constant function
    \[ \varphi = \sum_{x_1}\cdots\sum_{x_f} \max_{x_{f+1}}\cdots\max_{x_n}
        \prod_{S\in\calE}\psi_S(\mv x_S).
    \]
    Here, $\psi_S(\mv x_S) = \delta_{x_S \in R}$,
    where $R \in \atoms(\Phi)$
    is the atom corresponding to $S$.
    Note that $\sum$ and $\max$ are not commutative with one another.
    \label{ex:sharpCQ}
\end{ex}

\begin{ex}[$\qcq$ -- First Reduction]
    Let $\Phi$ be a first order formula of the form
    \[ \Phi(X_1,\dots,X_f) = Q_{f+1} X_{f+1} \cdots Q_n X_n 
    \left( \bigwedge_{R\in\atoms(\Phi)} R \right), \]
    where $Q_i \in \{\exists, \forall\}$, for $i>f$. 
    The problem is to compute the relation $\Phi$ on the free
    variables $X_1,\dots,X_f$.
    Then, we can reduce this problem to $\faq$ with the function
    \[ \varphi(x_1, \cdots, x_f) =
        \textstyle{\bigoplus^{(f+1)}_{x_{f+1}}\cdots\bigoplus^{(n)}_{x_n}}
        \prod_{S\in\calE}\psi_S(\mv x_S).
    \]
    Similar to Example~\ref{ex:sharpCQ}, 
    $\psi_S(\mv x_S) = \delta_{x_S \in R}$,
    where $R \in \atoms(\Phi)$ is the atom corresponding to $S$.
    And, $\bigoplus^{(i)} = \max$ if $Q_i = \exists$ and
    $\bigoplus^{(i)} = \min$ if $Q_i = \forall$.
\end{ex}

There is another way to reduce $\qcq$ to $\faq$.
It is this reduction that we will use later in the paper.

\begin{ex}[$\qcq$ -- Second Reduction]\label{ex:qcq}
    Let $\Phi$ be a first order formula of the form
    \[ \Phi(X_1,\dots,X_f) = Q_{f+1} X_{f+1} \cdots Q_n X_n 
    \left( \bigwedge_{R\in\atoms(\Phi)} R \right), \]
    where $Q_i \in \{\exists, \forall\}$, for $i>f$. 
    The problem is to compute the relation $\Phi$ on the free
    variables $X_1,\dots,X_f$.
    Then, we can reduce this problem to $\faq$ with the function
    \[ \varphi(x_1, \cdots, x_f) =
        \textstyle{\bigoplus^{(f+1)}_{x_{f+1}}\cdots\bigoplus^{(n)}_{x_n}}
        \prod_{S\in\calE}\psi_S(\mv x_S).
    \]
    Similar to Example~\ref{ex:sharpCQ}, 
    $\psi_S(\mv x_S) = \delta_{x_S \in R}$,
    where $R \in \atoms(\Phi)$ is the atom corresponding to $S$.
    And, 
    \[ \textstyle{\bigoplus^{(i)} = 
    \begin{cases}
    \max & \text{ if } Q_i = \exists, \\
    \times & \text{ if } Q_i = \forall.
    \end{cases}}
    \]
\end{ex}

What is interesting to note about the above reduction is that the product
occurs as an aggregate operator over (universally quantified) variables
and also as an aggregate operator over the input factors.

\begin{ex}[Example~\ref{ex:sqcq} revisited: $\sqcq$]
    This is similar to the above two examples.
    Let $\Phi$ be a first order formula of the form
    \[ \Phi(X_1,\dots,X_f) = Q_{f+1} X_{f+1} \cdots Q_n X_n 
    \left( \bigwedge_{R\in\atoms(\Phi)} R \right), \]
    where $Q_i \in \{\exists, \forall\}$, for $i>f$. 
    The problem is to {\em count} the number of tuples in relation $\Phi$ 
    on the free variables $X_1,\dots,X_f$.
    Then, we can reduce this problem to $\faq$ with the constant function
    \[ \varphi = \sum_{x_1} \cdots \sum_{x_f} 
        \textstyle{\bigoplus^{(f+1)}_{x_{f+1}}\cdots\bigoplus^{(n)}_{x_n}}
        \prod_{S\in\calE}\psi_S(\mv x_S).
    \]
    Where $\psi_S(\mv x_S) = \delta_{\mv x_S \in R}$,
    where $R \in \atoms(\Phi)$ is the atom corresponding to $S$.
    And, 
    \[ \textstyle{\bigoplus^{(i)} = 
    \begin{cases}
    \max & \text{ if } Q_i = \exists, \\
    \times & \text{ if } Q_i = \forall.
    \end{cases}}
    \]
\end{ex}

\section{Reductions from non-semiring to semiring}
\label{sec:appendix:noCS}

In practice, we often encounter queries which have the same format as \faq\ 
except that some of the aggregates $\oplus^{(i)}$ are neither product not 
semiring aggregates. Suppose that we want to compute the function
\[ \varphi: \prod_{i\in [f]} \Dom(X_i) \to \mathbf D\]
defined by
\[
    \varphi(\mv x_{[f]}) =
    \textstyle{\bigoplus^{(f+1)}_{x_{f+1}} \cdots
    \bigoplus^{(n)}_{x_{n}}}
    \bigotimes_{S\in\calE}\psi_S(\mv x_S).
\]
W.l.o.g. let $\oplus:=\oplus^{(n)}$ be neither a product nor a semiring aggregate (Otherwise, we could have eliminated $x_n$ as described in Section ~\ref{subsec:IO}). In many situations, it is possible to find a two-way mapping between $(\D, \oplus, \otimes)$ and some commutative semiring $(\bar \D, \bar\oplus, \bar \otimes)$ where $\bar \D$ is some extended domain and $\bar \oplus, \bar \otimes$ are extended versions of $\oplus, \otimes$ that fit the new domain $\bar \D$. If such a mapping exists, then we can transfer input factors into $\bar\D$, carry out the calculations over the semiring $(\bar \D, \bar\oplus, \bar \otimes)$, and then transfer the results back to the original domain $\D$, in order to bypass dealing with non-semirings.

Formally, suppose that there exists a function $\bar f: \D \to \bar\D$ and a function $f:\bar\D\to\D$ that satisfy the following conditions:
\be
\item For all $x_1, \ldots, x_n\in\D$ (where $n \geq 1$), $\bigoplus_{i\in[n]}x_i=f\left(\bar\bigoplus_{i\in[n]}\bar f(x_i)\right)$. \label{nsr:con1}
\item For all $x, y\in \D$, $\bar f(x \otimes y)=\bar f(x)\bar\otimes\bar f(y)$. \label{nsr:con2}
\item For all $\bar x, \bar y\in \bar\D$, $f(\bar x \bar\otimes \bar y)=f(\bar x)\otimes f(\bar y)$. \label{nsr:con3}
\ee
Note that for $n=1$, condition~\ref{nsr:con1} above reduces to $f(\bar f(x))=x$ for all $x \in \D$.
If the above conditions are met (Examples~\ref{ex:nsr:avg} and \ref{ex:nsr:unique} below), then variable elimination (Section~\ref{subsec:var:elim}) can take place as if $(\D, \oplus, \otimes)$ were a commutative semiring.
{\small
\begin{eqnarray}
   &&\varphi(\mv x_{[f]}) \\
    &=& \textstyle{\bigoplus^{(f+1)}_{x_{f+1}} \cdots
        \bigoplus^{(n-1)}_{x_{n-1}}
        f\left[\bar\bigoplus^{(n)}_{x_{n}}
        \bar\bigotimes_{S\in\calE}\bar f\left(\psi_S(\mv x_S)\right)\right]} \label{nsr:step1}\\
    &=& \textstyle{\bigoplus^{(f+1)}_{x_{f+1}} \cdots
        \bigoplus^{(n-1)}_{x_{n}}
        f\left[\Big(\bar\bigotimes_{S:n\notin S} \bar f\left(\psi_S(\mv x_S)\right)\Big)
        \bar\otimes \left(
        \bar\bigoplus^{(n)}_{x_{n}}
        \bar\bigotimes_{S:n\in S} \bar f\left(\psi_S(\mv x_S)\right)
        \right)\right]
        }\label{nsr:step2}\\
    &=& \textstyle{\bigoplus^{(f+1)}_{x_{f+1}} \cdots
        \bigoplus^{(n-1)}_{x_{n}}
        \bigotimes_{S:n\notin S} \psi_S(\mv x_S)
        \otimes \underbrace{\textstyle{f\left(
        \bar\bigoplus^{(n)}_{x_{n}}
        \bar\bigotimes_{S:n\in S} \bar f\left(\psi_S(\mv x_S)\right)
        \right)}}_{\text{new factor } \psi_{U_n-\{n\}}\text{ over }\D}
        }.\label{nsr:step3}
\end{eqnarray}
}

If only conditions \ref{nsr:con1} and \ref{nsr:con2} (but not \ref{nsr:con3}) are met (Example~\ref{ex:nsr:RmaxX}), then we can still apply steps~\eqref{nsr:step1} and \eqref{nsr:step2} (but not \eqref{nsr:step3}). In particular, factors $\psi_S(\mv x_S)$ will have to remain transformed $\bar f\left(\psi_S(\mv x_S)\right)$ into the new domain $\bar \D$, and the new factor will be over $\bar \D$ as well. We cannot transfer back to the original domain $\D$ until all products $\bar\otimes$ have been computed.

Let $\mv 0, \mv 1\in\D$ denote the identities for $\oplus, \otimes$ respectively, and $\bar{\mv 0}, \bar{\mv 1} \in \bar \D$ denote the identities for $\bar\oplus, \bar\otimes$ respectively. Many natural \faq\ algorithms rely on zero entries in order to save computation (e.g., \OI\ and \InsideOut\ in Sections~\ref{subsubsec:OI} and \ref{subsec:IO} respectively). Therefore, it is desired that $f$ and $\bar f$ satisfy the properties $\bar{\mv 0}=\bar f(\mv 0)$ and $\mv 0=f(\bar{\mv 0})$.

\begin{rmk}
Notice that conditions \ref{nsr:con1} and \ref{nsr:con3} above imply that $f(\bar{\mv 1})=\mv 1$. In particular, assuming $f(\bar{\mv 1})\neq \mv 1$, we have $f\left[\bar f(\mv 1)\bar \otimes \bar{\mv 1}\right]=f\left[\bar f(\mv 1)\right]=\mv 1$, while $f\left[\bar f(\mv 1)\right]\otimes f(\bar{\mv 1})=\mv 1\otimes f(\bar{\mv 1})=f(\bar{\mv 1})$.
\end{rmk}

Here are a few practical examples that can clarify the above abstract ideas (In all of those examples, we have $\bar{\mv 0}=\bar f(\mv 0)$, $\mv 0=f(\bar{\mv 0})$, $\bar{\mv 1}=\bar f(\mv 1)$, and $\mv 1=f(\bar{\mv 1})$).

\begin{ex}[Conditional average] Given a set of $n\geq 1$ numbers $x_1, \ldots, x_n\in \R$, suppose we are interested in computing the average of the numbers $x_i$ that satisfy a certain condition, e.g.~$x_i\neq 0$.
In particular, let $\avg(x_1, \ldots, x_n)$ denote the average of non-zero numbers among $x_1, \ldots, x_n$. (By convention, when all numbers are zeros, let $\avg$ be zero.)
\[\avg(x_1, \ldots, x_n):=
\begin{cases}
0 &\text{ if } x_1=x_2=\cdots=x_n=0,\\
\frac{x_1+\cdots+x_n}{\delta_{x_1\neq0}+\cdots+\delta_{x_n\neq0}} &\text{ otherwise}.
\end{cases}
\]
$(\R, \avg, \times)$ is not a semiring because $\functionname{avg}$ is not associative (e.g., $\avg(\avg(1, 2), 3)\neq \avg(1, \avg(2, 3))$). However, we can extend it into the commutative semiring $(\R\times \N, \bar\oplus, \bar\otimes)$ where $\bar\oplus$ and $\bar\otimes$ are defined as follows. For all $(a_1, b_1), (a_2, b_2)\in \R\times \N$,
\[(a_1, b_1)\bar\oplus(a_2, b_2):=(a_1+a_2, b_1+b_2),\]
\[(a_1, b_1)\bar\otimes(a_2, b_2):=(a_1\times a_2, b_1\times b_2).\]
Define $\bar f:\R\to \R\times\N$ such that for all $x \in \R$,
\[\bar f(x):=(x, \delta_{x\neq 0}).\]
Define $f:\R\times\N \to \R$ such that for all $(a, b)\in(\R, \N)$,
\[f(a, b):=
\begin{cases}
0 &\text{ if } b= 0,\\
\frac{a}{b} &\text{ otherwise}.
\end{cases}
\]
It is not hard to verify that conditions \ref{nsr:con1}, \ref{nsr:con2}, and \ref{nsr:con3} are met in this example.
Also, notice that $\mv 0=0$, $\mv 1=1$, $\bar{\mv 0}=(0, 0)$ and $\bar{\mv 1}=(1, 1)$.
\label{ex:nsr:avg}
\end{ex}

\begin{ex}[Uniqueness quantification ($\exists!$)] Given $n\geq 1$ Boolean variables $b_1, \ldots, b_n$, let $\unique(b_1, \ldots, b_n)$ denote the truth value of whether there is a unique $b_i$ whose value is $\true$.
\[\unique(b_1, \ldots, b_n):=(\exists! i\in[n] \suchthat b_i=\true).\]

The problem with $\unique$ is that it is not cumulative over $\{\true, \false\}$ (For example, $\unique(\true, \true, \true)=\false$ while $\unique(\unique(\true, \true), \true)=\true$.) In fact, binary $\unique$ reduces to logical $\xor$, while $n$-ary $\unique$ does not correspond to $\xor$ for $n>2$. Therefore, we cannot use the commutative semiring $(\{\true, \false\}, \xor, \wedge)$ to solve this \faq. However, we can extend the domain to become $\{0, 1, 2\}$ and define a commutative semiring $(\{0,1,2\}, \bar\oplus, \bar\otimes)$ such that for all $x, y\in\{0, 1, 2\}$,
\[x\bar\oplus y:=\min(x+y, 2),\]
\[x\bar\otimes y:=\min(x\times y, 2).\]
Define $\bar f:\{\true, \false\}\to\{0, 1, 2\}$ such that
\[\bar f(\true):=1,\]
\[\bar f(\false):=0.\]
Define $f:\{0, 1, 2\}\to\{\true, \false\}$ such that for all $x\in\{0, 1, 2\}$,
\[f(x):=(x=1).\]
Notice that conditions \ref{nsr:con1}, \ref{nsr:con2}, and \ref{nsr:con3} are met in this example. Also, notice that $\mv 0=\false$, $\mv 1=\true$, $\bar{\mv 0}=0$, and $\bar{\mv 1}=1$.
\label{ex:nsr:unique}
\end{ex}

\begin{ex}[$(\R, \max, \times)$]
$(\R, \max, \times)$ is not a semiring because $\max$ does not have an identity over $\R$. However, we can fix this by extending the domain to become $\D:=\R\cup\{\nan\}$ where $\nan$ is a special symbol having the following properties. For all $x \in \D$,
\[\max(x, \nan)=\max(\nan, x)=x,\]
\[\min(x, \nan)=\min(\nan, x)=x,\]
\[x\times \nan=\nan\times x=\nan.\]
Now, we have the $\max$ identity $\mv 0=\nan$ and the product identity identity $\mv 1=1$. However, $(\D, \max, \times)$ is still not a semiring because
$\times$ does not distribute over $\max$ in $\R$ (e.g., $-1\times \max(1, 2)\neq \max(-1, -2)$). However, we can extend it into the commutative semiring $(\bar\D, \bar\oplus, \bar\otimes)$ where
\[\bar \D:=\big\{(\nan, \nan)\big\}\cup\left\{(a, b)\in \R^2 \suchthat a\leq b\right\},\]
and for all $(a_1, b_1), (a_2, b_2)\in \bar \D$
\[(a_1, b_1)\bar\oplus(a_2, b_2):=(\min(a_1, a_2), \max(b_1, b_2)),\]
\[(a_1, b_1)\bar\otimes(a_2, b_2):=(\min(a_1a_2, a_1b_2, b_1a_2, b_1 b_2), \max(a_1a_2, a_1b_2, b_1a_2, b_1 b_2)).\]
Notice that $\bar{\mv 0}=(\nan, \nan)$ while $\bar{\mv 1}=(1, 1)$.

Define $\bar f:\D \to \bar \D$ such that for all $x\in\D$,
\[\bar f(x):=(x, x).\]
Define $f:\bar\D\to\D$ such that for all $(a, b)\in \bar\D$,
\[f(a, b):=b.\]
This example meets conditions \ref{nsr:con1} and \ref{nsr:con2} but \emph{not} \ref{nsr:con3}.
\label{ex:nsr:RmaxX}
\end{ex}

\section{Tree decompositions and variable elimination}
\label{app:sec:td}

In this section, we prove Proposition~\ref{prop:alpha:acyclicity},
Proposition~\ref{prop:beta:acyclicity}, and Lemma~\ref{lmm:g-width}. To this
end, we first describe ways to convert back and forth between a tree
decomposition and a vertex ordering of a given hypergraph.
We will also need the notion of a reduced (or non-redundant)
tree decomposition.

\paragraph*{Minimal tree decomposition}
The definition of tree decompositions
doesn't say anything about how large a
tree decomposition is, i.e. how many vertices $T$ has.
However, many tree decompositions are ``redundant'' in the sense that one
bag might be a subset of another. In that case, we can {\em reduce} the tree
decomposition as follows. Let $(T, \chi)$ be a tree decomposition of a
hypergraph $\calH$. Suppose there is a bag $B$ that is a subset of 
another bag $B'$. Note that there is a unique path $B=B_0,B_1,\dots, B_k=B'$ 
because $T$ is a tree. Then, $B$ has to be a subset of {\em every} bag on this 
path. We can thus remove $B$, connect all the neighbors of $B$, other than
$B_1$, to $B_1$.
It is easy to see that the new tree is a tree decomposition of the same
hypergraph still.
When no such bag removal is possible, we have a {\em reduced tree
decomposition} or a {\em non-redundant tree decomposition}. The following 
shows that reduced tree decompositions do not have too many nodes.

\bprop
Let $(T,\chi)$ be a reduced tree decomposition of a 
connected hypergraph $\calH$ on $n$
vertices, then $T$ has at most $n$ nodes.
\eprop
\bp
Any leaf bag $B$ of $T$ necessarily has a private vertex $v$, 
i.e. a vertex that does not belong to any other bag. 
Now, suppose we remove $v$ from $B$. If $B$ is a subset of its neighbor, then
we also remove $B$. Either way, what we end up with is a reduced tree
decomposition of a hypergraph on $n-1$ vertices. Induction completes the proof.
\ep

\paragraph*{Tree decomposition from vertex ordering.}
Given any vertex ordering $\sigma = (v_1,\dots,v_n)$ of a hypergraph
$\calH=(\calV,\calE)$, we construct a tree decomposition recursively as
follows. (Recall the hypergraph sequence defined in 
Definition~\ref{defn:elimination:hypergraph:sequence}.)
First, by induction, we construct a tree decomposition $T_{n-1}$ of the graph
$\calH^\sigma_{n-1}$ using the vertex ordering $(v_1,\dots,v_{n-1})$.
(The base case, i.e. the tree decomposition $T_1$ is trivial as the hypergraph
$\calH_1$ has only one vertex.)
Note that $U^\sigma_n-\{v_n\}$ {\em is} a hyperedge of $\calH^\sigma_{n-1}$.
Let $B$ be the bag of $T_{n-1}$ that contains this hyperedge.
Now, create a new bag $U_n$ and connect it to $B$.
Note that all hyperedges in $\partial^\sigma(v_n)$ are subsets of $U^\sigma_n$.
One can verify by induction that this is indeed a tree decomposition of 
$\calH^\sigma_n=\calH$.
The tree decomposition constructed this way has at most $n$
bags, but it is not necessarily non-redundant.
We reduce the tree decomposition to make it non-redundant,
and refer to the final tree decomposition as 
a {\em tree decomposition induced by the
vertex ordering}. The following proposition is straightforward by induction.

\bprop\label{prop:Uk-equal}
In a tree decomposition $(T, \chi)$ induced by the vertex ordering 
$\sigma=(v_1,\dots,v_n)$, every bag of $T$ is a set $U^\sigma_k$, for some
$k\in [n]$. Furthermore, $(T,\chi)$ is non-redundant.
\eprop

\paragraph*{GYO-elimination procedure.}
Next we explain why the vertex ordering is the reverse of
what typically is called an ``elimination order''.
The typical way to obtain a good vertex ordering from a tree decomposition
is the {\em GYO-elimination procedure} \cite{yu1984determining,
ontheuniversalrelation, Tarjan:1984:SLA:1169.1179}.
In this procedure, we repeatedly apply the following two operations on
a tree decomposition: (1) remove any bag that is a subset of another
bag, (2) remove any vertex that belongs to only one bag.
Typically this procedure is done by fixing arbitrarily a root bag and eliminating
bags and vertices from the leaves up to the root bag.

The reversed sequence of vertices that were removed is a vertex
ordering whose induced (fractional) width is as good as the (fractional)
width of the tree decomposition, as we show below.
One key aspect of the GYO-procedure is that we could have fixed any bag as the
root bag, and thus the vertices inside this bag can be eliminated last,
which means they will occur first in the resulting vertex ordering.

\paragraph*{Vertex ordering from tree decomposition.}
However, in this section for technical reasons we also describe a specific
realization of the basic GYO-elimination procedure.
Given a tree decomposition $(T,\chi)$ of $\calH=(\calV,\calE)$
we construct an ordering
of vertices $v_1,\dots,v_n$ as follows. First, we make $(T,\chi)$ a reduced
tree decomposition as described above. Second, we designate a node of $T$ as
the root node. For every vertex $v \in \calV$, let $t_v$ denote the node
highest in the tree such that $v \in \chi(t_v)$. We call the bag $\chi(t_v)$
the ``owner'' of of $v$, and $v$ a ``private vertex'' of $\chi(t_v)$.
We construct a vertex ordering as follows:
\be
 \item If $T$ is empty, return the empty sequence.
 \item Otherwise, take any leaf node $t$ of $T$, let $P_t$ denote the set of 
private vertices of $\chi(t)$. (Because $T$ is reduced, $P_t$ is
not empty!)
 \item Remove $t$ from $T$, resulting in a tree $T'$. Let $\sigma'$ denote the
vertex ordering of $\calV-P_t$ obtained from $T'$ by induction. Let $\sigma$
be obtained by appending vertices in $P_t$ to the end of $\sigma'$.
(Note that $T'$ still has the same root as $T$.)
\ee
The resulting vertex ordering is called a vertex ordering induced by the tree
decomposition. The following follows by induction.

\bprop \label{prop:vo:from:td}
Let $\sigma=(v_1,\dots,v_n)$ be a vertex ordering induced by a 
tree decomposition
$(T, \chi)$, then $U^\sigma_k$ is a subset of some bag of $T$, for
every $k\in [n]$.
Furthermore, the vertex ordering can be constructed in polynomial time.
\eprop

Now we are ready to characterize acyclicity and widths using vertex
orderings.

\bp[Proof of Proposition~\ref{prop:alpha:acyclicity}]
For the forward direction, assume $\calH$ is $\alpha$-acyclic. 
Then, there is a tree decomposition $(T,\chi)$ of $\calH$ in which 
every bag is a hyperedge. 
Let $\sigma=(v_1,\dots,v_n)$ be an induced vertex ordering of $(T,\chi)$. By
Proposition~\ref{prop:vo:from:td}, $U^\sigma_k \subseteq \chi(t)$ for some $t
\in V(T)$. Since $\chi(t)$ is a hyperedge of $\calH$, $\chi(t) \in \calE_k$ and
hence $U^\sigma_k = \chi(t) \in \partial^\sigma(v_k)$. 

Conversely, suppose there is a vertex ordering $\sigma$ for which
$U^\sigma_k \in \partial^\sigma(v_k)$ for all $k\in [n]$.
By induction, $\calH^\sigma_{n-1}$ is $\alpha$-acyclic. Thus, there is a
tree decomposition $T_{n-1}$ in which every bag of $T_{n-1}$ is a hyperedge of 
$\calH_{n-1}$. 
There is a hyperedge $B = U_n-\{v_n\}$ in $\calH_{n-1}$ which may not
be a hyperedge of $\calH_n$. 
If $B$ is indeed not a hyperedge of $\calH_n=\calH$, then we replace $B$
by $U_n$, add a bag for each set in $\partial(v_n)-\{U_n\}$, and connect those 
bags to the $U_n$-bag.
If $B$ {\em is already} a hyperedge of $\calH$, then
we create a new bag $B' = B \cup \{v_n\}$, connect it to $B$,
and connect the remaining bags in $\partial(v_n)$ to $B'$ as before.
Either way, we have just created a tree decomposition of $\calH$ in which every
bag is a hyperedge and vice versa.
\ep
 
Proposition~\ref{prop:beta:acyclicity} was shown in~\cite{nnrr}; we reproduce a
stand-alone proof here for completeness. 
To prove Proposition~\ref{prop:beta:acyclicity}, we use a different
characterization of $\beta$-acyclicity:

\bprop[See~\cite{DBLP:journals/jacm/Fagin83}]
A hypergraph $\calH = (\calV, \calE)$ is {\em $\beta$-acyclic}
if and only if there is {\bf no} sequence 
$$(F_1, u_1, F_2, u_2, \cdots, F_m,
u_m,
F_{m+1}=F_1)$$ with the following properties
\bi
 \item $m \geq 3$
 \item $u_1,\dots,u_m$ are distinct vertices of $\calH$
 \item $F_1, \dots, F_m$ are distinct hyperedges of $\calH$
 \item for every $i \in [m]$, $u_i \in F_i \cap F_{i+1}$, and $u_i \notin F_j$
 for every $j \in [m+1] -\{i,i+1\}$.
\ei
\label{prop:beta-acyclic}.
\eprop

\bp[Proof of Proposition~\ref{prop:beta:acyclicity}]
For the forward direction, suppose $\calH$ is $\beta$-acyclic.
A {\em nest point} of $\calH$ is a vertex $v \in \calH$ such that the
collection of hyperedges containing $v$ forms a nested sequence of subsets,
one contained in the next. 
Any $\beta$-acyclic hypergraph $\calH$ has at least two nest
points~\cite{brouwer-kolen-1980}.
Let $v_n$ be a nest point of $\calH=\calH_n$. Then, 
$\partial(v_n)$ is an inclusion chain and $U_n$ is the bottom element of 
$\partial(v_n)$.
Consequently, $\calH_{n-1}$ is precisely $\calH - \{v_n\}$, which is
$\beta$-acyclic. 
By induction there exists an elimination order
$v_1,\dots,v_{n-1}$ such that every collection $\partial(v_k)$ forms
a chain, $k\in [n-1]$,
Thus, the elimination order $v_1,\dots,v_n$ satisfies the desired
property.

Conversely, suppose there exists an ordering $\sigma=(v_1,\dots,v_n)$ of all 
vertices of
$\calH$ such that every collection $\partial^\sigma(v_k)$ is a chain. Assume to 
the contrary that
$\calH$ is not $\beta$-acyclic. Then, there is a sequence
$$(F_1,u_1,F_2,u_2,\dots,F_m,u_m,F_{m+1}=F_1)$$ satisfying the conditions stated
in Proposition~\ref{prop:beta-acyclic}.
Without loss of generality, suppose $u_m$ comes last in the vertex ordering
$\sigma$, and that $u_m = v_k$ for some $k$.
Then, the poset $\partial^\sigma(v_k)$ contains the set
$F_m \cap \{v_1,\dots,v_{k-1}\}$ and the set $F_1 \cap
\{v_1,\dots,v_{k-1}\}$. Since both $u_2$ and $u_{m-1}$ come before $u_m$ in the
ordering, we have
\begin{eqnarray*}
u_2 &\in& \left(F_1 \cap \{v_1,\dots,v_{k-1}\}\right) \setminus
        \left(F_m \cap \{v_1,\dots,v_{k-1}\}\right)\\
u_{m-1} &\in& \left(F_m \cap \{v_1,\dots,v_{k-1}\}\right) \setminus
        \left(F_1 \cap \{v_1,\dots,v_{k-1}\}\right).
\end{eqnarray*}
Consequently, $\partial^\sigma(v_k)$ is not a chain.
\ep

\bp[Proof of Lemma~\ref{lmm:g-width}]
For the forward direction, consider a tree decomposition 
$(T,\chi)$ with $g$-width $w$ and one
of its induced vertex orderings $\sigma = (v_1,\dots,v_n)$. 
By Proposition~\ref{prop:vo:from:td}, every set $U^\sigma_k$ is a subset of some 
bag $\chi(t)$, $t \in V(T)$. Thus, due to $g$'s monotonicity, 
$g(U_k) \leq g(\chi(t)) \leq w$.

Conversely, suppose there is a vertex ordering $\sigma = (v_1,\dots,v_n)$ 
for which
$g(U^\sigma_k) \leq w$ for all $k\in [n]$. Let $T$ be a tree decomposition induced by
this ordering. Then, by Proposition~\ref{prop:Uk-equal}, 
every bag in $T$ is some set $U_k$, which completes the proof.
\ep

\section{Missing details from the analysis of $\InsideOut$}
\label{app:sec:analysis}

\section{Quick applications of $\InsideOut$}
\label{app:sec:quick:applications}

We describe here some examples where expression \eqref{eqn:IO:runtime} is
``easy'' to minimize.
Note that to avoid cumbersome notation, up to re-indexing of variables,
we have stated \eqref{eqn:IO:runtime} with the assumption that the variable
ordering of the input expression is from $X_1$ to $X_n$. In the examples below,
the variable indices might be arranged
differently from the natural $1$ to $n$ order.

\begin{ex}[$\problemname{Matrix Chain Multiplication}$]
Consider the matrix multiplication problem and its reduction to 
$\faqcs$ described in Example~\ref{ex:matrix mult}. 
In this problem, the set of free variables is $F = \{1,n+1\}$. 
Let $v_2,\dots,v_n$ be an arbitrary permutation 
of $\{2,\dots,n\}$, and let $v_1 = 1, v_{n+1} = n+1$. 
Then, the problem can be re-stated as computing
the function
\[ \varphi(x_1,x_{n+1}) =
    \sum_{x_{v_2}} \sum_{x_{v_3}} \cdots \sum_{x_{v_n}} 
    \prod_{i=1}^n \psi_{v_i,v_{i+1}}(x_{v_i},x_{v_{i+1}}).
\]
And, expression \eqref{eqn:IO:runtime} becomes
\begin{equation}
    \tilde O \left(\sum_{k=2}^{n} \agm_{\calH_k}(U_k)+p_1 p_{n+1}\right).
    \label{eqn:PMM}
\end{equation}
Let us see how one can find a variable ordering $v_2,\dots,v_n$ to minimize
this expression.

Let $\ell = v_n \in \{2,\dots,n\}$.
Then, for this variable ordering
$U_n = \{\ell-1,\ell,\ell+1\}$. In fact, it is easy to see that
every set $U_k$ will have size $3$ during the elimination process;
and the number of hyperedges intersecting $U_k$ is at most $4$.
To fractionally cover $U_n$, we can 
bound
    \begin{eqnarray*}
        |\psi_{\ell,\ell+1}| &\leq& p_{\ell}p_{\ell+1}\\
        |\psi_{\ell-2,\ell-1 / U_n}| &\leq& p_{\ell-1}.
    \end{eqnarray*}
    Hence, we can bound the term $\agm_{\calH_n}(U_n)$
    in \eqref{eqn:PMM} to be
    $p_{\ell-1}\cdot p_\ell \cdot p_{\ell+1}$. 
    This is exactly the usual rough estimate of the cost of multiplying
    a $p_{\ell-1} \times p_\ell$-matrix with a 
    $p_{\ell} \times p_{\ell+1}$-matrix.

Inductively, finding the best variable ordering to minimize
\eqref{eqn:PMM} is {\em exactly} the same as finding the best sequence 
of matrix multiplications to minimize the overall multiplication cost
as set up in the textbook $\problemname{Matrix Chain Multiplication}$
problem \cite{MR2002e:68001}. 
This problem can be solved by dynamic programming!
\end{ex}

\begin{ex}[Matrix Vector Multiplication for structured matrices] We begin with the DFT matrix. Recall that we assume that $n=p^m$ for some constant prime $p$. Using the bound in~\eqref{eqn:PMM}, one can show that $\InsideOut$ runs in time $O(m^4p^m)=O(n\log^4{n})$, which is a $\log^3{n}$ factor off from the $O(n\log{n})$ runtime of FFT.
Next, we will show that our framework as a special case contains the message passing interpretation of the FFT algorithm of Aji and McEliece~\cite{aji-thesis,AM00}. In particular, we will need to do some pre-processing on the input factors before running $\InsideOut$. For every $0\le k<m$, define
\[\psi_{\cdot,Y_k}(x_0,\dots,x_{m-k-1},y_k)=\prod_{0\le j+k<m} \psi_{X_j,Y_k}(x_j,y_k).\]
Noting that 
the entire truth table representation of $\psi_{\cdot,Y_k}$ can be computed in time $O(mp^{m-k})$ and hence one can compute all the truth table of $\psi_{\cdot,Y_k}$ for $0\le k<m$ in time $O(mp^m)=O(n\log{n})$. Note that if we have these factors pre-computed, then we want to compute the following $\faqcs$ instance
\[\sum_{(y_0,\dots,y_{m-1})\in\Z_p^m} \psi_Y(y_0,\dots,y_{m-1})\cdot \prod_{k=0}^{m-1} \psi_{\cdot,Y_k}(x_0,\dots,x_{m-k-1},y_k).\]
Now consider the variable ordering $Y_{m-1},\dots,Y_0$ and note that $U_k$ when eliminating $Y_k$ (note we changed the notation from the default one), is given by $U_k=\{X_0,\dots,X_{m-k},Y_0,\dots,Y_k\}$. 
We further note that since $\D=\{0,1\}$ we do not need to use the indicator projections in $\InsideOut$. In particular, this implies that line~\ref{ln:intermediate1} in Algorithm~\ref{algo:IO} is computing
\[\sum_{Y_k\in\Z_p} \psi_{\cdot,Y_k} \cdot \psi_{U_{k+1}},\]
which can be done in $O(p^m)$ time. Since the above line is executed $m$ times, the overall run time is $O(mp^m)=O(n\log{n})$, as desired.

For the circulant matrix, we note that if one picks the variable ordering 
$$W_{m-1},\dots,W_0,Z_{m-1},\dots,Z_0,Y_{m-1},\dots,Y_0$$ 
then Algorithm~\ref{algo:IO} with this variable ordering corresponds to computing $\mv C\cdot \mv b$ as computing $\mv F\cdot \mv b$ first, then computing $\mv F\cdot \mv c$ next, then computing their component-wise product $\mv u = (\mv F\cdot \mv c)\cdot (\mv F\cdot \mv b)$ and then finally computing $\mv F^{-1}\cdot \mv u$. Note that each of the DFT computations we just saw is done in $O(n\log{n})$ leading to an overall runtime of $O(n\log{n})$ to compute $\mv C\cdot \mv b$.

Next, we consider the case when $\mv A$ is the Kronecker product $\mv D\otimes \mv E$. In this case for $\calH_{\mv D\otimes \mv E}$ consider the variable ordering $Y_1,Y_0$ and the corresponding $U_1=\{X_1,Y_0,Y_1\}$ and $U_0=\{X_0,X_1,Y_0\}$. To fractionally cover $U_1$, set $\lambda^{(1)}_{(X_1,Y_1)}=\lambda^{(1)}_{X_0,Y_0}=1$. Similarly to cover $U_0$, set $\lambda^{(0)}_{X_0,Y_0}=\lambda^{(0)}_{X_1,Y_1}=1$. Finally to cover $F=\{X_0,X_1\}$, set $\mu_{X_1,Y_1}=\mu_{X_0,Y_0}=1$. Then~\eqref{eqn:PMM} gives an overall runtime bound of $O(n_0^3)=O(n^{3/2})$.

We defer the discussion on the Khatri-Rao product case and now consider the case of $\mv A=\mv D \circ \mv E$. Now consider the run of Algorithm~\ref{algo:IO} with variable ordering $Y_0,Y_2,Y_1,Y_3$. The corresponding sets are $U_0=\{X_0,X_2,Y_0,Y_1,Y_2,Y_3\}, U_2=\{X_0,X_2,Y_1,Y_2,Y_3\}, U_1=\{X_0,X_1,X_2,X_3,Y_1,Y_3\}$ and $U_3=\{X_0,X_1,X_2,X_3,Y_3\}$. We note that it is enough to cover $U_0$ and $U_1$ since $U_2\subseteq U_0$ and $U_3\subset U_1$. To cover $U_0$, set $\lambda^{(0)}_{X_0,X_2,Y_0,Y_2}=\lambda^{(0)}_{X_1,X_3,Y_1,Y_3}=1$ and to cover $U_1$ set $\lambda^{(1)}_{X_0,X_2,Y_0,Y_2}=\lambda^{(1)}_{X_1,X_3,Y_1,Y_3}=1$. Finally, to cover $F=\{X_0,X_1,X_2,X_3\}$, set $\mu_{X_0,X_2,Y_0,Y_2}=\mu_{X_1,X_3,Y_1,Y_3}=1$. Then~\eqref{eqn:PMM} givens an overall runtime bound of $O(n_0^6)=O(n^{3/2})$.

Finally, we consider the case when $\mv A$ is the Khatri-Rao product $\mv A=\mv D\star\mv E$. Now consider the run of Algorithm~\ref{algo:IO} with variable ordering $Y_0,Y_1,Y_2$. The corresponding sets are $U_0=\{X_0,X_2,Y_0,Y_1,Y_2\}, U_1=\{X_0,X_1,X_2,Y_1,Y_2\}$ and $U_2=\{X_0,X_1,X_2,Y_2\}$. Since $U_2\subset U_1$, we  only need to cover $U_0$ and $U_1$. However, if we just try and use an edge cover and then~\eqref{eqn:PMM}, then we will get the trivial quadratic runtime. However, like in the case of the DFT, we can consider the $\InsideOut$ algorithm and argue an overall runtime of $O(n^{5/3})$. Since all the factors are represented using the truth table representation, Algorithm~\ref{algo:IO} would run in the same time even if we added in the projection of the factor $\psi_{X_0,Y_0,X_2,Y_2}$ onto $\{X_0,Y_0\}$ and the projection of $\psi_{X_1,Y_1,X_2,Y_2}$ on to $\{X_1,Y_1\}$. With these extra factors, we can cover $U_0$ with $\lambda^{(0)}_{X_0,X_2,Y_0,Y_2}=\lambda^{(0)}_{X_1,Y_1}=1$, $U_1$ with $\lambda^{(1)}_{X_1,X_2,Y_1,Y_2}=\lambda_{X_0,Y_0}=1$ and $F=\{X_0,X_1,X_2\}$ by $\mu_{X_0,Y_0}=\mu_{X_1,X_2,Y_1,Y_2}=1$. Then~\eqref{eqn:PMM} implies an overall runtime of $O(n_0^5)=O(n^{5/3})$, as desired.
\end{ex}

\section{More on characterizing $\EVO(\varphi)$ and approximating $\faqw(\varphi)$}
\label{app:sec:evo}

\subsection{Quick applications of $\faqcs$ without free variables (i.e. $\sumprod$)}
\label{app:subsec:quick:faqcs}

$\faqcs$ without free variables is the easiest special case of $\faq$ because 
all variable orderings are $\varphi$-equivalent. 
In this case, every variable aggregate is a semiring aggregate.
Hence, $K$ defined by~\eqref{eqn:the:set:K} reduces to $K = [n]$. Expression \eqref{eqn:simplified:IO:runtime} becomes
a simple quantity $\tilde O\left(N^{\faqw(\sigma)}+\repsize{\varphi}\right)$, where
\[ \faqw(\sigma) = \max_{k\in[n]} \rho^*_\calH(U_k). \]
In this case, $\faqw(\sigma)$ is exactly the induced fractional edge cover width
of $\sigma$ (Definition~\ref{defn:inducedFECWidth}).
Thanks to Lemma~\ref{lmm:g-width}, it follows that 
\[ \faqw(\varphi) = \fhtw(\calH). \] 

However, computing (a tree decomposition with) the optimal $\fhtw$ is $\np$-hard
\cite{Marx:2010:AFH:1721837.1721845,DBLP:journals/corr/FischlGP16}.
Marx~\cite{Marx:2010:AFH:1721837.1721845} introduced an approximation algorithm that, given a hypergraph $\calH$ with fractional hypertree width $\fhtw(\calH)$, computes a tree decomposition $(T,\chi)$ of $\calH$ (or equivalently a vertex ordering $\sigma$  of $\calH$, as discussed in Section~\ref{subsec:VO}) such that
the fractional hypertree width of $(T,\chi)$ is $O(\fhtw^3(\calH))$
(or equivalently  the induced fractional edge cover width of $\sigma$, given by Definition~\ref{defn:inducedFECWidth}, is $O(\fhtw^3(\calH))$)
and Marx's algorithm runs in time $O(|\calH|^{O(\fhtw^3(\calH))})$.
In particular, if $\fhtw(\calH)$ is bounded by a constant, then Marx's algorithm runs in polynomial time in $|\calH|$.
From this algorithm, we obtain the following corollaries,
which are essentially what was shown in Grohe and
Marx~\cite{Marx:2010:AFH:1721837.1721845} applied to a wider context.
(Recall from Table~\ref{tab:results} that in this paper we use $\tilde O$ to hide
a factor of $\log N$ where $N$ is the data size and a polynomial factor in the query size. Therefore when $\fhtw(\calH)$ is bounded by a constant, Marx's algorithm runs in $\tilde O(1)$ time.)

\bcor[Grohe-Marx~\cite{DBLP:journals/talg/GroheM14,Marx:2010:AFH:1721837.1721845}]
Let $\varphi$ be an $\faqcs$ query  without free variables over an arbitrary semiring.
Let $\calH$, the hypergraph of $\varphi$, have a fractional hypertree width $\fhtw(\calH)$ that is {\em bounded} by a constant. Then $\varphi$ can be solved 
in time 
\[\tilde O(N^{O(\fhtw^3(\calH))} + \repsize{\varphi}), \]
where $\repsize{\varphi}$ is the time needed to report the output,
and $N\geq 1$ is the input size in the listing representation.
\label{cor:grohe-marx}
\ecor

In most of our applications, $\repsize{\varphi} = \tilde O(1)$; however, as we have seen earlier in Example~\ref{ex:natural join} and shall visit again below for
the natural join problem, $\repsize{\varphi}$ can be a lot larger than $N^{\fhtw(\calH)}$.
We first describe several applications where $\repsize{\varphi} = \tilde O(1)$.

\bcor[Theorem 1 from \cite{pichler:2013}]
Quantifier-free $\scq$ is solvable in time 
$\tilde O(N^{O(\fhtw^3(\calH))})$ (under the assumption that $\fhtw(\calH)\leq c$ for some constant $c$).
In particular, $\scq$ is tractable for the class of conjunctive queries
with bounded fractional hypertree width.
\ecor

The following result in the probabilistic graphical model context might have
been known; but we could not find a paper or textbook that proves it.
(Note that $N^{\fhtw(\calH)} \leq D^{\tw(\calH)+1}$ if each input
domain is of size $D$. And, $N^{\fhtw(\calH)}$ can be {\em arbitrarily} smaller
than $D^{\tw(\calH)+1}$ for sparse inputs.)

\bcor[Partition function in $\pgm$]
The partition function in probabilistic graphical models can be computed
in time $\tilde O(N^{O(\fhtw^3(\calH)})$ (under the assumption that $\fhtw(\calH)\leq c$ for some constant $c$).
\ecor

For the natural join problem, there is the issue of output reporting.
We next work out what $\InsideOut$ does to the set semiring
formulation of the natural join problem (Example~\ref{ex:natural join}),
and briefly analyze the time it takes to report the output under this 
semiring.
We prove here the $\tilde O(N^{\fhtw(\calH)}+\repsize{\varphi})$-runtime of $\InsideOut$ for computing
joins as shown in Table~\ref{tab:results}.

\begin{ex}[$\InsideOut$ for joins under the set semiring]
To detail how $\InsideOut$ works with the $(2^{\calU}, \cup, \cap)$ semiring,
we describe its behavior using the familiar relational
algebra notations \cite{DBLP:books/aw/AbiteboulHV95}.
For this problem, the input factors $\psi_S$ are the input relations $R_S$;
the intermediate factors are materialized intermediate relations;
the indicator projections are precisely the relational algebra projections.
Following \eqref{eqn:maineqn}, 
the first intermediate factor we want to compute is
\[ R_{U_n} = \ 
    \left(\Join_{S\in\partial(n)} R_S\right)
    \Join
    \left( \Join_{S\notin\partial(n), S\cap U_n\neq\emptyset}
            \pi_{U_n}(R_S)
    \right)
\]
Then, we would like to ``marginalize out'' $X_n$ under the $\cup$ operator 
of the set semiring.
But we certainly do not want to perform set-union in a brute-force way because 
each tuple $\mv t_{U_n}$ is mapped to 
$D^{n-|U_n|}$ points in the output space $\calU$.
Luckily, all we have to do is to store $R_{U_n}$ using a 
B-tree or trie data structure in an attribute order such that 
$X_n$ comes {\em last}. 
This way, if we only traverse the trie on the first $|U_n|-1$ 
attributes,
we will have access to exactly the tuples in $R_{U_n-\{X_n\}}$.
Furthermore, if later we want to also visit the tuples in $R_{U_n}$ we can get down 
one more level in the trie.
Thus, a trie-like index that respects the variable ordering implicitly
computes the marginalization operation for us!

The next issue is how much time it takes to report the output.
Since our representation of the output set is implicit, the set $R_{U_1}$ 
is an implicit representation of the output.
Up until the point of computing $R_{U_1}$ we have only spent 
$\tilde O(N^{\faqw(\sigma)})$ time, but the total number of output tuples might 
be {\em a lot} larger than that.\footnote{Consider the {\em star} join query
$\phi=\ \Join_{i=1}^{n-1} R_i(X,Y_i)$. Since $\phi$ is acyclic, $\fhtw(\phi)=1$
but it is easy to come up with instances of the relations/factors such that the
output size is as large as $N^{n-1}$ (consider e.g. the case when each
$R_i=\{1\}\times [N]$).}
In order to actually report {\em all} output tuples, we will go
through each $x_1 \in R_{U_1}$, then each $x_2$ such that
$(x_1,x_2)$ aligns with $R_{U_2}$, and so forth, until $R_{U_n}$.
This is a series of semi-join reductions in disguise; and it roughly 
corresponds 
to the second and the third phases of Yannakakis algorithm 
\cite{dblp:conf/vldb/yannakakis81} performed at once.
The total amount of time to report all output tuples is 
   $O(n \repsize{\varphi} \log N)$, where $\repsize{\varphi}$ is the number of output tuples.
\label{ex:Yannakakis}
\end{ex}

\bcor[Relational join]
Given an optimal variable ordering, the $\InsideOut$ algorithm can solve the natural join query in time
\[\tilde O(N^{\fhtw(\calH)} + \repsize{\varphi})\]
where $\repsize{\varphi}$ is the output size.
Moreover, if an optimal ordering is not given and assuming that $\fhtw(\calH)\leq c$ for some constant $c$, the runtime becomes
\[\tilde O(N^{O(\fhtw^3(\calH))} + \repsize{\varphi}).\]
\label{cor:join}
\ecor

The above discussion assumes that we are presenting the output in the listing 
representations. 
There are other options which Section~\ref{sec:output-rep} discusses.

\subsection{Variable ordering for $\faq$ with two blocks of semiring aggregates}
\label{app:subsec:2blocks}

This section highlights some of the subtle problems that arise when there are
more than one variable aggregate. 
In particular, we consider the following
instance of the $\faq$ problem where there are no free variables, 
\begin{equation}\label{eqn:faq2blocks}
    \varphi =
    \textstyle{\bigoplus_{\mv x_L} 
    {\bar \bigoplus}_{\mv x_{[n]-L}}}
    \bigotimes_{S\in\calE}\psi_S(\mv x_S).
\end{equation}
The input $\faq$ query has two blocks of variable aggregates.
In the first block, for variables $X_i$ where $i\in L$, $\bigoplus$
is a semiring aggregate; and the second block, for variables $X_i$, $i\notin L$,
has $\bar \bigoplus$ as a functionally different semiring aggregate.
Example~\ref{ex:scq} showed that $\scq$ can be reduced to an $\faq$-instance of 
this form, where $\bigoplus = \sum$ and $\bar \bigoplus = \max$.
Note that $\bigoplus$ and $\bar\bigoplus$ being functionally different means
they are not commutative with one another, as shown in Proposition~\ref{prop:commute}.

\subsubsection{The precedence poset}

Define a poset called the {\em precedence poset} 
$P = ([n], \preceq)$ over the variables as follows.
Let $u \prec v$ for every pair $(u,v)$ in the same connected component 
of $\calH$ such that $u\in L$ and $v \notin L$. 
Let $\LE(P)$ denote the set of linear extensions of the precedence poset $P$.
We first show that $\LE(P)$ is ``sound''.

\bprop[$\LE(P) \subseteq \EVO(\varphi)$]
Suppose $\bigoplus$ and $\bar \bigoplus$ are not identical in the domain $\D$.
Given a query $\varphi$ with only two blocks of variable aggregates
as written in~\eqref{eqn:faq2blocks}, we have
$\LE(P) \subseteq \EVO(\varphi)$.
\label{prop:LE subseteq EVO 2 blocks case}
\eprop
\bp
Suppose $\calH$ has only one connected component. Then, every linear extension
of the poset $P$ has variables in $L$ listed before variables in $[n]-L$; the
soundness of $\LE(P)$ thus follows.
If $\calH$ has multiple connected components, then expression
\eqref{eqn:faq2blocks} factorizes into a product over those connected
components, where each factor in this product is an \faq-expression over the
variables in that connected component. So we are back to the case when there is
only one connected component.
This proves that $\LE(P)$ is sound.
\ep

In Section~\ref{subsec:evo:Kblocks}, we prove a more general result stating
that every $\varphi$-equivalent variable ordering $\sigma$ has the same $\faqw$
as some $\pi \in \LE(\varphi)$. In that sense, $\LE(P)$ is also
``complete'' and we only need to look in $\LE(P)$ to find a $\sigma$ minimizing
$\faqw(\sigma)$. In particular, we show that
\begin{equation}\label{eqn:LE(P) 2 blocks}
\faqw(\varphi) = \min_{\sigma \in \LE(P)} \faqw(\sigma). 
\end{equation}

We next explain why the precedence poset and its linear extensions help
find a good variable ordering.
In this case, since there is no free variable, expression \eqref{eqn:simplified:IO:runtime} 
becomes simply $\tilde O\left(N^{\faqw(\sigma)}\right)$. However, it is no longer clear how
$\faqw(\varphi)$ is related to $\fhtw(\calH)$. 
Here are some simple properties that we can observe straightaway:
\bi
 \item $\faqw(\varphi) \geq \fhtw(\calH)$, for any $L\subseteq [n]$.
 \item If $L = [n]$ or $L = \emptyset$, then $\faqw(\varphi) = \fhtw(\calH)$.
 \item When $\emptyset \subset L \subset [n]$, 
     the $\faq$-width $\faqw(\varphi)$
      can be arbitrarily far from $\fhtw(\calH)$.
     For example, consider the star graph $\calV = \{1,\dots,n\}$ with
     edges $(1,n), (2,n), \cdots, (n-1,n)$. In this case, $\fhtw(\calH) = 1$
     because the graph is acyclic. Now, let $L = [n-1]$, then every
     $\varphi$-equivalent variable ordering has $n$ as the last vertex,
     which means $\faqw(\sigma) = n-1$ for every $\varphi$-equivalent variable
     ordering.
\ei

\subsubsection{Relation to $L$-star size} 

Before proving our main result for this section, we analyze how the $\faq$-width
$\faqw(\varphi)$ is related to the notion
of width that was defined in Durand and Mengel \cite{DBLP:conf/icdt/DurandM13,DBLP:journals/jcss/DurandM14}
for dealing with the $\scq$ problem.
Durand and Mengel
\cite{DBLP:conf/icdt/DurandM13,DBLP:journals/jcss/DurandM14} introduced the notion of {\em $L$-star
size} of a hypergraph to characterize the complexity of counting solutions to
conjunctive queries. 

\bdefn[Independence Number $\alpha_\calF(B)$] Let $\calF$ be a set of hyperedges and $B\subseteq\bigcup_{S\in\calF} S$ be any set of vertices.
Then, the {\em independence number} $\alpha_\calF(B)$ is the maximum size
of a subset $I \subseteq B$ where no two vertices from $I$ belong to the same hyperedge in $\calF$.
\label{defn:alpha_F(B)}
\edefn

\bdefn[$L$-star size]
Let $\calH=(\calV, \calE)$ be a hypergraph, and $L\subseteq \calV$ be a subset of vertices.
For any connected component $C$ of $\calH - L$ with vertex set $\calV(C)$
and edge set $\calE(C)$, define
\begin{eqnarray}
    \bar \calE(C) &=& \{S \in \calE \suchthat S\cap \calV(C) \neq \emptyset \}\\
    U(C) &=& L \cap \left(\bigcup_{S\in \bar \calE(C)} S\right).
    \label{eqn:U(C)1}
\end{eqnarray}
Then, the {\em $L$-star size} of $\calH$, denoted by
$\Lss(\calH)$, is the maximum independence number
$\alpha_{\calE}(U(C))$ over all connected components $C$ of $\calH-L$,
i.e.
\[ \Lss(\calH) = \max \left\{
        \alpha_{\calE}(U(C)) \suchthat C \text{ is a connected component of }
\calH-L
\right\}
\]
\label{defn:Lss}
\edefn
To relate $\Lss$ to $\faqw$, we need a couple of technical tools.

\blmm
Let $(T,\chi)$ be a tree decomposition of a hypergraph $\calH=(\calV,\calE)$,
where the fractional hypertree width of $(T,\chi)$ is $w$. 
Let $L \subseteq \calV$ be an arbitrary set of vertices of $\calH$.
Then, there exists a tree decomposition $(T',\chi')$
of $\calH$ with the same fractional hypertree width
satisfying the following condition.
Let $\calV(C_1)$ and $\calV(C_2)$ be the vertex sets of
two {\em different} connected components of $\calH-L$.
Then, there is no bag in $(T',\chi')$ intersecting both $\calV(C_1)$
and $\calV(C_2)$. 
\label{lmm:separated TD}
\elmm
\bp
We modify $(T,\chi)$ gradually to satisfy the required condition without 
increasing $\rho^*_\calH(B)$ for any bag $B$ of the tree
decomposition. Fix an arbitrary root $r$ of $(T, \chi)$. 
While there still are two connected components $C_1$ and $C_2$ for which
the desired condition is violated, we do the following.

Let $B$ be the bag farthest from the root $r$ such that $B$ intersects both
$\calV(C_1)$ and $\calV(C_2)$. In particular, every child of $B$ intersects
at most one of the two sets $\calV(C_1)$ and $\calV(C_2)$.
Now, we split $B$ into two bags $B_1 = B \setminus \calV(C_2)$, 
and $B_2 = B \setminus \calV(C_1)$.
Remove $B$ and connect both $B_1$ and $B_2$ to the parent of $B$.
(If $B$ is already the root, then we connect $B_1$ and $B_2$ with an edge.)
Children of $B$ which intersect $\calV(C_i)$ are connected to $B_i$, for $i\in
\{1,2\}$. Children which do not intersect neither $\calV(C_i)$ are connected
arbitrarily to $B_1$ or $B_2$.
It is straightforward to verify that we still have a tree decomposition for 
$\calH$ and that $\rho^*_\calH(B_i) \leq \rho^*_\calH(B)$.
\ep

\bdefn[Fractional independence number $\alpha^*_\calF(B)$]
Let $\calF$ be a set of hyperedges and $B\subseteq\bigcup_{S\in\calF} S$ be a set of vertices.
Then, the {\em fractional independence number} $\alpha^*_\calF(B)$ is the optimal solution to the following linear program:
\begin{eqnarray}
\max && \sum _{v \in B} x_v\label{eqn:LP:alpha_F(B):frac}\\
\text{s.t.}&& \sum_{v\in S} x_v \leq 1, \quad \forall S\in \calF\nonumber\\
&& x_v \geq 0, \quad \forall v \in B.\nonumber
\end{eqnarray}
\label{defn:alpha_F(B):frac}
\edefn

\bprop
Given any set of hyperedges $\calF$ and any set of vertices $B\subseteq\bigcup_{S\in\calF} S$, we have
\[\alpha_\calF(B) \leq \alpha^*_\calF(B).\]
\label{prop:alpha:int<frac}
\eprop
\bp
From Definition~\ref{defn:alpha_F(B)}, $\alpha_\calF(B)$ is defined using the optimization problem~\eqref{eqn:LP:alpha_F(B):frac} but with the additional constraint
$x_v \in\{0, 1\}, \forall v \in B\nonumber$.
\ep

\bprop
Given any hypergraph $\calH=(\calV, \calE)$ and any set $B\subseteq V$, we have
\[\alpha^*_{\calE}(B) = \rho^*_\calH(B).\]
(Recall the definition of $\rho^*_\calH(B)$ from Section~\ref{subsec:agm:fractional}.)
\label{prop:frac-alpha=rho-star}
\eprop
\bp
\eqref{eqn:LP:alpha_F(B):frac} is the dual LP to \eqref{eqn:LP:rho-star}.
\ep

\bthm\label{thm:relation to Lss}
Let $\varphi$ be the $\faq$-query defined in \eqref{eqn:faq2blocks}.
For any hypergraph $\calH=(\calV,\calE)$ and any subset $L$ of vertices of
$\calH$, we have 
\bi
 \item[(i)] $\faqw(\varphi) \leq (1+\Lss(\calH)) \cdot \fhtw(\calH)$.
 \item[(ii)] $\faqw(\varphi) \geq \Lss(\calH)$.
\ei
\ethm
\bp
To prove (i), we show that there exists a $\varphi$-equivalent
vertex ordering $\sigma$ such that 
$\faqw(\sigma) \leq (1+\Lss(\calH)) \cdot \fhtw(\calH)$.
We do so by constructing a tree decomposition $(T,\chi)$ of $\calH$ for which
the fractional hypertree width of $(T,\chi)$ is at most
$(1+\Lss(\calH)) \cdot \fhtw(\calH)$, and for which there exists a
GYO-elimination order where vertices in $L$ are eliminated last.

The tree decomposition $(T,\chi)$ is constructed as follows.
Let $(T_L, \chi_L)$ be a tree decomposition of $\calH$ with fractional hypertree
width equal to $\fhtw(\calH)$. From Lemma~\ref{lmm:separated TD} we can assume
that every bag of $(T_L,\chi_L)$ intersects at most one set $\calV(C)$ for each
connected component $C$ of $\calH-L$.
A bag that intersects $\calV(C)$ is called a {\em $C$-bag}.
Then, adapting an argument in \cite{DBLP:conf/icdt/DurandM13},
for every $C$-bag $\chi_L(t)$ ($t\in T_L$), we amend $\chi_L(t)$ with
\[ \chi_L(t) = (\chi_L(t) \cup U(C)) \setminus \calV(C), \]
where $U(C)$ is defined as in Definition~\ref{defn:Lss}.
After this step is done, $(T_L,\chi_L)$ is a tree decomposition of the
restriction of $\calH$ on $L$. This is because the collection of all
$C$-bags form a connected subtree of $(T_L,\chi_L)$ before the amendments.

Next, for each connected component $C$ of $\calH - L$, we construct a tree 
decomposition $(T_C,\chi_C)$ with fractional hypertree width at most 
$\fhtw(\calH)$.
(This is possible because the $\fhtw$ of an {\em induced} subgraph is at
most the $\fhtw$ of the supergraph.)
Then, for each bag $\chi_C(t)$ of $(T_C,\chi_C)$ we set
\[ \chi_C(t) = \chi_C(t) \cup U(C). \]
Finally, we construct the tree decomposition $(T,\chi)$ by connecting the
tree decompositions $(T_C, \chi_C)$ to the tree decomposition $(T_L, \chi_L)$.
This is done by connecting an arbitrary bag of $(T_C,\chi_C)$ to a bag
of $(T_L, \chi_L)$ that used to be a $C$-bag.

We claim the following:

\begin{claim}
\label{clm:1}
$(T,\chi)$ is indeed a tree decomposition of $\calH$.
\end{claim}

\begin{claim}
\label{clm:2}
$(T, \chi)$ has fractional
hypertree width at most $\fhtw(\calH) \cdot (1+\Lss(\calH))$.
\end{claim}

Assuming the claims hold, let us complete the proof of part (i) of the
theorem.
We construct the $L$-prefixed variable ordering by showing that we can eliminate
vertices from $\calV-L$ {\em before} any vertex in $L$ using the GYO-elimination
procedure. 
Fix a connected component $C$ of $\calH-L$; then vertices in $\calV(C)$ 
all reside in the subtree $(T_C,\chi_C)$.
We now run the GYO-elimination procedure on this
subtree up to its root $r$, which was the node that was connected to the
center tree $(T_L,\chi_L)$. In this process, all vertices in $U(C)$ are not
eliminated since the neighbor of $r$ contains $U(C)$.
We inductively eliminated other $\calV(C')$ for other connected components
$C'$ of $\calH-L$. Part (i) of the theorem is thus proved.

We next prove Claim~\ref{clm:1}. Every hyperedge $S\in\calE$ is either 
completely contained in $L$, or $S\cap \calV(C) \neq\emptyset$ for some
$C$.
If $S\subseteq L$, then $S$ is a subset of some bag in $(T_L,\chi_L)$,
which means it is contained in some bag of the final tree 
decomposition $(T,\chi)$.
If $S\cap\calV(C) \neq\emptyset$, then $S\cap\calV(C)$ is contained in 
some bag of the original tree decomposition $(T_C,\chi_C)$. And so
$S$ is contained in some bag of $(T_C,\chi_C)$
after we amend each of those bags with the set $U(C)$.
To verify the running intersection property (RIP), fix a vertex $v\in\calV$.
If $v\in \calV(C)$ for some connected component $C$ of $\calH-L$, then
RIP holds for $v$ even after the amendments of bags in $(T_C,\chi_C)$.
If $v\in L$ then its RIP property holds also because $(T_L,\chi_L)$ is
a tree decomposition too.

Now, we prove Claim~\ref{clm:2}.
The crucial point to notice is that each bag is amended at most once
with $U(C)$ for some $C$. This is because a $C$-bag is not a $C'$-bag for
two different connected components $C$ and $C'$, thanks to 
Lemma~\ref{lmm:separated TD}.
From Lemma 17 of Durand and Mengel \cite{DBLP:journals/jcss/DurandM14}, 
each set $U(C)$ can be covered by at most
$\Lss(\calH)$ many $C$-bags. In particular, for each amended bag 
$\chi_C(t) = \chi_C(t) \cup U(C)$ we have
\begin{eqnarray*}
   \rho^*_\calH(\chi_C(t)) &\leq&
    \fhtw(\calH) + \rho^*_\calH(U(C)) \\
    &\leq&
    \fhtw(\calH) + \Lss(\calH) \cdot \fhtw(\calH)\\
    &=& (1 + \Lss(\calH)) \cdot \fhtw(\calH).
\end{eqnarray*}
Similarly, each amended bag $\chi_L(t)$ is also fractionally covered by
at most the same quantity,
which completes the proof of the claim.

Finally, we prove part (ii) of the theorem.
From Lemma~\ref{lmm:lowerbound2blocks} below, we have
\[\faqw(\varphi)\geq \max_{C} \rho^*_{\calH}(U(C))\]
where $C$ ranges overall connected components of $\calH-L$.
However by Propositions~\ref{prop:frac-alpha=rho-star} and~\ref{prop:alpha:int<frac},
\[\max_{C} \rho^*_{\calH}(U(C)) = \max_{C} \alpha^*_{\calE}(U(C)) \geq \max_{C} \alpha_{\calE}(U(C)) = \Lss(\calH).\]
\ep

From the above theorem along with that the fact that any $\faq$-query $\varphi$ of the form~\eqref{eqn:faq2blocks} satisfies $\faqw(\varphi)\geq \fhtw(\calH)$, we have
\bcor
\label{cor:bounded-faqw=bounded-Lss}
For any class of $\faq$-queries of the form~\eqref{eqn:faq2blocks}, $\faqw(\varphi)$ is bounded if and only if both $\Lss(\calH)$ and $\fhtw(\calH)$ are bounded.
\ecor

\subsubsection{Approximating $\faqw(\varphi)$}
\label{app:subsec:2blocks:approx}

We now prove a lowerbound on $\faqw(\varphi)$ that leads to an
approximation algorithm for computing $\faqw(\varphi)$ using an approximation
algorithm for $\fhtw(\calH)$ as a blackbox.
Recall that we are still considering the $\faq$-query $\varphi$ of the special
form \eqref{eqn:faq2blocks}. Some of the ideas developed in this section will
lead to the approximation algorithm for $\faqw(\varphi)$ for the general 
$\faq$-query case.

Let $\calH=(\calV,\calE)$ be a hypergraph and $L\subseteq\calV$ be a subset 
of vertices. 
For each connected component $C$ of $\calH-L$, let the sets
$\bar\calE(C)$ and $U(C)$ be as in Definition~\ref{defn:Lss}. 
Define the hypergraph $\calH_L = (L, \calE_L)$ where
\begin{equation}\label{eqn:calE_L}
 \calE_L = \left\{ S \suchthat S\in\calE \wedge S\subseteq L \right\} \cup
\left\{ U(C) \suchthat C \text{ is a connected component of } \calH - L
\right\}.
\end{equation}
The hypergraph $\calH_L$ above is basically the same as the {\em contraction} of $(\calH, L)$ defined earlier by Chen and Mengel~\cite{chen_et_al:LIPIcs:2015:4980}.

\blmm\label{lmm:lowerbound2blocks}
For any connected component $C$ of $\calH - L$ we have
\begin{eqnarray*}
    \faqw(\varphi) &\geq& \fhtw(C)\\
    \faqw(\varphi) &\geq& \rho^*_\calH(U(C)).
\end{eqnarray*}
(Note that $C$ itself is a hypergraph.) Furthermore,
\[ \faqw(\varphi) \geq \fhtw(\calH_L). \]
\elmm
\bp
For any connected component $C$ of $\calH-L$, it is easy to see that
\[ \faqw(\varphi) \geq \fhtw(\calH) \geq \fhtw(C). \]
Let $\sigma = (v_1,\dots,v_n)$ be an $L$-prefixed variable ordering of $\calH$ 
with $\faqw(\sigma)=\faqw(\varphi)$. (Such a variable ordering exists due to
\eqref{eqn:LE(P) 2 blocks}.)
For a connected component $C$ of $\calH-L$, let $k$ be the smallest integer 
such that $k>|L|$ and $v_k\in \calV(C)$. 
Then, the set $U_k$ is precisely $U(C) \cup \{v_k\}$.
(This is the reason why we chose the notation $U(C)$.)
To see this, note that each time we eliminate a vertex from $C$, we insert back a
hyperedge interconnecting all its neighbors to the next hypergraph in the
hypergraph sequence. And so by the time we reach $U_k$ all of the neighbors
of $C$ in $L$ are connected, which is the set $U(C)$.
It follows that 
\[ \faqw(\varphi) = \faqw(\sigma)\geq \rho^*_\calH(U_k) \geq \rho^*_\calH(U(C)).
\]

It remains to prove $\faqw(\varphi) \geq \fhtw(\calH_L)$.
From the above argument, for every connected component $C$ of $\calH-L$,
the set $U(C)$ is a hyperedge of $\calH^\sigma_{|L|}$, which is precisely
the graph $\calH_L$ defined above. 
We thus have
\begin{eqnarray*}
    \faqw(\varphi) &=&
    \min_{\sigma\in\LE(P)}\left(\max 
   \left\{ 
     \rho^*_\calH(U^\sigma_k) \suchthat 1 \leq k \leq n
   \right\}\right)\\
   &\geq&
   \min_{\sigma\in \LE(P)}\left(\max 
   \left\{ 
     \rho^*_\calH(U^\sigma_k) \suchthat 1 \leq k \leq |L| 
   \right\}\right)\\
   &\geq&
   \min_\tau \left(\max 
   \left\{ 
     \rho^*_{\calH_L}(U^\tau_k) \suchthat 1 \leq k \leq |L| 
   \right\}\right)\\
   &=& \fhtw(\calH_L),
\end{eqnarray*}
where $\tau$ is a variable ordering of the variables in $L$ (as opposed to
$\sigma$ which was a variable ordering of $[n]$).
\ep

\bthm\label{thm:faq2blocks}
Let $\varphi$ be any $\faq$ query of the form of \eqref{eqn:faq2blocks} whose hypergraph is $\calH$.
Suppose there is an approximation algorithm that,
given any hypergraph $\calH'$, outputs a tree decomposition 
of $\calH'$ with fractional hypertree width at
most $g(\fhtw(\calH'))$ in time $t(|\calH'|, \fhtw(\calH'))$ for some non-decreasing functions $g, t$. 
Then, we can in time $|\calH|\cdot t(|\calH|, \faqw(\varphi))$
compute a $\varphi$-equivalent vertex ordering $\sigma$ such that
\[ \faqw(\sigma) \leq \faqw(\varphi) + g(\faqw(\varphi)). \]
\ethm
\bp
We use the blackbox approximation algorithm to construct a tree decomposition 
$(T_C, \chi_C)$ for every connected component $C$ of $\calH - L$
and also construct a tree decomposition $(T_L,\chi_L)$ for $\calH_L = (L,
\calE_L)$. (Recall that $\calE_L$ was defined in \eqref{eqn:calE_L}.)
Then, we form a tree decomposition $(T,\chi)$ by connecting these tree 
decompositions together in the following way.
We add the set $U(C)$ to each bag of the tree decomposition
$(T_C,\chi_C)$, and arbitrarily connect any node of the tree $(T_C, \chi_C)$
to a bag $B$ of the tree $(T_L,\chi_L)$ for which $U(C)\subseteq B$.

From Lemma~\ref{lmm:lowerbound2blocks} and the fact that $g$ is non-decreasing,
to cover any bag of the sub-tree $(T_L,\chi_L)$ we need a fractional cover
number of at most 
\[ g(\fhtw(\calH_L)) \leq g(\faqw(\varphi)). \]
To cover any bag of the sub-tree $(T_C,\chi_C)$ (after $U(C)$ is added), we 
need a fractional cover number of at most 
\[ \rho^*_\calH(U(C)) + g(\fhtw(C)) \leq 
   \faqw(\varphi) + g(\faqw(\varphi)).  
\]
Finally, we obtain $\sigma$ by running the GYO-elimination procedure on the
combined tree decomposition $(T,\chi)$, making sure that variables in $L$
are eliminated {\em last}.
\ep

By applying the above theorem using the fractional hypertree width
approximation algorithm from Marx \cite{Marx:2010:AFH:1721837.1721845},
we obtain the following corollaries.
(Recall from Section~\ref{app:subsec:quick:faqcs} that Marx's algorithm runs in time $\poly(\calH)=\tilde O(1)$ when $\fhtw(\calH)$ is bounded by a constant.)
\bcor
Let $\varphi$ be any $\faq$ query
of the form of \eqref{eqn:faq2blocks}.
Suppose $\faqw(\varphi)\leq c$ for some constant $c$.
Then, in polynomial time we can compute a $\varphi$-equivalent vertex 
ordering $\sigma$ such that $\faqw(\sigma) = O(\faqw^3(\varphi))$.
In particular, we can solve the $\faq$ problem of the form \eqref{eqn:faq2blocks}
in time 
\[\tilde O\left(N^{O\bigl(\faqw^3(\varphi)\bigr)} \right). \]
\label{cor:faq-two-blocks}
\ecor

\bcor
Assuming $\faqw(\varphi)\leq c$ for some constant $c$, the $\scq$ problem is solvable in time 
\[ \tilde O\left(N^{O\bigl(\faqw^3(\varphi)\bigr)}\right).
\]
In particular, $\scq$ is tractable for the class of conjunctive queries
for which $\faqw(\varphi)$ is bounded.
\label{cor:scq}
\ecor

Note that due to part(i) of Theorem~\ref{thm:relation to Lss}, the above corollary 
implies the result by Durand and Mengel \cite{DBLP:conf/icdt/DurandM13}
that $\scq$ is solvable in time
$\tilde O(N^{\poly(\Lss(\calH), \fhtw(\calH))})$.
Moreover, Durand and Mengel~\cite{DBLP:conf/icdt/DurandM13} showed that bounded $\Lss(\calH)$ is necessary for the tractability of $\scq$ (assuming $\fpt\neq \wone$).
By part (ii) of Theorem~\ref{thm:relation to Lss}, this implies that bounded $\faqw(\varphi)$ is necessary for the tractability of $\scq$ as well (assuming $\fpt\neq \wone$).


%
%
%
%
%

\section{Factor representations}
\label{app:sec:representation}

\subsection{More on the factor oracle}
\label{app:subsec:oracle}

We argue that the listing representation (Definition~\ref{defn:listing:rep}) 
satisfies
both \Assumption and \ProdMar as long as we choose an appropriate data structure
to support listing. Furthermore, answering these queries from the data 
structure takes only at most $\tilde O(1)$ time. We will assume that 
the representation can depend on a given ordering $\sigma$ among the variables 
(we will see shortly why this latter requirement is needed).

Assume the non-$\mv 0$ elements of a factor $\psi_S$ are stored in a B-Tree 
data structure that respects $\sigma$. In particular, we will store the 
tuples $\mv z$ such that $\psi_S(\mv z)\neq \mv 0$ as follows. Except for the root 
node, all the nodes in one level correspond to the same variable, and 
the ordering of the variables (when sorted from the level closest to the root to 
the farthest) is consistent with $\sigma$. In other words, the children of the 
root node correspond to all $x_1$ such that 
$\psi_S(\cdot \suchthat x_1)\not\equiv \mv 0$ 
and are labeled with the corresponding value of $x_1$. (Here we assume that 
$1$ comes first in $\sigma$.) Further, these children are sorted in (say) 
increasing order of the $x_1$ values. Then for each child of such an $x_1$ 
we build the tree recursively for $\psi_S(\cdot \suchthat x_1)$. Finally, the 
leaves (which correspond to a vector $\mv z$ formed by concatenating the labels 
on the unique path from the leaf to the root) also store $\psi_S(\mv z)$: 
we will call this the {\em value} of the leaf. Note that it has to be the 
case that $\psi_S(\mv z)\neq \mv 0$.

We now argue why this listing representation satisfies the \Assumption. Assume 
w.l.o.g. that $\sigma=(1,2,\dots,n)$. Then given $\mv x_{[k]}$, we go down the 
path in the B-Tree labeled with $\mv x_{[k]}$. Say $u$ is the last node on 
this path. Then computing $x_{k+1}$ basically corresponds to figuring out 
where $y$ lies in the sorted list of values of all children of $u$. Thus, 
with a binary search we can return the desired result. Note that if the 
query did not respect the ordering $\sigma$, then this would not have 
been possible. 

We now argue why the listing representation satisfies the \ProdMar. Assume that 
under the ordering $\sigma$, $i$ comes last in $S$. Then one 
can perform product marginalization given $i$ and the B-Tree representation 
of $\psi_S$ as follows. Go through all leaves of $\psi_S$ and multiply the 
values of all leaves with the same parent, (pretend to) throw away the leaves 
and store the computed product as the value of the new leaf just constructed. 
(We don't really throw the leaves away because marginalizing just means we
pretend the depth of the trie/B-tree is one less.)
Again note that we need the query to respect the ordering $\sigma$.

In all our algorithms, we will have the case that all the queries respect the 
ordering $\sigma$ of the B-Trees. This is because the algorithm is given as 
an input a variable ordering $\sigma$ and all its queries respect this ordering. 
Then in essentially linear time one can construct the B-Tree representation 
of all factors with a simple pre-processing step.
(There is one caveat which is the indicator projections of the input factors; but
those can also be pre-processed in the same amount of time.)

\subsection{Truth table representation}
\label{subsec:truth table}

In the {\em truth-table representation}, each input factor $\psi_S$ is
represented using a table of $\prod_{i\in S}|\Dom(X_i)|$ many rows. 
Each row lists the parameters and the value of the function. The value can 
be implicit as in the set semiring.

One example is the (dense) matrix multiplication problem where each input
and output matrix is represented by listing out all of its entries.
Another example is text-book description of $\pgm$ inference, where 
{\em conditional probability tables} are often assumed to use the truth-table
representation.

The truth-table representation makes problems easier because the input sizes are
larger. However, it should be obvious that this representation is wasteful
when the input factors are {\em sparse}, in the sense that they might have many
$\mv 0$-valued entries.

It is easy to see that truth table representation can be converted to the listing format in linear time.

\subsection{Succinct Representation}
\label{sec:succ-rep}

This section considers four specific representations of
factors where the goal is to 
(i) have a more succinct representation of the
factors than the listing representation and 
(ii) have an effective representation on which one can run efficient 
algorithms without having to ``unpack'' the succinct representation 
(into the listing representation). 
Our main aim will be to explain how these representations are essentially 
encodings of each input factor as an output of another $\faq$ (or $\faqcs$) 
instance. In particular, we will use our notation as opposed to those used in 
existing work.

\paragraph{GDNFs.} We start with the generalized disjunctive normal form (GDNF)
for $\csp$s that was first considered by Chen and
Grohe~\cite{DBLP:journals/jcss/ChenG10a}. Recall that in the $\csp$ problem 
(Example~\ref{ex:csp}) the corresponding $\faqcs$ instance is to compute
\[ \varphi = \bigvee_{\mv x} \bigwedge_{S\in\calE} \psi_S(\mv x_S). \]
For the case when each of the $\psi_S$ is presented in the listing representation, we get back the Boolean conjunctive query. In the GDNF format for each $S\in\calE$, we have
\[\psi_S(\mv x_S)=\bigvee_{i=1}^{m_S} \bigwedge_{j\in S} \psi^{(i,j)}_S(\mv x_{\{j\}}).\]
For each $(i,j)\in [m_S]\times S$, $\psi^{(i,j)}_S$ is presented in the listing
representation. We note two things: 
(i) the GDNF
representation can bring down the representation size from 
$\prod_{j\in S} |\Dom(X_j)|$ (potentially) to 
$O(m_S\cdot\sum_{j\in S} |\Dom(X_j)|)$, which can be big savings and 
(ii) the effective $\faq$ problem that we need to solve is
\[ \varphi = \bigvee_{\mv x} \bigwedge_{S\in\calE} \bigvee_{i=1}^{m_S} \bigwedge_{j\in S} \psi^{(i,j)}_S(\mv x_{\{j\}}). \]

\paragraph{Decision Diagram Representation.} We now consider the decision
diagram representation of Chen and Grohe~\cite{DBLP:journals/jcss/ChenG10a},
which is a generalization of the well-studied ordered binary decision
diagrams or OBDDs. Again we consider the $\csp$ problem:
\[ \varphi = \bigvee_{\mv x} \bigwedge_{S\in\calE} \psi_S(\mv x_S). \]
In the decision diagram representation each $\psi_S$ is represented as follows.
For brevity we consider $S=\{1,\dots,s\}$:
\[ \psi_S(x_1,\dots,x_s) =
   \bigvee_{(y_0,y_1,\dots,y_s)}
   \bigwedge_{i=1}^s \psi_S^{(i)}(y_{i-1},x_i,y_{i}),
\]
where for each $S$ and $i$, $\psi_S^{(i)}$ is represented in the listing format. We make three remarks: (i) An equivalent way to think about the above representation (which is how it is presented in~\cite{DBLP:journals/jcss/ChenG10a} and makes the connection to OBDDs more apparent) is the following. All the tuples $\mv x=(x_1,\dots,x_s)$ such that $\psi_S(\mv x)=1$ correspond to the sequence of labels on all paths of length $s$ in the following layered graph (with $s+1$ layers): $ \psi_S^{(i)}(y_{i-1},x_i,y_{i})=1$ if and only is there is an edge from vertex $y_{i-1}$ (in layer $i-1$) to vertex $y_i$ (in layer $i$) that is labeled $x_i$; (ii) It is easy to see that a GDNF representation can be converted into a decision diagram of essentially the same size and it is shown in~\cite{DBLP:journals/jcss/ChenG10a} that the decision diagram representation can be exponentially smaller than any equivalent GDNF representation; (iii) 
Note that the effective $\faq$ problem that we need to solve is
\[ \varphi = \bigvee_{\mv x} \bigwedge_{S\in\calE} \bigvee_{(y^S_0,y^S_1,\dots,y^S_{|S|})}\bigwedge_{i=1}^{|S|} \psi_S^{(i)}(y^S_{i-1},x_{S[i]},y^S_{i}),\]
where $S[1],S[2],\dots,S[|S|]$ is some ordering of elements in $S$.

\paragraph{Factorized Databases.} Olteanu and Z\'{a}vodn\'{y}~\cite{OZ15} considered factorized  representations for conjunctive queries (Example~\ref{ex:conj}). Recall that we consider the following instance of $\faqcs$:
    \[ \varphi(\mv x_F) = \bigvee_{\mv x_{[n]-F}}
                          \bigwedge_{S\in\calE} \psi_S(\mv x_S). \]
In the factorized representation (called {\em f-representation} in~\cite{OZ15}), each $\psi_S$ is represented recursively as follows. In the general case either
\begin{equation}
\label{eq:f-rep-union}
\psi_S(\mv x)=\bigvee_{i=1}^{\ell} \psi_S^{\cup,i}(\mv x)
\end{equation}
or
\begin{equation}
\label{eq:f-rep-intersect}
\psi_S(\mv x)=\bigwedge_{i=1}^k \psi_{S_i}^{\times,i}(\mv x_{S_i}),
\end{equation}
where $S_1,\dots,S_k$ is a partition of $S$. In the base case we have factors of the form $\psi_{\{j\}}:\Dom(X_j)\to\{0,1\}$. We make five remarks: (i) An alternate formulation of $f$-representation is as follows. We can represent all the $\mv x$ such that $\psi_S(\mv x)=1$ in the following tree format. The leaves of the trees have single value for some attribute. Each of the internal nodes is either a `union' node (corresponding to~\eqref{eq:f-rep-union}) or an `Cartesian product' node (corresponding to~\eqref{eq:f-rep-intersect}) with the natural semantics attached to these nodes; (ii) It is easy to check that GDNFs are a special case of f-representations; (iii) A generalization of f-representation considered in~\cite{OZ15}, called {\em d-representations} is where one takes an f-representation and removes repeated sub-expressions: alternatively in the tree representation mentioned in (i), one `short-circuits' sub-trees that are repeated resulting in a DAG; (iv) The main contribution of~\cite{OZ15} is to show that if the factors are presented in a factorized representation, then the output of $\varphi$ can also be represented in the same format (and they prove tight bounds on the output size): we briefly touch on compressing the output in Section~\ref{sec:output-rep}; (v) One can think of $\varphi$ with the factors in a factorized representation as a big $\faq$ instance (where each $\psi_S$ is represented by an $\faq$ instance itself over the Boolean semiring, where this instance is determined by the recursive definition of $\psi_S$ above).

\paragraph{Fast Matrix Vector Multiplication.} We saw in Example~\ref{ex:dft} that for the DFT matrix $\mv F$ we can represent the factor corresponding to the matrix more succinctly. We will now use the $\faqcs$ formulation to describe a generic way to talk about matrices $\mv A$ for which we can get a sub-quadratic matrix vector multiplication algorithm: further, our algorithm is the same. The only thing that changes is the representation of the $\mv A$ and the elimination order. Next, we provide more details on this claim.

Recall that we are aiming to compute $\mv A\cdot\mv b$. Assuming $n=n_0^m$ for  integers $m,n_0\ge 1$, the most general way to represent an $n\times n$ matrix $\mv A=(A_{i,j})$ is via the following factor
\[\psi_{\mv A}(x_0,\dots,x_{m-1},y_0,\dots,y_{m-1})=A_{(x_0,\dots,x_{m-1}),(y_0,\dots,y_{m-1})},\]
where we think of $(x_0,\dots,x_{m-1}),(y_0,\dots,y_{m-1})$ as integers in the range $[0,n-1]$. Now assume that there is a hypergraph $\calH_{\mv A}=(\{X_0,\dots,X_{m-1},Y_0,\dots,Y_{m-1}\},\calE_{\mv A})$ such that
\begin{equation}
\label{eq:A-factor}
\psi_{\mv A}(x_0,\dots,x_{m-1},y_0,\dots,y_{m-1})=\prod_{e\in \calE_{\mv A}} \psi_e((\mv x,\mv y)_e),
\end{equation}
where we use $(\mv x,\mv y)_e$ as a short hand for the projection of $\{x_0,\dots,x_{m-1},y_0,\dots,y_{m-1}\}$ onto $e$ (i.e. if $X_j\in e$ then we pick $x_j$ and if $Y_j\in e$ then we pick $y_j$). With this notation in place, recall we are trying to solve the problem:
\begin{equation}
\label{eq:Ab-factorized}
\phi_{\mv A\cdot\mv b}(x_0,\dots,x_{m-1}) = \sum_{(y_0,\dots,y_{m-1})\in \Z_{n_0}^m} \psi_{\mv b}(y_0,\dots,y_{m-1})\cdot \prod_{e\in \calE_{\mv A}} \psi_e((\mv x,\mv y)_e).
\end{equation}

Next we present some specific instantiations of the $\faqcs$ instance from~\eqref{eq:A-factor}.

We begin with DFT and related matrices. In particular, recall the DFT matrix 
$$\mv F=\left(f_{x,y}=e^{i2\pi\frac{x\cdot y}{n}}\right)_{0\le x,y<n}$$
and from Example~\ref{ex:dft} that in this case we have the hypergraph 
$\calH_{\mv F}=(\calV_{\mv F},\calE_{\mv F})$ with 

 \begin{eqnarray*} 
     \calV_{\mv F}&=&\{X_0,\dots,X_{m-1},Y_0,\dots,Y_{m-1}\},\\
     \calE_{\mv F}&=&\{(Y_0,\dots,Y_{m-1})\}\cup
\{(X_j,Y_k)\}_{0\le j+k<m}.
 \end{eqnarray*}


Consider the $n\times n$ circulant matrix $\mv C$ where the first column is say the vector $(c_0,\dots,c_{n-1})$ and the rest of the columns are a cyclic shift of the previous column. It can be shown that $\mv C\cdot \mv b$ for a vector $\mv b$ is the same as the {\em convolution} of $\mv c=(c_0,\dots,c_{n-1})$ and $\mv b$. In other words,
\[\mv C\cdot \mv b=\mv F^{-1}\cdot (\mv F\cdot \mv c)\cdot (\mv F\cdot \mv b).\]
Note that the above is equivalent to the following $\faqcs$ instance:
\begin{multline*}
   \phi_{\mv C\cdot \mv b}(\mv x)=\sum_{\mv y\in\F_p^m} \sum_{\mv z\in \F_p^m}
   \sum_{\mv w\in \F_p^m} b_{\mv w}\cdot c_{\mv z}\cdot  \prod_{0\le j+k< m}
   \psi_{Y_j,Z_k}(y_j,z_k) \\
   \cdot
\prod_{0\le j+k< m} \psi_{Y_j,W_k}(y_j,w_k)\cdot \prod_{0\le j+k<m}
\psi_{X_j,Y_k}(-x_j,y_k).\end{multline*}


Next, we consider the case when the matrix $\mv A$ is itself a product of two or more matrices. For simplicity, we will only consider the case when $\mv A$ is a product of two square matrices $\mv D$ and $\mv E$ (of the same order). Both of these restrictions can be removed but we focus on these special cases since our goal is to show how our framework gives a uniform way to measure the efficiency of our algorithm for matrix vector multiplication with structured matrices $\mv A$.

We begin with the Kronecker product. Given two $n_0\times n_0$ matrices $\mv D$ and $\mv E$, their Kronecker product $\mv A=\mv D\otimes\mv E$ is defined as follows: for every $(x_0,x_1),(y_0,y_1)\in \Z_{n_0}^2$, $A[(x_0,x_1),(y_0,y_1)]=E[x_0,y_0]\cdot F[x_1,y_1]$. In this case, we have the hypergraph $\calH_{\mv D\otimes\mv E}=(\{X_0,X_1,Y_0,Y_1\},\calE_{\mv D\otimes\mv E})$, where
\[\calE_{\mv D\otimes\mv E}=\{(X_1,Y_1),(Y_0,Y_1),(X_0,Y_0)\}.\]


Next, we consider the Khatri-Rao product~\cite{khatri-rao}. We think of  $\mv E$ and $\mv F$ as collection of $n_0^2$ matrices $\mv E^{x_2,y_2}$ and $\mv F^{x_2,y_2}$ each of which are $n_0\times n_0$ matrices. The Khatri-Rao product $\mv A=\mv E\star\mv F$ is defined as follows: for every $(x_0,x_1,x_2),(y_0,y_1,y_2)\in\Z_{n_0}^3$, $A[(x_0,x_1,x_2),(y_0,y_1,y_2)]=E^{x_2,y_2}[x_0,y_0]\cdot F^{x_2,y_2}[x_1,y_1]$. In this case, we have the hypergraph \linebreak $\calH_{\mv E\star\mv F}=$$(\{X_0,X_1,X_2,Y_0,Y_1,Y_2\},\calE_{\mv E\star\mv F})$, where
\[\calE_{\mv E\star\mv F}=\{(X_0,X_2,Y_0,Y_2), (X_1,X_2,Y_1,Y_2), (Y_0,Y_1,Y_2)\}.\]

Finally, we consider the Tracy-Singh product~\cite{tracy-singh}. We think of  $\mv E$ and $\mv F$ as collection of $n_0^2$ matrices $\mv E^{x_2,y_2}$ and $\mv F^{x_3,y_3}$ each of which are $n_0\times n_0$ matrices.  The Tracy-Singh  product $\mv A=\mv E\circ\mv F$ is defined as follows: for every $(x_0,x_1,x_2,x_3),(y_0,y_1,y_2,y_3)\in\Z_{n_0}^4$, $A[(x_0,x_1,x_2,x_3),(y_0,y_1,y_2,y_3)]=E^{x_2,y_2}[x_0,y_0]\cdot F^{x_3,y_3}[x_1,y_1]$. In this case, we have the hypergraph $\calH_{\mv E\circ\mv F}=(\{X_0,X_1,X_2,X_3,Y_0,Y_1,Y_2,Y_2\},\calE_{\mv E\circ\mv F})$, where
\[\calE_{\mv E\circ\mv F}=\{(X_0,X_2,Y_0,Y_2), (X_1,X_3,Y_1,Y_3), (Y_0,Y_1,Y_2,Y_3)\}.\]

We conclude by noting that if one is willing to go beyond semirings and look at fields (in particular, all the matrices above are over complex numbers), then one can do faster matrix vector multiplication for a wider family of matrices~\cite{P00,OS00}. Since we are interested primarily in semirings in this paper, we do not further explore this connection.

\subsection{Representations of sparse tables in $\pgm$s}
We now summarize the two main succinct representation of factors in $\pgm$S (Example~\ref{ex:pgm}): the first one corresponds to the listing representation (Definition~\ref{def:listing}) and the second one is similar to the decision diagram representation (Section~\ref{sec:succ-rep}).

\paragraph{Sparse tables.} This is the listing representation: only list inputs $\mv y$ in representation of $\psi_S$ such that $\psi_S(\mv y)\neq 0$. This is called {\em evidence shrinking} in~\cite{DBLP:journals/ijar/Darwiche96}.

\paragraph{Algebraic Decision Diagrams.} Algebraic Decision Diagrams (or ADDs) were introduced in Bahar et al.~\cite{DBLP:journals/fmsd/BaharFGHMPS97}. This is a succinct representation that is very closely related to the decision diagrams from Section~\ref{sec:succ-rep}, as we shall see shortly. (Bahar et al. defined ADDs, considered some of their basic properties, and analyzed  (standard) algorithms for matrix multiplication, shortest path problems and linear algebra when the inputs for the problems were represented as ADDs.)

We will present the definition of ADD for representing a single factor over bits (larger domain elements can be represented as bits and the generalization to multiple factors is straightforward). In particular consider the case when $\psi:\{0,1\}^n \to \D$. An ADD representation for $\psi$ is a DAG $G$ with the sinks (i.e. vertices in $G$ with no outgoing edges) being labeled with values from $\D$. Further, each non-sink node is labeled with one of the $n$ variables such that it is consistent with an ordering $\sigma$ of the $n$ variables. (By consistent we mean that if $(u,v)$ is a directed edge and they are labeled $\ell(u)$ and $\ell(v)$, then $\sigma(\ell(u))<\sigma(\ell(v))$.) Further, every non-sink node has two outgoing edges: one labeled $0$ and the other $1$. Given an ADD for a factor $\psi$ one can compute $\psi(\mv x)$ for an input $\mv x=(x_1,\dots,x_n)$ in the natural way. Start with the root $r$ and take the branch corresponding to $x_{\ell(r)}$ and continue recursively and stop when you reach a sink $s$. The value of $\psi(\mv x)$ is the label of $s$ (recall that the sinks are labeled with elements from $\D$).

The ADD representation is very similar to the decision diagram representation presented in Section~\ref{sec:succ-rep}. The main difference is that unlike the decision diagram case where the underlying DAG $G$ is a layered graph, in an ADD representation $G$ might not be a layered graph. In particular, if we put all vertices in $G$ with the same label in a layer then there can be edges that connect layers that are not right next to each other. However, if we are willing to tolerate a blowup of $O(n)$ in the representation size, then we can indeed convert a $G$ for any ADD representation into a layered graph $G'$ as follows. For any edge $(w_i,w_j)$ between layer $i$ and $j$ (for $j>i+1$),\footnote{W.l.o.g. we assume that the ordering $\sigma$ is $1,\dots,n$.} we replace the edge with a dummy path $(w_i,w_{i+1}), (w_{i+1},w_{i+2}), \dots, (w_{j-1},w_j)$, where $w_{i+1},\dots,w_{j-1}$ are new nodes in $G'$. Further, $(w_i,w_{i+1})$ has the same label as $(w_i,w_j)$ while the rest of the new edges have a label of both $0$ and $1$ (more precisely, there are two parallel edges one labeled $0$ and the other labeled $1$). Finally, 
we add an $(n+1)$th layer and connect all sinks to distinct vertices in this layer. (Again if the sink is in layer $i$ for $i<n$, then we'll actually need a `dummy' path of appropriate length). Let $L$ be the function that maps each vertex in the $(n+1)$th layer to the corresponding sink label. Given the ADD in the layered form $G'$, one can represent $\psi$ as before (with suitable modifications since $\psi$ is no longer binary as in Section~\ref{sec:succ-rep}):
\[\psi(x_1,\dots,x_n)=\sum_{y_0,\dots,y_{s+1}} \left(\prod_{i=1}^n \psi^{(i)}(y_{i-1},x_i,y_i)\right)\cdot \psi^{(n+1)}(y_n,y_{n+1})\cdot L(y_{n+1}),\]
where $\psi^{(i)}$ encodes the edges between layer $(i-1)$ and layer $i$ (for $i\in [n+1]$). In the definition above these $\psi^{(i)}$ are binary functions and we use the listing representation for each $\psi^{(i)}$.

If the blowup by a factor of $n$ when going from the original ADD representation to the layered representation is not acceptable, then there are potential ways to mitigate this blowup:
\begin{enumerate}
\item If there are $s$ sinks in total, then one can convert $G$ into a layered graph $G''$ with an {\em additive} blowup of $O(sn+n^3)$. To see this note that in the construction of $G'$ above, each sink leads to a dummy path of length $O(n)$, which justifies the $O(sn)$ term. Next note that for every edge between layers $i$ and $j$ (for $j>i+1$), the dummy path is the same except for the label on the first edge. In other words, all these dummy paths can have the same common suffix path of length $j-i-1$. Note that there are $O(n^3)$ distinct such suffixes, which justifies the $O(n^3)$ term. This additive factor is advantageous over the multiplicative blowup of $n$ when $G$ has many more vertices when compared to $n$ and $s$.
\item Alternatively, we can encode the ADD as an $\faq$ instance by essentially encoding the Boolean function suggested by the ADD (modulo the labeling of the sink). More formally, let $r$ be the root of $G$ and let $c_0$ and $c_1$ be its two outgoing neighbors. Then it is easy to see that
\[\psi_G(x_1,\dots,x_n)=\psi_{\ell(r)}(x_{\ell(r)})\cdot \psi_{G_{c_1}}(x_{\ell(r)+1},\dots,x_n) + \bar{\psi}_{\ell(r)}(x_{\ell(r)})\cdot \psi_{G_{c_0}}(x_{\ell(r)+1},\dots,x_n),\]
where $\psi_{\ell(r)}(x_{\ell(r)})=x_{\ell(r)}$ and $\bar{\psi}_{\ell(r)}(x_{\ell(r)})=1-x_{\ell(r)}$ and $\psi_{G_{c_1}}, \psi_{G_{c_0}}$ are defined recursively. Finally, for a sink $u$, $\psi_{G_u}(\mv x)=L(u)$. The advantage of this representation is that there is only an $O(1)$ blowup in the representation size, but now the size of the resulting $\faq$ instance is the same as the size of the original as opposed to our earlier representation where the size of the resulting $\faq$ instance only depends on $n$ and $s$ (but not on the size of $G$).
\end{enumerate}

Gogate and Domingos~\cite{DBLP:conf/uai/GogateD13} present message passing algorithms that work with both the sparse table and ADD representation. Unlike this work, which focuses on exact computation of $\varphi$, the work of~\cite{DBLP:conf/uai/GogateD13} presents algorithms that approximate $\varphi$.

\end{document}